\documentclass{ws-ijmpb}

\usepackage{graphicx}
\usepackage{amssymb}
\usepackage{amsfonts}
\usepackage{bm}

\catcode`\@=11

\newtoks\@stequation

\def\subequations{\refstepcounter{equation}%
  \edef\@savedequation{\the\c@equation}%
  \@stequation=\expandafter{\theequation}%   %only want \theequation
  \edef\@savedtheequation{\the\@stequation}% % expanded once
 \edef\oldtheequation{\theequation}%
  \setcounter{equation}{0}%
  \def\theequation{\oldtheequation\alph{equation}}}

\def\endsubequations{\setcounter{equation}{\@savedequation}%
  \@stequation=\expandafter{\@savedtheequation}%
  \edef\theequation{\the\@stequation}\global\@ignoretrue}
\catcode`\@=12

\newcommand{\bsubequations}{\begin{subequations}}
\newcommand{\esubequations}{\end{subequations}}

\newcommand{\bbf}{}

\newcommand{\qqph}{\qquad \phantom{.}}
\newcommand{\qph}{\quad \phantom{.}}  \newcommand{\irm}{{ i}} 
            \newcommand{\ie}{{i.e.}}  % that is
            \newcommand{\eg}{{e.g.}}  % for example
            \newcommand{\pdag}{{\phantom{\dagger}}}
            \newcommand{\ext}{{\rm ext}}

            \newcommand{\here}{{\rm here}}
            \newcommand{\there}{{\rm there}}
            \newcommand{\class}{{{\rm cl}}}
            
            \newcommand{\rw}{{\rm rw}}
            
            \newcommand{\Keldysh}{{\rm Keldysh}}
            
            \newcommand{\AAK}{{\rm AAK}}
            \newcommand{\gAAK}{{\gamma_\varphi^\AAK}}
            \newcommand{\self}{{\rm self}}

            \newcommand{\vertex}{{\rm vert}}
            \newcommand{\leading}{{\rm leading}}

            \newcommand{\full}{{\rm full}}
            \newcommand{\bare}{{\rm bare}}
            \newcommand{\free}{{\rm free}}
            \newcommand{\cqp}{{\rm cqp}}

            \newcommand{\ns}{{{\rm (ns)}}}
            \newcommand{\bard}{{{\bar d}}}  \newcommand{\nns}{{{\rm ns}}}

            \newcommand{\DC}{{\rm DC}}  \newcommand{\rreal}{{\rm real}}
            \newcommand{\iimag}{{\rm imag}}
            \newcommand{\nud}{\nu} % {\nu_d}
            \newcommand{\nav}{\langle n
              \rangle} % {\langle n_d \rangle}
            \newcommand{\Dd}{D} % {D_d}
            \newcommand{\Drude}{{\rm Drude}}
            \newcommand{\WL}{{\rm WL}}
            \newcommand{\RPA}{{\rm RPA}}
            \newcommand{\firstprinciples}{}%{\rm 1.princ}}
            \newcommand{\inflfunct}{}%{{\rm infl.\ funct}}
            \newcommand{\dressed}{{\rm dressed}}
            \newcommand{\GZ}{{\rm GZ}}
            
            \newcommand{\tot}{{\rm tot}}
            \newcommand{\inter}{{\rm int}}
            \newcommand{\recoiling}{{\rm rec}}
            \newcommand{\eff}{{\rm eff}}
            \newcommand{\toh}{{\textstyle{\frac{1}{2}}}}
            
            \newcommand{\toq}{{\textstyle{\frac{1}{4}}}}
            \newcommand{\Vol}{{\rm Vol}}
            \newcommand{\tauphi}{{\tau_\varphi}}
            \newcommand{\tauH}{{\tau_H}}
            
            \newcommand{\gammaphi}{{\gamma_\varphi}}
            \newcommand{\Pauli}{{(\tilde \delta - 2 \tilde \rho^0)}}
            \newcommand{\gammaH}{{\gamma_H}}

            \newcommand{\nonint}{{\rm nonint}}
            \newcommand{\Hartree}{{\rm Hartree}}
            \newcommand{\imp}{{\rm imp}}
            \newcommand{\tauel}{{\tau_{\rm el}}}
            \newcommand{\lel}{{l_{\rm el}}}
            \newcommand{\kF}{k_{\rm F}}
            
            \newcommand{\eF}{\varepsilon_{\rm F}}
            \newcommand{\ve}{\varepsilon}
            \newcommand{\vF}{v_{\rm F}}
            
            \newcommand{\dis}{{\rm dis}}

            \newcommand{\Tr}{{\rm Tr}} \newcommand{\Sch}{{ S}}
            \newcommand{\ddd}{{ d}}
            \newcommand{\Eq}[1]{Eq.~(\ref{#1})} % Eq. (#) Equation
\newcommand{\Eqs}[1]{Eqs.~(\ref{#1})}      % Eqs. (#) Equations
\newcommand{\Sec}[1]{Sec.~\ref{#1}}        % Sec. (#)  Section
\newcommand{\bmrho}{{\bm{\rho}}}
\newcommand{\gpR}{{\gamma_\varphi^{R,\self}}}
\newcommand{\gpI}{{\gamma_\varphi^{I,\self}}}
\newcommand{\bari}{{\bar \imath}}
\newcommand{\barj}{{\bar \jmath}}
\newcommand{\ibari}{{i \bar \imath}}

\newcommand{\barii}{{\bar \imath i}}
\newcommand{\iiijjj}{{i \! j}}
\newcommand{\ji}{{ji}}

\newcommand{\bmR}{{\bm{R}}}
\newcommand{\bmE}{{\bm{E}}}

\newcommand{\bmr}{{\bm{r}}}
\newcommand{\bmj}{{\bm{j}}}
\newcommand{\bmA}{{\bm{A}}}
\newcommand{\bmp}{{\bm{p}}}
\newcommand{\bmk}{{\bm{k}}}
\newcommand{\bmq}{{\bm{q}}}

\newcommand{\bmQ}{{\bm{Q}}}

\newcommand{\bmP}{{\bm{P}}}

\newcommand{\bmK}{{\bm{K}}}
\newcommand{\bnabla}{{\bm{\nabla}}}
\newcommand{\bmrone}{{\bm{r}}_1}
\newcommand{\bmrtwo}{{\bm{r}}_2}

\newcommand{\bmpsi}{{\hat {\bm{\psi}}}}
\newcommand{\unbmpsi}{{\hat {\underline{\bm{\psi}}}}}
\newcommand{\bmone}{{\bm{1}}}
\newcommand{\bmgamma}{{{\bm{\gamma}} }}

\newcommand{\G}{{\tilde G}}
\newcommand{\Uz}{{{\tilde U}^0}}
\newcommand{\U}{{\tilde U}}
\newcommand{\GR}{{\tilde G^R}}
\newcommand{\GA}{{\tilde G^A}}
\newcommand{\GK}{{\tilde G^K}}

\newcommand{\hbG}{{\bm{{\widehat {\tilde G}}}}}
\newcommand{\chbG}{{\bm{{\underline {\widehat {\tilde G}}}}}}
\newcommand{\bG}{{\bm{{\tilde G}}}}
\newcommand{\cbG}{{\bm {\underline {\tilde G}}}}
\newcommand{\cbSigma}{{\bm {\underline {\tilde \Sigma}}}}
\newcommand{\bL}{{\bm{L}}}
\newcommand{\btL}{{\bm{{\tilde L}}}}
\newcommand{\LL}{{\tilde {\cal L}}}
\newcommand{\tW}{\widetilde W}
\newcommand{\bW}{\overline W}

\newcommand{\JJbm}{{\bm{{\cal J}}}}
\newcommand{\bLL}{{\bm{{\LL}}}}
\newcommand{\bcL}{{\bm{{\underline \LL}}}}
\newcommand{\AAA}{\tilde {\cal A}}

\newcommand{\btau}{{\bm{\tau}}}
\newcommand{\ttau}{{\widetilde \tau}}

\newcommand{\bbtau}{{\bar \tau}}
\newcommand{\bbrho}{{\bar \bmrho}}
\newcommand{\unbmP}{{\underline{\bm{P}}}}

\newcommand{\unV}{{\bm {\underline {V}}}}
\newcommand{\unv}{{\bm {\underline {v}}}}

\newcommand{\bcG}{{\overline {\cal G}}}
\newcommand{\tcP}{{\tilde {\cal P}}}
\newcommand{\bcP}{{\overline {\cal P}}}
\newcommand{\tcG}{{\tilde {\cal G}}}
\newcommand{\tcC}{{\tilde {\cal C}}}

\newcommand{\bSigma}{{\overline \Sigma}}
\newcommand{\tSigma}{{\tilde \Sigma}}

\newcommand{\bcC}{{{\overline {\cal C}}}}
\newcommand{\bcD}{{\overline {\cal D}}}
\newcommand{\bcCqw}{{{\overline {\cal C}^0_\bmq (\omega)}}}
\newcommand{\bcDqw}{{{\overline {\cal D}^0_\bmq (\omega)}}}
\newcommand{\bcCqwself}{{{\overline {\cal C}^\self_\bmq (\omega)}}}
\newcommand{\bPiqw}{{{\overline \Pi_\bmq (\omega)}}}
\newcommand{\bGqw}{{{\overline \Gamma_\bmq (\omega)}}}

\newcommand{\bLRbbqw}{{{\overline {\cal L}^R_\bbmq (\bomega)}}}
\newcommand{\bLAbbqw}{{{\overline {\cal L}^A_\bbmq (\bomega)}}}
\newcommand{\bLKbbqw}{{{\overline {\cal L}^K_\bbmq (\bomega)}}}

\newcommand{\bomega}{{\bar \omega}}
\newcommand{\tomega}{{\widetilde \omega}}

\newcommand{\bbmq}{{\bar \bmq}}

\newcommand{\Fint}{F \hspace{-10pt} \int}
\newcommand{\Bint}{B \hspace{-10.5pt} \int}

\begin{document}

\markboth{Jan von Delft} {Influence functional calculation of
  decoherence in weak localization}

%%%%%%%%%%%%%%%%%%%%% Publisher's Area please ignore %%%%%%%%%%%%%%%
%
\catchline{}{}{}{}{}
%
%%%%%%%%%%%%%%%%%%%%%%%%%%%%%%%%%%%%%%%%%%%%%%%%%%%%%%%%%%%%%%%%%%%%

\title{INFLUENCE FUNCTIONAL FOR DECOHERENCE OF INTERACTING ELECTRONS
  IN DISORDERED CONDUCTORS}

\author{JAN VON DELFT}

\address{
Fakult\"at f\"ur Physik, Arnold Sommerfeld Center for Theoretical
    Physics, \\ and Center for {Nano}Science 
\\Ludwig-Maximilians-Universit\"at
M\"unchen \\ Theresienstr. 37, 80333, M\"unchen, Germany
\\
vondelft@lmu.de
}

\maketitle

\begin{history}
  \received{November 16, 2007}
%\revised{Day Month Year}
%\accepted{(Day Month Year)}
%\comby{(xxxxxxxxxx)}
\end{history}

\begin{abstract}
  In this review, we rederive the controversial influence functional
  approach of Golubev and Zaikin (GZ) for interacting electrons in
  disordered metals in a way that allows us to show its equivalence,
  before disorder averaging, to diagrammatic Keldysh perturbation
  theory.  By representing a certain Pauli factor $\Pauli$ occuring in
  GZ's effective action in the frequency domain (instead of the time
  domain, as GZ do), we also achieve a more accurate treatment of
  recoil effects.
%The main improvement relative to GZ's formulation, achieved by
%  representing a certain Pauli factor $\Pauli$ in their effective
%  action in the frequency domain, is that recoil effects are now
%  treated correctly.  In this paper we show that this improved
  With this change, GZ's approach reproduces, in a remarkably simple
  way, the standard, generally accepted result for the decoherence
  rate. -- The main text and appendices~A.1 to A.3 of the present
  review are comparatively brief, and have been published
  previously; %\cite{Granada};
  for convenience, they are included here again (with minor
  revisions). The bulk of the review is contained in several
  additional, lengthy appendices containing the relevant technical
  details.
\end{abstract}


\section{Introduction}

A few years ago, Golubev and Zaikin (GZ) developed an influence
functional approach for describing interacting fermions in a
disordered conductor\cite{GZ1,GZ2,GZ3,GZS,%
  GolubevZaikin02,GolubevHereroZaikina02}. 
Their key idea was as follows: to understand how
the diffusive behavior of a given electron is affected by its
interactions with other electrons in the system, which constitute its
effective environment, the latter should be integrated out, leading to
an influence functional, denoted by $e^{-{1 \over \hbar}
(i \tilde S_R + \tilde
  S_I)}$, in the path integral $\int \! \tilde {\cal D}^\prime \bmR $
describing its dynamics. To derive the effective action $(i \tilde S_R
+ \tilde S_I)$, GZ devised a strategy which, when implemented with
sufficient care, \emph{properly incorporates the Pauli principle} --
this is essential, since both the particle and its environment
originate from the same system of indistinghuishable fermions, a
feature which makes the present problem conceptually interesting and
sets it apart from all other applications of influence functionals
that we are aware of.

GZ used their new approach to calculate the electron decoherence rate
$\gammaphi (T)$ in disordered conductors, as extracted from the
magnetoconductance in the weak localization regime, and found it to be
finite at zero temperature\cite{GZ1,GZ2,GZ3,GZS,%
  GolubevZaikin02,GolubevHereroZaikina02}, $\gamma_\varphi^\GZ (T \to
0) = \gamma^{0,\GZ}_\varphi$, in apparent agreement with some
experiments\cite{MW}.  However, this result contradicts the standard
view, based on the work of Altshuler, Aronov and Khmelnitskii
(AAK)\cite{AAK82}, that $ \gamma_\varphi^\AAK (T \to 0) = 0$, and
hence elicited a considerable controversy\cite{controversy}.  GZ's
work was widely
questioned,\cite{EriksenHedegard,VavilovAmbegaokar98,KirkpatrickBelitz01,%
  Imry02,vonDelftJapan02,Marquardt02}, with the most detailed and
vigorous critique coming from Aleiner, Altshuler and Gershenzon
(AAG)\cite{AAG98} and Aleiner, Altshuler and Vavilov
(AAV)\cite{AAV01,AAV02}, but GZ rejected each
critique\cite{GZ3,GZS,GolubevZaikin02,controversy} with equal vigor.
It is important to emphasize that the debate here was about a
well-defined theoretical model, and not about experiments which do or
do not support GZ's claim.

The fact that GZ's final results for $\gamma_\varphi^\GZ (T)$ have
been questioned, however, does not imply that their influence
functional approach, as such, is fundamentally flawed. To the
contrary, we show in this review that it is sound in principle, and
that \emph{ the standard result $\gamma_\varphi^\AAK (T) $ can be
  reproduced using GZ's method,} provided that it is applied with
slightly more care to correctly account for recoil effects (\ie\ the
fact that the energy of an electron changes when it absorbs or emits a
photon).  We believe that this finding conclusively resolves the
controversy in favor of AAK and company; hopefully, it will also serve
to revive appreciation for the merits of GZ's influence functional
approach.

The premise for understanding how $\gAAK$ can be reproduced with GZ's
methods was that we had carried out a painfully detailed analysis and
rederivation GZ's approach, as set forth by them in two lengthy papers
from 1999 and 2000, henceforth referred to as GZ99\cite{GZ2} and
GZ00\cite{GZ3}. Our aim was to establish to what extent their method
is related to the standard Keldysh diagrammatic approach.  As it
turned out, the two methods are essentially equivalent, and GZ
obtained unconventional results only because a certain ``Pauli
factor'' $(\tilde \delta - 2 \tilde \rho^0)$ occuring in $\tilde S_R$
was not treated sufficiently carefully, where $\tilde \rho^0$ is the
single-particle density matrix. That their treatment of this Pauli
factor was dubious had of course been understood and emphasized
before: first and foremost it was correctly pointed out by
AAG\cite{AAG98} that GZ's treatment of the Pauli factor caused their
expression for $\gamma_\varphi^\GZ$ to aquire an artificial
ultraviolet divergence, which then produces the term
$\gamma_\varphi^{0, \GZ}$, whereas no such divergence is present in
diagrammatic calculations.  GZ's treatment of $\Pauli$ was also
criticized, in various related contexts, by several other
authors\cite{EriksenHedegard,%
  VavilovAmbegaokar98,vonDelftJapan02,Marquardt02,AAV01}.  However,
none of these works (including our own\cite{vonDelftJapan02}, which,
in retrospect, missed the main point, namely recoil) had attempted to
diagnose the nature of the Pauli factor problem \emph{with sufficient
  precision to allow a successful remedy to be devised within the
  influence functional framework}.

 This will be done in the present review.  Working in the time domain,
 GZ represent $(\tilde \delta - 2 \tilde \rho^0 (t)) $ as $1 - 2 n_0
 \bigr [ \tilde h_0 (t)/2T\bigr]$, where $n_0$ is the Fermi function
 and $\tilde h_0 (t)$ the free part of the electron energy.  GZ
 assumed that $\tilde h_0 (t)$ does not change during the diffusive
 motion, because scattering off impurities is elastic.  Our diagnosis
 is that this assumption \emph{unintentionally neglects recoil
   effects} (as first pointed out by Eriksen and
 Hedegard\cite{EriksenHedegard}), because the energy of an electron
 actually does change at each interaction vertex, \ie\ each time it
 emits or absorbs a photon.  The remedy (not found by Eriksen and
 Hedegard) is to transform from the time to the frequency domain, in
 which $\Pauli$ is represented by $1 - 2 n_0 [\hbar (\bar \ve -
 \bomega)] = \tanh[\hbar (\bar \ve-\bomega)/2T]$, where $\hbar
 \bomega$ is the energy change experienced by an electron with energy
 $\hbar \bar \ve$ at an interaction vertex.  Remarkably, this simple
 change of representation from the time to the frequency domain is
 sufficient to recover $\gAAK$.  Moreover, the ensuing calculation is
 free of ultraviolet or infrared divergencies, and no cut-offs of any
 kind have to be introduced by hand.
 
 The main text of the present review has two central aims: firstly, to
 concisely explain the nature of the Pauli factor problem and its
 remedy; and secondly, to present a transparent calculation of
 $\gammaphi$, using only a few lines of simple algebra.  (Actually, we
 shall only present a ``rough'' version of the calculation here, which
 reproduces the qualitative behavior of $\gAAK (T)$; an improved
 version, which achieves quantitative agreement with AAK's result for
 the magnetoconductance [with an error of at most 4\% for quasi-1-D
 wires], has been published in 
a separate analysis by Marquardt, von Delft,
Smith and Ambegaokar\cite{MarquardtAmbegaokar04}.
The latter consists of two parts, 
referred to as MDSA-I and DMSA-II below, which use alternative
routes to arrive at conclusions that fully  confirm
the analysis of this review.)
 
 We have made an effort to keep the main text reasonably short and to
 the point; once one accepts its starting point
 [\Eqs{eq:sigmageneraldefinePI-MAINtext} to \Eq{subeq:ingredients}],
 the rest of the discussion can easily be followed step by step.
 Thus, as far as possible, the main text avoids technical details of
 interest only to the experts.  These have been included in a set of
 five lengthy and very detailed appendices, B to F, in the belief that
 when dealing with a controversy, \emph{all} relevant details should
 be publicly accessible to those interested in ``the fine print''.
For the benefit of those readers (presumably the majority)
 with no time or inclination to read lengthy appendices, a concise
 appendix~A summarizes (without derivations)
 the main steps and approximations involved in obtaining the influence
 functional.
 
 The main text and appendices A.1 to A.3 have already been published
 previously\cite{Granada}, but for convenience are included here again
 (with minor revisions, and an extra sketch in
 Fig.~\ref{fig:Keldyshvertices}), filling the first 23 pages.  The
 content of the remaining appendices is as follows: In
 App.~\ref{sec:interchangingaverages} we address GZ's claim that a
 strictly nonperturbative approach is needed for obtaining
 $\gamma_\varphi$, and explain why we disagree (as do many
 others\cite{AAG98,AAV01,AAV02}). In App.~B, we rederive the influence
 functional and effective action of GZ, following their general
 strategy in spirit, but introducing some improvements. The most
 important differences are: (i) instead of using the
 coordinate-momentum path integral $\int \! {\cal D} \bmR \int {\cal
   D} \bmP$ of GZ, we use a ``coordinates-only'' version $\int \!
 \tilde {\cal D}^\prime \bmR$, since this enables the Pauli factor to
 be treated more accurately; and (ii), we are careful to perform
 thermal weigthing at an initial time $t_0 \to - \infty$ (which GZ do
 not do), which is essential for obtaining properly energy-averaged
 expressions and for reproducing perturbative results: the standard
 diagrammatic Keldysh perturbation expansion for the Cooperon in
 powers of the interaction propagator is generated if, \emph{before
   disorder averaging}, the influence functional is expanded in powers
 of $(i \tilde S_R + \tilde S_I)/ \hbar$. In App.~C we review how a
 general path integral expression derived for the conductivity in
 App.~B can be rewritten in terms of the familiar Cooperon propagator,
 and thereby related to the standard relations familiar from
 diagrammatic perturbation theory.  In particular, we review the
 Fourier transforms required to obtain a path integral $\tilde
 P^\ve_\eff (\tau)$ properly depending on both the energy variable
 $\hbar \ve$ relevant for thermal weighting and the propagation time
 $\tau$ needed to traverse the closed paths governing weak
 localization.  Appendix~D gives an explicit time-slicing definition
 of the ``coordinates-only'' path integral $\int \! \tilde {\cal
   D}^\prime \bmR$ used in App.~B.  Finally, for reference purposes,
 we collect in Apps.~E and~F some standard material on the
 diagrammatic technique (although this is bread-and-butter knowledge
 for experts in diagrammatic methods and available elsewere, it is
 useful to have it summarized here in a notation consistent with the
 rest of our analysis).  App.~E summarizes the standard Keldysh
 approach in a way that emphasizes the analogy to our influence
 functional approach, and App.~F collects some standard and well-known
 results used for diagrammatic disorder averaging.  Disorder averaging
 is discussed last for a good reason: one of the appealing features of
 the influence functional approach is that most of the analysis can be
 performed \emph{before} disorder averaging, which, if at all, only
 has to be performed at the very end.

\section{Main Results of Influence Functional Approach}
\label{sec:mainresults}

We begin by summarizing the main result of GZ's influence functional
approach.  Our notations and also the content of some of our formulas
are not identical to those of GZ, and in fact differ from their's in
important respects.  Nevertheless, we shall refer to them as ``GZ's
results'', since we have (re)derived them (see App.~B
for details) in the spirit of GZ's approach.

\begin{figure}[t]
{\includegraphics[clip,width=0.98\linewidth]{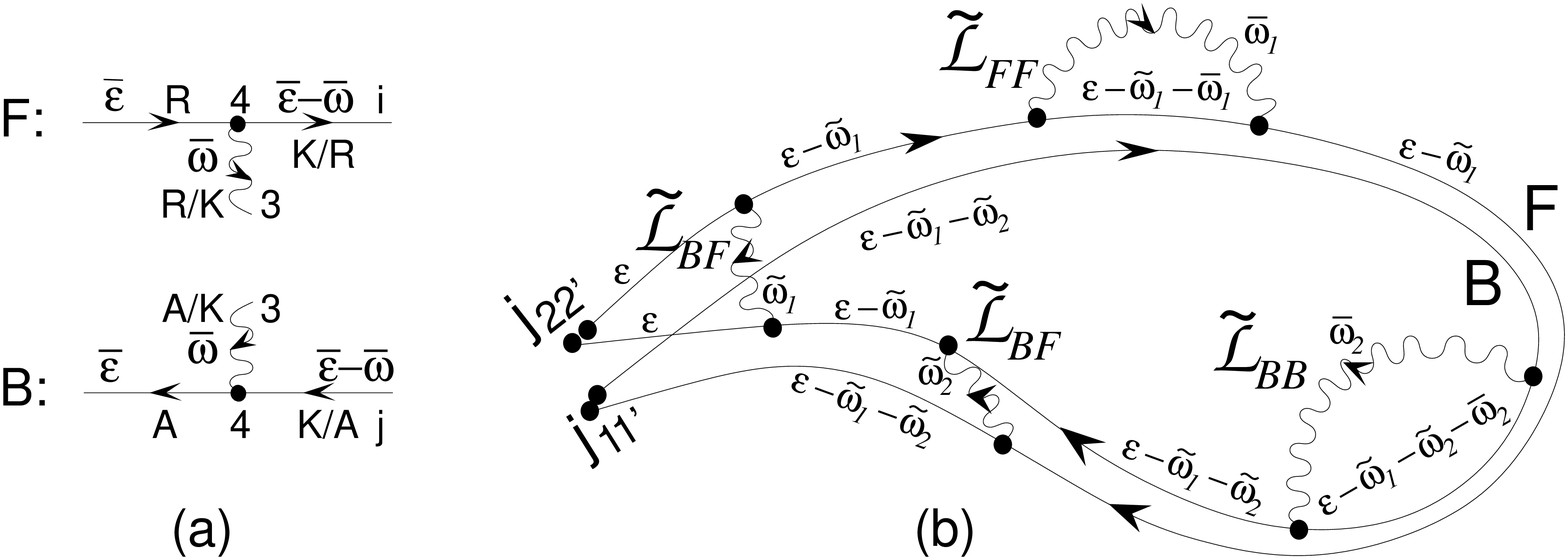}}%
\caption{
  (a) Structure of vertices on the forward or backward contours of
  Keldysh perturbation theory. F: the combinations $\G^K_{i_F
    4_F}\tilde {\cal L}^{R}_{34_{F}}$ and $\G^R_{i_F 4_F} \tilde {\cal
    L}^{K}_{34_{F}}$ occur if vertex 4 lies on the upper forward
  contour. B: the combinations $\tilde {\cal L}^{A}_{4_B 3} \G^K_{4_B
    j_B}$ and $ \tilde {\cal L}^{K}_{4_B 3} \G^A_{4_B j_B}$ occur if
  vertex 4 lies on the lower contour.  Arrows point from the second to
  first indices of propagators.  (b) Sketch of a pair of time-reversed
  paths connecting the points at which the current operators
  $\bmj_{11'} \! \cdot \! \bmj_{\,22'}$ act [cf.\ 
  \Eq{eq:sigmageneraldefinePI-MAINtext-a}], decorated by several
  (wavy) interaction propagators $\LL^{R/A/K}_{aa'} (\omega)$. In the
  Keldysh formalism, the electron lines represent the electron
  propagators $\G^{R/A}(\omega)$ or $\G^K (\omega) = \tanh(\hbar
  \omega/2T) [\G^R - \G^A](\omega)$. The effective action defined in
  \Eqs{eq:SIR-LIR-aa-main} to (\ref{eq:recallLCD}) in effect neglects
  the frequency transfers $\omega_i$ in the arguments of all retarded
  and advanced electron Green's functions [$\G^{R/A} (\ve - \omega_i -
  \dots) \to \G^{R/A} (\ve ) $], but, for every occurence of the
  combination $\LL^{R/A} (\omega_i) \G^K (\ve - \omega_i)$, retains it
  in the factor $\tanh[\hbar (\ve -\omega_i) / \hbar]$ of the
  accompanying $\GK$ function. The latter prescription
%As discussed in
%  Sec.~\ref{sec:Paulifactorshort} or~\ref{sec:ruleofthumb}, 
  ensures that a crucial feature of the Keldysh approach is retained
  in the influence functional formalism, too, namely that all
  integrals $\int d \omega_i$ over frequency transfer variables are
  limited to the range $|\hbar \omega_i| \lesssim T$ [which is why the
  neglect of $\omega_i$ in $\G^{R/A} (\ve - \omega_i - \dots )$ is
  justified]. In contrast, GZ also neglect the $- \omega_i$ in
  $\tanh[\hbar (\ve -\omega_i) / \hbar]$ [see
  Sec.~\ref{sec:GZ-classical-paths}], which amounts to neglecting
  recoil. As as a result, their $\int d \omega_i$ integrals are no
  longer limited to $|\hbar \omega_i| \lesssim T$, \ie\ artificial
  ultraviolet divergencies occur, which produce GZ's
  temperature-independent contribution $\gamma_{\varphi}^{0,\GZ}$ to
  the decoherence rate [see \Eq{eq:GZwrongresult}]. Thus,
  $\gamma_{\varphi}^{0,\GZ}$ is an artefact of GZ's neglect of recoil,
  as is their claimed ``decoherence at zero temperature''.}
\label{fig:Keldyshvertices}
\end{figure}

The Kubo formula represents the DC conductivity $\sigma_\DC$ in terms
of a retarded current-current correlator $\langle [ \hat \bmj (1),
\hat \bmj (2) ] \rangle$.  This correlator can (within various
approximations discussed in App.~B.5.6, B.5.7, B.6.3 and
\ref{sec:thermalaveragingAppA}) be expressed as follows in terms of a
path integral $\tilde P^\ve_\eff$ representing the propagation of a
pair of electrons with average energy $\hbar \ve$, thermally averaged
over energies:
\bsubequations
\label{eq:sigmageneraldefinePI-MAINtext}
\begin{eqnarray}
\phantom{.} \hspace{-.5cm}
\label{eq:sigmageneraldefinePI-MAINtext-a}
\sigma_{\DC} \!\! & =  & \!\! % \nonumber
  {2 \over d}
\! \int \! \!  dx_2 \, \bmj_{11'} \! \cdot \! \bmj_{\,22'}
\!\!  \int (d \varepsilon) [ - n' (\hbar \varepsilon )] \,
\int_0^\infty \!\! d \tau  \,
\tilde P^{1 2', \ve}_{21', \eff} (\tau) \; ,
\\
\tilde P^{1 2', \ve}_{21', \eff} (\tau)\!\!  & = &  \!\!
  \Fint_{\bmR^F (-{ \tau \over 2}) =\bmr_{2'}}^{\bmR^F ({ \tau \over
  2}) = \bmr_1}
  \Bint_{\bmR^B (-{ \tau \over 2}) = \bmr_{2}}^{\bmR^B ({ \tau \over
  2}) = \bmr_{1'}}
\Bigl. \widetilde {\cal D}' \bmR \,
\, e^{{1 \over \hbar}  [i(\tilde  S_0^F - \tilde S_0^B) -( i \tilde
  S_R + \tilde S_I)] (\tau)} \; . 
\label{eq:sigmageneraldefinePI-MAINtext-b}
\end{eqnarray}
\esubequations The propagator $\tilde P^{1 2', \ve}_{21', \eff}
(\tau)$, defined for a given impurity configuration, is written in
terms of a forward and backward path integral ${\displaystyle \Fint
  \Bint \widetilde {\cal D}' \! \bmR }$ between the specified initial
and final coordinates and times. It gives the amplitude for a pair of
electron trajectories, with average energy $\hbar \ve$, to propagate
from $\bmr_{2'}$ at time $- \toh \tau$ to $\bmr_{1}$ at $\toh \tau $
or from $\bmr_{1'}$ at time $ \toh \tau$ to $\bmr_2$ at $- \toh \tau$,
respectively.  [The sense in which both $\tau$ and $\varepsilon$ can
be specified at the same time is discussed in
App.~\ref{sec:thermalaveragingAppA}, and in more detail in
App.~\ref{sec:definefullCooperon}, \Eqs{eq:Pfixedenergy-a} to
(\ref{eq:fixenergytoEFmoduloT})]. We shall call these the forward and
backward paths, respectively, using an index $a = F,B$ to distinghuish
them.  $\tilde S_0^a = \tilde S_0^{F/B}$ are the corresponding free
actions, which determine which paths will dominate the path integral.
The weak localization correction to the conductivity, $\sigma^{\rm
  WL}_\DC$, arises from the ``Cooperon'' contributions to
$\sigma_\DC$, illustrated in Fig.~\ref{fig:Keldyshvertices}(b), for
which the coordinates $\bmr_1$, $\bmr_1'$, $\bmr_2$ and $\bmr_2'$ all
lie close together, and which feature self-returning random walks
through the disordered potential landscape for pairs of paths
$\bmR^{F/B}$, with path $B$ being the time-reversed version of path
$F$, \ie\ $\bmR^F (t_3) = \bmR^B (- t_3)$ for $t_3 \in (-\toh \tau,
\toh \tau)$.  The effect of the other electrons on this propagation is
encoded in the influence functional $e^{- (i \tilde S_R + \tilde
  S_I)/\hbar }$ occuring in \Eq{eq:sigmageneraldefinePI-MAINtext-b}.  The
effective action $i \tilde S_R + \tilde S_I$ turns out to have the
form [for a more explicit version, see \Eq{eq:defineSiRA} in App.~A;
or, for an equivalent but more compact representation, see
\Eqs{eq:Seff} and (\ref{subeq:LAA'FT}) of
Sec.~\ref{sec:alteffaction}]:
%\bsubequations
\begin{eqnarray}
  \label{eq:SIR-LIR-aa-main}
\Biggl\{ \! \! \begin{array}{c}
 i \tilde S_R (\tau)
\rule[-2mm]{0mm}{0mm} \\
 \tilde S_I (\tau)
\end{array} \! \! \Biggr\}   =  - \toh i  \sum_{a,a'= F,B}  s_a
\int_{-{\tau \over 2}}^{\tau \over 2} d t_{3_a}
\int_{-{\tau \over 2}}^{t_{3_a}} d t_{4_{a'}}
\Biggl \{ \!\! \!  \begin{array}{c}
\phantom{s_{a'}}
\tilde {\cal L}^{a'}_{{3_a} 4_{a'}}
\rule[-2mm]{0mm}{0mm}
\\
 s_{a'}  \tilde {\cal L}^K_{{3_a} 4_{a'}}
\end{array} \!\!\! \Biggr\} \; .
\end{eqnarray}
Here $s_a$ stands for $s_{F/B} = \pm 1$, and the shorthand $\tilde
{\cal L}_{3_a 4_a'} = \tilde {\cal L} \bigl[t_{3_a} - t_{4_{a'}},
\bmR^a (t_{3_a}) - \bmR^{a'} (t_{4_{a'}})\bigr]$ describes, in the
coordinate-time representation, an interaction propagator linking two
vertices on contours $a$ and $a'$. It will be convenient below to
Fourier transform to the momentum-freqency representation, where the
propagators $\overline {\cal L}^K$ and $\overline {\cal L}^{a'}$ can be
represented as follows [$(d \bomega) (d \bbmq) \equiv {(d \bomega \, d
  \bbmq)/(2 \pi)^4}$]:
\bsubequations
  \label{subeq:defineLKRA}
\begin{eqnarray}
  \label{eq:defineLK}
\tilde {\cal L}^K_{3_a 4_{a'}}
\!\! & \equiv & \!\!
\int (d \bomega) (d \bbmq)
e^{i   \left(\bbmq \cdot \left[\bmR^a (t_{3_a}) -
\bmR^{a'} (t_{4_{a'}}) \right] - \bomega (t_{3_a} - t_{4_{a'}}) \right)}
\overline {\cal L}^K_\bbmq (\bomega) \, ,
\\
  \label{eq:defineLRArealspace}
\tilde {\cal L}^{a'}_{3_a 4_{a'}}
\!\! & \equiv & \!\!
 \left\{
\begin{array}{ll}
\bigl[\Pauli \tilde {\cal L}^R \bigr]_{3_a 4_F} & \qquad  \mbox{if}
\quad a' = F \; ,
\\
\bigl[ \tilde {\cal L}^A \Pauli \bigr]_{4_B 3_a } & \qquad  \mbox{if}
\quad a' = B \; ,
\end{array}
\right.
\\   \label{eq:defineLRArealomega}
\!\! & \equiv & \!\!
\int (d \bomega) (d \bbmq)
e^{i s_{a'}  \left(\bbmq \cdot \left[\bmR^a (t_{3_a}) -
\bmR^{a'} (t_{4_{a'}}) \right] - \bomega (t_{3_a} - t_{4_{a'}}) \right)}
\overline {\cal L}^{a'}_\bbmq (\bomega) \, . \qqph
\end{eqnarray}
\esubequations
[Note the sign $s_{a'}$ in the Fourier exponential in
\Eq{eq:defineLRArealomega}; it reflects the opposite order of indices
in \Eq{eq:defineLRArealspace}, namely 34 for $F$ vs. 43 for $B$.]
Here $\tilde {\cal L}^K$ is the Keldysh interaction propagator, while
$\tilde {\cal L}^{F/B}$, to be used when time $t_{4_{a'}}$ lies on the
forward or backward contours, respectively, represent ``effective''
retarded or advanced propagators, modified by a ``Pauli factor''
$(\tilde \delta - 2 \tilde \rho^0)$ (involving a Dirac-delta $\tilde
\delta_\iiijjj$ and single-particle density matrix $\tilde \rho^0_\iiijjj$ in
coordinate space), the precise meaning of which will be discussed
below.  $\overline {\cal L}^{K,R,A}_\bbmq (\bomega) $ denote the Fourier transforms
of the standard Keldysh, retarded, or advanced interaction
propatators.  For the screened Coulomb interaction in the unitary
limit, they are given by
\bsubequations
\label{subeq:ingredients}
\begin{eqnarray}
  \label{eq:recallLCD}
  \bLRbbqw & = & [\bLAbbqw]^\ast =
- {E^0_\bbmq - i \bomega\over   2 \nud E^0_\bbmq} =
- {[\bcD^0_\bbmq (\bomega)]^{-1} \over
2 \nud E^0_\bbmq} \; ,
\\
\bLKbbqw
& = & 2 \, i \coth (\hbar \bomega / 2T) \, {\rm Im}
[\bLRbbqw] \; , \\
\bcC^0_\bbmq (\bomega) & = & {1 \over E_\bbmq - i \bomega}
\; ,
\qquad
\bcD^0_\bbmq (\bomega) = {1 \over E_\bbmq^0 - i \bomega }
\; , \qqph
\\
\label{eq:defineEq}
E^0_\bbmq  & =  & \Dd \bbmq^2 \; , \qquad
E_\bbmq = \Dd \bbmq^2 + \gamma_H \; ,
\end{eqnarray}
where, for later reference, we have also listed the Fourier transforms
of the bare
diffuson $\bcD^0 $ and Cooperon $\bcC^0 $ (where $\gammaH$ is
the dephasing rate of the latter in the presence of a magnetic field,
$\Dd$ the diffusion constant and $\nu$ the density of states per spin).
Finally, $\overline {\cal L}^{a'}_\bbmq (\bomega)$ in
\Eq{eq:defineLRArealomega} is defined as
\begin{eqnarray}
  \label{eq:modifiedRAtanh}
\overline {\cal L}^{F/B}_\bbmq (\bomega) =
\tanh[\hbar (\ve - \bomega)/ 2 T] \,
   \overline {\cal L}^{R/A}_\bbmq (\bomega) \; ,
\end{eqnarray}
\esubequations
where $\hbar \ve$ is the same energy as that occuring in the thermal
weighting factor $[ - n' (\hbar \varepsilon )] $ in 
\Eq{eq:sigmageneraldefinePI-MAINtext-a}.

Via the influence functional, the effective action
(\ref{eq:SIR-LIR-aa-main}) concisely incorporates the effects of
interactions into the path integral approach.  $\tilde S_I$ describes
the \emph{classical} part of the effective environment, and if one
would replace the factor $\coth(\hbar \bomega/ 2T ) $ in $\tilde {\cal
  L}^K_\bbmq (\bomega)$ by $2 T / \hbar \bomega$ (as is possible for
high temperatures) it corresponds to the contribution calculated by
AAK\cite{AAK82}.  With $\tilde S_R$, GZ succeeded to additionally
also include the quantum part of the environment, and in particular,
via the Pauli factor $\Pauli$, to properly account for the Pauli
principle.

Casual readers are asked to simply accept the above equations as
starting point for the remainder of this review, and perhaps glance
through App.~A to get an idea of the main steps and approximations
involved in deriving them.  Those interested in a detailed derivation
are referred to App.~B (where $\tilde S_{R/I}$ are obtained in
Sec.~B.5.8).  It is also shown there [Sec.~B.6] that the standard
results of diagrammatic Keldysh perturbation theory can readily be
reproduced from the above formalism by expanding the influence
functional $e^{- (i \tilde S_R + \tilde S_I)/ \hbar}$ in powers of $(i
\tilde S_R + \tilde S_I)/\hbar$.  For present purposes, simply note
that such an equivalence is entirely plausible in light of the fact
that our effective action (\ref{eq:SIR-LIR-aa-main}) is linear in the
effective interaction propagators $\tilde {\cal L}$, a structure
typical for generating functionals for Feynman diagrams.

\section{Origin of the Pauli Factor}
\label{sec:Paulifactorshort}

The occurence of the Pauli factor $\Pauli$
in $\tilde S_R$ was first found by GZ in precisely
the form displayed in the position-time representation
of the effective action used in \Eq{eq:SIR-LIR-aa-main}. However,
their subsequent treatment of this factor differs from ours,
in a way that will be described below. In particular,
they did not represent this factor
in the frequency representation, as in our \Eq{eq:modifiedRAtanh},
and this is the most important difference between our analysis
and theirs.

The origin of the Pauli factor in the form given by our
\Eq{eq:modifiedRAtanh} can easily be understood if one is familiar
with the structure of Keldysh perturbation theory.  [For a detailed
discussion, see Sec.~B.6.2.]  First recall two exact relations for the
noninteraction Keldysh electron propagator: in the coordinate-time
representation, it contains a Pauli factor,
\bsubequations
\label{subeq:Keldyshid}
\begin{eqnarray}
\G^K_\iiijjj \! = \!  \int \! \! d x_k \,(\G^R - \G^A)_{ik} \Pauli_{kj}
\! = \!  \int \! d x_k \,
\Pauli_{ik} (\G^R - \G^A)_{kj} \hspace{-5mm} \phantom{.}
\nonumber \\
\label{eq:Keldyshid}
\end{eqnarray}
 which turns into a $\tanh$ in the coordinate
frequency representation:
\begin{eqnarray}
  \label{eq:Keldyshtanh}
  \G^K_\iiijjj (\bomega) = \tanh(\hbar \bomega/2T)
\Bigl[\G^R_\iiijjj (\bomega)  - \G^A_\iiijjj (\bomega)  \Bigr] \; .
\end{eqnarray}
\esubequations
Now, in the Keldysh approach, retarded or advanced interaction
propagators always occur [see Fig.~\ref{fig:Keldyshvertices}(a)] together
with Keldysh electron propagators, in the combinations $\G^K_{i_F
  4_F}\tilde {\cal L}^{R}_{34_{F}}$ or $ \tilde {\cal L}^{A}_{4_B 3}
\G^K_{4_B j_B}$, where the indices denote coordinates and times.
[Likewise, the Keldysh interaction propagators always come in the
combinations $\G^R_{i_F 4_F} \tilde {\cal L}^{K}_{34_{F}}$ or $\tilde
{\cal L}^{K}_{4_B 3} \G^A_{4_B j_B}$.]  In the momentum-frequency
representation, the combinations involving $\G^K$ therefore turn into
$\overline {\cal L}^{R/A}_\bbmq 
(\bomega) \bigl[ \bar G^R - \bar G^A \bigr]_{\bmq
  - \bbmq} (\bar \ve - \bomega) \, \tanh[\hbar (\bar \ve - \bomega)/ 2
T] $. Thus, \emph{in the frequency representation the Pauli factor is
  represented as} $\tanh[\hbar (\bar \ve - \bomega)/ 2 T]$.  Here the
variable $\hbar \bar \ve$ represents the energy of the electron line
on the upper (or lower) Keldysh contour before it enters (or after it
leaves) an interaction vertex at which its energy decreases (or
increases) by $\hbar \bomega$ [see Fig.~\ref{fig:Keldyshvertices}(a)].
The subtraction of $\bomega$ in the argument of $\tanh$ thus reflects
the physics of recoil: emitting or absorbing a photon causes the
electron energy to change by $\hbar \bomega$, and it is this changed
energy $\hbar (\bar \ve - \bomega)$ that enters the Fermi functions
for the relevant final or initial states.

Of course, in Keldysh perturbation theory, $\hbar \bar \ve$ will have
different values from one vertex to the next, reflecting the history
of energy changes of an electron line as it proceeds through a Feynman
diagram [as illustrated in Fig.~\ref{fig:Keldyshvertices}(b)].  It is
possible to neglect this complication in the influence functional
approach, if one so chooses, by always using one and the same energy
in \Eq{eq:modifiedRAtanh}, which then should be chosen to be the same
as that occuring in the thermal weighting factor $[ - n' (\hbar
\varepsilon )] $, \ie\ $\hbar \bar \ve = \hbar \ve$.  This
approximation, which we shall henceforth adopt, is expected to work
well if the relevant physics is dominated by low frequencies, at which
energy transfers between the two contours are sufficiently small
[$\hbar (\bar \ve - \ve ) \ll T$, so that the electron ``sees''
essentially the same Fermi function throughout its motion. [For a
detailed discussion of this point, see App.~\ref{sec:ruleofthumb}.]

Though the origin and neccessity of the Pauli factor is eminently
clear when seen in conjunction with Keldysh perturbation theory, it is
a rather nontrivial matter to derive it cleanly in the functional
integral approach [indeed, this is the main reason for the length of
our appendices!]. The fact that GZ got it completely right in the
position-time representation of \Eq{eq:SIR-LIR-aa-main} is, in our
opinion, a significant and important achievement.  It is regrettable
that they did not proceed to consider the frequency representation
(\ref{eq:modifiedRAtanh}), too, which in our opinion is more useful.

\section{Calculating $\tauphi$ $\grave {{\rm a}}$ la GZ}
\label{sec:GZ-classical-paths}

To calculate the decoherence rate $\gammaphi = 1/\tauphi$, one has to
find the long-time decay of the Cooperon contribution to the
propagator $\tilde P^\ve_\eff (\tau)$ of
\Eq{eq:sigmageneraldefinePI-MAINtext}. To do this, GZ proceeded as
follows: using a saddle-point approximation for the path integral for
the Cooperon, they replaced the sum over all pairs of self-returning
paths $\bmR^{F/B}(t_{3_{F/B}})$ by just the contribution $\langle e^{-
  {1 \over \hbar} (i \tilde S_R + \tilde S_I)(\tau)} \rangle_\rw$ of
the classical ``random walk'' paths $\bmR_\rw (t)$ picked out by the
classical actions $\tilde S_0^a$, namely $\bmR^F(t_{3_F}) = \bmR_\rw
(t_{3_F})$ and $\bmR^B(t_{3_B}) = \bmR_\rw (- t_{3_B})$, for which the
paths on the forward and backward Keldysh contours are
\emph{time-reversed} partners.  The subscript ``rw'' indicates that
each such classical path is a self-returning \underline{r}andom
\underline{w}alk through the given disorder potential landscape, and
$\langle \; \; \rangle_\rw$ means averaging over all such paths.
Next, in the spirit of Chakravarty and
Schmid\cite{ChakravartySchmid86}, they replace the average of the
exponent over all time-reversed pairs of self-returning random walks,
by the exponent of the average, $e^{- F(\tau)}$, where $F(\tau) = {1
  \over \hbar} \langle i \tilde S_R + \tilde S_I \rangle_\rw $ (cf.\
Eq.  (67) of GZ99\cite{GZ2}). This amounts to expanding the exponent
to first order, then averaging, and then reexponentiating.  The
function $F(\tau)$ thus defined increases with time, starting from
$F(0) = 0$, and the decoherence time $\tauphi$ can be defined as the
time at which it becomes of order one, \ie\ $F(\tauphi) \approx 1$.

To evaluate $\langle i \tilde S_R + \tilde S_I \rangle_\rw $, GZ
Fourier transform the functions $\tilde {\cal L}_{3_a 4_a'} = \tilde
{\cal L} [t_{34}, \bmR^a (t_3) - \bmR^{a'} (t_4)]$ occuring in $\tilde
S_{R/I}$, and average the Fourier exponents 
using\cite{ChakravartySchmid86} the distribution function for diffusive
motion, which gives probability that a random walk that passes point
$\bmR_\rw (t_4)$ at time $t_4$ will pass point $\bmR_\rw (t_3)$ at
time $t_3$, \ie\ that it covers a distance $\bmR = \bmR_\rw (t_3) -
\bmR_\rw (t_4)$ in time $|t_{34}|$:
\begin{eqnarray}  \nonumber
\Bigl\langle
e^{i \bbmq \cdot [\bmR_\rw (t_3)- \bmR_\rw (t_4)]}
\Bigr\rangle_\rw \!\! & \simeq  & \!\!
\int \!d^\bard \bmR \left(\frac{\pi}{ D |t_{34}|} \right)^{\bard/2}
e^{-\bmR^2/ (4 D |t_{34}|) } \,
e^{{i} \bbmq \cdot \bmR} \qqph
\\ \label{eq:impurity-average-of-eikr-A}
\!\! &=& \!\!
e^{- \bbmq^2 D |t_{34}|} \to \tilde C^0_\bbmq (|t_{34}|)
= e^{- E_\bbmq |t_{34}|} \; .
\end{eqnarray}
(Here $t_{34} = t_3 - t_4$.)  The arrow in the second line makes
explicit that if we also account for the fact that such time-reversed
pairs of paths are dephased by a magnetic field, by adding a factor
$e^{-\gammaH |t_{34}|}$, the result is simply equal to the bare
Cooperon in the momentum-time representation.

Actually, the above way of averaging is somewhat inaccurate, as was
pointed out to us by Florian Marquardt: it neglects the fact that the
diffusive trajectories between $t_3$ and $t_4$ are part of a larger,
\emph{self-returning} trajectory, starting and ending at $\bmr_1
\simeq \bmr_2$ at times $\mp \toh \tau$.  It is actually not difficult
to include this fact, see MDSA-I\cite{MarquardtAmbegaokar04}, and this
turns out to quantitatively improve the numerical prefactor for
$\tauphi$ (\eg\ in \Eq{eq:F1dexplicitfinal} below).  However, for the
sake of simplicity, we shall here be content with using
\Eq{eq:impurity-average-of-eikr-A}, as GZ did.

Finally, GZ also assumed that the Pauli factor $\Pauli$ in $\tilde S_R$
remains unchanged throughout the diffusive motion: they use a
coordinate-momentum path integral $\int{\cal D} \! \bmR \int \! {\cal
  D} \bmP$ [instead of our coordinates-only version $\int \!
\widetilde {\cal D}' \bmR$], in which $(\tilde \delta - 2 \tilde
\rho^0) $ is replaced by $[1 - 2 n_0 (\tilde h_0)] = \tanh (\tilde
h_0/2 T)$, and the free-electron energy $\tilde h_0 \bigl[\bmR (t_a),
\bmP(t_a)\bigr]$ is argued to be unchanged throughout the diffusive motion,
since impurity scattering is elastic [cf.\ p.~9205 of GZ99\cite{GZ2}:
``$n$ depends only on the energy and not on time because the energy is
conserved along the classical path'']. Indeed, this is true
\emph{between} the two interaction events at times $t_3$ and $t_4$, so
that the averaging of \Eq{eq:impurity-average-of-eikr-A} \emph{is}
permissible. However, as emphasized above, the full trajectory
stretches from $- \toh \tau$ to $t_4$ to $t_3$ to $\toh \tau$, and the
electron energy \emph{does} change, by $\pm \hbar \bomega$, at the
interaction vertices at $t_4$ and $t_3$. Thus, \emph{GZ's assumption of a
time-independent Pauli factor neglects recoil effects}. As argued in
the previous section, these can straightforwardly taken into account
using \Eq{eq:modifiedRAtanh}, which we shall use below. In contrast,
GZ's assumption of time-independent $n$ amounts dropping the $- \hbar
\bomega$ in our $\tanh[\hbar (\ve - \bomega)/ 2 T] $ function.

If one uses GZ's assumptions to average \Eq{eq:SIR-LIR-aa-main}, but
uses the proper $\tanh[\hbar (\ve - \bomega)/ 2 T] $ function, one
readily arrives at
\begin{eqnarray}
\label{eq:averageSRIcl}
 \left\{ \! \! \begin{array}{c}
\langle i \tilde S_R \rangle_\rw
\rule[-3mm]{0mm}{0mm}
\\
\langle  \tilde S_I \rangle_\rw
\end{array} \! \! \right\}    =
2 {\rm Re} \left[
- \toh i  \int (d \bomega) (d \bbmq)
  \left\{ \! \! \begin{array}{c}
\overline {\cal L}^{F}_\bbmq (\bomega)
\rule[-3mm]{0mm}{0mm}
\\
 \bLKbbqw \,
\end{array} \! \! \right\}
\Bigl[ f^\self  - f^\vertex   \Bigr]\!
(\tau)
\right] ,
\end{eqnarray}
where $f^\self - f^\vertex$ are the
first and second terms of the double time integral
\begin{eqnarray}
  \label{eq:doubletimeintegral}
\int_{-{\tau \over 2}}^{\tau \over 2} d t_{3}
\int_{-{\tau \over 2}}^{t_{3}} d t_{4} \,
e^{- i \bomega t_{34} }
\Bigl\langle
e^{i \bmq \cdot [\bmR_\rw (t_3)- \bmR_\rw (t_4)]}
- e^{i \bmq \cdot [\bmR_\rw (- t_3)- \bmR_\rw (t_4)]}
\Bigr\rangle_\rw \! ,
\end{eqnarray}
corresponding to self-energy ($a=a'= F$) and vertex ($a \neq a'= F$)
contributions, and the $2{\rm  Re} [\; \; ]$ in
\Eq{eq:averageSRIcl} comes from adding the contributions of $a' = F$
and $B$. Performing the integrals in \Eq{eq:doubletimeintegral}, we
find
\bsubequations
  \label{subeq:averageSRIcl}
\begin{eqnarray}
  \label{subeq:averageSRIclself}
f^\self (\tau)  & = &
\bcC^0_\bbmq (-\bomega)
\tau \; + \;
\bigl[ \bcC^0_\bbmq (-\bomega) \bigr]^2 \,
\Bigl[e^{- \tau (E_\bbmq + i \bomega)} - 1 \Bigr] \;  , \qqph
\\
  \label{subeq:averageSRIclvert}
 f^\vertex (\tau)  & = &
\bcC^0_\bbmq (\bomega)  \biggl[
 {e^{-i \bomega \tau} -1  \over -i \bomega} +
{e^{-E_\bbmq \tau} - 1 \over E_\bbmq}
 \biggr]  \; .
\end{eqnarray}
\esubequations
Of all terms in \Eqs{subeq:averageSRIcl}, the first term of $f^\self
$, which is linear in $\tau$, clearly grows most rapidly, and hence
dominates the leading long-time behavior.  Denoting the associated
contribution to \Eq{eq:averageSRIcl} by ${1 \over \hbar} \langle i
\tilde S_R/ \tilde S_I \rangle^{\leading, \self}_\rw \equiv \tau
\gamma_\varphi^{R/I, \self}$, the corresponding rates
$\gamma_\varphi^{R/I,\self}$ obtained from \Eqs{eq:averageSRIcl} and
(\ref{subeq:averageSRIcl}) are:
\bsubequations
  \label{eq:finalselfenergy}
  \begin{eqnarray}
    \label{eq:finalSigmaR}
\gamma_\varphi^{R, \self} & = & {1 \over \hbar}
\int (d \bomega ) (d \bbmq )
\tanh \left[{\hbar (\ve - \bomega) \over 2T} \right] \,
 2 {\rm Re} \left[ {  \toh i (E_\bbmq^0 - i \bomega )
\over 2 \nud E_\bbmq^0 ( E_\bbmq + i \bomega) } \right]  , \qqph
\\
    \label{eq:finalSigmaI}
  \gamma_\varphi^{I, \self}  & = &
{1 \over \hbar}
\int (d \bomega ) (d \bbmq )
\coth \left[{\hbar \bomega \over 2T} \right] \,
 2 {\rm Re} \left[ { \bomega \over  2 \nu E_\bbmq^0
(E_\bbmq + i \bomega) } \right]  .
  \end{eqnarray}
\esubequations
Let us compare these results to those of GZ, henceforth
using $\gamma_H = 0$. Firstly, both our
$\gamma_\varphi^{I,\self}$ and $\gamma_\varphi^{R, \self}$ are
nonzero. In contrast, in their analysis GZ concluded that $\langle
\tilde S_R \rangle_\rw = 0$.  The reason for the latter result is,
evidently, their neglect of recoil effects: indeed, if we drop the $-
\hbar \bomega$ from the $\tanh$-factor of \Eq{eq:finalSigmaR}, we
would find $\gamma_\varphi^R = 0$ and thereby recover GZ's result,
since the real part of the factor in square brackets is odd in
$\bomega$.

Secondly and as expected, we note that \Eq{eq:finalSigmaI} for
$\gamma_\varphi^{I, \self}$ agrees with that of GZ, as given by their
equation (71) of GZ99\cite{GZ2} for $1/\tauphi$, \ie\
$\gamma_\varphi^{I,\self} = \gamma_\varphi^\GZ$. [To see the
equivalence explicitly, use \Eq{eq:RIvsLRA-main}.]  Noting that the $\int \! d
\bomega$-integral in \Eq{eq:finalSigmaI} evidently diverges for large
$\bomega$, GZ cut off this divergence at $ 1/\tauel$ (arguing that the
diffusive approximation only holds for time-scales longer than
$\tauel$, the elastic scattering time). For example, for
quasi-1-dimensional wires, for which $\int (d \bbmq) = a^{-2} \int
dq/(2 \pi)$ can be used ($a^2$ being the cross section, so that
$\sigma_1 = a^2 \sigma^\Drude_\DC$ is the conductivity per unit
length, with $\sigma^\Drude_\DC = 2 e^2 \nud \Dd$), they obtain (cf.\
(76) of GZ99\cite{GZ2}):
\begin{eqnarray}
{1 \over  \tau_\varphi^\GZ } \simeq  {e^2 \sqrt{2\Dd} \over \hbar \sigma_1}
\int_{1 \over \tau^\GZ_\varphi}^{1 \over \tauel}
{ (d \bomega) \over \omega^{1/2} } \coth \left[ \hbar \bomega
  \over 2 T \right] \; \simeq \;
{e^2 \over \pi \hbar \sigma_1} \sqrt {2 D \over \tauel}
\left[{2 T \sqrt {\tauel \tau_\varphi^\GZ } \over \hbar}
 + 1 \right] \; .
% \hspace{-1cm} \phantom{.} 
%\nonumber
%\\
  \qqph \label{eq:GZwrongresult}
\end{eqnarray}
[The use of a self-consistently-determined lower frequency cut-off is
explained in Sec.~\ref{sec:vertex}]. Thus, they obtained a
temperature-independent contribution $\gamma_\varphi^{0,\GZ}$ from the
+1 term, which is the result that ingited the controversy.

However, we thirdly observe that, due to the special form of the
retarded interaction propagator in the unitary limit, the real parts
of the last factors in square brackets of \Eqs{eq:finalSigmaR} and
(\ref{eq:finalSigmaI}) are actually \emph{equal} (for $\gammaH = 0$).
Thus, the ultraviolet divergence of $\gpI$ is \emph{cancelled} by a
similar divergence of $\gamma_\varphi^{R,\self}$.  Consequently, the
total decoherence rate coming from self-energy terms,
$\gamma_\varphi^\self = \gamma_\varphi^{I,\self} +
\gamma_\varphi^{R,\self}$, is free of ultraviolet divergencies.  Thus
we conclude that the contribution $\gamma_\varphi^{0,\GZ}$ found by GZ
is an artefact of their neglect of recoil, as is their claimed
``decoherence at zero temperature''.

\section{Dyson Equation and Cooperon Self Energy}
\label{sec:DysonCooperonSelfenergy}

The above results for $\gpR + \gpI$ turn out to agree completely with
those of a standard calculation of the Cooperon self energy
$\tSigma$ using
diagrammatic impurity averaging [details of which are summarized in
Appendix~F]. We shall now summarize
how this comes about.

Calculating $ \tSigma$ is an elementary excercise within diagrammatic
perturbation theory, first performed by Fukuyama and 
Abrahams\cite{FukuyamaAbrahams83}.
However, to facilitate comparison with the influence functional
results derived above, we proceed differently: We have derived
[Sec.~B.6.1] a general expression\cite{erratum}, before impurity
averaging, for the Cooperon self-energy of the form $\tSigma =
\sum_{aa'} \left[ \tSigma^{I}_{aa'} + \tSigma^{R}_{aa'} \right]$,
which keeps track of which terms originate from $i \tilde S_R$ or
$\tilde S_I$, and which contours $a,a'=F/B$ the vertices sit on.  This
expression agrees, as expected, with that of Keldysh perturbation
theory, before disorder averaging; it is given by
\Eq{eq:selfenergies-explicit-main} and illustrated by
Fig.~\ref{fig:cooperonselfenergy} in App.~A.  We then disorder average
using standard diagrammatic techniques. For reference purposes, some
details of this straightforward excercise are collected in
Appendix~F.2.

For present purposes, we shall consider only the ``self-energy
contributions'' ($a=a'$) to the Cooperon self energy, and neglect the
``vertex contributions'' ($a \neq a'$), since above we likewise
extracted $\gamma_\varphi^{R/I}$ from the self-energy contributions to
the effective action, $\langle \tilde S_{R/I} \rangle^{\leading,
  \self}_\rw $.  After impurity averaging, the Cooperon then
satisfies a Dyson equation of standard form,
%[Fig.~\ref{fig:Hikamiboxes}(a)],
$ \bcCqwself  =  \bcCqw + \bcCqw  \, \bSigma_\bmq^\self (\omega) \,
  \bcCqwself$,
with standard solution:
%\bsubequations
%\label{CooperonDyson}
\begin{eqnarray}
\label{CooperonDysona}
  \bcCqwself & = &
\label{CooperonDysonb}
{1 \over E_\bmq - i \omega - \bSigma_\bmq^\self (\omega) }
\; ,
 \end{eqnarray}
where $\overline \Sigma^{R/I,\self} = \sum_{a} \overline \Sigma^{R/I,
  \self}_{aa}$, with  $\overline \Sigma^{ R/I,\self }_{\bmq,FF}
(\omega) = \left[\overline \Sigma^{ R/I,\self }_{\bmq,BB} (-\omega)
\right]^\ast$, and
\bsubequations
\label{subeq:selfenergySELF}
\begin{eqnarray}
%\nonumber %\label{eq:selfenergy-selfI}
\overline  \Sigma^{ I,\self }_{\bmq, FF} (\omega) \!\!
 & \equiv &   - {1 \over \hbar} \int (d \bomega) (d \bbmq)
\coth \left[{\hbar \bomega \over 2T} \right]
{\rm Im} \bigl[ \bLRbbqw \bigr] \,
 \bcC^0_{\bmq - \bbmq}( \omega - \bomega) \; , \qqph
\end{eqnarray}
\begin{eqnarray}
  \label{eq:selfenergy-selfR}
\overline   \Sigma^{R,\self }_{\bmq,FF} (\omega) \!\!
 & \equiv & \!\!
{ 1 \over \hbar } \int (d \bomega) (d \bbmq)
\Biggl\{
\tanh \Bigl[{\hbar (\varepsilon + \toh \omega - \bomega) \over 2T}\Bigr] \,
\toh i  \bLRbbqw
\qqph \label{subeq:selfenergygorydetail}
\\
& & 
\nonumber \phantom{.} \hspace{-0.5cm}
\times
 \left[ \bcC^0_{\bmq - \bbmq} ( \omega - \bomega) +
\bigl[ \bcD^0_\bbmq(\bomega) \bigr]^2
\Bigl(
\bigl[ \bcC^0_{\bmq} (\omega ) \bigr]^{-1}
+ \bigl[ \bcD^0_{\bbmq} (\bomega) \bigr]^{-1}
\Bigr)
\right] \Biggr\} .
\end{eqnarray}
\esubequations
In \Eq{subeq:selfenergygorydetail}, the terms proportional to $\bigl(
\bcD^0 \bigr)^2 \bigl[ \bigl(\bcC^0 \bigr)^{-1} + \bigl(\bcD^0
\bigr)^{-1} \bigr]$ stem from the so-called Hikami contributions, for
which an electron line changes from $\G^{R/A}$ to $\G^{A/R}$ to
$\G^{R/A}$ at the two interaction vertices. As correctly emphasized by
AAG\cite{AAG98} and AAV\cite{AAV01}, such terms are missed by GZ's
approach of averaging only over time-reversed pairs of paths, since
they stem from paths that are not time-reversed pairs.

Now, the standard way to define a decoherence rate for a Cooperon of
the form (\ref{CooperonDysonb}) is as the ``mass'' term that survives
in the denominator when $\omega = E_\bmq = 0$, \ie\
$\gamma_\varphi^\self = - \bSigma^\self_{\bm{0}} (0) = - 2 \textrm {Re}
\left[ \bSigma^{I+R, \self}_{{\bm{0}}, FF} (0)\right]$.
%= \gamma^{I,\self}_\varphi + \gamma^{R,\self}_\varphi $.
In this limit the contribution of the Hikami terms vanishes
identically, as is easily seen by using the last of
\Eqs{eq:recallLCD}, and noting that ${\rm Re} [ i (\bcD^0)^{-1}
(\bcD^0)^2 (\bcD^0)^{-1} ] = {\rm Re} [i] = 0$.  (The realization
of this fact came to us as a surprise, since AAG and AAV had argued
that GZ's main mistake was their neglect of Hikami 
terms\cite{AAG98,AAV01}, thereby implying that the contribution of these
terms is not zero, but essential.) The remaining (non-Hikami) terms of
\Eq{subeq:selfenergygorydetail} agree with the result for $\tilde
\Sigma$ of AAV\cite{AAV01} and reproduce \Eqs{eq:finalselfenergy}
given above, in other words:
\begin{eqnarray}
  \label{eq:Dyson=<S>}
\gamma^\self_\varphi = [ - \bSigma^\self_{\bm{0}} (0)]  =
  {1 \over \tau \,\hbar} \langle i
\tilde S_R +  \tilde S_I \rangle^{\leading, \self}_\rw  \; .
\end{eqnarray}
Thus, the Cooperon mass term $- \bSigma^\self_{\bm{0}} (0)$ agrees
identically with the coefficient of $\tau$ in the leading terms of the
averaged effective action of the influence functional.  This is
no coincidence: it simply reflects the fact that averaging in
the exponent amounts to reexponentiating the \emph{average of the
  first order term} of an expansion of the exponential, while in
calculating the self energy one of course \emph{also} averages the
first order term of the Dyson equation.  It is noteworthy, though, that
for the problem at hand, where the unitary limit of 
the interaction propagator is considered, 
 it suffices to perform this average
exclusively over pairs of time-reversed paths --- more complicated
paths are evidently not needed, in contrast to the expectations voiced
by AAG\cite{AAG98} and AAV\cite{AAV01}.

The latter expectations do apply, however, if one consideres forms of
the interaction propagator $ \bLRbbqw$ more general than the unitary
limit of (\ref{eq:recallLCD}) (\ie\ not proportional to $\bigl[
\bcD^0_\bbmq (\bomega)]^{-1})$.  Then, the Hikami contribution to
$\gamma_\varphi^\self = - \bSigma^\self_{\bm{0}} (0)$ indeed does not
vanish; instead, by noting that for $\omega = \bmq = \gammaH = 0$ the
second line of \Eq{subeq:selfenergygorydetail} can always be written
as $2 {\rm Re} \bigl[ \bcD^0_{\bbmq} (\bomega)\bigr]$, we obtain
\begin{eqnarray}
\nonumber
\gamma_\varphi^\self
 & = &    {1 \over \hbar} \int (d \bomega) (d \bbmq)
\left\{ \coth \Bigl[{\hbar \bomega \over 2T} \Bigr]
 + \tanh \Bigl[{\hbar (\varepsilon  - \bomega) \over
 2T}\Bigr] \right\} 
\\   \label{eq:gammaphigeneral}
 & & \qquad \times {\rm Im} \bigl[ \bLRbbqw \bigr] \,
{2   E^0_\bbmq \over 
(E^0_\bbmq)^2 + \bomega^2} \; ,
\end{eqnarray}
which is the form given by AAV\cite{AAV01}.

\section{Vertex Contributions}
\label{sec:vertex}

\Eq{eq:finalSigmaI} for $\gamma^{I,\self}_\varphi$ has the deficiency
that its frequency integral is \emph{infrared} divergent (for $\bomega
\to 0$) for the quasi-1 and 2-dimensional cases, as becomes explicit
once its $\bbmq$-integral has been performed [as in
\Eq{eq:GZwrongresult}].  This problem is often  dealt with by arguing
that small-frequency environmental fluctuations that are slower than
the typical time scale of the diffusive trajectories are, from the
point of view of the diffusing electron, indistuingishable from a
static field and hence cannot contribute to decoherence. Thus, a
low-frequency cutoff $\gamma_\varphi$ is inserted by hand into
\Eqs{eq:finalselfenergy} [\ie\ $\int_0 d \bar \omega \to
\int_{\gammaphi} d \bar \omega$], and $\gamma_\varphi$
determined selfconsistently. This procedure was motivated in quite
some detail by AAG\cite{AAG98}, and also adopted by GZ in
GZ99\cite{GZ2} [see \Eq{eq:GZwrongresult} above]. However, as
emphasized by GZ in a subsequent paper, GZ00\cite{GZ3}, it has the serious
drawback that it does not necessarily reproduce the correct functional
form for the Cooperon in the time domain; \eg, in $\bard = 1$
dimensions, the Cooperon is known\cite{AAK82} to decay as $e^{-a
  (\tau/\tauphi)^{3/2}}$, \ie\ with a nontrivial power in the
exponent, whereas a ``Cooperon mass'' would simply give
$e^{-\tau /\tauphi}$.  

A cheap fix for this problem would be to take the above idea of a
self-consistent infrared cutoff one step further, arguing that
the Cooperon will decay as $e^{- \tau \gamma_\varphi^\self
(\tau)}$, where $\gamma_\varphi^\self (\tau)$ is  a
\emph{time-dependent} decoherence rate, whose time-dependence
enters via a time-dependent infrared cutoff. Concretely, using
\Eqs{subeq:selfenergySELF} and (\ref{eq:finalselfenergy}), one would write
\begin{eqnarray}
\nonumber \gamma_\varphi^\self (\tau)  & = & 2 \int_{1/
\tau}^\infty (d \bomega ) \, \bomega \left\{ \coth \left[ {\hbar
\bomega \over 2T}\right]
 +  \toh \sum_{s = \pm} s \tanh \left[{\hbar (\ve - s \bomega) \over 2T} \right]
% - \tanh \left[{\hbar (\ve + \bomega) \over 2T} \right] \right)
\right\}
 \\
\label{eq:cheapfix}
& & \times \int {(d \bbmq)  \over \hbar  \nu
} { 1 \over  (\Dd \bbmq^2)^2 +  \bomega^2 } \; . \qqph
\end{eqnarray}
It is straightforward to check [using steps analogous to those
used below to obtain \Eq{eq:F1dexplicitfinal}]  that in $\bard =
1$ dimensions, the leading long-time dependence is
$\gamma_\varphi^\self (\tau) \propto \tau^{1/2}$, so that this
cheap fix does indeed produce the desired $e^{- a (\tau /
\tauphi)^{3/2}}$ behavior.

The merits of this admittedly rather ad hoc cheap fix can be
checked by doing a better calculation: It is well-known that the
proper way to cure the infrared problems is to include ``vertex
contributions'', having interactions vertices on opposite
contours.  In fact, the original calculation of AAK\cite{AAK82}
in effect did just that.  Likewise, although GZ neglected vertex
contributions in GZ99\cite{GZ2}, they subsequently included them
in GZ00\cite{GZ3}, exploiting the fact that in the influence functional
approach this is as straightforward as calculating the
self-energy terms: one simply has to include the contributions to
$\langle i\tilde S_R / \tilde S_I \rangle_\rw $ of the vertex
function $- f^\vertex$ in \Eq{eq:averageSRIcl}, too. The leading
contribution comes from the first term in
\Eq{subeq:averageSRIclvert}, to be called $\langle i \tilde S_R/
\tilde S_I \rangle^{\leading, \vertex}_\rw $, which gives a
contribution identical to $\langle i \tilde S_R/ \tilde S_I
\rangle^{\leading, \self}_\rw $, but multiplied by an extra
factor of $- {\sin (\bomega \tau) \over \bomega \tau}$ under the
integral. Thus, if we collect all contributions to
\Eq{eq:averageSRIcl} that have been termed ``leading'', our final
result for the averaged effective action is ${
  1 \over \hbar} \langle i \tilde S_R + \tilde S_I
\rangle^{\leading}_\rw \equiv F_\bard (\tau) $, with
\begin{eqnarray}
\nonumber
F_\bard (\tau)  & = & \tau
\int (d \bomega ) \, \bomega \left\{
\coth \left[ {\hbar \bomega \over 2T}\right]
 +  \tanh \left[{\hbar (\ve - \bomega) \over 2T} \right] \right\}
\left(1 -  {\sin (\bomega \tau) \over \bomega \tau}
\right) \\
  \label{eq:finalgammaphi}
& & \times \int {(d \bbmq)  \over \hbar  \nu }
{ 1 \over  (\Dd  \bbmq^2)^2 +  \bomega^2 }
\; . \qqph
\end{eqnarray}
This is our main result: an expression for the decoherence function
$F_\bard (\tau)$ that is both ultraviolet and infrared convergent (as
will be checked below), due to the $(\coth + \tanh)$ and $(1 -
\sin)$-combinations, respectively.  Comparing this to
\Eqs{eq:cheapfix}, we note that $F_\bard (\tau)$ has precisely the
same form as $\tau \gamma_\varphi^\self (\tau)$, except that the
infrared cutoff now occurs in the $\int (d \bomega)$ integrals through
the $(1- \sin)$ combination. Thus, the result of including vertex
contributions fully confirms the validity of using the cheap fix
replacement $\int_0 (d \bomega) \to \int_{1/ \tau }(d \bomega)$, the
only difference being that the cutoff function is smooth instead of
sharp (which will somewhat change the numerical prefactor of
$\tauphi$).

It turns out to be possible to also obtain \Eq{eq:finalgammaphi} [and
in addition \emph{all} the ``subleading'' terms of
\Eq{eq:averageSRIcl}] by purely diagrammatic means: to this end, one
has to set up and solve a Bethe-Salpeter equation. This is a
Dyson-type equation, but with interaction lines transferring energies
between the upper and lower contours, so that a more general Cooperon
$\bcC^\ve_\bmq (\Omega_1, \Omega_2)$, with three frequency variables,
is needed.  Such an analysis will be published 
in DMSA-II\cite{MarquardtAmbegaokar04}.

To wrap up our rederivation of standard results, let us perform the
integrals in \Eq{eq:finalgammaphi} for $F_\bard (\tau)$ for the
quasi-1-dimensional case $\bard =1$.  The $\int (d \bbmq)$-integral
yields $\bomega^{-3/2} \sqrt {\Dd/2} / (\sigma_1 \hbar/e^2)$.  To do the
frequency integral, we note that since the $(\coth +
\tanh)$-combination constrains the relevant frequencies to be $|\hbar
\bomega| \lesssim T$, the integral is dominated by the small-frequency
limit of the integrand, in which $ \coth (\hbar \bomega / 2T) \simeq
2T/\hbar \bomega$, whereas $\tanh$, making a subleading contribution,
can be neglected.  The frequency integral then readily yields
\begin{eqnarray}
  \label{eq:F1dexplicitfinal}
  F_1(\tau) & = &
{4 \over 3 \sqrt \pi}
{T \tau / \hbar  \over
 g_1 (\sqrt{\Dd \tau}) }
\equiv  {4 \over 3 \sqrt \pi}
\left (\tau \over \tau_\varphi \right)^{3/2} \; ,
\end{eqnarray}
so that we correctly obtain the known $e^{-a (\tau/\tauphi)^{3/2}}$
decay for the Cooperon.  Here $g_\bard (L) = (\hbar /e^2) \sigma_\bard
L^{\bard-2}$ represents the dimensionless conductance, which is $\gg
1$ for good conductors.  The second equality in
\Eq{eq:F1dexplicitfinal} defines $\tauphi$, where we have exploited
the fact that the dependence of $ F_1$ on $\tau$ is a simple
$\tau^{3/2}$ power law, which we made dimensionless by introducing the
decoherence time $\tau_\varphi$. [Following AAG\cite{AAG98}, we
purposefully arranged numerical prefactors such that none occur in the
final \Eq{eq:definetauphig} for $\tauphi$ below.]  Setting $\tau =
\tau_\varphi$ in \Eq{eq:F1dexplicitfinal} we obtain the
self-consistency relation and solution (cf.\ Eq.~(2.38a) of AAG\cite{AAG98}):
\begin{eqnarray}
  \label{eq:definetauphig}
  {1 \over \tauphi } =  {T / \hbar \over g_\bard (\sqrt{\Dd
  \tauphi}) } \;  , \qquad \Rightarrow \qquad
\tauphi = \left( {\hbar^2 \sigma_1 \over T e^2 \sqrt \Dd } \right)^{2/3}
  \; .
\end{eqnarray}
The second relation is the celebrated result of AAK,
which diverges for $T \to 0$.
This completes our recalculation of $\gamma_\varphi^\AAK$ using
GZ's influence functional approach.

\Eq{eq:F1dexplicitfinal} can be used to calculate the
magnetoconductance for $\bard = 1$ via
\begin{eqnarray}
\label{eq:sigma(H)}
  \sigma_\DC^\WL (H) = - {\sigma_\DC^\Drude \over \pi \nu \hbar}
\int_0^\infty d \tau \, \tilde C^0_{\bmr = 0} (\tau) \, e^{-F_1 (\tau)} \; .
\end{eqnarray}
(Here, of course, we have to use $\gammaH \neq0 $ in $\tilde C^0_{\bmr
  = 0} (\tau)$.  Comparing the result to AAK's result for the
magnetoconductance (featuring an ${\rm Ai'}$ function for $\bard =
1$), one finds qualitatively correct behavior, but deviations of up to
20\% for small magnetic fields $H$.  The reason is that our
calculation was not sufficiently accurate to obtain the correct
numerical prefactor in \Eq{eq:F1dexplicitfinal}.  [GZ did not attempt
to calculate it accurately, either].  It turns 
out (see MDSA-I\cite{MarquardtAmbegaokar04}) 
that if the averaging over random walks
of \Eq{eq:impurity-average-of-eikr-A} is done more accurately,
following Marquardt's suggestion of ensuring that the random walks are
\emph{self-returning}, the prefactor changes in such a way that the
magnetoconductance agrees with that of AAK to within an error of at
most 4\%. Another improvement that occurs for this more accurate
calculation is that the results are well-behaved also for finite
$\gammaH$, which is not the case for our present \Eq{eq:finalSigmaR}:
for $\gammaH \neq 0$, the real part of the square brackets contains a
term proportional to $\gammaH / E_\bbmq^0$, which contains an infrared
divergence as $\bbmq \to 0$. This problem disappears if
the averaging over paths is performed more 
accurately, see MDSA-I\cite{MarquardtAmbegaokar04}.

\section{Discussion and Summary}

We have shown [in Apps.~B to D, as summarized in App.~A]
that GZ's influence functional approach to interacting fermions is
sound in principle, and that standard results from Keldysh
diagrammatic perturbation theory can be extracted from it, such as the
Feynman rules, the first order terms of a perturbation expansion in
the interaction, and the Cooperon self energy.

Having established the equivalence between the two aproaches in
general terms, we were able to identify precisely why GZ's treatment
of the Pauli factor $\Pauli$ occuring $ \tilde S_R$ was problematic:
representing it in the time domain as $\tanh[\tilde h_0 (t)/2T]$, they
assumed it not to change during diffusive motion along time-reversed
paths. However, they thereby neglected the physics of recoil, \ie\
energy changes of the diffusing electrons by emission or absorption of
photons. As a result, GZ's calculation yielded the
result % $\gamma_\varphi^{R,\GZ} =
$\langle i \tilde S_R^\GZ \rangle_\rw
= 0$.  The ultraviolet divergence in %$\gamma_\varphi^I =
$\langle \tilde S_I^\GZ \rangle_\rw $, which in diagrammatic
approaches is cancelled by terms involving a $\tanh$ function, was
thus left uncancelled, and instead was cut off at $\bomega \simeq 1 /
\tauel$, leading to the conclusion that $\gamma_\varphi^\GZ (T \to 0)$
is finite.

In this review, we have shown that the physics of recoil can be
included very simply by passing from the time to the frequency
representation, in which $\Pauli$ is represented by $\tanh [ \hbar
(\ve - \bomega)/2T]$.  Then $ \langle i \tilde S_R \rangle_\rw$ is
found \emph{not} to equal to zero; instead, it cancels the ultraviolet
divergence of $\langle \tilde S_I \rangle_\rw$, so that the total rate
$\gammaphi = \gamma_\varphi^{I} + \gamma_\varphi^{R}$ reproduces the
classical result $\gamma_\varphi^\AAK$, which goes to zero for $T \to
0$.  Interestingly, to obtain this result it was sufficient to average
only over pairs of time-reversed paths; more complicated paths, such
as represented by Hikami terms, are evidently not needed.  (However,
this simplification is somewhat fortuitous, since it occurs only when
considering the unitary limit of the interaction propagator; for more
general forms of the latter, the contribution of Hikami terms
\emph{is} essential, as emphasized by AAG and AAV\cite{AAG98,AAV01}.)

The fact that the standard result for $\gammaphi$ \emph{can} be
reproduced from the influence functional approach is satisfying, since
this approach is appealingly clear and simple, not only conceptually,
but also for calculating $\gammaphi$.  Indeed, once the form of the
influence functional (\ref{eq:SIR-LIR-aa-main}) has been properly
derived (wherein lies the hard work), the calculation of $\langle i
\tilde S_R + \tilde S_I \rangle_\rw$ requires little more than
knowledge of the distribution function for a random walk and can be
presented in just a few lines [Sec.\ref{sec:GZ-classical-paths}];
indeed, the algebra needed for the key steps [evaluating
\Eq{eq:averageSRIcl} to get the first terms of
(\ref{subeq:averageSRIcl}), then finding (\ref{eq:finalselfenergy})
and (\ref{eq:finalgammaphi})] involves just a couple of pages.

We expect that the approach should be similarly useful for the
calculation of other physical quantities governed by the long-time,
low-frequency behavior of the Cooperon, provided that one can
establish unambiguously that it suffices to include the contributions
of time-reversed paths only --- because Hikami-like terms, though
derivable from the influence functional approach too, can not easily
be evaluated in it; for the latter task, diagrammatic impurity
averaging still seems to be the only reliable tool.

\section*{Acknowledgements}

I dedicate this
  review to Vinay Ambegaokar on the occasion of his 70th birthday. He
  raised and sustained my interested in the present subject by telling
  me in 1998: ``I believe GZ have a problem with detailed balance'',
  which turned out to be right on the mark, in that recoil and
  detailed balance go hand in hand.  I thank D.  Golubev and A.
  Zaikin, and, in equal measure, I. Aleiner, B.  Altshuler, M.
  Vavilov, I. Gornyi, R. Smith and F. Marquardt, for countless patient
  and constructive discussions, which taught me many details and
  subtleties of the influence functional and diagrammatic approaches,
  and without which I would never have been able to reach the
  conclusions presented above.  I also acknowledge illuminating
  discussions with J.  Imry, P.  Kopietz, J. Kroha, A.  Mirlin, H.
  Monien, A. Rosch, I.  Smolyarenko, G. Sch\"on, P.  W\"olfle and A.
  Zawadowski. Finally, I acknowledge the hospitality of the centers
  for theoretical physics in Trieste, Santa Barbara, Aspen, 
  Dresden and the Newton Institute in Cambridge, where some of this
  work was performed. This research was supported in part by
 SFB631 of the DFG, and by the National Science Foundation under
Grant No. PHY99-07949.

%\appendix

%\appeqn

\appendix{Outline of GZ's Influence Functional Approach}

Without dwelling on details of derivations, we outline in this
appendix how the influence functional presented in
Sec.~\ref{sec:mainresults} is derived. (A similar summary is contained
in a previous paper by this 
author\cite{vonDelftJapan02,erratum}; however, it is
incomplete, in that we have introduced important improvements since.)
Before we start, let us point out the two main differences between our
formulation and that of GZ:

\noindent (i)
GZ  formulated the Cooperon propagator in terms
of a coordinate-momentum path integral $\int{\cal D} \! \bmR \int \! {\cal
  D} \bmP$, in which $(\tilde \delta - 2 \tilde
\rho^0) $ is represented as $[1 - 2 n_0 (\tilde h_0)] = \tanh (\tilde
h_0/2 T)$, where the free-electron energy
$\tilde h_0 \bigl[\bmR (t_a), \bmP(t_a)\bigr]$ depends on position and
momentum. This formulation makes it difficult to
treat the Pauli factor with sufficient accuracy
to include recoil. In contrast, we achieve
the latter by using a  coordinates-only version $\int \!
\widetilde {\cal D}' \bmR$, in which exact relations
between noninteracting Green's functions make
an accurate treatment of  the Pauli factor possible,
upon Fourier-transforming the effective action to the
frequency domain.

\noindent (ii)
GZ effectively performed thermal weighting at an initial time $t_0$
that is not sent to $- \infty$, but (in the notation of the main text)
is set to $t_0 = - \tau/2$; with the latter choice, it is impossible
to correctly reproduce the first (or higher) order terms of a
perturbation expansion. GZ's claim in GZ00\cite{GZ3} that they have
reproduced these is incorrect (see end of App.~C.3),
 since their time integrals have $-\tau/2$ as the
lower limit, whereas in the Keldysh approach
they run from $- \infty$ to $+ \infty$.
We have found that with some (but not much) extra effort it \emph{is}
possible to properly take the limit $t_0 \to - \infty$, to correctly
recover the first order perturbation terms [App.~C.3] and to express
the conductivity in a form containing thermal weighting in the energy
domain explicitly in the form of a factor $\int (d \ve) [-n_0^\prime
(\hbar \ve)] \tilde P^\ve$, where $\tilde P^\ve$ is an
energy-dependent path integral, obtained by suitable Fourier
transformation [App.~C.4].

\subappendix{Outline of Derivation of Influence Functional}
\label{app:GZoutline}

Consider a disordered system of interacting fermions,
 with Hamiltonian $\hat H = \hat H_0 + \hat H_\irm$:
\bsubequations
\begin{eqnarray}
  \hat H_0 & =  \int d x \,
\hat \psi^\dag (x)  h_0 (x) \hat \psi (x) \; ,
\\ %  \label{eq:defHint} \nonumber
\hat H_\irm  & =  {e^2 \over 2} \!\!
 \int \!\!   d x_1 \, d x_2 \,
: \! \hat \psi^\dag (x_1) \hat \psi (x_1) \! :
\tilde V^\inter_{12}
: \! \hat \psi^\dag (x_2) \hat \psi (x_2) \! :
\qquad \phantom{.}
%\\
%\hat n^\pdag_{\iiijjj } & =
%\hat \psi^\dag (x_j) \hat \psi (x_i) \; ,
%\\ \label{eq:defrho0}
% \langle \hat O \rangle_0  & = \Tr \{ \hat O \, \hat \rho_0 \}  ,
%\quad  \hat \rho_0 =  e^{- \beta \hat H_0} /
%   \{ \Tr e^{- \beta \hat H_0} \}  . \qquad \phantom{.}
\end{eqnarray}
\esubequations
Here $\int dx = \sum_\sigma \int d \bmr$,  and
$\hat \psi
(x) \equiv \hat \psi (\bmr, \sigma) $ is the
electron field operator for creating a spin-$\sigma$ electron at position
$\bmr$, with the following expansion in terms of the exact
eigenfunctions $\psi_\lambda(x)$ of
$h_0 (x) = \frac{- \hbar^2 }{2 m} \bnabla_{\bmr}^2 +
V^\pdag_\imp (\bmr) - \mu$:
\begin{eqnarray}
\label{eq:definepsifieldsmain}
  \hat \psi (x)  =  \sum_\lambda \psi_\lambda (x) \hat c_{\lambda} ,
  \quad  \mbox{[} h_0 (x) - \xi_\lambda \mbox{]}
\psi_\lambda (x) = 0.
\end{eqnarray}
The interaction potential $\tilde V^\inter_{12} = \tilde V^\inter
(|\bmr_1 - \bmr_2|)$ acts between the normal-ordered densities at
$\bmr_1$ and $\bmr_2$. The Kubo formula for the DC conductivity of a
$d$-dimensional conductor gives
%in either of  the forms
\bsubequations
%\label{eq:sigmaDC-final}
\begin{eqnarray}
%\nonumber
\sigma_{\rm DC} & = & -  \mbox{Re} \left[   \lim_{\omega_0 \to 0}
{1 \over d \omega_0} \sum_{\sigma_1}
\int \! dx_2 \,
    \bmj_{11'} \cdot   \bmj_{22'}
    \tilde {J}_{11',22} (\omega_0)\Big|_{x_1 = x_{1'}}
\right] , \,
\\
%\end{eqnarray}
%\begin{eqnarray}
\label{eq:defineGomega0}
\tilde {   J}_{11',22'} (\omega_0) &= &
 \! \int_{- \infty}^\infty
\!\!  dt_{12} e^{i \omega_0 t_{12}}
 \theta (t_{12}) \, \tilde {   C}_{[11',22']} \; ,
\\
%\end{eqnarray}
%\begin{eqnarray}
%\nonumber %\label{eq:define-C-firsttime-a}
\label{eq:define-C-firsttime-aAA}
\tilde {\cal C}_{[11',22']}  & \equiv &  {1 \over \hbar}
\langle [\hat \psi^\dag (t_1,x_{1'}) \hat \psi (t_1,x_{1}), \hat  \psi^\dag
(t_2,x_{2'}) \hat \psi (t_2,x_{2}) ] \rangle_H , \qquad \phantom{.}
\end{eqnarray}
\esubequations
where $ \bmj_{11'}  \equiv {- i e \hbar \over 2 m}
( \bnabla_1 - \bnabla_{1'})$ and a  uniform applied
electric field $\bm{E}(\omega_0)$ was
represented using a uniform, time-dependent vector potential,
$ \bm{E}(\omega_0) = i \omega_0 \bm{A} (\omega_0) $.
%Both the time dependence and the statistical weighting in the
%generalized density-density commutator $\tilde {   C}_{[11',22']} $
%are governed by the full Hamiltonian $\hat H$.  Note that the indices
%1 and $1'$ refer to the same time $t_1$, and likewise for 2, $2'$ and
%$t_2$.
A path integral representation for $\tilde { C}_{[11',22']} $ can be
derived using the following strategy, adapted from 
GZ99\cite{GZ2}: \label{p:Alistofsteps}
(1) introduce a source term into the Hamiltonian, in
which an artificial source field $\tilde v_{2'2}$ couples to $ \hat
\psi^\dag (t_2,x_{2'}) \hat \psi (t_2,x_{2})$, and write $\tilde {
  C}_{[11',22']} $ as the linear response to the source field $\tilde
v_{22'}$ of the single-particle density matrix $\tilde
{\bm{\rho}}_{11'} = $ $\langle \hat \psi^\dag (t_1,x_{1'}) \hat \psi
(t_1,x_{1}) \rangle_H$.  (2) Decouple the interaction using a
Hubbard-Stratonovitch transformation, thereby introducing a functional
integral $\langle \dots \rangle_V$ 
 over real scalar fields $V_{F/B}$, the so-called
``interaction fields'', defined on the forward and backward Keldysh
contours, respectively; these then constitute a dynamic, dissipative
environment with which the electrons interact.  (3) Derive an equation
of motion for $\tilde \rho_{11'}^V$, the single-particle density
matrix for a given, fixed configuration of the fields $V_{F/B}$, and
linearize it in $\tilde v_{2'2}$, to obtain an equation of motion for
the linear response $\delta \tilde \rho_{11'}^V (t)$ to the source
field.  (4) Formally integrate this equation of motion by introducing
a path integral $\int \widetilde {\cal D}' (\bmR )$ over the
coordinates  of the single degree of freedom associated
with the single-particle density matrix $\delta \tilde \rho^V_{11'}$.
(5) Use the RPA-approximation to bring the effective action $S_V$ that
governs the dynamics of the fields $V_{F/B}$ into a quadratic form.
(6) Neglect the effect of the interaction on the single-particle
density matrix whereever it occurs in the exponents occuring under the
path integral $\int \widetilde {\cal D}' \bmR $, i.e.\ replace $\tilde
\rho^V_{\iiijjj}$ there by the free single-particle density matrix
\begin{eqnarray}
\label{eq:exact-rho0-a-main}
\tilde  \rho^0_\iiijjj   =
 \langle \hat \psi^\dag (x_j) \hat \psi (x_i) \rangle_0
=  \sum_\lambda \psi^\ast_\lambda (x_j) \psi_\lambda (x_i) \, n_0
 (\xi_\lambda)
\; ,
\end{eqnarray}
where thermal averaging is performed using $\langle \hat
O\rangle_0 = \Tr e[^{- \beta \hat H_0} \hat O ] / \Tr [e^{- \beta \hat
  H_0}]$.  (7) Perform the functional integral 
$\langle \dots \rangle_V$ (which steps (5) and
(6) have rendered Gaussian) over the fields $V_{F/B}$; the environment
is thereby integrated out, and its effects on the dynamics of the
single particle are encoded in an influence functional of the form
$e^{-(i \bar S_R + \bar S_I)}$.  The final result of this strategy is
that $\bmj_{22'} \cdot \bmj_{11'} \, \tilde {\cal C}_{[11',22']}$ can
be written as [cf. (II.49)]
\begin{eqnarray}
\label{eq:intermediatej22Jrho-main}
%\nonumber
\phantom{.} \hspace{-0.8cm}
\int \!\! dx_{2} \, \bmj_{22'} \cdot \bmj_{11'}
\tilde {\cal C}_{[11',22']} &  = &
%\\ \nonumber & = &
\int  \!\! d x_{0_F,\bar 0_B} \, \tilde \rho^0_{0_F \bar  0_B}
\Fint_{0_F}^{1_F}  \!\!\!
\Bint_{\bar 0_B}^{1'_B} \!\!\!
\widetilde {\cal D}' (\bmR ) \quad
\\ \nonumber
\phantom{.} \hspace{-0.8cm}
& & \times
{1 \over \hbar} \Biggl\{
\Bigl[ \hat \bmj (t_{2_F})  - \hat \bmj (t_{2_B}) \Bigr]
 \hat \bmj (t_1)
e^{-[i \tilde S_R  + \tilde S_I]
(t_1,t_0)/\hbar}
\end{eqnarray}
% \Biggr.
%\\
%& &
%\Biggl.
where ${\displaystyle \Fint \Bint \widetilde {\cal D}' \! (\bmR )}$ is
used as a shorthand for the following forward and backward path
integral between the specified initial and final coordinates and
times:
\begin{eqnarray}
\nonumber
\Fint_{j_F}^{i_F}
\Bint_{\barj_B}^{\bari_B}
\widetilde {\cal D}' \!  (\bmR) \dots
& \equiv & \int_{\bmR^F (t^F_j)  = \bmr^F_j}^{\bmR^F (t^F_i) =
\bmr^F_i}  \widetilde {\cal D}' \!  \bmR^F (t^F_3)
\, e^{i \tilde  S_0^F (t^F_i, t^F_j)  / \hbar}
\\
& & \phantom{.}
\hspace{-0.6cm} \times  \int_{\bmR^B (t^B_j)  = \bmr^B_\barj}^{\bmR^B (t^B_i) =
\bmr^B_\bari} \widetilde {\cal D}' \!  \bmR^B(t_3^B) \,
e^{- i  \tilde S_0^B (t^B_i, t^B_j)/ \hbar} \dots \;
\qquad \phantom{.}
\end{eqnarray}
The complex weighting functional $e^{i (\tilde S_0^F - \tilde S_0^B)}$
occuring  therein involves  the  action for a single, free electron.
Expression (\ref{eq:intermediatej22Jrho-main}) has a simple interpretation:
thermal averaging with $\tilde \rho_{0 \bar 0}^0$ at time $t_0$ (for
which we take the limit $ \to
- \infty$) is followed by propagation in the presence of interactions
(described by $e^{-[i \tilde S_R + \tilde S_I]}$) from time $t_0$ up
to time $t_1$, with insertions of current vertices $\hat \bmj
(t_{2_a})$ at time $t_2$ on either the upper or lower Keldysh contour,
and $\hat \bmj (t_1)$ at the final time $t_1$.

For the purpose of calculating the Cooperon propagator, we now make
the following approximation in \Eq{eq:intermediatej22Jrho-main}
[referred to as ``approximation (ii)'' in App.~B]: For the first or
second terms, for which the current vertex occurs at time
$t_{2_{\tilde a}}$ on contour $\tilde a = F$ or $B$ respectively, we
neglect all interaction vertices that occur on the \emph{same} contour
$\tilde a$ at earlier times $t_{3_{ \tilde a}}$ or $t_{4_{\tilde a}}
\in [t_0 , t_{2_{\tilde a}}]$; however, for the opposite contour containing no
current vertex, we include interaction vertices for \emph{all} times
$\in [t_0, t_1]$, with $t_0 \to - \infty$.  [This turns out to be
essential to obtain, after Fourier transforming, the proper thermal
weighting factor $[-n_0^\prime (\hbar \ve)]$ occuring in
\Eq{eq:sigmageneraldefinePI-MAINtext-a}, see App.~C.4.]  The rationale
for this approximation is that, in diagrammatic language, this
approximation retains only those diagrams for which \emph{both}
current vertices $\bmj_{2 2'}$ and $\bmj_{1 1'}$ are always sandwiched
between a $\GR$- and a $\GA$-function; these are the ones relevant for
the Cooperon. The contributions thereby neglected correspond to the
so-called ``interaction corrections''.  [If one so chooses, they
latter \emph{can} be kept track of, though.]

This approximation (ii) is much weaker than the one used by GZ at a
similar point in their calculation: to simplify the thermal weighting
factor describing the initial distribution of electrons, namely to
obtain the explicit factor $\rho_0$ in Eq.~(49) of GZ99\cite{GZ2}, they
set $t_0 \to t_2$ (their $t'$ corresponds to our $t_2$), and thereby
perform thermal weighting at time $t_2$, instead of at $-\infty$.  As
a consequence, in their analysis all time integrals have $t_2$ as
lower limit, which means that (contrary to their claims in GZ00\cite{GZ3})
they did not correctly reproduce the Keldysh first order perturbation
expansion for $\tilde {\cal C}_{[11',22']}$, in which all time
integrals run to $- \infty$.  A detailed discussion of this matter is
given at the end of App.~C.3. [Contrary to our initial expectations,
but in agreement with those of GZ, it turns out, though, that the
choice of $t_0$ does not have any implications for the calculation of
$\tauphi$, which does not depend on whether one chooses $t_0 = t_2$ or
sends it to $- \infty$.]

Having made the above approximation (ii), the effective action $(i
\tilde S_R + \tilde S_I)$ occuring in \Eq{eq:intermediatej22Jrho-main} is
found to have the following form (we use the notation $i \tilde
S_R/ \tilde S_I$ to write two equations with similar structure in one
line, and upper or lower terms in curly brackets refer to the first or
second case):
\begin{eqnarray}
\label{eq:defineSiRA}
[i \tilde S_R/ \tilde S_I] (t_1, t_0)
& \equiv &
 \sum_{aa'} \int_{t_0}^{t_1} d t_{3} \int_{t_0}^{t_1} d t_{4}
 \,
(i \tilde L^R / \tilde L^I)_{3_a 4_{a'}}  \; ,
\qqph
\end{eqnarray}
\bsubequations
    \label{eq:LtildesA}
\begin{eqnarray}
\label{eq:LtildesFF-main}
 (i \tilde L^R / \tilde L^I)_{3_F 4_F}  & = &
-  \toh i \:\, \theta_{34} \,  \tilde \delta_{3_F \bar 3_F}
\biggl \{  \begin{array}{c}
%2 \tilde \rho^0)_{4_F \bar 4_F}
\!\! [\tilde \delta -
2 \tilde \rho^0 ]_{4_F \bar 4_F}
\rule[-2mm]{0mm}{0mm}
\\
\, \tilde \delta_{4_F \bar 4_F}
\end{array} \!\! \biggr\}
\, \tilde {\cal L}^{R/K}_{\bar 3_F \bar   4_F}
\; ,
\end{eqnarray}
\begin{eqnarray}
%\rule[-7mm]{0mm}{0mm}
%\\
\label{eq:LtildesBF-main}
 (i \tilde L^R / \tilde L^I)_{3_B 4_F}  & = &
\phantom{-}  \toh i \:  \, \theta_{34} \,
\biggl \{  \begin{array}{c}
%  (\tilde \delta - 2 \tilde \rho^0)_{4_F \bar 4_F}
\!\!
[\tilde \delta - 2 \tilde \rho^0 ]_{4_F \bar 4_F}
\rule[-2mm]{0mm}{0mm}
\\
 \tilde \delta_{4_F \bar 4_F}
\end{array} \!\! \biggr\}
\, \tilde {\cal L}^{R/K}_{\bar 3_B \bar   4_F} \,
\tilde \delta_{\bar 3_B 3_B} \; ,
\end{eqnarray}
\begin{eqnarray}
%  \rule[-7mm]{0mm}{0mm}
%\\
 \label{eq:LtildesFB-main}
(i \tilde L^R / \tilde L^I)_{3_F 4_B}  & = &
\mp \toh i \:  \, \theta_{34} \, \tilde \delta_{3_F \bar 3_F}
\, \tilde {\cal L}^{A/K}_{ \bar   4_B \bar 3_F}
\,
\biggl \{  \begin{array}{c}
%(\tilde \delta - 2 \tilde \rho^0)_{\bar 4_B 4_B}
\!\!
[\tilde \delta -  2 \tilde \rho^0 ]_{\bar 4_B 4_B}
\rule[-2mm]{0mm}{0mm}
\\
\tilde \delta_{\bar 4_B 4_B}
\end{array} \!\! \biggr\}
\; ,
\end{eqnarray}
%\rule[-7mm]{0mm}{0mm}
%\\
\begin{eqnarray}
\label{eq:LtildesBB-main}
  (i \tilde L^R / \tilde L^I)_{3_B 4_B}  & = &
\pm  \toh i \; \, \theta_{34} \,
 \tilde {\cal L}^{A/K}_{\bar 4_B \bar   3_B} \,
 \tilde \delta_{\bar 3_B 3_B} \,
\biggl \{  \begin{array}{c}
%  (\tilde \delta - 2 \tilde \rho^0)_{\bar 4_B 4_B}
\!\!
[\tilde \delta -  2 \tilde \rho^0 ]_{\bar 4_B 4_B}
\rule[-2mm]{0mm}{0mm}
\\
\tilde \delta_{\bar 4_B 4_B}
\end{array} \!\! \biggr\} \; .
\end{eqnarray}
\esubequations
Here $\tilde \delta_{\bari i} = \delta_{\sigma_\bari \sigma_i} \delta
(\bmr_\bari - \bmr_i)$ and $(\tilde {\cal L}^{R,A,K})_{\bari_a
  \barj_{a'}} = (\tilde {\cal L}^{R,A,K}) \bigl( t_{i_a}-t_{j_{a'}},
\bmr^a_{\bari} (t_{i_a})- \bmr^{a'}_\barj (t_{j_{a'}}) \bigr) $ are
the standard retarded, advanced and Keldysh interaction propagators.
For each occurrence in \Eqs{eq:LtildesA} of a pair of indices, one
without bar, one with, e.g.\ $4_{a}$ and $\bar 4_{a}$, the
corresponding coordinates $x_4^{a}$ and $x^{a}_{\bar 4}$ are
both associated with the \emph{same} time $t_4$, and integrated over,
$\int d x^{a}_4 d x^{a}_{\bar 4}$, in the path integral $\int
{\cal D}' \! (\bmR )$. (This somewhat unusual aspect of the
``coordinates-only'' path integral used in our approach is discussed
in explicit detail in App.~D.4; it is needed to account for the fact
that the density-matrix $\tilde \rho^0$ is non-local in space, and
arises upon explicitly performing the $\int {\cal D} \bmP$ momentum
path integral in GZ's formulation.) The $\tilde \delta_{\bari i}$
functions on the right hand side of \Eqs{eq:LtildesA} will kill one of
these double coordinate integrations at time $t_i$.

\Eqs{eq:defineSiRA} and (\ref{eq:LtildesA}) are the main result of our
rederivation of the influence functional approach. They are identical
in structure (including signs and prefactors) to the corresponding
expressions derived by GZ (Eqs.~(68) and (69) of GZ99\cite{GZ2}),
as can be verified by using the relations
\begin{eqnarray}
\label{eq:RIvsLRA-main}
- e^2 \tilde R_\iiijjj = \tilde {\cal
  L}^R_\iiijjj =  \tilde {\cal L}^A_{ji}, \qquad
e^2 \tilde I_\iiijjj = e^2 \tilde I_{ji} = - \toh i \tilde {\cal L}^K_\iiijjj ,
\qquad \phantom{.}
\end{eqnarray}
to relate our interaction propagators $\tilde {\cal L}_\iiijjj$ to the
functions $R_\iiijjj$ and $I_\iiijjj$ used by GZ.  However, whereas Eqs.~(68)
and (69) of GZ99\cite{GZ2} are written in a mixed coordinate-momentum
representation in which it is difficult to treat the Pauli factors
$\Pauli$ sufficiently accurately, our expressions (\ref{eq:LtildesA})
are formulated in a coordinates-only version.  Formally, the two
representations are fully equivalent. The key advantage of the latter,
though, is that passing to a coordinate-frequency representation
(which can be done \emph{before} disorder averaging, allows us to
sort out the fate of $\Pauli$, as discussed in
Sec.~A.\ref{sec:Paulifactorshort} [and extensively in App.~B.6.2].

\subappendix{Cooperon Self Energy before Disorder Averaging}

\begin{figure}[htbp]
  \centering
  \includegraphics[width=\linewidth]{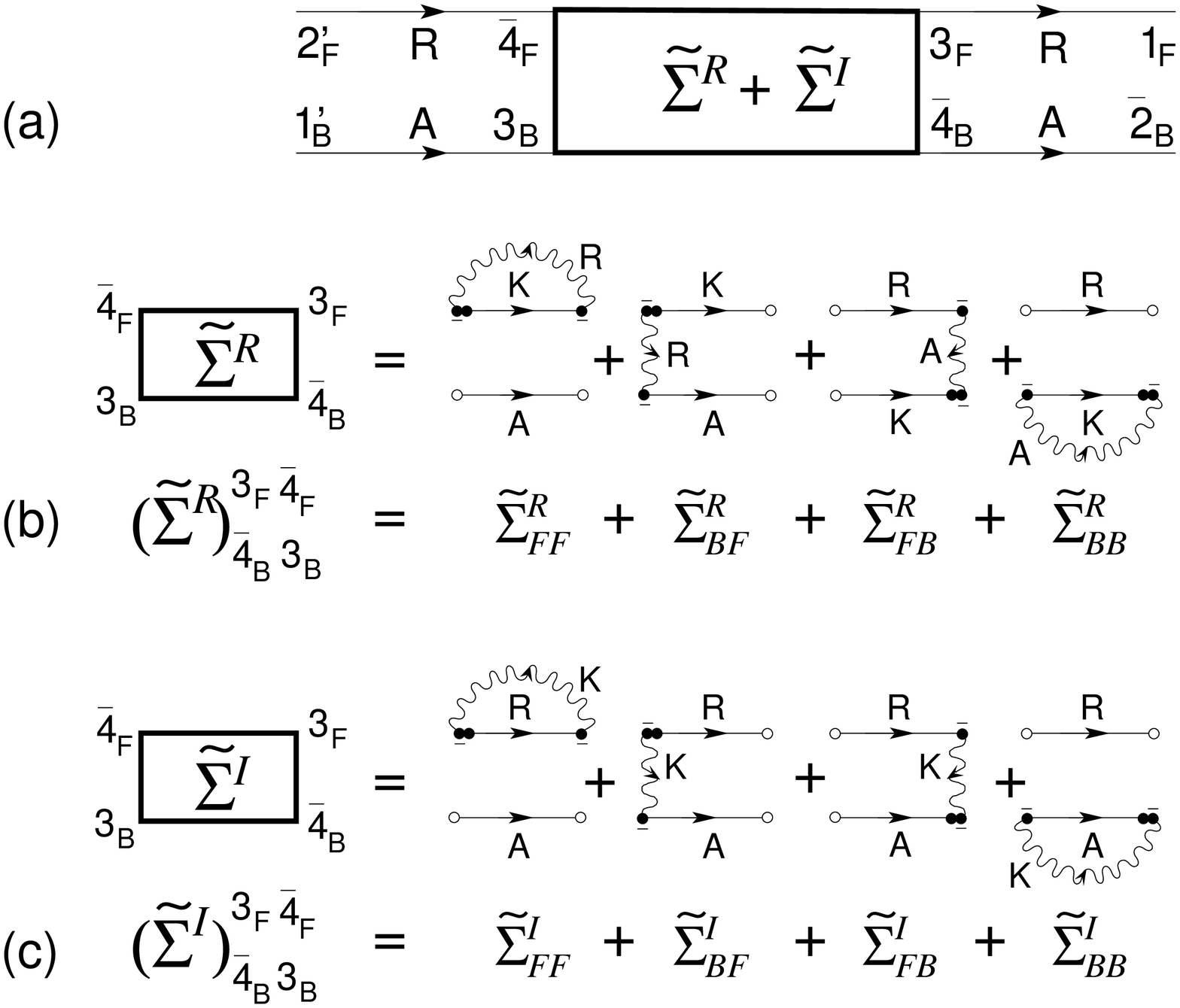}
  \caption{First order contributions to the irreducible self energy
    of the Cooperon, illustrating 
    Eqs.~(\protect\ref{eq:selfenergies-explicit-main}).
    The arrows associated with each factor $\tilde G_\iiijjj$ or $\tilde
    {\cal L}_\iiijjj$ in
    Eqs.~(\protect\ref{eq:selfenergies-explicit-main}) are drawn to
    point from the second index to the first ($j$ to $i$).  Filled
    double dots denote the occurence of a factor $(\tilde \delta - 2
    \tilde \rho)_{4_F \bar 4_F}$ on the upper contour or $(\tilde
    \delta - 2 \tilde \rho)_{\bar 4_B 4_B}$ on the lower contour. Bars
    on filled dots are used to indicate the barred indices to which
    the interaction lines depicting $\tilde {\cal L}_{\bari \barj}$ are
    connected.  Both filled and open single dots indicate a delta
    function $\tilde \delta$; the open dots stand for delta functions
    that have been inserted to exhaust dummy integrations, as
    discussed after Eqs.~(\protect\ref{eq:LtildesA}) 
    [and, in more detail, in  Sec.~6.1].
    The diagrams in (b) and (c) coincide precisely with those obtained
    by standard Keldysh diagrammatic perturbation theory for the
    Cooperon self energy, as depicted, \eg, in Fig.~2 of
    Ref.~\protect\cite{AAV01}. (There, impurity lines needed for impurity
    averaging are also depicted; in the present figure, impurity
    averaging has not yet been performed.) }
  \label{fig:cooperonselfenergy}
\end{figure}

From the formalism outlined above, it is possible to recover the
standard results of diagrammatic Keldysh perturbation theory,
\emph{before disorder averaging}, by expanding the path integral
(\ref{eq:intermediatej22Jrho-main}) in powers of the effective action
${1 \over \hbar} (i
\tilde S_R + \tilde S_I$). For example, using \Eqs{eq:LtildesA}
[and being sufficiently careful with signs, see App.~B.6.1] one
readily obtains the following expressions for Cooperon self energy
$\tilde \Sigma^{R/I} = \sum_{aa'} \tilde \Sigma^{R/I}_{a a'}$,
summarized diagrammatically in Fig.~\ref{fig:cooperonselfenergy}:
\bsubequations
  \label{eq:selfenergies-explicit-main}
  \begin{eqnarray}
\left( \tilde \Sigma^{R/I}_{FF} \right)^{3_F \bar   4_F}_{\bar 4_B 3_B} &=&
- {i \hbar \over 2} \;
%(-  \toh i \hbar ) \;
(\tilde G^{K/R})^{3_F \bar 4_F} \tilde G^A_{\bar 4_B 3_B}
 ( \tilde {\cal L}^R/ \tilde {\cal L}^K)^{3_F \bar 4_F}
\; , \quad \rule[-6mm]{0mm}{0mm}
\\
\left( \tilde \Sigma^{R/I}_{BF} \right)^{3_F \bar
 4_F}_{\bar 4_B 3_B} &=&
- {i \hbar \over 2} \;
%(-  \toh i \hbar ) \;
  (\tilde G^{K/R})^{3_F \bar 4_F} \tilde G^A_{\bar 4_B 3_B}
 ( \tilde {\cal L}^R/ \toh \tilde {\cal L}^K)_{3_B}^{\;\;\;\;  \bar 4_F}
 ,   \rule[-6mm]{0mm}{0mm}
\\
 \left( \tilde \Sigma^{R/I}_{FB} \right)^{3_F \bar
  4_F}_{\bar 4_B 3_B} &=&
- {i \hbar \over 2} \;
%(-  \toh  i \hbar ) \;
\tilde G^{R, 3_F \bar 4_F} (\tilde G^{K/A})_{\bar 4_B 3_B}
( \tilde {\cal L}^A/ \toh \tilde {\cal L}^K)_{\bar 4_B}^{\; \; \; \,  3_F}
 ,   \rule[-6mm]{0mm}{0mm}
\\
\left( \tilde \Sigma^{R/I}_{BB} \right)^{3_F \bar
  4_F}_{\bar 4_B 3_B} &=&
- {i \hbar \over 2} \;
%(-  \toh  i \hbar ) \;
\tilde G^{R, 3_F \bar 4_F}
(\tilde G^{K/A})_{\bar 4_B 3_B}
( \tilde {\cal L}^A/ \tilde {\cal L}^K)_{\bar 4_B 3_B}
 \; .  \rule[-1mm]{0mm}{0mm}
      \end{eqnarray}
\esubequations
To obtain this, we exploited the fact that every vertex occuring in
the effective action is sandwidched between retarded propagators if it
sits on the upper contour, and advanced ones on the lower contour. The
Keldysh functions arise from using some exact identities, valid
(before impurity averaging) in the coordinate-time representation:
depending on whether a vertex at time $t_{4_a'}$ sits on the forward
(time-ordered) or backward (anti-time-ordered) contour ($a' = F/B$),
the factor $(\tilde \delta - 2 \tilde \rho^0) \tilde {\cal L}^{R/A}$
occuring in $\tilde L^R_{a a'}$ is sandwidched as follows (on the left
hand sides below, a coordinate integration $\int dx_{4_{a'}}$ over the
un-barred variable at vertex 4 is implied):
\bsubequations
 \label{eq:1-2rho->tanhindicesF-main}
\begin{eqnarray}
%\nonumber
\phantom{.} \hspace{-.7cm}
\Big[\tilde G^R_{i_F 4_F}
(\tilde \delta - 2 \tilde \rho^0)_{4_F \bar 4_F} \Bigr]
\tilde {\cal L}^R_{3 \bar 4_F} \,
\G^R_{\bar 4_F j_F} \! &  \to  &  \! \phantom{-}
\G^K_{i_F \bar 4_F} (\bar \ve - \bomega) \,
\tilde {\cal L}^R_{3 \bar 4_F} (\bomega) \,
 %\tanh \Bigl[ {\hbar (\ve - \bomega) \over 2 T} \Bigr]
\G^R_{\bar 4_F j_F} (\bar \ve) ,
%\tanh[\;]
\qqph
% \hspace{-2cm} \phantom{.}
\\
%\nonumber
\phantom{.} \hspace{-.7cm}
\G^A_{\barj_B \bar 4_B} \, \tilde {\cal L}^A_{\bar 4_B 3}
\Bigl[(\tilde \delta - 2 \tilde \rho^0)_{\bar 4_B 4_B}
 \G^A_{4_B \bari_B} \Bigr]
\!  &  \to & \! -
%\tanh \Bigl[ {\hbar (\bar \ve - \bomega) \over 2 T} \Bigr]
\G^A_{\barj_B \bar 4_B} (\bar \ve) \, \tilde {\cal L}^A_{\bar 4_B 3} (\bomega)
\, \G^K_{\bar 4_B \bari_B} (\bar \ve - \bomega)
\qqph
% \hspace{-2cm} \phantom{.}\\
 % \label{eq:1-2rho->tanhindicesF}
\end{eqnarray}
\esubequations
The left- and right-hand sides are written in the time and frequency
domains, respectively.  To obtain Keldysh functions from the left-hand
side expressions, we exploit the fact that the upper or lower contours
are time- or anti-time-ordered to add an extra $- \G^{A/R} = 0$, and
then exploited \Eq{eq:Keldyshid} to obtain a factor $ \pm \G^K $ (see
Sec.~B.6.2).

\subappendix{Thermal Averaging}
\label{sec:thermalaveragingAppA}

It remains to figure out how the thermal weighting in 
\Eq{eq:sigmageneraldefinePI-MAINtext-a} can be derived from our
general path integral expression \Eq{eq:intermediatej22Jrho-main}.
This is a standard, if nontrivial, excercise in Fourier
transformation, carried out (along the lines of a similar analysis by
AAK\cite{AAK82}) in % App.~B.6.4 and 
App.~C.4. The result is an equation for the conductivity similar to
but more general than \Eq{eq:sigmageneraldefinePI-MAINtext}, with
$\int_0^\infty \!\! d \tau \, \tilde P^{1 2', \ve}_{21', \eff} (\tau)$
replaced by $\int_0^\infty \!\! d \tau_{12} \, \tilde P^{1 2',
  \ve}_{21'} (\tau_{12})$, involving a slightly more complicated path integral
[\Eq{eq:Pfixedenergy-a}], defined as
\begin{eqnarray}
\label{eq:Pfixedenergy-amaintext}
\tilde P^{1 2', \ve}_{21'} (\tau)
& \!\! = \!\! & \!\!  
\int_{-\infty}^\infty \!   d \bar \tau \, 
e^{i \bar \tau \ve } \;
\Fint_{\bmR^F (-{\tau \over 2}  - {\bar \tau \over 2})
 = \bmr_{2'}}^{\bmR^F ({\tau \over 2} ) = \bmr_1}
\Bint_{\bmR^B (-{\tau \over 2} +  {\bar \tau \over 2}) 
= \bmr_2}^{\bmR^B ({\tau \over 2}) = \bmr_{1'}}
\Bigl. \widetilde {\cal D}' (\bmR) \,
 e^{-[i \tilde S_R + \tilde S_I]/\hbar} \; .
\qqph \qph
\end{eqnarray}
{\bbf  Note that the duration of the forward and backward paths differs by a
time $\bar \tau$, in contrast to the path integral
(\ref{eq:sigmageneraldefinePI-MAINtext-b}) used in the main text.  The
combination $\int d\ve \int d \bar \tau$ of integrals from
\Eqs{eq:sigmageneraldefinePI-MAINtext-a} and
(\ref{eq:Pfixedenergy-amaintext}) have the effect of 
fixing\cite{ChakravartySchmid86} the average energy of the forward and
backward trajectories to be close to the Fermi energy, with energy
spread of roughly $\pm T$ (see App.~\ref{sec:definefullCooperon} for a
detailed discussion).  This energy $\ve$ is the same as the one that
in perturbative calculations shows up in the $\tanh [ \hbar (\ve -
\bomega)/2T]$-factors of the Keldysh electron Green's functions
$\tilde G^K (\ve - \bomega)$, which play a role in determining the
phase space available for electrons to get scattered upon absorbing or
emitting a noise quantum.  In App.~\ref{sec:definefullCooperon} we
argue that the simplest way to keep track of this in the influence
functional approach is to replace \Eq{eq:Pfixedenergy-amaintext} by
\Eq{eq:sigmageneraldefinePI-MAINtext-b}, which mimicks the effect of
the former's integral $\int d \bar \tau e^{i \ve \bar \tau}$ by using
(i) forward and backward paths of \emph{equal} duration $\tau$ and 
(ii) an effective action whose time integration boundaries are fixed at $\pm
\tau/2$, but which \emph{depends explicitly on the average propagation
  energy $\ve$} [via \Eqs{eq:SIR-LIR-aa-main},
(\ref{eq:modifiedRAtanh}), or equivalently \Eqs{eq:Seff},
(\ref{subeq:LAA'FT})].

Note that GZ's approach in effect employs the same simplification,
since they likewise have no $\int d \bar \tau e^{i \ve \bar \tau}$
integral and use forward and backward paths of equal duration $\tau$.
Their effective action depends on the average energy $\ve$, too, via
the $\tanh[\hbar \ve/2T]$-factor in their $\tilde S_R$. However,
lacking the $- \bomega$ recoil shift, their $\tanh$-terms turn out to
yield zero after averaging over random walks, so that $\langle i
\tilde S_R^\GZ \rangle_\rw \simeq 0$.}

\subappendix{Perturbative vs. Nonperturbative Methods}
\label{sec:interchangingaverages}

{\bbf We conclude this overview-style appendix with some general
  comments on whether it is sufficient to calculate $\tauphi$
  perturbatively, as we contend (in agreement with 
others\cite{AAG98,AAV01,AAV02}), or whether a truly nonperturbative
  approach is needed, as GZ have argued in GZ00\cite{GZ3}. 
  We have made an effort to keep the discussion  as nontechnical
  as possible and accessible to casual readers that have not
  studied    App.~\ref{sec:rederivation} in detail, although we will on 
  occasion refer to results from the latter.

  In GZ's influence functional approach, the decoherence time is
  defined as the scale at which the function $F(\tau) ={1 \over \hbar}
  \langle i
  \tilde S_R + \tilde S_I \rangle_\rw$, which in their theory is
  \emph{linear} in the interaction propagators $\tilde R/\tilde I$,
  becomes of order one.  This means that $\tauphi$ is the crossover
  scale between the regimes where perturbation theory is rigorously
  valid or breaks down, $F (\tau) \ll 1$ or $\gg 1$, respectively. To
  determine this scale, we contend that it is sufficient to calculate
  $F(\tau)$ \emph{perturbatively} (assuming, strictly speaking,
  $F(\tau) \ll 1$), and then to enquire for what time the perturbative
  result so obtained ceases to be small, setting $F(\tauphi) \simeq
  1$.  (This is analogous to the fact that the crossover scales $T_K$
  or $T_c$, the Kondo temperature in the Kondo problem or the critical
  temperature in the theory of superconductivity, can be calculated
  perturbatively as the scales where perturbation theory breaks down.)
  An accurate knowledge of $F(\tau)$ for $\tau \gtrsim \tauphi$ would be
  needed only if we desired to accurately include exponentially small
  ($e^{-F(\tau)} \ll 1$) contributions to weak localization, which is
  usually deemed not worth the effort.  (In contrast, for the Kondo
  problem or superconductivity, nonperturbative treatments \emph{are}
  worth the effort, because the phenomena of interest become strong in
  the nonperturbative regimes.)

  GZ have argued in GZ00\cite{GZ3} that a perturbative treatment of weak
  localization is insufficient, because according to them it fails to
  disentangle the effects of preexponent and exponent in an Ansatz for
  the Cooperon of the general form $C(\tau) = A(\tau) e^{-F(\tau)}$:
  when this is expanded in powers of the interaction, both $A$ and $F$
  contribute to the first-order term $C^{(1)}$. The influence
  functional approach avoids this problem by very naturally generating
  a general expression for the function $F$ in the exponent -- which
  in GZ's approach turns out to be \emph{linear} in the interaction
  propagator [\Eq{eq:SIR-LIR-aa-main}, or \Eq{eq:Ltildes}]. 
  However, the problem of disentangling
  the exponent from the preexponent is easily avoided in the
  diagrammatic approach, too, by calculating not the Cooperon itself,
  but its \emph{self energy}, to linear order in the interaction;
  Fourier transforming the resulting Cooperon $C(\omega)$ into the
  time domain, this automatically yields an expression of the form $A
  (\tau) e^{-F(\tau)}$, again with $F$ \emph{linear} in the
  interaction propagator. [The prefactor arises from wave-function
  renormalization effects, see DMSA-II\cite{MarquardtAmbegaokar04},
  Eq.~(14a).]  Since both the influence functional and diagrammatic
  strategies yield results for which the exponent $F$ is linear in the
  interaction (and contains contributions with a similar $\coth + \tanh$
  structure), it is reasonable to expect that if both approaches are
  implemented with sufficient care, their answers for $F$ should agree
  completely.

  They do agree, in fact, if the recoil-incorporating effective action
  proposed in this work and featuring $\tanh[ \hbar (\ve
\mp \bomega)/2T]$-factors is used.  (This agreement is demonstrated
  explicitly in DMSA-II\cite{MarquardtAmbegaokar04}.)  But they
  differ if GZ's procedure is followed without modification, 
  leading to their no-recoil $\tanh[ \hbar \ve/2T]$-factors. It is
  important and instructive, therefore, to identify at which point of
  the derivation of the influence functional approach the need for a
  modification of GZ's approach first manifests itself. We shall now
  argue that this point is reached when the order in which two
  distinct averaging procedures are performed, over paths $\bmR^a$ and
  fields $V$, is tacitly interchanged, an aspect that has not been
  emphasized in the preceding sections.

%\bsubequations
%\label{subeq:orderofaverages}
%\begin{eqnarray}
%\label{eq:pathsfirst}
%  {\rm First Principles: \; } & &
%% \bigl\langle \langle \dots \rangle_\paths \bigr\rangle_V \; ,
%\bigl\langle \langle \, \cdot \,  \rangle_\paths \bigr\rangle_V \; , 
%\\
%\label{eq:Vfirst} 
%  {\rm Influence Functional: \; } & & 
%  \bigl\langle \langle  \, \cdot \, \rangle_V \bigr \rangle_\paths \; . 
%\qqph 
%\end{eqnarray}
%\esubequations

To be concrete, let us focus on an intermediate stage of GZ's
first principles 
calculation of the weak localization contribution $\sigma_\DC^\WL$
to the conducitivity. Following the enumeration of steps used in
App.~\ref{app:GZoutline}, p.~\pageref{p:Alistofsteps}, 
this stage is reached after steps (1) to (6) [or
according to the enumeration of App.~\ref{sec:roadmap}, 
p.~\pageref{p:Blistofsteps}, after steps {\bf (A)} to {\bf (G)}],
resulting in the following expressions [the first
of which corresponds to \Eq{eq:sigmaDCreal}]:
\bsubequations
\label{subeq:Jfirstprinciples}
\begin{eqnarray}
\label{eq:sigmareal-AppA}
\sigma_{\DC,\rreal} & = &
\sum_{\sigma_1}  {1 \over d}
\int \!  dx_2 \, \bmj_{11'} \! \cdot \! \bmj_{\,22'}
\, 
\tilde {J}_{12',21'}^\prime (0)  \; , 
\\
\tilde {J}_{12',21'}^{\prime \firstprinciples} (0) & =  &
\tilde {J}_{12',21'}^{\prime \, \free} (0)  \, 
\Bigl\langle \Bigl \langle e^{i (\tilde S_V^F - \tilde S_V^B)/\hbar}
\Bigr \rangle_\cqp
\Bigr \rangle_V \; , 
  \label{eq:Jfirstprinciples}  
\\
\label{eq:SFBV}
\tilde S_V^{F/B} & = & \int_{ -{\tau \over 2} \mp {\bar \tau \over 2}}^{
  {\tau \over 2}} d t_3 \, 
\tilde h_V^{F/B} \bigl(t_3, \bmR^{F/B} (t_3) \bigr)  \; , 
\\
\label{eq:<<>_cp>_V}
\Bigl \langle \dots 
\Bigr \rangle_\cqp
& = & 
\left[ \tilde {J}_{12',21'}^{\prime \, \free} (0) \right]^{-1}
\int (d \varepsilon) [ - n' (\hbar \varepsilon )] 
\int_{-\infty}^\infty \!   d \bar \tau \, 
e^{i \bar \tau \ve } \; 
\label{eq:definecqp}
\\
& & \hspace{-2cm} \times 
\Fint_{\bmR^F (-{\tau \over 2}  - {\bar \tau \over 2})
  = \bmr_{2'}}^{\bmR^F ({\tau \over 2} ) = \bmr_1} 
%\nonumber \\
%& & 
%\hspace{1.5cm} 
%\times 
\Bint_{\bmR^B (-{\tau \over 2} +  {\bar \tau \over 2}) 
  = \bmr_2}^{\bmR^B ({\tau \over 2}) = \bmr_{1'}}
\Bigl. \widetilde {\cal D}' (\bmR) \,
e^{i \left[\tilde  S_0^F ({\tau \over 2}, -{\tau \over 2} 
 - {\bar \tau \over 2}) 
- \tilde S_0^B ({\tau \over 2}, -{\tau \over 2}  + {\bar \tau
    \over 2}) \right]/ \hbar} \dots \;
\nonumber
  \end{eqnarray}
  \esubequations The correlator $\tilde {J}_{12',21'}^{\prime
    \firstprinciples} (0) $ originates from $\tilde {\cal
    C}_{[11',22']}$ of \Eq{eq:define-C-firsttime-aAA}. It has here
  been expressed [starting from \Eqs{eq:C1122afterderivatives} to
  (\ref{eq:defJ(F/B)}), and using the results of
  \Eqs{eq:sigmageneraldefinePI}, (\ref{eq:defineJprimetimes-a}) and
  (\ref{eq:Pfixedenergy-a})] as a double average $\langle \langle
  \dots \rangle_\cqp \rangle_V$ over a pair of 
  phase factors $e^{i(\tilde S_V^F - \tilde S_V^B)}$,
  which describe the influence of interactions,
  represented by fluctuating fields
  $V_{F/B} (t_3, \bmr_3)$, on a pair of closed
quantum-mechanical paths (cqp). 
[The detailed form (\ref{eq:SFBV}) 
of the phase factors follow from
  \Eq{eq:def-path-integral-pureRmaintext}, with $\tilde h^a_V$ given
  by (\ref{eq:tildehBF}); see also \Eq{eq:shorthandtildehfunctionR},
  and the discussion thereafter].
%over $V$ [using
%\Eq{eq:V-average}] at the end. 
%These expressions instruct us to first
%pick out a 
\Eq{eq:Jfirstprinciples}   instructs us to first pick out
a specific configuration of the fields $V_{F/B} (t_3, \bmr_3)$, then 
to calculate the average $\langle \dots \rangle_\cqp$ of this
phase factor over all closed quantum-mechanical paths with
boundary conditions specified in \Eq{eq:definecqp} [as obtained from
\Eq{eq:Pfixedenergy-a}], and to evaluate the average over
all field configurations in the end.  Thus, for a given $V$, the set
of paths making the dominant contribution will depend on $V$. 
%The same is true for diagrammatic perturbation theory, which can be
%derived from \Eqs{subeq:Jfirstprinciples} by expanding the phase
%factors in powers of $V$: if, for a given diagram and a given
%field configuration one would evaluate the electron Green's 
%functions semiclassically,  the set
%of paths making the dominant contribution will depend on $V$.  
%[Of course, in practice, the diagrams ]

Now, the next step of GZ's strategy [step (7) according to
App.~\ref{app:GZoutline}, or step {\bf (H)} according to
App.~\ref{sec:roadmap}], is to perform the average $\langle \dots
\rangle_V$ over the interaction fields. To carry out this step, GZ
(tacitly) interchange the order of averages [as do we in
App.~\ref{sec:coordinatePI}], in effect replacing
\Eq{eq:Jfirstprinciples} by 
\bsubequations
  \label{subeq:<<>_V>_cp}
\begin{eqnarray}
  \label{eq:<<>_V>_cp}
  \tilde {J}_{12',21'}^{\prime \inflfunct} (0) & =  &
  \Bigl\langle 
  \Bigl \langle e^{i (\tilde S_V^F - \tilde S_V^B)/\hbar} \Bigr \rangle_V
  \Bigr \rangle_\cqp  \; 
  \stackrel{\RPA}{\simeq} \;
  \Bigl\langle e^{ -  \tilde S_\eff / \hbar} \Bigr \rangle_\cqp \; , 
  \\
\tilde S_\eff [\bmR^a] & = & 
   {1 \over 2 \hbar} \bigl \langle (\tilde S_V^F - \tilde S_V^B)
  (\tilde S_V^F - \tilde S_V^B) \bigr \rangle_V \; = \; 
  i \tilde S_R + \tilde S_I  \; . 
\end{eqnarray}
\esubequations \Eq{eq:<<>_V>_cp} instructs us to first pick out a
specific pair of paths $\bmR^{F/B}(t_3)$, and then to calculate the
influence functional $\langle e^{i (\tilde S_V^F - \tilde S_V^B)/
  \hbar} \rangle_V$ which describes how the chosen pair of paths are
effected on average by interactions. Within the RPA approximation, the
$\langle \dots \rangle_V$ average can now be done exactly, yielding an
effective action $\tilde S_\eff$ that is linear in interaction
correlators $\tilde R/\tilde I$ [\Eq{eq:Ltildes}].  The sum over all
closed quantum paths is to be performed at the end.

%When introducing the influence functional $\tilde {\cal F}$, however,
%the order of averages is opposite: According to the path integral
%expression (\ref{eq:JrhoRonly}) for $\langle \tilde J^V\rangle_V$,
%% and defining the influence
%%  functional $\tilde {\cal F}_{(t_1,t_0)} [\bmR^a ]$ of
%%  \Eq{eq:tildeFtildeB-define}, 
%we are instructed to first pick out a given pair of paths
%$\bmR^{F/B}(t_3)$, then to calculate the influence functional $\tilde
%{\cal F}_{(t_1,t_0)} [\bmR^a ]$ [\Eqs{eq:tildeFtildeB-define}] which
%describe how the chosen pair of paths are effected on average by
%interactions, and in the end to obtain the interaction-modified time
%evolution of the density matrix by summing over all paths in the path
%integral (\ref{eq:JrhoRonly}).  Thus, the order of averages has
%tacitly been changed to $ \langle \langle \, \cdot \, \rangle_V \bigr
%\rangle_\paths $ [\Eq{eq:Vfirst}].

Now, this seemingly innocuous change in the order of averages is
without consequence only if both averages are performed exactly, as is
possible in an order-for-order perturbation expansion (or, to all
orders, for exactly solvable models such as the Caldeira-Leggett
model). However, this is not the case in GZ's theory (or our version
thereof), which proceeds to use the semiclassical approximation
of replacing the sum over all closed
quantum paths by a sum over only the saddle point paths that extremize
the action.  In principle, this can be done in at least two different
ways, which we indicate schematically as follows:
\bsubequations
  \label{subeq:semiclassics}
  \begin{eqnarray}
     \Bigl\langle e^{ - {1 \over \hbar} 
\tilde S_\eff [\bmr^a]} \Bigr \rangle_\cqp 
% \longrightarrow
%    \left\{\begin{array}{lll} 
    \label{eq:semiclassics-GZ}
        & \stackrel{\GZ}{\longrightarrow} & 
        \Bigl\langle e^{ - {1 \over \hbar} 
          \tilde S^\GZ_\eff [\bmr^a_\bare]} \Bigr
        \rangle_\bare  % & 
        \simeq
        e^{ - {1 \over \hbar}  
          \langle \tilde S^\GZ_\eff [\bmr^a_\bare] \rangle_\bare }
         \; , 
        % &  %\quad (\GZ)
        \rule[-4mm]{0mm}{0mm}
\\
     \Bigl\langle e^{ - {1 \over \hbar} 
       \tilde S_\eff [\bmr^a]} \Bigr \rangle_\cqp 
   \label{eq:semiclassics-ideally} 
        &  \stackrel{{\rm ideally}}{\longrightarrow} & 
\Bigl\langle e^{ - {1 \over \hbar} \tilde S_\eff [\bmr^a_\dressed]} 
\Bigr \rangle_\dressed  
         \; .  % &  %\simeq
%e^{ - \langle \tilde S^\GZ_\eff [\bmr^a_\dressed] \rangle_\dressed
%} 
%  & \quad {\rm (ideally)} \qqph
%        \rule[-6mm]{0mm}{0mm}
% \end{array} \right.
 \end{eqnarray}
%\\
%    \label{eq:semiclassics-JvD}
%%        \stackrel{\JvD}{\longrightarrow} &
%    \Bigl\langle e^{ - \tilde S_\eff [\bmr^a_\recoiling]} 
%    \Bigr \rangle^\recoiling_\class  & (\JvD)\\
%\end{array} \right.
%  \end{eqnarray}
 \esubequations 
Here the subscripts bare/dressed indicate that the sums over the
 paths on the right side of \Eq{subeq:semiclassics} are taken only
 over those paths, with boundary conditions as specified in
 \Eq{eq:definecqp}, which extremize the bare action $\tilde S_0 =
 \tilde S_0^F - \tilde S_0^B$ (``bare'' paths,
 \Eq{eq:semiclassics-GZ}, 
used by GZ), or the full action $\tilde
 S_\tot = \tilde S_0 + i \tilde S_\eff$ (``dressed'' paths,
 \Eq{eq:semiclassics-ideally}, 
discussed below). On the far right of
 \Eq{eq:semiclassics-GZ} 
we indicated a further (uncontroversial) approximation,
 used by GZ and others when an exponential is to be averaged over bare
 paths, namely to lift the average into the exponent.  
%In practice,
% [\eg\ Sec.~\ref{sec:GZ-classical-paths}], the average $\langle \dots
% \rangle_\bare$ is performed 
%semiclassically as an
% average $\langle \dots \rangle_\rw$ over all diffusive paths, treated
% as random walks (rw), with appropriate boundary conditions, while
% fluctuations about these semiclassical paths are neglected.  (Note
% that the step of ``impurity averaging'' is never carried out
% explicitly; it enters only implicitly, in the sense that $\langle
% \langle \dots \rangle_\bare \rangle_\imp = \langle \dots
% \rangle_\rw$.)
In practice,
 [\eg\ Sec.~\ref{sec:GZ-classical-paths}], the averages 
$\langle \dots  \rangle_\bare$ on the r.h.s.\ of 
\Eq{eq:semiclassics-GZ} are replaced by
$\langle \langle \dots \rangle_\bare \rangle_\dis  \to  \langle \dots
 \rangle_\rw$, where the latter average is over
diffusive random walks (rw) with appropriate boundary conditions.
In other words, $ \langle \dots  \rangle_\bare$ 
is approximated by considering only semiclassical trajectories 
in a disordered potential landscape 
(while fluctuations about these semiclassical paths are neglected), 
and, after after implicit disorder averaging, these
semiclassical trajectories are treated
as random walks. 

 Ideally, one would of course prefer to average over dressed paths
 [\Eq{eq:semiclassics-ideally}], which ``know'' about the effects of
 interactions due to the role that $i \tilde S_\eff$ plays in
 determining the saddle point trajectories. (Even more ideally, one would
 also take into account fluctuations about these dressed paths).  
 In such a calculation, the $i\tilde
 S_I$ term in $i \tilde S_\eff$ would cause the dressed paths $\tilde
 \bmR_\dressed^a$ to acquire an imaginary component (we thank Igor
 Gornyi for alerting us to this fact), implying that the contributions
 of the two terms in $(i \tilde S_R + \tilde S_I)[\bmR^a_\dressed]$
 can partially cancel, even though both $ \tilde S_R $ and $ \tilde
 S_I $ are purely real functionals of their arguments (GZ overlooked
 the possibility of such a partial cancellation, because they
 considered only bare paths, see below.  Marquardt\cite{Marquardt02}
 has illustrated how such a partial cancellation occurs in the
 Caldeira-Leggett model).  Note that such a dressed-path procedure
 would require only that $\tilde S_\tot[\bmR^a_\dressed] / \hbar \gg
 1$, and would not require $\tilde S_\eff[\bmR^a_\dressed] / \hbar
 $ to be small.  Indeed, its results would be nonperturbative in the
 interaction correlator $\tilde R/\tilde I$, since $\tilde S_\eff$,
 though linear in $\tilde R/\tilde I$, is a non-linear functional of
 $\bmR^a_\dressed$, which itself is nonlinear in $\tilde R/\tilde I$
 (as illustrated explicitly in the Caldeira-Leggett model, where all
 these nonlinear functions can be evaluated explicitly).

 However, in the present theory, using fully dressed paths is not
 technically feasible. Therefore, GZ made the standard and seemingly
 natural choice of averaging purely over \emph{bare} paths.  Indeed, they
 write (just before Eq.~(61) of GZ99\cite{GZ2}): ``In the zero-order
 approximation one can neglect the terms $\tilde S_R$ and $\tilde S_I$
 describing the effect of Coulomb interaction'' so that ``the path
 integral is dominated by the saddle-point trajectories for the action
 $\tilde S_0$''. In other words, bare paths don't ``know'' about the
 interactions \emph{at all}. Consequently, GZ used an effective action
 $\tilde S^\GZ_\eff$ obtained by treating the Pauli factor $\tilde
 \delta - 2 \tilde \rho$ in $\tilde S_\eff$ as time-independent
 (arguing that the energy argument of the corresponding Fermi function
 is conserved during propagation), essentially replacing it by $\tanh[
 \hbar \ve/2T]$. [See p.~9205 of GZ99\cite{GZ2}: ``$n$ depends only on the
 energy and not on time because the energy is conserved along the
 classical path''.]

 Once the approximation of using purely bare paths has been made, the
 effective action $\tilde S^\GZ_\eff [\bmR^a_\bare]$ is \emph{linear}
 in the interaction propagators (since $\bmR^a_\bare$ is now
 independent of the interaction).  This implies in our view that GZ's
 results are \emph{purely perturbative}.  GZ dispute this
 characterization, calling their approach ``nonperturbative'' because
 in their view it does not require $\tilde S^\GZ_\eff [\bmR^a_\bare]
 \lesssim \hbar$, only $\tilde S^\GZ_\eff [\bmR^a_\bare] \ll \tilde
 S_0 [\bmR^a_\bare] $. We disagree, contending that GZ do need the
 former condition, because without it, their use of purely bare paths
 would not be be justified: a semiclassical treatment requires the
 evaluation of the action $\tilde S_\tot [\bmR^a]$ to be accurate to
 within $\hbar$, implying that the effects of $\tilde S_\eff$ on the
 semiclassical paths can be neglected only if $\tilde S^\GZ_\eff
 [\bmR^a_\bare] \lesssim \hbar$.  [Note, also, that an approach that
 reliably evaluates $\langle \tilde S_\eff \rangle $ in the regime
 where the result is $\lesssim \hbar$ would yield the function
 $F(\tau)$ in the regime where it is $\lesssim 1$, which is entirely
 sufficient to reliably extract $\tauphi$, as argued in the second
 paragraph of this subsection.]

While it is a matter of somewhat empty semantics whether an approach
using purely bare paths can be called nonperturbative or not, the
validity of such an approach can be subjected to a hard test: does the
result which this approach produces for $F(\tau)$ \emph{after} the average
$\langle \dots \rangle_\bare$ has been performed agree, 
in the regime $F(\tau) \ll 1 $, with that obtained from Keldysh
perturbation theory?  (GZ's claim in GZ00\cite{GZ3} that
their approach agrees with Keldysh perturbation theory, is true only if
the perturbation expansion is performed \emph{before} averaging over
paths, see Secs.~\ref{sec:selfenergy} and
\ref{app:1storderperturbation}).  The answer is no:
%Actually, we contend that using purely bare paths is insufficiently accurate
%even when $\tilde S_\eff [\bmR^a_\bare] \lesssim 1$. First, GZ actually
%point out themselves that the interactions \emph{do} modify the paths.
%In fact, using their effective action $\tilde S_\eff^GZ$, they derive a
%Langevin-type equation of motion for the relevant paths which includes
%the effect of interactions (Eq.~(98) of GZ99\cite{GZ2}) and shows that the
%dressed paths do differ from the bare one. Second and more
%tellingly, comparison to perturbation theory 
The perturbation expansion obtained by expanding the first-principles
expression (\ref{subeq:Jfirstprinciples}) in powers of $(\tilde S_V^F
- \tilde S_V^B)/ \hbar$ shows unambiguously that the paths arising in
the perturbation expansion, \emph{do} know about the interactions (in
contrast to bare paths): energy conservation induces recoil at each
interaction vertex, so that the electron frequencies incident and
leaving a vertex differ from each other, $\bar \ve$ vs.\ $\bar \ve \mp
\bomega$, in a way relevant for the Pauli factors, which depend on
$\bar \ve \mp \bomega$.  This effect is negligible for retarded and
advanced electron propagators, which depend only on the combination
$\hbar (\bar \ve \mp \bomega) - \xi_\bmp \pm i \hbar / \tauel$, with
$\xi_\bmp = \bmP^2/2m$ [\Eq{eq:GRAdisavepe}], since there energy
shifts by $\hbar |\bomega| \lesssim \hbar/\tauel$ are negligible.
However, it is \emph{not} negligible for the Keldysh propagator, which
contains fermion functions of the form $[e^{[\hbar (\bar \ve \mp
  \bomega) - \ve_F]/T} + 1]^{-1}$ [\Eq{eq:GKsummary}], in which the
largeness of $\hbar \bar \ve$ is \emph{cancelled} by that of $\ve_F$.
Thus, interaction events with recoil energies of order $\hbar
|\bomega| \gtrsim T$ strongly change the value of the Fermi function
which specifies the phase space available for a given transition.
Since these recoil effects are present in the original order $\langle
\langle \dots \rangle_\cqp \rangle_V$ of doing the average but absent
in the switched order $\langle e^{-\tilde S_\eff^\GZ/ \hbar}
\rangle_\bare$ if GZ's version of the effective action is used,
something is clearly amiss in the latter approach.

The main assertion of our own work is that for the purpose of
describing decoherence in weak localization, recoil effects
\emph{can} be taken into account in the influence functional
approach provided GZ's use of bare paths is supplemented by the 
use of an effective action that keeps track of recoil (rec):
  \begin{eqnarray}
    \label{eq:semiclassics-JvD} 
    \Bigl\langle e^{ - {1 \over \hbar} 
\tilde S_\eff [\bmr^a]} \Bigr \rangle_\cqp 
%\longrightarrow
        \stackrel{{\rm JvD}}{\longrightarrow} 
\Bigl\langle e^{ - {1 \over \hbar} 
\tilde S^\recoiling_\eff [\bmr^a_\bare]} \Bigr
\rangle_\bare  
\; \simeq \; 
e^{ - {1 \over \hbar} 
\langle \tilde S^\recoiling_\eff [\bmr^a_\bare] \rangle_\bare} 
\; . 
% \qquad  (\JvD) \; . \qqph 
\end{eqnarray}
[In practice, we perform the averages 
over bare paths on the r.h.s.\ the same way as GZ do,\cite{GZ2,GZ3} 
i.e.\ using an
average over semiclassical diffusive random walks
(and neglecting fluctuations about these), 
$\langle \langle \dots \rangle_\bare \rangle_\dis  \to  \langle \dots
\rangle_\rw$.] 
Here $\tilde S^\recoiling_\eff$ is the effective action obtained by
representing $(\tilde \delta - 2 \tilde \rho)$ by $\tanh[ \hbar (\ve
\mp \bomega)/2T]$-factors [as made explicit in
\Eq{subeq:generalizedruleofthumbA}].  The result for $\tilde
S_\eff^\recoiling [\bmR^a_\bare]$ so obtained
[\Eq{eq:SIR-LIR-aa-main}, or \Eq{eq:Seff} with (\ref{subeq:LAA'FT})]
is linear in the interaction propagators $\tilde R / \tilde I$, just
as GZ's effective action is, but in contrast to the latter, an
expansion of $e^{- {{1 \over \hbar} \langle \tilde
    S_\eff^\recoiling}\rangle_\bare}$ to first order in the exponent
yields results consistent with the Keldysh perturbation expansion also
if the average over paths in ${1 \over \hbar} \langle \tilde
S_\eff^\recoiling \rangle_\bare$ is performed explicitly first and the
exponential expanded only \emph{thereafter}.  Moreover, the results
for $F (\tau)$ so obtained agree fully with those from a diagrammatic
Bethe-Salpeter calculation of the Cooperon (see
DMSA-II\cite{MarquardtAmbegaokar04}).  A crucial ingredient for
ensuring this agreement is that the ultraviolet divergencies arising
in each of the two terms in $ \langle (i \tilde S^\recoiling_R +
\tilde S^\recoiling_I)[\bmR^a_\bare]\rangle_\bare$ cancel each other
[\Sec{sec:GZ-classical-paths}].  This cancellation is possible because
the functional $\tilde S^\recoiling_R [\bmR^a_\bare]$, despite using
only bare paths, is not purely real, thereby capturing an essential
feature of $\tilde S^{\phantom{R}}_\eff [\bmR^a_\dressed]$ that is not
present in $\tilde S^\GZ_\eff [\bmR^a_\bare]$.

Since GZ contend that their approach is nonperturbative, they reject
arguments based on perturbation theory, defending their use of purely
bare paths by evoking only the standard semiclassical approximation.
But the need to keep track of recoil arises within the latter
framework, too, in a way very similar to that described above: The
standard condition for the validity of the semiclassical approximation
is that the propagation energies and momenta of the quantum particle
that is to be described semiclassically should be much larger than the
typical frequencies and wave numbers characterizing the potential
landscape which it is moving in, so that the latter appears
``smooth''.  If one were to consider a single noninteracting electron
propagating with energy $\hbar \bar \ve \simeq \ve_F$ through a
disordered potential landscape, this implies the conditions $\ve_F \gg
\hbar/\tauel $ (or $k_F \gg 1/ \lel$), which certainly are satisfied
in the regime of weak localization. However, GZ's theory for an
electron propagating through and interacting with a Fermi sea of other
electrons shows that the propagation energy enters not only in the
free part of the action, but also in the Fermi function $[e^{ (\hbar
  \bar \ve - \ve_F)/T} + 1]^{-1}$ arising from the Pauli factor
$\tilde \delta - 2 \tilde \rho$ in $\tilde S_R$.  To ensure that this
factor is treated accurately, the standard semiclassical condition
$\ve_F \gg \hbar / \tauel $ evoked by GZ is not sufficient, since
inside the Fermi function the largeness of $\hbar \bar \ve$ is
\emph{cancelled} by the largeness of $\ve_F$.  Thus,
interaction-induced changes in $\bar \ve$ of order $\bomega \lesssim
1/\tauel$ will produce strong changes in the Fermi function between
$\simeq 0$ and $\simeq 1$.  These changes need to be kept track of. As
argued above, this can be accomplished by the recoil contributions
$\mp \bomega$ in our $\tanh[ \hbar (\ve \mp \bomega)/2T]$-factors.
% In other words, by using $\tilde S^\recoiling_\eff [\bmR^a_\bare]$,
% we accurately keep track of the energy changes \emph{at} each
% interaction vertex, and evoke the semiclassical approximation only
% for the propagation at constant energy \emph{between} interaction
% vertices.

}

\appendix{Derivation of Influence Functional Approach}

\label{sec:rederivation}

\label{sec:Kubo}
\label{sec:notations}

In this appendix, we rederive the influence functional approach of GZ,
with the aim of establishing clearly (i) how far it can be taken
without any approximations, and (ii) what the consequences are of the
approximations that they eventually do make. We generally follow the
strategy they have chosen to use, but the details of our notations and
derivations deviate from GZ's whenever we believe that greater
compactness, clarity or generality can thereby be achieved.  The most
important difference is that instead of using the coordinate-momentum
path integral $\int {\cal D} (\bmR \bmP)$ of GZ, we use a
coordinate-only version $\int {\cal D}^\prime \bmR$, since this enables the
Pauli factor to be treated more accurately.

The outline of this appendix is as follows. After a summary of our
notational conventions, Secs.~\ref{sec:model} to \ref{sec:hs} define
the model and decouple the interaction using a Hubbard-Stratonovich
transformation within a Keldysh framework.  In Sec.~\ref{sec:roadmap},
we summarize GZ's procedure for deriving their influence functional
approach, and in Sec.~\ref{sec:repeatGZ} repeat their steps in
explicit detail, though with some changes. Finally,
Sec.~\ref{sec:IFvsPT} establishes a link between the influence
functional so derived and Keldysh perturbation theory, and discusses
the fate of the Pauli factor.

\subsection*{Notational Conventions}

We begin by summarizing, for ease
of reference,  some notational conventions to
be used throughout: We shall use the shorthands $x \equiv (\bmr,
\sigma)$ for electron position and spin, and $
\label{eq:shorthand-psi-int} \int dx \equiv \sum_\sigma \int d \bmr.  $
Operators will generally carry hats (e.g.\ $\hat H_0$), and the
subscripts {\small $S$}, {\small $H$} and {\small $I$} will distinguish
operators in the Schr\"odinger, Heisenberg or interaction pictures,
respectively.  For $c$-number fields, the shorthand $V_i \equiv V_i
(t_i) \equiv V (t_i, \bmr_i)$ will often be used, i.e.\ the
time argument, when not displayed explicitly, will be understood to be
$t_i$.  $c$-number functions depending on two different coordinates,
i.e.\ coordinate-space matrices, will generally carry tildes [e.g.\
$\tilde \rho_{i \! j} = \rho (x_i, x_j)$], and their Fourier
transforms w.r.t.\ $\bmr_i - \bmr_j$ will carry bars, e.g.\
\begin{eqnarray}  \label{eq:FT-conventions}
\bar \rho(\bmR_{\iiijjj}, \bmp) \equiv \int d \bmr_{\iiijjj}
e^{-{i} \bmp \cdot \bmr_{\iiijjj}}
 \,  \tilde \rho (\bmR_{\iiijjj} + \toh \bmr_{\iiijjj},
\bmR_{\iiijjj} - \toh \bmr_{\iiijjj}) \;,
\end{eqnarray}
where $\bmR_{\iiijjj}$ and $\bmr_{\iiijjj}$ will generally
denote center-of-mass and relative coordinates,
\begin{equation}\label{eq:defineRijrij}
  \bmR_{\iiijjj} = (\bmr_i + \bmr_j)/2 \, , \qquad
  \bmr_{\iiijjj} =  \bmr_i - \bmr_j \, .
\end{equation}
We do not set $\hbar = 1$ but display it explicitly throughout.
Hence, the variable $\bmp$ in \Eq{eq:FT-conventions}
(and likewise $\bmk$, $\bmq$ below) denotes a wave-number,
with units of inverse length, not a momentum; the corresponding
momenta will always be denoted by capital letters:
\begin{eqnarray}
  \label{eq:P=hbarp}
  \bmP = \hbar \bmp \; \qquad \bmK = \hbar \bmk, \qquad
\bmQ = \hbar \bmq \; .
\end{eqnarray}

For correlation functions, the shorthand $\tilde G_{\iiijjj} \equiv \tilde
G_{\iiijjj} (t_{\iiijjj}) \equiv G (t_{\iiijjj}; x_i, x_j)$ will often be used,
i.e.\ the time argument, when not displayed explicitly, will be
understood to be $t_{\iiijjj} = t_i - t_j$. [For the step function, we use
$\theta_\iiijjj \equiv \theta (t_\iiijjj)$.]  The corresponding frequency
Fourier transform w.r.t.\ $t_{\iiijjj}$ will be denoted by
\begin{eqnarray}\label{eq:define-frequency-FT}
%  \tilde G_{\iiijjj} (\omega) & = & \int d t_{\iiijjj}
%    e^{i \omega t_{\iiijjj}} \tilde G_{\iiijjj} (t_{\iiijjj}) \; , \qquad
    \tilde G_{\iiijjj} (t_{\iiijjj})  =   \int (d \omega)  e^{- i \omega
    t_{\iiijjj}}  \tilde G_{\iiijjj} (\omega) \; , \qquad 
  \int (d \omega)  \equiv \int {d \omega \over 2 \pi} \; , 
\qqph
\end{eqnarray}
where $\omega$ has units of inverse time.  If coordinate-space
subscripts are not displayed explicitly, they are understood to be
summed over, e.g.\ $[\tilde G(t) \tilde G(t')]_{\iiijjj} \equiv \int dx_k
\tilde G_{ik}(t) \tilde G_{kj}(t')$.  We distinguish forward and
backward parts of the Keldysh contour by an index $a = F,B$ [GZ use
$a=1,2$ instead], and use boldface for matrices in Keldysh space,
e.g.\ $\bG_{\iiijjj}$.

A pair of indices such as $\scriptstyle {ii'}$, appearing
once without prime and once with, will denote independent coordinates $x_i$
and $x_{i'}$ referring to the same time (\ie\ $t_i = t_{i'}$ is
to be understood), which are, however, to be set equal at the
very end of the calculation, after being differentiated upon, \ie\
\begin{eqnarray}
  \label{eq:indexindex'}
(\bnabla_{i} - \bnabla_{i'} ) \tilde \rho_{ii'} \equiv
\left[ (\bnabla_i - \bnabla_{i'}) \tilde \rho_{i
    i'}\right]_{x_i = x_{i'}} \; .
  \end{eqnarray}
  We shall often encounter double summations over coordinates
  referring to the same time. For such coordinates we shall use the
  index pair $\scriptstyle {i \bari}$, one without bar and one with,
take it to be understood that $t_{\bari} \equiv t_i$, and 
  denote the double summation by
\begin{eqnarray}
  \label{eq:integratenobarwithbar}
  \int \! d x_{i, \bari} \equiv \sum_{\sigma_i} \int \! d \bmr_i d
\bmr_{\bari} \; .
\end{eqnarray}

When taking the limit of infinite volume,
we shall use the shorthand notation
\bsubequations
    \label{eq:shorthandmomentaintegrals}
\begin{eqnarray}
  \label{eq:shorthandmomentaintegralsdp}
  \int (d \bmp) & \equiv & \int {d \bmp \over (2 \pi)^d}
 =  \lim_{\Vol \to \infty} {1 \over \Vol} \sum_{\bmp} \; ,
\\
\bar \delta^{(d)} (\bmp - \bmp) & \equiv &
\lim_{\Vol \to \infty} \delta_{\bmp , \bmp'} \Vol \; ,
\end{eqnarray}
\esubequations
so that $\int (d\bmp) \bar \delta (\bmp) = 1$, \ie, $\bar \delta
(\bmp) $ equals $(2 \pi)^d$ times a $d$-dimensional Dirac delta
function.  If the integrand under $\int (d \bmp)$ depends only on the
energy $\xi_{\bmp} = \bmP^2 /2m - \eF$ and if it decays at least as fast as
$1/\xi^2_{\bmp}$ for $\xi_{\bmp} \to \infty$, we shall use $\int (d
\bmp) \to \nud \int \! d \xi_{\bmp}$. Here $\nud$ denotes the density
of states per spin at the Fermi surface, which  in $d = 3$ or 2
dimensions is given by
% Book 6, p. 133
\begin{eqnarray}
  \label{eq:densityofstatesnd}
\nud = {m \over 2 \pi \hbar^2 } \left({\kF \over \pi }
\right)^{d-2} \;
= \; {d\, \nav \over 2 \eF } \; ,
\end{eqnarray}
where $\nav = \int_{k < \kF} (d \bmk) = {\pi \over 2 d} \left({ \kF
    \over \pi} \right)^d $ is the average electron density per spin.
The \emph{purely} 1-dimensional case
$d=1$ will not be considered here; nevertheless, $d=3$
or 2 of course include the case that a sample is quasi 1-dimensional,
in the sense that only one of its dimensions is larger than the
phasebreaking length, $L \gtrsim \sqrt {\Dd \tauphi}$,
where  $\Dd = \vF^2 \tauel/d$ is the diffusion constant.

For quasi-$\bard$-dimensional diffusion, the actually measured (bare)
DC conductivity $\sigma_\bard$ is related to the Drude conductivity
$\sigma^\Drude_\DC = 2 e^2 \nu \Dd$ by an extra factor $a^{d -
  \bard}$, which accounts for the sample's transverse directions along
which motion is not diffusive ($d$ = 3 or 2 is the actual dimension of
the sample, $\bard$= 3, 2 or 1 the effective dimension for diffusive
motion).  Hence, it is customary to define [cf.\ AAK\cite{AAK82},
after Eq.~(5)] $\sigma_\bard = \sigma^\Drude_\DC a^{d - \bard}$ as the
conductivity per unit length and unit area of a 3D sample (for $\bard
= 3$), or the inverse square resistance of a film (for $\bard = 2$),
or the inverse resistance per unit length of a wire (for $\bard = 1$).
Likewise, the weak localization correction to the conductivity is
often expressed in terms of these actual conductivities by defining
$\sigma^\WL_\bard \equiv \sigma^\WL_\DC a^{d - \bard}$.

The fact that the weak localization correction is small compared to
the Drude term is often made explicit by writing the prefactor of the
Cooperon term as $\sigma^\Drude_\DC / g_\bard (L_H)$ [see
\Eq{eq:sigmaDSdimensionless}], where $L_H = \sqrt{D \tauH}$ is the
magnetic length and 
\begin{eqnarray}
  \label{eq:dimconductance}
g_{\bard} (L) \equiv (\hbar / e^2) \, \sigma_{\bard} \, L^{\bard - 2} \; ,   
\end{eqnarray}
is the so-called dimensionless conductance, defined as the
conductance, in units of $e^2 / \hbar$, of a rectangular
($d$-dimensional) block with volume $a^{d- \bard} L^\bard$, measured
along one of the ``long'' directions (of length $L$).

For good conductors, $g_\bard (L) = (\pi^{1- d}/d) (a
\kF)^{d-\bard} (\lel \kF)^{\bard -1} (L/\lel)^{\bard - 2}$ is large
whenever $L$ is large: we may assume $\lel \kF \gg 1$ and $a \kF \gg
1$ throughout, thus for $\bard = 3$ or $2$, any length $L \ge \lel$
implies $g_\bard(L) \gg 1$; for $\bard = 1$ the function $g_1 (L)$
likewise starts out being $\gg 1$ for $L \simeq \lel$, but decreases
with increasing $L$; nevertheless, it reaches $g_1 (L_\ast) = 1$ only
when $L$ exceeds the very large length scale $L_\ast = (a \kF)^{d-1}
\lel \gg \lel$.

\subappendix{The Model and Kubo Formula} 
\label{sec:model}

Following GZ, we consider a disordered system of interacting fermions,
described by the Hamiltonian $\hat H = \hat H_0 + \hat H_\irm$, where
\begin{eqnarray}
  \hat H_0 & = & \int d x \,
\hat \psi_S^\dag (x)  h_0 (x) \hat \psi^\pdag_S (x) \; ,
% \\
  \qquad 
h_0 (x) =  \frac{- \hbar^2 }{2 m} \bnabla_{\bmr}^2 +
V^\pdag_\imp (\bmr) - \mu \; , \qqph \qph
\\ \label{eq:defHint}
\hat H_\irm  & = & {e^2 \over 2}
 \int \!\!   d x_1 \, d x_2 \,
: \! \hat n_{11S} \! : \,
\tilde V^\inter_{12}
: \! \hat n_{22S} \! : % \over |\bmr_1 - \bmr_2|}
\, ,
%\\
%V^\inter_{12} &  = & % V^\inter (|\bmr_1 - \bmr_2|)
%\int {d^3 \bmq \over (2 \pi)^3} \, e^{i (\bmr_1 - \bmr_2) \cdot \bmq}
%\, V^\inter (\bmq) \; ,
\\
\label{normalordering}
: \! \hat n_{\iiijjj S} \! : &=& \hat n_{\iiijjj S} -
\langle \hat n_{\iiijjj S} \rangle_0 \, , \qquad
\hat n^\pdag_{\iiijjj S} \equiv
\hat \psi_S^\dag (x_j) \hat \psi^\pdag_S (x_i) \; ,
\\ \label{eq:defrho0}
 \langle \hat O \rangle_0  & =& \Tr \{ \hat O \, \hat \rho_0 \}  ,
\quad  \hat \rho_0 =  e^{- \beta \hat H_0} /
   \{ \Tr e^{- \beta \hat H_0} \}  .
\end{eqnarray}
$V_\imp (\bmr)$ is the disorder potential.  We shall assume that the
interaction potential $\tilde V^\inter_{12} = \tilde V^\inter (|\bmr_1
- \bmr_2|)$, which guarantees that its Fourier transform in $\bar d$
effective dimensions, $\bar V^\inter_{\bar d} (\bmq) = \bar
V^\inter_{\bar d} (|\bmq|) = a^{3-d} \int d^d r \, e^{-i \bbmq \cdot \bmr} 
\tilde V^\inter (r)$, is real.  For example, for the Coulomb interaction,
they are given by $\tilde V^\inter_{12} = { 1 \over |\bmr_1 -
  \bmr_2|}$, and 
\begin{eqnarray}
  \label{eq:screenedCoulomb}
V^{(3)}_\bbmq  =  { 4 \pi  \over  \bbmq^2} , 
\qquad V^{(2)}_\bbmq  =  {a\,   2 \pi \over  |\bbmq| } , 
\qquad 
V^{(1)}_\bbmq  = a^2  \ln (\bbmq^2 a^2) \; . 
%\bar V^\inter (\bmq) = {4 \pi \over \lambda_0^2 + q^2} \; .
\end{eqnarray}
\Eq{normalordering} corresponds to a
normal-ordering prescription which subtracts the expectation value
w.r.t.\ the free density matrix $\hat \rho_0$.
The second-quantized 
electron field  $\hat \psi^\pdag_S
(x) \equiv \hat \psi^\pdag_S (\bmr, \sigma) $ 
(in the Schr\"odinger picture) destroys
a spin-$\sigma$ electron at position
$\bmr$, and can be expanded as follows in terms of the exact
eigenfunctions $\psi_\lambda(x)$ of $h_0(x)$, with eigenvalues 
$\xi_\lambda$:
\begin{eqnarray}
\label{eq:definepsifields}
  \hat \psi_S (x) & = & \sum_\lambda \psi_\lambda (x) \, \hat c_{\lambda S} ,
  \quad  \bigl[ h_0 (x) - \xi_\lambda \bigr]
 \psi_\lambda (x) = 0. \quad \phantom{.}
\end{eqnarray}

The current density operator
% can be expressed as [this is (GZ-II.48)]
has the form
\begin{eqnarray}
  \label{eq:current-define}
\hat \JJbm_{\!\! S }
(t_1, \bmrone)  & = &
\sum_{\sigma_1} \left[  \bmj_{11'} -
  { e^2 \over m} \bmA (t_1, \bmrone)
\right]\hat n^\pdag_{11'S} \; , \quad 
%\quad \phantom{.}
%\\
%\hat \bmj_S (x_1) &=&
%\bmj_{11'} \, \hat n^\pdag_{11'S}  , \quad
%     \bmj_{11'} \equiv % &  \equiv &
%{- i e \hbar \over 2 m}
%( \bnabla_1 - \bnabla_{1'})   ,
%\phantom{.}
\end{eqnarray}
where $\bmA$ is the vector
potential, $     \bmj_{11'} \equiv % &  \equiv &
(- i e \hbar/  2 m) ( \bnabla_1 - \bnabla_{1'}) $, and the
convention of \Eq{eq:indexindex'} was used for the indices
$\scriptstyle {1 1'}$.  An external electric field,
%\begin{equation}
%\label{eq:defineE}
$  \bmE (t_2, \bmr)  = - \bnabla V^\pdag_\ext (t_2, \bmr) -
\partial_{t_2} \bmA (t_2, \bmr)$,
%\end{equation}
switched on at time $t'_0$, is described by the
perturbation\footnote{
\label{e<0} 
We use $e < 0$ for the electron charge, as do AAG\cite{AAG98},
whereas GZ use $-e < 0$, hence our potentials are related to GZ's by a
minus sign: $ e V_\ext^{\rm here} = - e V^\GZ_x$, and likewise $e
V^{\rm here}_{F} = - e V^\GZ_1$, $ e V^{\rm here}_{B} = - e V^\GZ_2$
for the potentials introduced in \Eq{eq:hatV} below.}
\bsubequations  \label{eq:externalperturbation}
\begin{eqnarray}
\label{eq:Vexternal}
  \hat H_\ext (t_2)  &  = &
 \theta (t_2-t'_0) \int d  x_2 \, \hat h^\ext_S (t_2, x_2) \; ,
 \\
\hat h^\ext_S (t_2, x_2) &=&
  h^\ext_{22'}   \, \hat n^\pdag_{22' S} \; , \qquad
% \\
   h^\ext_{22'}   \equiv  e V^\pdag_\ext (t_2, \bmr_2) -
   \bmA (t_2,\bmr_2) \cdot \bmj_{22'} \; .
   \quad \phantom{.} \qqph
\end{eqnarray}
\esubequations
According to the Kubo formula, % (\ref{eq:Kubo}),
the linear response of the current density to this perturbation
is 
\begin{eqnarray}
\label{eq:Kubo-for-current}
 \langle \delta \hat \JJbm_{\!\! H} (t_1, \bmr_1) \rangle
 =   \sum_{\sigma_1}
\Biggl[
- {e^2 \over m} \bmA (t_1, \bmr_1) \langle \hat n^\pdag_{11S} \rangle
%\Biggr.
% \\ & & \Biggl. \phantom{.} \hspace{5mm} \nonumber
 - i \int^{t_1}_{t'_0} \!\! dt_2 \! \int \! dx_2 \,
  h^\ext_{22'} \, \bmj_{11'} \,
\tilde {\cal C}_{[11',22']}
%\langle [ \hat n^\pdag_{11'H} (t), \hat n^\pdag_{22'H} (t') ] \rangle
\Biggr] \; .  \qph
\end{eqnarray}
The first term of \Eq{eq:Kubo-for-current} is the diamagnetic
contribution, $\nav = 2 \nud \eF/d $ being the average electron
density per spin [cf.\ \Eq{eq:densityofstatesnd}]. In the second
term, the correlator 
\begin{eqnarray}
\label{eq:define-C-firsttime}
 \tilde {\cal C}_{[11',22']} 
\equiv {1 \over
  \hbar} \langle [ \hat n^\pdag_{11'H} , \hat n^\pdag_{22'H} ]
\rangle \; 
\end{eqnarray}
is to be evaluated with $\hat H_\ext$ set to zero, where $\hat B_H (t)
= e^{i \hat H t/\hbar } \hat B_S e^{-i \hat H t/ \hbar}$ describes
time evolution in the Heisenberg picture, and thermal averaging is
defined by $ \langle \hat O \rangle = \Tr \{ \hat O \, \hat \rho_H
\}$, where $\hat \rho_H= e^{- \beta \hat H} / \{ \Tr e^{- \beta \hat
  H} \} $ is the full equilibrium density matrix.

The DC conductivity is defined via the low-frequency limit of the
current response to a spatially homogeneous applied AC field $\bmE
(t_2) = \int (d \tilde \omega_0) e^{- i \tilde \omega_0 t_2} \bmE
(\tilde \omega_0)$. For a $d$-dimensional isotropic sample it can be
written as
\begin{equation}
\label{eq:DCconductivity} \sigma_\DC = \lim_{\omega_0 \to 0} {1
\over d} {\partial \phantom{\bmE \bmE} \over \partial \bmE (\omega_0) } \cdot
\langle \delta \hat \JJbm (\omega_0) \rangle  \; ,
\end{equation}
where $\bmE (\omega_0)$ can be represented by either of the
choices (related by a gauge transformation)
\bsubequations
\label{eq:representing-E}
\begin{eqnarray}
\label{eq:representing-Ea} \bmA & =&  0\; ,
\phantom{\bmE } %\omega_0)}
  V^\pdag_\ext (\omega_0, \bmr_2)
 = - \bmr_2 \cdot  \bmE (\omega_0)\; ,
\qquad \phantom{.} \\
\label{eq:representing-Eb} \bmA (\omega_0) & = & \frac{
 \bmE (\omega_0)}{i \omega_0} \; , \phantom{0} \phantom{\bmr_2)}
 V^\pdag_\ext = 0 \; .
\end{eqnarray}
\esubequations \noindent
GZ use choice (\ref{eq:representing-Ea}) (but note our
footnote~\ref{e<0}), AAG use choice (\ref{eq:representing-Eb}).
Taking the limit $t'_0 \to - \infty$, one then readily finds from
Eqs.~(\ref{eq:DCconductivity}) and (\ref{eq:Kubo-for-current}) that
$\sigma_\DC$ can be written in either of the following forms,
depending on whether the electric field is represented using a scalar
or a vector potential:
\bsubequations
\label{eq:sigmaDC-final}
\begin{eqnarray}
\label{eq:sigmaGZ}
\sigma_{\rm DC} \! \! & = & \! \! - {e \over d } \sum_{\sigma_1}
%\sum_{\alpha}
\! \int \! \! dx_2 \,
    \bmj_{11'} \! \cdot \!  \bmr_2
    \lim_{\omega_0 \to 0}
    \tilde {   J}_{12',21'} (\omega_0) \, , 
\\
\sigma_{\rm DC} \! \! & = &  \!\!    \lim_{\omega_0 \to 0} {1 \over \omega_0}
\sum_{\sigma_1} \left[ {1 \over d }
\! \int \! \!  dx_2 \, \bmj_{11'} \! \cdot \! \bmj_{\,22'}
\, \tilde {J}_{12',21'} (\omega_0)  
\label{eq:sigmaAAG}
+ {i  e^2 \langle \hat n_{11 H} \rangle
 \over m} \right] , \qquad \phantom{.} \qph
%\end{eqnarray}
\end{eqnarray}
\esubequations
where we have introduced the retarded correlator [with $\theta_{12}
\equiv \theta (t_{12})$]
\begin{eqnarray}
\label{eq:defineJomega0}
\tilde {J}_{12',21'} (\omega_0) &=&
 \! \int_{- \infty}^\infty
\!\!  dt_{12} e^{i \omega_0 t_{12}}
\, \theta_{12}
 \, \tilde {   C}_{[11',22']} \; .
 \label{eq:defineGR1221}
\end{eqnarray}
Sometimes it is covenient to average the coordinate $\bmr_1$ over the
volume $\Vol$, in which case one should replace $\sum_{\sigma_1}$ in 
\Eq{eq:sigmaDC-final} by $\int {d x_1 \over \Vol} $. 

\subappendix{Keldysh Approach with Source Fields}
\label{sec:sources}

We now use the Keldysh real-time approach
to rewrite $ \tilde {\cal C}_{[11',22']}$
in terms of correlators whose dynamical and
statistical properties are governed entirely by $\hat H_0$:
% (this functional may be represented as a path integral,
%and/or expanded in powers of $\hat H_\irm$).
First, thermal weighting is done in the infinite past using the
\emph{free} density matrix $\hat \rho_0$, and then the interaction is
turned on adiabatically. For arbitrary operators $\hat A_H$ and $\hat
B_H$, this amounts to the replacement
\begin{eqnarray}
  \label{eq:Keldysh}
&&   \langle \hat A_H (t_1) \hat B_H (t_2) \rangle \to
  \langle \hat A_H (t_1-t_0) \hat B_H (t_2-t_0) \rangle_0 \; ,
\end{eqnarray}
where the initial time $t_0$ is sent to $- \infty$ so that all disturbances
associated with switching on the interactions have decayed in the infinite
past (the limit $t_0 \to - \infty$,  will be understood but
not displayed below).
Second, the time evolution of all operators is expressed in
the interaction representation, using the familiar
operator identity
\bsubequations
\begin{eqnarray}
  \label{eq:Uint}
% \hat U^\pdag_H (t-t_0)
e^{-i H (t_i-t_0)/\hbar } & = &  e^{- i \hat H_0 (t_i-t_0)/ \hbar }
   \hat U^\pdag_I (t_i,t_0) \; ,
\\
\label{eq:UT}
  \hat  U^\pdag_I (t_i,t_j) & = &
{\cal T} e^{-{i\over \hbar} \int_{t_j}^{t_i} \ddd  t_3
   \hat H_{iI} ( t_3) } \; ,
\\
  \label{eq:AH0}
\hat A_I (t_i) &=& e^{i \hat H_0 (t_i - t_0)/\hbar} \hat A_\Sch
   e^{- i \hat H_0 (t_i-t_0)/\hbar} \; ,
\end{eqnarray}
\esubequations \noindent
where ${\cal T}$ is the time-ordering operator (the anti-time-ordering
operator, needed for $\hat U_I^\dag$, will be denoted by $\overline {\cal
  T}$). The correlator
$\tilde {\cal C}_{[11',22']}$  then becomes
\begin{eqnarray}
  \label{eq:C-interaction} \tilde {\cal C}_{[11',22']}  & = & 
%\\ \nonumber
% & &  \hspace{-0.5cm} 
{1 \over \hbar} \langle \hat U_I^\dag (t_1,t_0)
\hat n_{11'I} (t_1)
\hat U^\pdag_I (t_1,t_2) \hat  n_{22'I} (t_2)
\hat U^\pdag_I (t_2,t_0) \rangle_0 \qqph 
\\ \nonumber
& & % \hspace{-0.5cm} 
-   {1 \over \hbar} \langle \hat U_I^\dag (t_2,t_0)
\hat n_{22'I} (t_2) \hat U^\dag_I (t_1,t_2) \hat  n_{11'I} (t_1)
\hat U^\pdag_I (t_1,t_0) \rangle_0 \; .
\end{eqnarray}
This expression can be recovered via functional derivates from
the following construction:
\bsubequations
\label{eq:functional-integrals}
\begin{eqnarray}
\label{C-1122-derivative}  \tilde {\cal C}_{[11',22']} & = &
i  \left. \frac{ \delta \tilde {\bm{\rho}}^\pdag_{11'} (t_1,t_0)}{
\delta \tilde v^\pdag_{2'2} }\right|_{\tilde v= 0} \; ,
\\
\tilde {\bm{\rho}}^\pdag_{11'} (t_1,t_0)  & \equiv &
\label{eq:rho-interaction} {\langle \hat U_{IB}^\dag (t_1,t_0) \hat
n^\pdag_{11'I}(t_1)
 \hat U^\pdag_{IF} (t_1 , t_0) \rangle_0 \over
\langle \hat U_{IB}^\dag (t_1,t_0)
 \hat U^\pdag_{IF} (t_1 , t_0) \rangle_0 } \, , \qquad \phantom{.}
\\
\label{eq:UT-sources}
  \hat  U^\pdag_{Ia} (t_1,t_0) & \equiv &
{\cal T} e^{-{i\over \hbar} \int_{t_0}^{t_1} d  t_3
   \left[\hat H_{\irm I} ( t_3) + \hat v_I (t_3) \right]} \; ,
\\
\label{eq:defvhat}
 \hat v_I (t_3) & \equiv &
\int \! d x_{3,\bar 3} \, \tilde v^\pdag_{\bar 3 3} (t_3)  \,
\hat n_{3 \bar 3 I} (t_1) \; .
\end{eqnarray}
\esubequations \noindent
%Evidently, $ \bar \rho$
%depends on the initial time $t_0$ too, but this dependence will not be
%displayed explicitly.
The index $a = F,B$ will be used to distinguish propagation associated
with $U^\pdag_I$ or $U^\dag_I$ in \Eq{eq:C-interaction}, i.e.\ with
the forward or backward parts of the Keldysh contour, respectively.
Since $\hat U_{IB}^\dagger \hat U_{IF} = 1$, the denominator in
\Eq{eq:rho-interaction}, included for later convenience, in fact
equals unity.  $\tilde {\bm{\rho}}^\pdag_{11'} (t_1,t_0) = \tilde
{\bm{\rho}} (t_1,t_0; x_1,x_{1'})$ is the \emph{reduced
  single-particle density matrix}.  We call it ``reduced'', since the
thermal average $\langle \; \rangle_0$ in \Eq{eq:rho-interaction}
traces out all electron degrees of freedom but one, to be called the
``singled-out electron'', for which the others constitute an effective
environment.  Note that we have defined $\tilde
{\bm{\rho}}^\pdag_{11'} (t_1,t_0)$ in the presence of a source
term\footnote{For our purposes it turns out to be sufficient to use
  the \emph{same} source term $\hat v$ and source field $\tilde
  v_{\bar 3 3}$ on the forward and backward contour; to calculate
  correlators more general than $\tilde {\bm{\rho}}^\pdag_{11'}$, one
  would introduce a separate source term $\hat v^a$ and corresponding
  source fields $\tilde v^a_{\bar 3 3}$ for each of the forward and
  backward contours, $a = F/B$. The corresponding generalizations
  below are straightforward.}  to generalize this to $\hat v (t_3)$,
defined by \Eq{eq:defvhat} [which uses the conventions of
\Eq{eq:integratenobarwithbar}] on the interval $t_3 \in [t_0, t_1]$ in
terms of a real $c$-number ``source field'' $\tilde v^\pdag_{\bar 3 3}
(t_3) = \tilde v (t_3; x_{\bar 3},x_3)$ that couples to the (not
normal-ordered) operator $\hat n_{3 \bar 3 I} (t_3)$. The source field
is devoid of physical meaning, and is introduced merely as a
mathematical device to generate $\tilde {\cal C}_{[11',22']}$ via
functional differentiation. For $\tilde v = 0$, our reduced density
matrix $\tilde {\bm{\rho}}^\pdag_{12} (t,t_0)$ corresponds to
$\bm{\rho} (t; \bmrone, \bmrtwo)$ of (GZ-II.20) of GZ, who simply call
it ``density matrix''.

In the usual Keldysh approach, all time integrals involving the
interaction extend from $- \infty$ to $\infty$.  This can also be
achieved in the present approach, if desired, by inserting a factor of
$1 = \hat U^\dagger_{IB} ( t_\infty, t_1) \hat U^\pdag_{IF} (t_\infty, t_1)$
just to the left or right of $\hat n_{11'I} (t_1) $ in the first or
second lines of \Eq{eq:C-interaction}, respectively, and taking the
limit $t_\infty \to \infty$, $t_0 \to - \infty$.  However, the actual
value chosen for $t_\infty$ does not matter, and in the present
approach, it is actually simplest to use $t_\infty = t_1$.

\subappendix{Hubbard-Stratonovich Transformation}
\label{sec:hs}

Following GZ, we now decouple the interaction term $\hat H_\irm$ in
$\hat U_{Ia}$ using a Hubbard-Stratonovich transformation that
introduces a path-integral over a further pair of real,
spin-independent $c$-number fields $V_a(t_3, \bmr_3)$:
\bsubequations
  \label{eq:Hubbard-Stratonovitch-all}
\begin{eqnarray}
  \label{eq:Hubbard-Stratonovitch}
  \hat U^\pdag_{Ia} (t_1, t_0) & =  &
{ \int {\cal D} V_{a} (t_3, \bmr_3)
 \, \hat {\cal U}_a (t_1,t_0) \, e^{i S_V^{0a} (t_1,t_0)/ \hbar}
\over
\int {\cal D} V_{a} (t_3,\bmr_3) \,   e^{i S_V^{0a} (t_1,t_0)/\hbar } } \; ,
\qquad \phantom{.}
\\
\label{eq:defineSa}
S_V^{0a} (t_1,t_0)  & = &
\!  \int_{t_0}^{t_1} \!\! d t_3 \! \int \! \! { d \bmq \over (2 \pi)^3}
{\bar V_a (t_3, -\bmq) \, \bar V_a (t_3, \bmq) \over 2 \, \bar V^\inter (\bmq)}
\; , \qquad \phantom{.}
\\
\label{eq:Uadefine}
 \hat  {\cal U}_a (t_1,t_0)
& = & {\cal T} e^{-{i\over \hbar}
\int_{t_0}^{t_1} d t_3 \left[ \hat V_a (t_3)
+ \hat v_I (t_3) \right]}
\; ,
\\
\label{eq:hatV}
\hat V_a (t_3)   & = &
  \int d x_3 \,e V_{a} (t_3,\bmr_3) \, : \! \hat n_{33I} \! : (t_3) \; .
\end{eqnarray}
\esubequations \noindent
%Here the index $a = F$ (or $B$) is used to distinguish objects associated
%with the forward (or backward) parts of the Keldysh contour, \ie\ the parts
%associated with propagation forward (or backward) in time with $\hat {\cal
%  U}^\pdag_{IF}$ (or $\hat {\cal U}^\dag_{IB}$).
The fields $V_a(t_3, \bmr_3)$ and their Fourier transforms $\bar
V_a(t_3, \bmq)$ are defined on the interval $t_3 \in [t_0, t_1]$ on
the upper or lower Keldysh contour for $a = F$ or $B$, respectively
(\ie\ the time argument of $V_a$ is understood to carry an implicit
index $\scriptstyle a$).  By using \Eqs{eq:Hubbard-Stratonovitch} to
rewrite all $\hat U_{Ia}$ in \Eq{eq:rho-interaction} in terms of $\hat
{\cal U}_a$, the reduced density matrix can be expressed as
\bsubequations
\label{eq:reddenmat-Hub}
\begin{eqnarray} \label{eq:densitymHS}
 \tilde {\bm{\rho}}^\pdag_{11'}  (t_1,t_0)  & = &
\langle \tilde \rho^\pdag_{11'} (t_1,t_0)
\rangle_V^\pdag \; , \\
  \label{eq:define-densitymatrix}
  \tilde \rho^\pdag_{11'} (t_1,t_0) & \equiv &
{  {\langle
\hat {\cal U}_B^\dag(t_1,t_0) \, \hat n^\pdag_{11'I} (t_1) \, \hat
{\cal U}_F (t_1,t_0) \rangle_0 } \over Z (t_1,t_0)}  \; ,
\qquad \phantom{.}
\\
Z (t_1,t_0) & \equiv  &  \langle
\hat {\cal U}_B^\dag(t_1,t_0)  \, \hat {\cal U}_F (t_1,t_0) \rangle_0   \; ,
\\
\label{eq:V-average}
  \langle {\cal F}[V_a]  \rangle_V^\pdag & \equiv &
  {\int {\cal D} V_F   \int {\cal D} V_B \,
  {\cal F} [V_a] \, e^{ i S_V^\tot (t_1,t_0)/ \hbar }}
\over
  \int {\cal D} V_F   \int {\cal D} V_B \,
  e^{ i S_V^\tot (t_1,t_0)/ \hbar}
\; ,
\\
\label{eq:S[V_a]}
i S_V^\tot  (t_1,t_0) & \equiv & i(S_V^{0F} - S_V^{0B}) + \hbar \ln Z  \; .
\end{eqnarray}
\esubequations \noindent
In \Eq{eq:densitymHS}, the reduced density matrix $\tilde
{\bm{\rho}}^\pdag_{11'}$ is expressed as a functional average, over all
configurations of the fields $V_a$, of the functional $\tilde
\rho^\pdag_{11'}(t_1,t_0)$. The latter, defined in \Eq{eq:define-densitymatrix}
(and called $\rho_V$ by GZ), is the reduced density matrix corresponding to a
particular configuration of the fields $V_a$.  For any such functional ${\cal
  F}[V_a]$, the functional average is defined by the functional integral
(\ref{eq:V-average}), with an effective action $S_V^\tot$ given by
(\ref{eq:S[V_a]}).  Note that $S_V^\tot$, via its dependence
on $Z$, depends on the source field $\tilde v$.

%\subappendix{Effective Action for Hubbard-Stratonovich Field}

\subappendix{Roadmap for GZ's Strategy}
\label{sec:roadmap}

If, in \Eq{eq:define-densitymatrix} for $\tilde
\rho^\pdag_{11'}(t_1,t_0)$, the evolution operators $\hat {\cal U}_a$
are expanded in powers of the $\hat V_a$'s, the standard Keldysh
perturbation expansion for these correlators would result (as
recapitulated in App.~\ref{app:Keldysh}).  The approach of 
AAG\cite{AAG98} amounts to doing just such an expansion to order $\hat
V_a^2$.  However, such a perturbation expansion has infrared
divergencies which are cured only when the leading divergencies are
summed to all orders (or by introducing an infrared cut-off by hand,
such as an external magnetic field, as done by AAG).  At present, no
exact way of summing the entire perturbation series is known. Already
in 1982, AAK\cite{AAK82} were able to perform a summation of the
leading infrared divergencies by treating $\hat V_a$ as a classical
field; this indeed cured the infrared problems, but neglects the
quantum nature of $\hat V_a$, hence corrections to AAK's calculation
are to be expected at sufficiently low temperatures.

GZ attempted to proceed both beyond AAK's calculation (by including
quantum corrections) and beyond perturbation theory (by summing an
infinite subset of the perturbation series). The essence of their idea
was to integrate out all electron degrees of freedom but one, 
the ``singled-out electron'', thereby
deriving an influence functional describing the effect of the other
electrons (an effective dissipative environment) on the diffusive
motion of the singled-out electron.  To this end, they adopted the
following
strategy,  which we shall repeat below in our own notation:\\
{\bf (A)} \label{p:Blistofsteps}
An exact equation of motion is derived for $\tilde
\rho^\pdag_{11'} (t,t_0)$ [(GZ-II.24), our
(\ref{eq:finalresultfordrhodt})].
\\
{\bf (B)} From this, another exact equation of motion is derived for
the linear response $\delta \tilde \rho^\pdag_{11'} (t_1,t_0)$ to the
source field $\tilde v$ [(GZ-II.39), our
(\ref{eq:GZlinearresponseeomd+inhom})], together with the form of the
effective Hamiltonian $\tilde H_\iiijjj^a$ [(GZ-II.40), our
(\ref{eq:GZlinearresponseall})]
which governs the dynamics of $\delta \tilde \rho^\pdag_{11'} (t_1,t_0)$.   \\
{\bf (C)} This second equation of motion is integrated exactly
[(GZ-II.41), our (\ref{eq:GZlinearresponse-rhovv})] in terms of
effective evolution matrix functions $\tilde U^a_{\iiijjj}(t,t')$
[(GZ-II.42), our (\ref{eq:definetildeU})].
\\
{\bf (D)} A functional derivative of $\delta \tilde \rho^\pdag_{11'}
(t_1,t_0)$ w.r.t.\ to the source field $\tilde v$ is taken to obtain
an expression for the conductivity [(GZ-II.49), our
(\ref{eq:finalGZsigma}) or (\ref{subeq:sigmaDCreal})], which involves
a functional average of the form $\langle \tilde U^F \tilde \rho^0
\tilde U^B \rangle_V$
%integral $\int {\cal D} V_a$
over the fields $V_a$ [\Eqs{eq:defJ(F/B)}], where $\tilde \rho^0
\equiv \tilde \rho_{V_a = 0}$ is the (initial) density matrix in the
absence of interactions.  The purpose of the subsequent steps {\bf
  (E)} to {\bf
  (G)} is to facilitate the evaluation of this functional average.  \\
{\bf (E)} The evolution functions $\tilde U_{\iiijjj}^a$ introduced in
{\bf (C)} are represented as path integrals over the degrees of
freedom of a single electron, whose Hamiltonian depends on the fields
$V_a$.  We shall use a coordinate-space-only path integral $\int
\widetilde {\cal D}' (\bmR)$
[\Eq{eq:def-path-integral-pureRmaintext}], thereby deviating somewhat
from GZ, who use position-momentum space integrals $\int {\cal D}
\bmR^a {\cal D} \bmP^a$ [(GZ-II.44), our \Eq{eq:def-path-integral}].
The relation between GZ's position-momen\-tum and our coordinates-only
path integrals is explained in time-slicing detail in
App.~\ref{app:path-integral}.
\\
{\bf (F)} The action $S_V^\tot$ (more specifically, the term $\ln Z$)
that governs the weights of different configurations of the fields
$V_a$ in the functional average $\langle \tilde U^F \tilde \rho^0
\tilde U^B \rangle_V$, is expanded to second order in $\hat V_a$,
corresponding to the standard RPA approximation
[(GZ-II.30), our Eqs.~(\ref{eq:RPA-approx}) and (\ref{subeq:VAV})]. \\
{\bf (G)} The density matrix $\tilde \rho^\pdag_{ii'} (t_i,t_0)$,
whereever it still occurs, is approximated by its noninteracting
($V_a=0$) version $\tilde \rho^0_{ii'}$.
%{\bf (Is the following
%  necessary; what small parameter controlls this replacement?)}  This
%is equivalent to evaluating $\tilde \rho^\pdag_{ii'} (t_i,t_0)$ for
%$t_0 = t_i$:
%\begin{eqnarray}
%\label{eq:GZmainapproximation-general}
%\tilde \rho^\pdag_{ii'} (t_i,t_0) &
%\; \stackrel{{\rm GZ}}{\longrightarrow} &  \tilde \rho_{ii'}^0 \; \equiv \;
%\tilde \rho^\pdag_{ii'} (t_i,t_0)_{V_a=0} \\
%&  = &
%\tilde \rho^\pdag_{ii'} (t_i,t_i) \; .
%\end{eqnarray}
[GZ make this approximation twice: (i) in the propagators $\tilde
U^a_\iiijjj$, to obtain (GZ-II.43), and (ii) in the initial-time thermal
averaging, to obtain (GZ-II.49); we use the analogue of (i)
[\Sec{sec:rho->rho0}], but do not need (ii).]
\\
{\bf (H)} The functional average $\langle \; \rangle_V$, which through
the approximations {\bf (F)} and {\bf (G)} has been reduced to a
Gaussian functional integral, is performed to yield an effective
action $i \tilde S_R + \tilde S_I$ [(GZ-II.54), (GZ-II.55), our
(\ref{eq:Gaussian-Integral-Ronly}), (\ref{eq:defineSiR})].  This
effective action determines the influence functional of the
environment (the other electrons) on the singled-out electron.
\\
\indent In GZ's paper the above steps are presented in a somewhat
different order: approximation {\bf (F)} is discussed already after
{\bf (B)}, and approximation {\bf (G)} is made directly after {\bf
  (C)}.  We prefer to carry out the steps in the order stipulated
above, because this allows us to postpone each approximation to the
latest possible stage.

The results derived by the above steps are used in
Secs.~\ref{sec:IFvsPT} and \ref{app:1storderperturbation} to make
contact with diagrammatic perturbation theory, and in the main text
[\Sec{sec:GZ-classical-paths}] to calculate the decoherence time. For
the latter, we continue to follow GZ's approach in spirit, but use a
more careful treatment of a certain ``Pauli factor''; remarkably, this
turns out to lead to AAK's result for $\tauphi$ instead of GZ's.
Although the details of this calculation are presented in the main
text, we shall now summarize them here, too, in order that the present
brief overview of GZ's strategy be complete.
\\
{\bf (I)} The term $i \tilde S_R$ in the effective action turns out to
depend on a certain ``Pauli factor'' $(\tilde \delta - 2 \tilde
\rho^0)$, which we treat differently from GZ: In their
position-momentum path integral it is represented as $[1 - 2 n_0
(\tilde h_0)]$, where $\tilde h_0 (\bmR(t), \bmP(t))$ is the
single-particle energy of the singled-out electron, which GZ assume to
remain constant during the diffusive motion. In our opinion, this
assumption neglects recoil effects associated with electron-electron
interactions [see \Sec{sec:ruleofthumb}]. Therefore, we instead use a
Fourier representation of the effective action, in which the Pauli
factor is represented as $[1 - 2 n_0 (\hbar (\ve - \bomega))]$
[\Eq{eq:1-2rho->tanh}], where $\hbar \ve$ corresponds to GZ's $\tilde
h_0$, and $\bomega$ is the frequency transfer upon emission or
absorption of a photon.
\\
{\bf (J)} The path integrals $\int \widetilde {\cal D} \bmR'$ for the
singled-out electron are performed in the saddle point approximation,
meaning that only the contributions of pairs of time-reversed
diffusive (or ``random walk'')
paths are retained. \\
{\bf (K)} The average of the influence functional over all such random
walk paths, namely $\langle e^{- (i \tilde S_R + \tilde S_I)}
\rangle_\rw$, is approximated by the exponentiating the average of the
effective action, $e^{- \langle i \tilde S_R + \tilde S_I \rangle_\rw}
$ [(GZ-II.67), our \Sec{sec:GZ-classical-paths}].
\\
{\bf (M)} The exponent $F (\tau) = \langle i \tilde S_R + \tilde S_I
\rangle_\rw $, a growing function of time, is evaluated by Fourier
transforming the effective action into the frequency-momentum domain
and averaging the Fourier exponents, using $\langle e^{i \bbmq \cdot
  [\bmR (t_3)- \bmR (t_4)]} \rangle_\rw \simeq e^{- \bbmq^2 D
  |t_{34}|} $ [our Eq.~(\ref{eq:impurity-average-of-eikr-A})].
\\
{\bf (L)} The resulting function $F(\tau)$ is used to identify the
decoherence time as the time for which $F(\tauphi)$ becomes of order
unity [(GZ-II.67), (GZ-II.70), or (GZ-III.6) (GZ-III.22), or our
Eq.~(\ref{eq:F1dexplicitfinal})].

\subappendix{Repeating GZ's Strategy in Detail}
\label{sec:repeatGZ}

The remainder of this appendix is devoted to a detailed discussion of
steps {\bf (A)} to {\bf (I)}, using our own notation.
 
\subsubappendix{Exact Equation of Motion for \mbox{$\tilde \rho_{ii'} (t,t_0)$}}

To derive an exact equation of motion for $\tilde \rho^\pdag_{ii'} (t,t_0)$,
we start from the simple relations
\bsubequations
\label{eq:time-derivatives}
\begin{eqnarray}
  \label{eq:eom-psi}
  i \hbar \partial_t \hat \psi^\pdag_I (t, x)
 &=& h_0 (x)  \hat \psi^\pdag_I (t, x) \; ,
\\
  \label{eq:eom-U}
  i \hbar \partial_t \hat {\cal U}^\pdag_{a} (t)
&=& \left[ \hat V_{a} (t) + \hat v_I (t) \right]
     \hat {\cal U}^\pdag_{a} (t) \; .
\end{eqnarray}
\esubequations \noindent
Since all functions in \Eqs{eq:time-derivatives} are evaluated at the same
time $t$, as are all other functions needed below up
to \Eq{eq:newV's},
we shall suppress the time argument below and use the shorthand notation
\bsubequations
\label{eq:shorthands}
\begin{eqnarray}
  \label{eq:shorthand-1}
&&
\tilde \rho^\pdag_{ii'} = \tilde \rho^\pdag_{ii'} (t,t_0) , \qquad \; \,
\hat {\cal U}_a   \equiv \hat {\cal U}_a (t,t_0) \; ,
\qquad
\hat n^\pdag_{ii'} = \hat n^\pdag_{ii' I} (t) \; ,
\\
&& h_{0i}  \equiv h_0 (x_i) , \qquad \, \quad \; 
V^\pdag_{ai}  \equiv V_a (t,\bmr_i) \, , \qquad 
  \label{eq:shorthand-3}
\tilde \delta_{ii'} = \delta^{(d)}
( \bmr_i - \bmr_{i'}) \delta_{\sigma_i \sigma_{i'}} \; . 
\qquad \phantom{.} \qqph
  \end{eqnarray}
\esubequations \noindent From \Eqs{eq:time-derivatives},
we then readily find
\bsubequations
\label{eq:rho-derivatives}
\begin{eqnarray}
  \label{eq:drhodt}
   i \hbar \partial_t  \tilde \rho^\pdag_{ii'}   &=&
 Z^{-1}
\left[ i \hbar \partial_t \langle \hat {\cal U}_B^\dag
 \, \hat n^\pdag_{ii'} \, \hat {\cal U}_F \rangle_0
- \tilde \rho^\pdag_{ii'} i \partial_t %\langle \hat U^\dag_B {\cal U}_F \rangle_0
Z \right] \; ,
\\
 \label{eq:ZdtZ}
 i \hbar \partial_t   Z   &=&
%- Z^{-1} \int d \bmrthree e  \left[ V^\pdag_{Fk} - V^\pdag_{Bk} \right]
%\langle \hat U^\dag_B \hat \psi^\dag_k
  \int d x_k ( e V^\pdag_{Fk} -  e V^\pdag_{Bk} )
\langle \hat {\cal U}_B^\dag  : \!  \hat n^\pdag_{kk} \! :
 \hat {\cal U}_F \rangle_0  \; ,
\\ \nonumber 
 i \hbar \partial_t   \langle \hat {\cal U}_B^\dag
\,  \hat n^\pdag_{ii'} \, \hat {\cal U}_F \rangle_0 &=&
(h_{0i} - h_{0i'})
\langle \hat {\cal U}_B^\dag \,  \hat n^\pdag_{ii'} \, \hat {\cal U}_F
\rangle_0 \\
&  &  \nonumber 
  \phantom{.} \hspace{-2cm} +
   \int \! d x_k  \langle \hat {\cal U}_B^\dag \left[
 \hat n^\pdag_{ii'}  (e V^\pdag_{Fk} : \! \hat n^\pdag_{kk} \! :  )
 - (e V^\pdag_{Bk} : \! \hat n^\pdag_{kk} \! : ) \hat n^\pdag_{ii'}
\right] \hat {\cal U}_F \rangle_0 \qqph
\\ &  &
  \phantom{.} \hspace{-2cm} +
 \int \! d x_{k, \bar k}  v^\pdag_{{\bar k}k}
\langle \hat {\cal U}_B^\dag
[   \hat n^\pdag_{ii'}  ,  \hat n^\pdag_{k{\bar k}} ]
\hat {\cal U}_F \rangle_0 \; .
\end{eqnarray}
\esubequations \noindent
\Eq{eq:drhodt} can be brought into the form
\begin{eqnarray}
\label{eq:finalresultfordrhodt}
 i \hbar \partial_t \tilde \rho^\pdag_{ii'}
&  = &  \left[ h_{0i} + e V^\pdag_{Fi}\right]    \tilde \rho^\pdag_{ii'}
- \tilde \rho^\pdag_{ii'} \left[ h_{0i'} + e V^\pdag_{Bi'} \right]
\\ \nonumber && %\quad
-  \int d x_k  \,
\left[ \tilde \rho^\pdag_{ik}  (e V^\pdag_{Fk} - e V^\pdag_{Bk})
\tilde \rho^\pdag_{ki'} + \tilde \rho^\pdag_{ik}  \tilde v^\pdag_{ki'} -
\tilde v^\pdag_{ik} \tilde \rho^\pdag_{ki'} \right]
\end{eqnarray}
by using the identities
\begin{eqnarray}
\nonumber  \label{eq:identities-psipsi}
&&  \hat n^\pdag_{ii'} \hat n^\pdag_{k{\bar k}} =
  \hat \psi^\dag_{i'} \left[ \tilde \delta^\pdag_{i{\bar k}} -
\hat \psi^\dag_{\bar k} \hat \psi_i \right] \hat \psi_k
 =   \hat n^\pdag_{k{i'}} \tilde \delta^\pdag_{i{\bar k}} +
\hat \psi^\dag_{i'} \hat \psi^\dag_{\bar k} \hat \psi_k  \hat \psi_i
\\
\nonumber
&&  \hat n^\pdag_{k{\bar k}} \hat n^\pdag_{ii'} =
  \hat \psi^\dag_{\bar k} \left[ \tilde \delta^\pdag_{k{i'}} - \hat \psi^\dag_{i'} \hat \psi_k \right]
\hat \psi_i
 =    \hat n^\pdag_{i{\bar k}} \tilde \delta^\pdag_{k{i'}} +
 \hat \psi^\dag_{i'} \hat \psi^\dag_{\bar k} \hat \psi_k   \hat \psi_i \, ,
\\
  \label{eq:Wick}
&&  Z^{-1}
\langle \hat {\cal U}_B^\dag \hat \psi^\dag_{i'}
\hat \psi_{\bar k}^\dag \hat \psi_k \hat \psi_i \hat
  {\cal U}_F \rangle_0  =  \tilde \rho^\pdag_{ii'} \tilde \rho^\pdag_{k{\bar k}} - \tilde \rho^\pdag_{i{\bar k}}
\tilde \rho^\pdag_{k{i'}} \,  .
\end{eqnarray}
The last of these can be checked by expanding both sides in powers of
$\hat V_a$, and evaluating each term in the expansion using Wick's
theorem.  Since $\hat V_a$ is quadratic in $\hat \psi$'s, one readily
finds that the combinatorial factor for each topologically distinct
diagram is just equal to 1, and that the left- and right-hand sides of
\Eq{eq:Wick} generate precisely the same set of topologically distinct
diagrams.

\Eq{eq:finalresultfordrhodt} is the desired equation of motion for
$\tilde \rho^\pdag_{ii'}$. [It reduces to (GZ-II.24) upon setting the
source fields to zero, $\tilde v^\pdag = 0$ and recalling our
footnote~\ref{e<0}.]  The term on the right-hand side of
\Eq{eq:finalresultfordrhodt} that contains a term \emph{quadratic} in
$\tilde \rho$, coupling to $e (V_F - V_B)$, will be seen below to be
responsible for enforcing the Pauli principle.  Note that
\Eq{eq:finalresultfordrhodt} contains \emph{only} $c$-number functions
(no hats occur).  Hence the order of factors in products does not
matter as long as their subscripts are displayed explicitly (the
derivatives contained in the functional operator $h_{0{i'}}$ should be
understood to act on index ${i'}$ of $\tilde \rho^\pdag_{ii'}$ even if
we write them in the order $\tilde \rho^\pdag_{ii'} h_{0{i'}}$).
Nevertheless, the subscripts do imply that the products have the
structure of matrix multiplication in coordinate space; we hence chose
to write the factors in an order that is suggestive of this matrix
multiplication.  [This order conforms to that used by GZ, in whose
notation the coordinate indices are not displayed, but are implicit.]

\subsubappendix{Equation of Motion for $\delta
\tilde \rho_{ii'}(t,t_0)$ }

Next, we expand the reduced density
matrix to linear order in the source field (which is sufficient
for a linear respone calculation of the conductivity) by writing
$\tilde \rho^\pdag_{ii'} = \tilde \rho_{ii'}^\ns + \delta \tilde
\rho^\pdag_{ii'}$, where the superscript (ns) denotes ``no
sources'' and $\delta \tilde \rho^\pdag_{ii'}$ is linear in
$\tilde v$.  It satisfies the following equation of motion, found
by expanding \Eq{eq:finalresultfordrhodt},
\bsubequations
 \label{eq:GZlinearresponseeomd+inhom}
 \begin{eqnarray}
 i \hbar \partial_t \delta \tilde \rho^\pdag_{i i'} & = &
 \tilde D_{i i'}  +  \label{eq:GZlinearresponseeom-F}
\!\!  \int \! d x_{\bari}
 \tilde H^F_{i \bari}   \delta \tilde \rho^\pdag_{\bari i'}  -
\! \! \int \! d x_{{\bari'}}  \,
\delta \tilde \rho^\pdag_{i \bari'}   \tilde H^B_{\bari' \! i'} ,
\qquad \phantom{.}
\\
\label{eq:tildeDH}
\phantom{.} \hspace{-0.5cm}
\tilde {   D}_{ii'}  &  \equiv  &
\!  \int \! d x_{\bari}
\tilde v^\pdag_{i \bari}
\tilde \rho^\ns_{\bari i'} -
 \! \int \! d x_{{\bari'}}  \,
\tilde \rho_{i {\bari'}}^\ns \tilde v^\pdag_{{\bari'} i'} \; ,
\end{eqnarray}
\esubequations
where the  effective Hamiltonians
$\tilde H^F$ and $\tilde H^B$ are defined as follows:\footnote{
Note  that $\tilde H^a_{\ibari}$, like $h_{0i}$,
is a $c$-number functional operator -- the derivatives contained in
$h_{0i} \tilde \delta^\pdag_{\ibari}$ get ``transferred'' onto the
function it multiplies:
\begin{eqnarray}
%\label{eq:how-del-acts}
 \int \! d x_\bari \,
(\bnabla^2_i \tilde \delta^\pdag_{\ibari}) \delta \tilde
\rho^\pdag_{\bari k}
= \bnabla^2_{i} \delta \tilde \rho^\pdag_{ik} \; ,
\quad
 \int \! d x_\bari  \,
\delta \tilde \rho^\pdag_{\ibari} (\bnabla^2_\bari \tilde
\delta^\pdag_{\bari k})
= \bnabla^2_{k}  \delta \tilde \rho^\pdag_{\bari k} \; . \nonumber 
\end{eqnarray}
}
\bsubequations
\label{eq:GZlinearresponseall}
\begin{eqnarray}
\label{eq:tildeHa}
&&
\phantom{.} \hspace{-0.5cm}
\tilde H^F_{\ibari}  \: \equiv \: h_{0 i} \, \tilde \delta^\pdag_{\ibari} +
\tilde h^F_{V\ibari} \; ,
\quad  \; \;
\tilde H^B_{\barii} \,  \equiv \,  \tilde \delta^\pdag_{\barii} \, h_{0 i} +
\tilde h^B_{V\barii} \; , \rule[-4mm]{0mm}{0mm} \qqph
\\
\label{eq:tildeHF} \label{eq:tildeHB}
\label{eq:tildehBF}
&&
\phantom{.} \hspace{-0.7cm}
\begin{array}{rclcl}
\tilde h^F_{V\ibari} & = &
{\displaystyle \tilde \delta_{\ibari} \, e V_{F \bari} - \tilde \rho_\ibari
(e V_{F \bari} - e V_{B \bari}) }
& = &
{\displaystyle
\!\!\! \sum_{\alpha = \pm}
\tilde w^{F\alpha}_{\ibari}  V_{\alpha \bari}} \, ,
\rule[-6mm]{0mm}{0mm}
\\
%\qquad
\tilde h^B_{V\barii} & = &
{\displaystyle
e V_{B \bari} \, \tilde \delta_{\barii}
+ (eV_{F \bari} - eV_{B \bari})  \tilde \rho_\barii }
& = &
{\displaystyle
\!\!\! \sum_{\alpha = \pm}
 V_{\alpha \bari} \tilde w^{B\alpha}_{\barii}  } \, ,
\end{array}
\qph
\\
\label{eq:wvertices}
& &
\phantom{.} \hspace{-0.5cm}
\tilde w^{a +}_\ibari \equiv  e \, \tilde \delta_\ibari \; , \qquad
\tilde w^{a -}_\ibari \equiv s_a
\toh e (\tilde \delta_\ibari - 2 \tilde \rho^\ns_{\ibari} ) . \qquad \phantom{.}
%\tilde h^B_{V\barii} & \equiv & e V_{+\bari} \, \tilde \delta^\pdag_{\barii}
% -   e V_{- \bari} \, \toh (1-2 \tilde \rho^\ns)_{\barii}
%. \qquad \phantom{.}
 \end{eqnarray}
\esubequations \noindent
[Eqs.(\ref{eq:GZlinearresponseall}) correspond to (GZ-II.39,40); their
$-e V_x (\bmr_i)$ corresponds to our $\tilde v_{ii}$.]  In
\Eqs{eq:GZlinearresponseeomd+inhom}, the combination of indices
$\scriptstyle{ i \bari}$ or $\scriptstyle {\bari' \! i'}$, one without
bar, one with, will always refer to two independent position indices
associated with the \emph{same} time (\ie\ $t_\bari \equiv t_i$).  The
Hamiltonians $\tilde H^F$ and $\tilde H^B$ are associated with
propagation along the upper and lower Keldysh contours, which is why
in \Eq{eq:GZlinearresponseeom-F} they are contracted from the
left or right with the first or second index of $\delta \tilde
\rho^\pdag_{i i'}$.
In \Eq{eq:wvertices} for the vertices $\tilde w^{a \alpha}_\ibari$ and
elsewhere below, the symbol $s_a$ stands for ``sign of $a$'',
with $s_{F/B}  = \pm 1$.
The fields $V_{\alpha \bari} = V_{\alpha \bari} (t_\bari, \bmr_\bari)$
(with $\alpha = \pm$) occuring in \Eq{eq:tildehBF} are defined as
symmetric and antisymmetric linear combinations of the fields
$V_{a \bari}$ (\ie\ the time and coordinate arguments of $V_{+ \bari}$
and $V_{- \bari}$ on  the upper and lower
Keldysh contours are \emph{both} equal to ($t_\bari$, $\bmr_\bari$)]:
\begin{eqnarray}
  \label{eq:newV's}
 { V_{+\bari} \choose V_{-\bari} }
\equiv \left(\begin{array}{cc}
  1/2   & 1/2 \\
  1 & -1
\end{array} \right)
{ V_{F \bari} \choose V_{B \bari} }\; .
\end{eqnarray}

Since both $H^F_{ \ibari}$ and $H^B_{\barii}$ depend, through
$V_{\alpha \bari}$, on \emph{both} $V_{F \bar i}$ and $V_{B \bari}$,
crossterms will occur below that link the forward and backward Keldysh
contours.  Note  that the field $V_{\alpha \bari}$, which
shall always carry a ``barred'' index below, is contracted with the
second index of $w^{F\alpha}_{\ibari}$ in $\tilde h^{F}_{i \bari}$ or
the first index of $\tilde w^{B\alpha}_{\barii}$ in $\tilde
h^{F}_{\bari i}$, respectively. Thus $V_-$ and $\tilde w^{a-}$ ``do not
commute'', which will be important below. The factor $(\tilde \delta - 2
\tilde \rho)$ in $\tilde w^{a-}$ will be seen below to account for the
Pauli principle.  
%Note also that $\tilde H^a_{\ibari}$ is not hermitian (i.e.\ $\tilde
%H^{a \dagger}_{\ibari} \equiv \tilde H^{a \ast}_{\bari i} \neq \tilde H^{a
%\pdag}_{\ibari}$), because the arguments of the field $V_{- }$ are
%$\bmr_\bari$ or $\bmr_i$ (if instead they had been  $(\bmr_i +
%\bmr_\bari)/2$, $\tilde H^a_{\ibari}$ \emph{would} have been hermitian).

All functions occuring in \Eqs{eq:GZlinearresponseall} depend on
the same time argument $t$, which we henceforth display explicitly
again.  It is worth emphasizing that, through their dependence on
$\tilde \rho^{\rm (ns)}_{\ibari} (t,t_0)$, the expressions
$\delta \tilde \rho^\pdag_{ii'}$, $\tilde D^\pdag_{\ibari}$, $\tilde
H^a_{\ibari}$ and $\tilde U^a_{\iiijjj}$ [defined below in
\Eqs{eq:definetildeU}] all explicitly depend on the initial time
$t_0$, too, although, for notational brevity, this $t_0$-dependence
will be displayed below only for $\tilde \rho^{\rm ns}_{\ibari}
(t,t_0)$.

\subsubappendix{Exact Expression for $\delta \tilde \rho_{ii'} (t,t_0)$}

The formally exact solution of \Eqs{eq:GZlinearresponseeomd+inhom} can
be written in the form
\begin{eqnarray}
  \delta \tilde \rho^\pdag_{ii'} (t)  & = & - i \! \int_{t'_0}^t
\! \! dt' \! \!
\int \! \! d x_{k, \bar k}  \,
\tilde U^{F}_{ik} (t,t')  \tilde {   D}_{k \bar k} (t')
 \tilde U^{B}_{\bar k i'} (t',t)  ,
\label{eq:GZlinearresponse-rhovv}
\end{eqnarray}
where  the functions $\tilde U^F_{\iiijjj}(t,t')$ and
$\tilde U^{B}_{ji}(t',t)$ are defined by the requirements that
\bsubequations
\label{eq:tildeUconditions}
\begin{eqnarray}
\label{eq:tildeU-bound}
\tilde U^F_{\iiijjj}(t,t) & = & \tilde U^B_{ji}(t,t) \, =  \,
   \tilde \delta^\pdag_{\iiijjj} \; ,
\\
\label{eq:tildeU-diff-F}
 i \hbar \partial_t \tilde U^F_{\iiijjj}(t,t') & = & \phantom{-} \int \! dx_\bari \,
\tilde H^F_{i\bari} (t) \tilde U^F_{\bari j} (t,t') \; ,
\\
\label{eq:tildeU-diff-B}
 i \hbar \partial_t \tilde U^{B}_{ji}(t',t) & = & - \int \! dx_\bari \,
\tilde U^{B\pdag}_{j\bari} (t',t) \, \tilde H^B_{\barii} (t)  \; .
\end{eqnarray}
\esubequations \noindent
\Eqs{eq:tildeUconditions} are fulfilled by time-ordered
exponentials% \cite{UFUBUaallequal},
\bsubequations
\label{eq:definetildeU}
\begin{eqnarray}
\label{eq:Ftime-ordered-exponential}
\tilde U^F_{\iiijjj}(t,t')    & = &
\Bigl[ {\cal T} e^{-{i \over \hbar} \int_{t'}^t d t_3  \tilde H^F (t_3)
}\Bigr]_\iiijjj \; ,
\\ & & \phantom{.} \hspace{-2cm}
\nonumber \label{eq:Ftime-ordered-power-series}
\equiv \tilde \delta^\pdag_{\iiijjj}
-{i \over \hbar} \int_{t'}^t \! \! d t_3 \, \tilde H^F_{\iiijjj} (t_3)
- {1 \over \hbar^2}
\int_{t'}^t \! \! d t_3 \! \int_{t'}^{t_3} \! \! d t_4 \!
  \int \! d x_k \, \tilde H^F_{ik} (t_3) \tilde H^F_{kj} (t_4) + \dots
\\
\tilde U^{B}_{ji}(t',t)
\label{eq:Btime-ordered-exponential}
& = & \Bigl[ \overline {\cal T} e^{{i\over \hbar}
 \int_{t'}^t d t_3  \tilde H^B (t_3) } \Bigr]_{ji} \; ,
\\ \nonumber \label{eq:Btime-ordered-power-series} 
& & \phantom{.} \hspace{-2cm}
\equiv \tilde \delta^\pdag_{\iiijjj}
+ { i \over \hbar} \int_{t'}^t \! \! d t_3 \, \tilde H^B_{ji} (t_3)
 - { 1\over \hbar^2}
\int_{t'}^t \! \! d t_4 \! \int_{t'}^{t_4} \! \! d t_3 \!
  \int \! d x_k \, \tilde H^B_{jk} (t_3) \tilde H^B_{ki} (t_4) + \dots
\end{eqnarray}
\esubequations \noindent
where we always take $t>t'$, and where each ``internal'' product of
two factors $\tilde H^a_{ik} \tilde H^a_{kj}$ that arises when
expanding the exponential involves a further coordinate integral $\int
dx_k$.  [Below, we shall often suppress time arguments and use the
short-hand $\tilde U^a_\iiijjj \equiv \tilde U^a_\iiijjj (t_i, t_j)$ and
likewise $\tilde U^a_{\bari \! j} \equiv \tilde U^a_{\bari \! j} (t_i,
t_j)$.]  Note that the time-ordered exponentials
(\ref{eq:Ftime-ordered-exponential}) and
(\ref{eq:Btime-ordered-exponential}) for $\tilde U^F_\iiijjj$ and $\tilde
U^B_\iiijjj$ are \emph{defined} in terms of
the power series expansions indicated 
above;
%(\ref{eq:Ftime-ordered-power-series}) and
%(\ref{eq:Btime-ordered-power-series}) 
the same is true for all path integral representations of
$\tilde U^F_\iiijjj$ and $\tilde U^B_\iiijjj$ to be used below.  Note also
that $\tilde U^a_{\iiijjj}$ is spin-diagonal, since this is the case for
$\tilde H_{\ibari}^a(t) = \delta_{\sigma_i \sigma_\bari} \tilde H^a
(t, \bmr_i, \bmr_\bari)$ and $\tilde \rho_{\ibari}(t,t_0) =
\delta_{\sigma_i \sigma_\bari} \tilde \rho (t,t_0; \bmr_i,
\bmr_\bari)$.

\Eq{eq:GZlinearresponse-rhovv} corresponds to GZ's exact
Eq.~(GZ-II.41).  Note that the procedure by which
it was obtained, namely, first differentiating $\tilde \rho$ and then
integrating $\delta \tilde \rho$ w.r.t.\ $t$, has produced a result in which
the reduced density matrix $\rho^{\rm (ns)}$ \emph{appears in the
  exponent}, via its occurence in $\tilde H^a_{\iiijjj}$ and $\tilde U^a_{\iiijjj}$.
Accordingly, the effective action to be derived below
will likewise depend on $\rho^{\rm (ns)}$.
%Approximations involving $  \rho^{\rm (ns)}$ thus have ``exponentially
%severe consequences''.

Let us now also derive an equation of motion for the time evolution of
the density matrix in the \emph{absence} of source fields, $\tilde
\rho^\ns_{\iiijjj'}$, since we need it in \Eq{eq:GZlinearresponse-rhovv},
where it enters via the $\tilde D$ of \Eq{eq:tildeDH}.  [This point is
not discussed by GZ, who simply replace $\tilde \rho^\ns_{\iiijjj'}$ in
\Eq{eq:GZlinearresponse-rhovv} by $\tilde \rho^0_{\iiijjj'}$, as discussed
in Section~\ref{sec:rho->rho0}.]  Evidently, the desired equation of
motion for $\tilde \rho^\ns_{\iiijjj'}$ is the $\tilde v = 0$ version of
that of $\tilde \rho_{\iiijjj'}$, namely
\Eq{eq:finalresultfordrhodt}${}_{\tilde v = 0}$, which can, in analogy
to \Eq{eq:GZlinearresponseeom-F} (without its first term), be
rewritten as
 \begin{eqnarray}
 \label{eq:GZlinearresponseeomd-ns}
 i \hbar \partial_t  \tilde \rho^\ns_{\iiijjj'} & = &
 \int \! d x_{\bari}
 \tilde H^{\prime F}_{i \bari}  \tilde \rho^\ns_{\bari j'}  -
\! \! \int \! d x_{{\barj'}}  \,
\tilde \rho^\ns_{i{\barj'}}   \tilde H^{\prime B}_{{\barj'} j'} .
\qquad \phantom{.}
\end{eqnarray}
Here the primed Hamiltonians $\tilde H^{\prime a}_{\ibari}$ are
defined by equations identical to \Eqs{eq:GZlinearresponseall} for the unprimed
ones, except that the vertices $\tilde w^{a \alpha}_\ibari $ of
\Eq{eq:wvertices} are replaced by primed vertices $\tilde w'^{ a
  \alpha}_\ibari $ that are defined as follows:\footnote{The reason for the extra $y^a$ in front of $\tilde \rho^\ns_{\ibari}$
for $\tilde w'^{a -}_\ibari$, which is the only difference compared to
$\tilde w^{a -}_\ibari $ of \Eq{eq:wvertices}, is as follows: The
linear response equation of motion for $\delta \tilde \rho^\pdag_{i
  i'}$ contains two \emph{different} contributions that are quadratic
in $\tilde \rho$, namely $\tilde \rho^\ns_{ik} e V^\pdag_{-k} \delta
\tilde \rho^\pdag_{ki'}$ and $\delta \tilde \rho^\pdag_{ik} e
V^\pdag_{-k} \tilde \rho^\ns_{ki'}$, which in
\Eq{eq:GZlinearresponseeom-F} were grouped with the first and second
terms of respectively.  In contrast, the equation of motion for $
\tilde \rho^\ns_{\iiijjj'}$ turns out to contain on the right-hand side
just one type of term quadratic in $\tilde \rho$, namely $\tilde
\rho^\ns_{ik} e V^\pdag_{-k} \tilde \rho^\ns_{ki'}$, with total weight
1.  By using $y^F + y^B = 1$ in \Eq{eq:definetildewa-prime}, we have
distributed this term with weights $y^F$ and $y^B$ among the two terms
on the right-hand side of \Eq{eq:GZlinearresponseeomd-ns}.}
% Book 7, p. 18
\begin{eqnarray}
  \label{eq:definetildewa-prime}\tilde w'^{a +}_\ibari
\equiv e \, \tilde \delta_{\ibari} \; , \qquad
\tilde w'^{a -}_\ibari \equiv s_a
\toh e ( \tilde \delta_\ibari - y^a 2  \tilde \rho^\ns_{\ibari} )
\;  . \qquad
\phantom{.}
\end{eqnarray}
Here the $y^{F/B} \in [0,1]$ are (arbitrary) real numbers, with $y^F +
y^B \equiv 1$. It will turn out below to be convenient to let the
choice of values for $y^{\tilde a}$ depend on which contour the
current vertex at time $t_{2_{\tilde a}}$ is located: if it is on
contour $F/B$, we shall choose $y^{F/B} = 0 = 1 - y^{B/F}$ (compactly:
$y^{\tilde a} = 0$ for $\hat \bmj (t_{2_{\tilde a}})$ on contour
$\tilde a$; Fig.~\ref{fig:J122221} below shows an example with $\tilde
a = F$).
%[The factor $y^a$ is the only difference\cite{missingfactoror2}
%compared to $\tilde w^{a -}_\ibari $ of \Eq{eq:wvertices}.]  
% The reason
%for not making a definite choice for $y^a$ at this point is that it
%will be convenient later to pick $y^F =0$, $y^B =1$ on some occasions,
%but $y^B = 0$, $y^F =1$, on others (namely, for all terms for which
%the current vertex at time $t_2$ sits on contour $\tilde a (=F/B)$, we
%shall choose $y^{\tilde a} = 0$).
  The solution of \Eq{eq:GZlinearresponseeomd-ns} can be expressed as
\bsubequations
  \label{subeq:tilderhonsUU-ns}
\begin{eqnarray}
  \label{eq:tilderhonsUU-ns}
  \tilde \rho^\ns_{\iiijjj'} (t,t_0) & = & \tilde U^{\prime F}_{ik} (t,t_0)
\tilde \rho^0_{k \bar k} \tilde U^{\prime B}_{\bar k \! j'} (t_0, t)
\; ,
\qquad \mbox{with} \quad y^F + y^B =1 \; ,
\\
  \label{eq:tilderhonsUU-ns-twochoices}
& = & \left\{
\begin{array}{ll}
\Bigl.
\tilde U^{\prime F}_{ik}(t,t_0) \Bigr|_{y^F = 0}
\tilde \rho^0_{k \bar k}
\tilde U^{B}_{\bar k \! j'} (t_0, t) & \; \mbox{if} \quad y^F = 0,  \;
y^B =1 \; , \rule[-6mm]{0mm}{0mm}
\\
\Bigl. \tilde U^{F}_{ik} (t,t_0)
\tilde \rho^0_{k \bar k}
\tilde U^{\prime B}_{\bar k \! j'} (t_0, t) \Bigr|_{y^B=0}
& \; \mbox{if} \quad y^B =0,
\; y^F = 1  \; ,
\end{array}
\right.
\end{eqnarray}
\esubequations
%where both versions of \Eq{eq:tilderhonsUU-ns-twochoices} will be used
%below, depending on which is more convenient. 
The primed propagators $\tilde U^{\prime a}_\iiijjj$ are defined
analogously to \Eqs{eq:definetildeU}, but with $\tilde H \to \tilde
H'$ everywhere.  In \Eqs{subeq:tilderhonsUU-ns} we have implemented
the standard initial condition for the Keldysh approach, namely that
at time $t = t_0$, the density matrix was free, \ie\ $\tilde
\rho_{\iiijjj'}^\ns (t_0, t_0) \equiv \tilde \rho_{\iiijjj'}^0$.  Below [cf.\ 
\Eq{eq:defineJ1231-00}], we shall insert
\Eqs{eq:tilderhonsUU-ns-twochoices}, with $t_0 \to - \infty$, into
\Eq{eq:GZlinearresponse-rhovv}, where it enters via the $\tilde D$ of
\Eq{eq:tildeDH}, to ensure that thermal averaging is done in the
infinite past. This is an important improvement relative to an
approximation used by GZ, who simply replace $\tilde \rho^\ns_{\iiijjj'}
(t,t_0) $ in \Eq{eq:GZlinearresponse-rhovv} by $\tilde \rho^0_{\iiijjj'}$;
they thereby effectively perform thermal averaging with a
nonequilibrium initial density matrix, as discussed in
Section~\ref{sec:rho->rho0}.

The way in which $\tilde U^F_{\iiijjj}$, $\tilde U^{B}_\iiijjj$ and $\tilde
\rho^\ns_\iiijjj$ differ from their free versions is evidently through
their dependence on the fields $V_a$ and the density matrix $\tilde
\rho_{\iiijjj}$ in Eqs.~(\ref{eq:GZlinearresponseall}). Let us now briefly
discuss their free versions.  First, in the absence of all
interactions the expression for the reduced density matrix $\tilde
\rho^\ns_\iiijjj$ reduces to the form 
%[see App.~\ref{app:explicit-expressions} for details]
\bsubequations
\label{eq:exact-rho0}
\begin{eqnarray}
\label{eq:exact-rho0-a}
\tilde  \rho^0_\iiijjj  & = &
 \sum_\lambda  \psi_\lambda (x_i) \, 
\psi_\lambda^\ast (x_j) \, n_0 (\xi_\lambda)
= (\tilde  \rho^{0}_{ji } )^\ast \; , 
%\\
%&=& \label{eq:rho-U-relation} \int \frac{d \omega}{2 \pi} n_0
%(\hbar \omega) \, \tilde U_\iiijjj^0 (\omega) \; ,
\end{eqnarray}
\esubequations
where $n_0 (\xi_\lambda) = [e^{\xi_\lambda /T} + 1]^{-1}$ is the Fermi
function, and $\psi_\lambda (x_i)$ are the exact single-particle
eigenfunctions of $\tilde h_{0i}$, with eigenvalues $\xi_\lambda$
[cf.\ Eq.~(\ref{eq:definepsifields})]. Next, let $\tilde U^{0a}_{\iiijjj}$
denote the propagator to which $\tilde U^a_{\iiijjj}$ reduces in the
absence of interactions, \ie\ for $V_{ai} = 0$ in
Eqs.~(\ref{eq:GZlinearresponseall}) [so that $\tilde H_{\iiijjj}^a =
h_{0i} \tilde \delta_{\iiijjj}$]. Its explicit form is easily found by
constructing an object satisfying the defining
\Eqs{eq:tildeUconditions} for $V_{ai} = 0$; the result is independent
of whether $a=F$ or $B$, and given by:
%\bsubequations  \label{eq:exact-U0}
\begin{eqnarray}
\tilde  U^{0a}_{\iiijjj}  \equiv \tilde  U^{0}_{\iiijjj}
&= & \sum_\lambda \psi_\lambda (x_i) \, \psi^\ast_\lambda (x_j)  \,
e^{-i \xi_\lambda t_\iiijjj / \hbar }
\label{eq:exact-U0-RA}
 \; = \;  i  \hbar (\tilde G^R_{\iiijjj} - \tilde G^A_{\iiijjj})\; , \qqph
\end{eqnarray}
where $\tilde G^{R/A}_{\iiijjj} = \pm \theta (\pm t_\iiijjj) (\tilde G^>_{\iiijjj}
- \tilde G^<_{\iiijjj})\; $ are the standard free retarded and advanced
electron Green's functions, with 
\begin{eqnarray}
\label{eq:defineG<} \label{eq:defineG>}
\mp i \hbar \,  \G_{\iiijjj}^{</>}  & \!\!  \equiv \!\! &  
\left\{ \!\! \begin{array}{c}
 \langle \hat \psi_I^\dag
(t_j,x_j)  \hat \psi^\pdag_I (t_i,x_i) \rangle_0 \rule[-3mm]{0mm}{0mm}
\\
\langle \hat \psi^\pdag_I (t_i,x_i) \hat \psi_I^\dag (t_j,x_j)
\rangle_0 \end{array} \!\! \right\}
= \sum_\lambda \psi_\lambda (x_i) \psi_\lambda^\ast (x_j) 
e^{-i \xi_\lambda t /\hbar} n_0 (\pm \xi_\lambda)  \; . \qqph \qph
%  \G_{\iiijjj}^<  & \equiv  &  {i\over \hbar}
% \langle \hat \psi_I^\dag
%(t_j,x_j)  \hat \psi^\pdag_I (t_i,x_i) \rangle_0 \; , 
%%\\
%\label{eq:defineG>}
%\quad   \G_{\iiijjj}^> \equiv % & \equiv &
%-{i\over \hbar}
%\langle \hat \psi^\pdag_I (t_i,x_i) \hat \psi_I^\dag (t_j,x_j)
%\rangle_0 \; . 
\end{eqnarray}
 It follows that for a given time
order, as occurs under a time-ordered integral, $\tilde U^{0a}_\iiijjj$ is
equal to \emph{either} a retarded \emph{or} an advanced Green's
function; \eg, for $t_i > t_j$, we have
$  \tilde U^{0F}_{\iiijjj} = i \hbar \tilde G^R_\iiijjj$ and 
$   \tilde U^{0B}_{ji} = - i \hbar \tilde G^A_{ji} $.
%\begin{eqnarray}
%  \label{eq:UabecomesRA}
%  \tilde U^{0F}_{\iiijjj} = i \hbar \tilde G^R_\iiijjj, \qquad
%  \tilde U^{0B}_{ji} = - i \hbar \tilde G^A_{ji} .
%\end{eqnarray}
%\esubequations
Nevertheless, it will be useful to generally retain both terms in
\Eq{eq:exact-U0-RA}, because that allows expressions involving the
free reduced density matrix to be simplified by Fourier transforming
from the time to the frequency domain: for example, denoting the
frequency Fourier transform of $\tilde U^0_{\iiijjj} (t)$ by $\tilde
U_{\iiijjj}^0 (\omega) $, we immediately find the exceedingly useful
relations:
\bsubequations
  \label{subeq:G<GK}
\begin{eqnarray}
  \label{eq:rhoUomegamaintext}
\int d x_{\bari}  \, 
\tilde  \rho^0_{i \bari}\,  \tilde U_{\bari j}^0 (\omega) & = &  
 n_0 (\hbar \omega) \, \tilde U_\iiijjj^0 (\omega) = - i \hbar \tilde
 G^<_\iiijjj (\omega) \; , 
\\
  \label{eq:GKsummary}
\int d x_{\bari}  \, 
[\tilde \delta - 2 \tilde  \rho^0]_{i \bari}\,  \tilde U_{\bari j}^0 (\omega) & = &  
[1 - 2 n_0 (\hbar \omega/2T)]
\, \tilde U_\iiijjj^0 (\omega) =  i \hbar \tilde
 G^K_\iiijjj (\omega)  , \qqph
\end{eqnarray}
\esubequations
where $ G^K_\iiijjj = \tilde G^>_{\iiijjj} + \tilde G^<_{\iiijjj} $ is the Keldysh
function. Note, in particular, that by passing to the frequency
represantion, the Pauli factor $(\tilde \delta - 2 \tilde \rho^0)$ in
\Eq{eq:GKsummary} gets mapped onto $[1 - 2 n_0 (\hbar \omega/2T)] =
\tanh (\hbar \omega/2T) $, a fact that will be very useful in
\Sec{sec:ruleofthumb} below.

For future reference, we note that when the matrix propagators $\tilde
U^F_{\iiijjj}$ and $\tilde U^{B}_{\iiijjj}$ [with $t > t'$] are expanded in
powers of $\tilde h^a_V$ [\ie\ in powers of the fields $V_\alpha$, see
Eqs.~(\ref{eq:tildeHF})], they take the form of time-ordered or
anti-time-ordered power series, respectively:
\begin{eqnarray}
 \left. \begin{array}{r}
 \tilde U^{F}_{\iiijjj} (t,t') \rule[-4mm]{0mm}{0mm} \\
 \tilde U^{B}_{ji} (t',t)
\end{array} \right\} \! \!\! & = & \!\!\!
  \sum_{N=0}^\infty { 1 \over \hbar^N}
 \int_{t_j}^{t_i} \!\! d t_1
\!\! \int_{t_j}^{t_1} \! \!d t_2
 \dots
\!\! \int_{t_j}^{t_{N-1}} \! \! \!\! d t_N
\!\! \int \! dx_{1,\bar1} dx_{2,\bar 2} \dots dx_{N, \bar N}
\nonumber 
\\ \label{eq:PIexpandedinfullmaintext}
& &  \times
\left\{ \!\! \begin{array}{l} (-i)^N
\tilde U^{0F}_{i1} \tilde h^F_{V 1 \bar 1}  \tilde U^{0F}_{\bar 12}
\dots \tilde h^F_{V N \bar N}  \tilde U^{0F}_{\bar Nj} 
\rule[-4mm]{0mm}{0mm}
\\
(+i)^N
\tilde U^{0B}_{j \bar N} \tilde h^B_{V \bar N N}
\dots \tilde U^{0B}_{2 \bar 1} \tilde h^B_{V \bar 1 1} \tilde U^{0B}_{1i}
\end{array} \right. \; . \quad \hspace{-8cm} \phantom{.}
\end{eqnarray}
These expansions [illustrated in App.~D by the third
row of Fig.~\ref{fig:timeslice}] are alternative but equivalent to 
those of \Eqs{eq:definetildeU}, and, just as the latter, 
can be regarded as formal \emph{definitions} of
$\tilde U^a_\iiijjj$, and of all path-integral representations thereof to
be used below.  Note that for each occurrence of a ``vertex'' $ \tilde
h^{F}_{V l_F \bar l_F}$ or $ \tilde h^{B}_{V \bar l_B l_B}$, the
vertex coordinates $x_{l_a}$ and $x_{\bar l_a}$ are both associated
with the \emph{same} time $t_l$, and both are integrated over in $\int
d x_{l, \bar l}$ [cf.\ \Eq{eq:integratenobarwithbar}]. 
This need for a double position integral at each vertex is a 
direct consequence of the fact that the effective
Hamiltonians $\tilde H^a_\iiijjj$ of \Eqs{eq:GZlinearresponseall}
are nonlocal in space. Since the
integrals in \Eq{eq:PIexpandedinfullmaintext} are time-ordered, each
$\U^{0F}$ occuring in $\U^F$ can be replaced by $i \hbar \G^R$, and
each $\U^{0B}$ in $\U^B$ by $- i \hbar \G^A$ [see
\Eq{eq:PIexpandedinfullGRGA}]. Indeed, the latter replacements are, in
effect, used in the path integral representation of $\U^a$ to be
introduced below [\Eq{eq:def-path-integral-pureRmaintext}].  We have
nevertheless chosen to write \Eq{eq:PIexpandedinfullmaintext} in terms
of $\U^{0a}$ functions, as a reminder that the density matrices
occuring in the interaction vertices $\tilde h^a_{V}$ can be converted
to Fermi functions using \Eqs{subeq:G<GK}.

%
%\bsubequations
%\begin{eqnarray}
%\label{eq:expandtildeU-F}
%\tilde U^F_{\iiijjj} (t_i,t_j)
%& = &
%\sum_{N=0}^\infty {(-i)^N \over \hbar^N}
%\int_{t_j}^{t_i} \!\! d t_1
%\int_{t_j}^{t_1} \! \!d t_2
% \dots
%\int_{t_j}^{t_{N-1}} \! \! d t_N
%\int \! dx_{1,\bar1} dx_{2,\bar 2} \dots dx_{N, \bar N} \,
%\tilde U^{0F}_{i1} \tilde h^F_{V 1 \bar 1}  \tilde U^{0F}_{\bar 12}
%\dots \tilde h^F_{V N \bar N}  \tilde U^{0F}_{\bar Nj} \; ,
%\\
%\label{eq:expandtildeU-B}
%    \label{eq:Uaexpansion-F}
%\tilde U^B_{ji} (t_j,t_i)
%& = &
%\sum_{N=0}^\infty {i^N \over \hbar^N}
%\int_{t_j}^{t_i} \!\! d t_1
%\int_{t_j}^{t_1} \! \!d t_2
% \dots
%\int_{t_j}^{t_{N-1}} \! \! d t_N
%\int \! dx_{1,\bar1} dx_{2,\bar 2} \dots dx_{N, \bar N} \,
%\tilde U^{0B}_{j \bar N} \tilde h^B_{V \bar N N}
%\dots \tilde U^{0B}_{2 \bar 1} \tilde h^B_{V \bar 1 1}
%\tilde U^{0B}_{1i} \; .
%\end{eqnarray}
%\esubequations

\subsubappendix{Exact Expression for Conductivity}

The density-density commutator $\tilde {\cal C}_{[11',22']}$
[\Eqs{eq:define-C-firsttime}, (\ref{eq:functional-integrals}),
(\ref{eq:densitymHS}) and (\ref{eq:GZlinearresponse-rhovv})] can now
be obtained by taking the functional derivative of $ \delta \tilde
\rho^\pdag_{12} (t) $ with respect to the source field $\tilde v$
[occuring in (\ref{eq:GZlinearresponse-rhovv}) via $\tilde D$ of
(\ref{eq:tildeDH})].  Henceforth writing $t \equiv t_1$ and $t' \equiv
t_2$, the result can be written as
\begin{eqnarray}
  \label{eq:C1122afterderivatives}
    \tilde {\cal C}_{[11',22']} (t_1 - t_2)
& = &  
\sum_{\tilde a = F,B} \tilde J^{(\tilde a)}_{12',21'} (t_1, t_2; t_0)
\; + \;     \tilde {\cal C}_{[11',22']}^\Hartree \; ,
\end{eqnarray}
where $\tilde {\cal C}_{[11',22']}^\Hartree $ is a contribution
irrelevant for weak localization, which will be dropped
 henceforth.\footnote{\label{f:Hartree}
  The term $ \tilde {\cal C}_{[11',22']}^\Hartree $ in 
\Eq{eq:C1122afterderivatives} has the form
\bsubequations
\begin{eqnarray}
\nonumber %  \label{eq:C1122disconnected}
\tilde {\cal C}_{[11',22']}^\Hartree  % (t_1 - t_2)
 & =  &  
% & & \phantom{.} \hspace{-2cm} \nonumber
i  \langle \tilde {\bm{\rho}}^\pdag_{12} (t_1,t_0)
{ \delta \ln Z \over \delta \tilde v_{2'2} (t_2)}
\rangle_{V, {\rm ns}} 
%\\ \nonumber & & 
\; -  \;
i \langle \tilde {\bm{\rho}}^\pdag_{12} (t_1,t_0) \rangle_{V, {\rm ns}}
\, \langle { \delta \ln Z \over \delta \tilde v_{2'2} (t_2)}
\rangle_{V, {\rm ns}} , \qqph \qph
\\   \nonumber % \phantom{.} \hspace{-4cm}
{ \delta \ln Z \over \delta \tilde v_{2'2} (t_2)} &  = & 
-i {1 \over \hbar Z}
\Bigl[ \langle \hat {\cal U}_B^\dagger (t_1,t_0)
\hat {\cal U}^\pdag_F(t_1, t_2) \hat n_{22'I} (t_2)
\hat {\cal U}^\pdag_F(t_2,t_0) \rangle_0 
\\  \nonumber 
& & \qqph -
\langle \hat {\cal U}_B^\dagger (t_2,t_0) \hat n_{22'I} (t_2)
\hat {\cal U}_B^\dagger
(t_1, t_2)  \hat {\cal U}^\pdag_F(t_2,t_0) \rangle_0
 \Bigr] \; ,
\end{eqnarray}
\esubequations
and arises since the effective action $S_V^\tot$ of
Eq.~(\ref{eq:S[V_a]}) in the functional average (\ref{eq:V-average})
depends, via $\ln Z$, on $\tilde v$ too.  $ \tilde {\cal
  C}_{[11',22']}^\Hartree $ corresponds to (GZ-II.47) and is neglected
by GZ [see the discussion after (GZ-II.47)], because in the absence of
interactions, it vanishes entirely, and hence does not contribute to
the weak localization correction to the conductivity (in other words,
$ \tilde {\cal C}_{[11',22']}^\Hartree $ is irrelevant to the question
how this correction is affected by interactions). We shall not
consider it further either, since in diagrammatic terms it corresponds
to Hartree contributions to the electron Green's functions, which
merely renormalize the magnitude of the conductivity (and were
neglected by AAG\cite{AAG98}, too).  }
The $ \tilde J^{(\tilde a)}$'s are defined in terms of the correlator
\bsubequations
\label{subeq:defineJ1231}
\begin{eqnarray}
   \label{eq:defineJ1231}
 \tilde {   J}^V_{12', 2 \bar 2' , {\bar 2}1'} (t_1, t_2;t_0) & \equiv &
{1 \over \hbar}
 \tilde U^F_{12'} (t_1, t_2) \tilde \rho^\ns_{2 \bar 2'} (t_2,t_0)
 \tilde U^{B}_{\bar 2 1'} (t_2, t_1)\;  \qqph 
\\
&  &  \phantom{.} \hspace{-3.0cm} 
\label{eq:defineJ1231-00}
= {1 \over \hbar} \int d x_{0, \bar 0} \,
 \tilde U^F_{12'} (t_1, t_2) \,
 \tilde U'^F_{20} (t_2, t_0) \,
\tilde \rho^0_{0 \bar 0}  \,
 \tilde U'^{B}_{\bar 0 \bar 2'} (t_0, t_2) \,
 \tilde U^{B}_{\bar 2 1'} (t_2, t_1)\; , \qqph 
\end{eqnarray}
\esubequations
[the second line follows via \Eq{eq:tilderhonsUU-ns}] by:
%for $\tilde \rho^\ns_{2 \bar 2'} (t_2,t_0)$], by:
\bsubequations
\label{eq:defJ(F/B)}
\begin{eqnarray}
\label{eq:defJ(F)}
 \tilde J^{(F)}_{12',21'} (t_1, t_2; t_0) & \equiv &
\phantom{-}
  \int \!  {d x_{\bar 2} } \,
\Bigl\langle \tilde J^V_{12',2 \bar 2,{\bar 2}1'} (t_1, t_2; t_0)
\Bigr\rangle_{V, {\rm ns}}  \\ \nonumber 
&  = & 
\phantom{-} 
 \int  \! d x_{0, \bar 0} \, {1  \over \hbar}
\Bigl\langle
\tilde U^F_{12'}\,   \tilde U^{\prime F}_{20}
 \, \tilde \rho^0_{0\bar0}
\, \tilde U^B_{\bar 0 1'}
\Bigr\rangle^{y^F=0}_{V,{\rm ns}} \; ,
\end{eqnarray}
\begin{eqnarray}
%\\
\label{eq:defJ(B)}
 \tilde J^{(B)}_{12',21'} (t_1, t_2; t_0) & \equiv &
-   \int \! {d x_{\bar 2} } \,
\Bigl \langle  \tilde J^V_{1 \bar 2, \bar 2 2',2 1'}  (t_1, t_2; t_0)
\Bigr\rangle_{V, {\rm ns}}  \\ \nonumber
&  = & 
-  \int \!  d x_{0, \bar 0}
{1 \over \hbar} \, \Bigl\langle
\tilde U^{F}_{1 0}
\, \tilde \rho^0_{0 \bar 0} \,
\tilde U'^B_{\bar 0 2'}
\, \tilde U^B_{2 1'}
\Bigr\rangle^{y^B=0}_{V,{\rm ns}} \; .
 \qquad  \phantom{.}
\end{eqnarray}
\esubequations
%and the correlators $\tilde J(t_1, t_2; t_0)$ and $\tilde
%J^{(F/B)}(t_1, t_2; t_0)$ are to be considered only for $t_0 < t_2 <
%t_1$.
$\tilde J^{(F)}$ [illustrated in Fig.~\ref{fig:J122221}] and $\tilde
J^{(B)}$ denote correlators that have a current vertex inserted on the
forward or backward Keldysh contours, respectively.  As a notational
reminder, the indices 2, $2'$, and $\bar 2$ here all refer to the same
time, $t_2$ in this case, and after performing the derivatives
implicit in $\bmj_{11'}$ and $h^\ext_{22'}$, we have to set $2 = 2'$.
However, $x_{\bar 2}$ in \Eqs{eq:defJ(F/B)} is an independent
integration variable.  The subscript ns (for ``no sources'') in
$\langle \; \rangle_{V, {\rm ns}}$ indicates that, following the
prescription of \Eq{C-1122-derivative}, all remaining $\tilde
v$-dependencies are to be dropped henceforth by setting $\tilde v =
0$.  The second lines of \Eqs{eq:defJ(F)} and (\ref{eq:defJ(B)}),
in which we set $y^{\tilde a} = 0$ for the correlator $\tilde
J^{(\tilde a)}$ containing the current vertex on contour $\tilde a$,
follow from \Eq{eq:defineJ1231-00} for $\tilde J^V_{12',2 \bar 2',
  \bar 21'}$ by using the first or second line of
\Eq{eq:tilderhonsUU-ns-twochoices} for $ \tilde U^{\prime F} \, \tilde
\rho^0 \, \tilde U'^B$, respectively [thereby conveniently avoiding
primed propagators $\tilde U'$ under the $x_{\bar 2}$-integrals on the
``other'' contour $\tilde a' \neq \tilde a$, which thus have the form
$\int dx_{\bar 2} \tilde U^{ \tilde a'}_{i \bar 2} \, \tilde U^{\tilde
  a'}_{\bar 2 j} = \tilde U^{\tilde a'}_\iiijjj$; the latter composition
rule follows from \Eq{eq:exact-U0-RA} and the completeness of the
wavefunctions $\psi_\lambda (x_i)$ occuring therein.]

%\begin{figure}[h!!]%[htbp]
\begin{figure}[!ht]
\includegraphics[width=\linewidth]{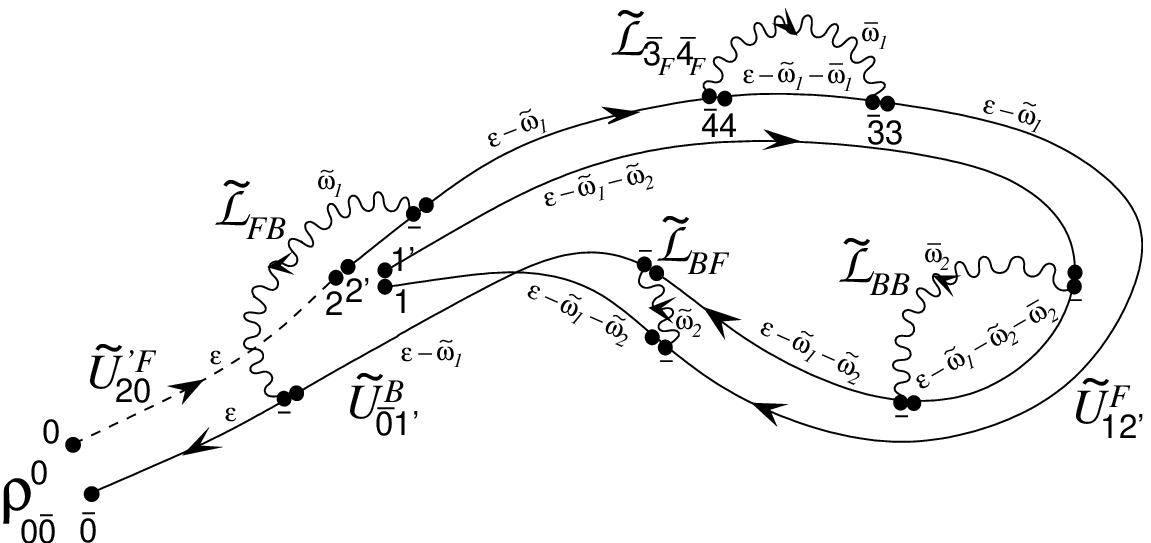}
\caption{ A pair of backward (B) and forward (F) paths contributing to
  $ \tilde { J}^{(F)}_{12', 21'} (t_1, t_2;t_0)$, with $t_1 > t_2 >
  t_0$.  There are two ways to view this figure: (i) Ignore the wavy
  interaction lines, double dot vertices and frequency assignments;
  then this figure illustrates the second line of \Eq{eq:defJ(F)}, and
  the solid or dashed lines represent the full unprimed ($\tilde
  U^{F/B}_\iiijjj$) or primed ($\tilde {U'}^{F}_\iiijjj)$ propagators,
  respectively.  (ii) Imagine the propagators $\tilde U^{F/B}$ and
  $\tilde U'^{F}$ to have been expanded in powers of the interaction
  [as in \Eq{eq:PIexpandedinfullmaintext}]. This generates a forward
  and backward backbone of free propagators $\tilde U^{0 F}_{\bari j}$
  or $\tilde U^{0B}_{j \bari}$ (represented by either solid or dashed
  lines, which now have identical meanings), which are respectively
  decorated by the vertices $\tilde h^F_{V \ibari}$ and $\tilde h^B_{V
    \barii}$ (represented by a pair of dots; both dots are associated
  with the same time, but the one drawn on the side of earlier times
  is distinguished by a bar; the origin of this convention is
  explained in App.~\ref{app:path-integral},
  Fig.~\ref{fig:timeslice}).  The vertices generate, after averaging
  over the fields $V_{\alpha \bari}$, the wavy interaction lines
  $\tilde {\cal L}_{\bari a \barj_{a'}}$, connected to the barred
  dots. [The interaction lines are labelled according to
  \Eq{eq:Ltildes} below: $\LL_{a a'}$ stands for $\LL^K_{\bari_a
    \barj_{a'}}$, $\LL^R_{\bari_a \barj_{a'}}$ or $\LL^A_{\barj_{a'}
    \bari_a}$, if generated by $\langle V_{+ \bari_a} V_{+ \barj_{a'}}
  \rangle_V$, $\langle V_{+ \bari_a} V_{- \barj_{a'}} \rangle_V$, or
  $\langle V_{- \bari_a} V_{+ \barj_{a'}} \rangle_V$, respectively,
  cf.\ \Eq{eq:RIvsLRA}.]  For both cases (i) and (ii), arrows are
  drawn to point from the second index to the first index of each of
  $\tilde U^{ F}_{\bari j}$, $\tilde U^{B}_{j \bari}$ and
  $\LL^{K/R/A}_{\bari_a \barj_{a'}}$.  Thus, they point from later to
  earlier times along the backward Keldysh contour, and from earlier
  to later times along the forward Keldysh contour (\ie\ they form a
  continous loop, starting on the backward contour from $t_1$
  backwards to $t_0 = -\infty$, then continuing on the forward contour
  from $t_0 = -\infty$ forwards via $t_2$ to $t_1$). Finally, the
  frequencies label the interaction correlators $\LL_{aa'}(\omega)$
  and Green's functions $\G^{R/A}(\omega)$ and $\G^K(\omega)
  = \tanh(\hbar \omega/2T) [\G^R - \G^A](\omega)$ that arise (before
  disorder averaging) upon Fourier transforming from the time to the
  frequency domain, as for \Eqs{subeq:Jprime1storder} or
  (\ref{subeq:Cooperonenergytimetime}) below. The effective action
  defined in \Eqs{eq:SIR-LIR-aa-main} to (\ref{eq:recallLCD}) of the
  main text neglects the frequency transfers $\omega_i$ in the
  arguments of all retarded and advanced electron Green's functions
  [$\G^{R/A} (\ve - \omega_i - \dots) \to \G^{R/A} (\ve ) $], but, for
  every $\LL^{R/A} (\omega_i) \G^K (\ve - \omega_i)$, retains it in
  the factor $\tanh[\hbar (\ve -\omega_i) / \hbar]$ of the
  accompanying $\GK$ function. [As discussed in
  Sec.~\ref{sec:Paulifactorshort} or~\ref{sec:ruleofthumb}, this is
  justified by the fact that all integrals over frequency transfer
  variables are limited by Fermi factors to the range $|\hbar
  \omega_i| \lesssim T$].  }
\label{fig:J122221}
\end{figure}

Inserting \Eq{eq:C1122afterderivatives} for $ \tilde {\cal
C}_{[11',22']}$ into \Eq{eq:sigmaGZ},
the expression for $\sigma_\DC$ that results
upon representing the applied field in terms of a scalar potential,
and then relabelling $x_2 \leftrightarrow x_{\bar 2}$ in 
the term containing $\tilde J^{(B)}$, we find:
%\bsubequations
%  \label{eq:finalcurrentpert}
%\begin{eqnarray}
%  \label{eq:finalcurrentpert-a}
%\langle \delta \hat \JJbm (t, \bmrone) \rangle
%&  = &  \sum_{\sigma_1}
%\Biggl\{
%- {e^2 \over m} \bmA (t, \bmrone) \langle \hat n^\pdag_{11S} \rangle
%% \Biggr.  \\ & & \Biggl. \phantom{.} \hspace{-15mm} \nonumber
% - i \int^t_{t'_0} \! dt' \! \int \! dx_{2, \bar 2} \,
% \left[  h^\ext_{22'} -  h^\ext_{33'} \right] (t')
%  j_{11'}^\nu \,  \langle \tilde {   J}^\pdag_{12',31'} (t_1, t_2)
%\tilde \rho^\ns_{23'} (t',t_0) \rangle_{V, {\rm ns}}^\pdag  \Biggr\} \; ,
%\\
%  \label{eq:finalcurrentpert-b}
% \tilde {   J}^\pdag_{12',1'3} (t_1, t_2) & \equiv &
% \tilde U^F_{12'} (t_1, t_2) \tilde U^{B}_{31'} (t',t)\; .
%\end{eqnarray}
%\esubequations \noindent
\begin{eqnarray}
\label{eq:finalGZsigma}
\sigma_{\rm DC} & = & 
{e^2 \over   2 \, m \, d }\sum_{ \sigma_1}
 \!  \int^{t_1}_{-\infty} \! dt_2 \! \int \! dx_{2, \bar 2} \,
 \\ \nonumber & & \qquad 
\left[\bmr_2 - \bmr_{\bar 2} \right] \cdot
 (\bnabla_{1} -  \bnabla_{1'}) 
\langle   \tilde J^V_{12', 2 \bar 2, \bar 2 1'}(t_1,  t_2; t_0)
\rangle_{V, {\rm ns}}^\pdag \; . \qqph 
\end{eqnarray}
\Eq{eq:finalGZsigma} for the DC conductivity is analogous (but, as
discussed below, not identical) to (GZ-II.49) [the factor $ \tilde
U^F_{12'} (t_1, t_2) \tilde U^{B}_{\bar 2 1'} (t_2, t_1)$ which occurs
in our $J_{12', 2 \bar 2, \bar 2 1'}(t_1, t_2; t_0)$ is the analogue of
the function $J (t_1,t_2;\bmr_1, \bmr'_1; \bmr_2, \bmr_{\bar 2})$
occuring in Eqs.~(GZ-II.49) and (GZ-II.50)].  In deriving
\Eq{eq:finalGZsigma}, no approximations have been
  made, apart from not displaying the Hartree terms
 [cf.\ footnote~\ref{f:Hartree}].

  Instead of (\ref{eq:sigmaGZ}) and (\ref{eq:finalGZsigma}), it will
  be more convenient for our purposes to use \Eq{eq:sigmaAAG} as
  alternative expression for $\sigma_{\rm DC}$, derived by
  representing the applied external field via a vector potential. The
  correlator $ \tilde { J}_{12',21'} (\omega_0)$ occuring therein can
  [via \Eq{eq:C1122afterderivatives}] be expressed as:
\begin{eqnarray}
\label{sub:finalJ1221omegadefine}
 & &  
\tilde J_{12',21'} (\omega_0)
\equiv  
\int_{- \infty}^\infty dt_{12} \,
e^{i \omega_0 t_{12}} \,
\lim_{t_0 \to -\infty}
\sum_{\tilde a = F, B} \theta_{12} 
\tilde J^{(\tilde a)}_{12',21'} (t_1, t_2; t_0) \; . 
\end{eqnarray}
%We begin by identifying the real and imaginary contributions 
%to \Eq{eq:sigmaAAG} for $\sigma_\DC$. 
Since $ \tilde J_{12',21'}$ stems from the commutator $\tilde
{\cal C}_{[11',22']} $ [\Eq{eq:define-C-firsttime}], whose
terms satisfy
$\langle \hat n^\pdag_{11'H} \hat n^\pdag_{22'H} \rangle = \langle
\hat n^\pdag_{22'H} \hat n^\pdag_{11'H} \rangle^\ast $,
the correlators  $\tilde J^{(a)}$  satisfy
\begin{eqnarray}
\nonumber
 \tilde J^{(B)}_{12',21'} (t_1, t_2; t_0) = % &  = &
-  \tilde J^{(F)\ast}_{1'2,2'1} (t_1, t_2; t_0)  \; ,
\quad
 \tilde J^{(B)}_{12',21'} (\omega_0) = % &  = &
- \Bigl[ \tilde J^{(F)}_{1'2,2'1} (- \omega_0) \Bigr]^\ast \!\!  . 
\end{eqnarray}
The first of these [which implies the second]
is manifestly obeyed by \Eqs{eq:defJ(F/B)}.
Taylor-expanding
\Eq{eq:sigmaAAG} using 
%\begin{eqnarray}
%  \label{eq:taylorexpandJ}
$\tilde J (\omega_0) =
%\sum_a \tilde J^{(a)} (\omega_0) = 
 \tilde  J (0) +
\omega_0 \tilde  J^\prime (0) + \dots  $,   
%\end{eqnarray}
and  separating $\sigma_\DC = \sigma_{\DC,\rreal} +
i \sigma_{\DC,\iimag} $ into real and imaginary parts,
we obtain
\bsubequations
 \label{subeq:sigmaDCreal}
 \begin{eqnarray}
 \label{eq:sigmaDCreal}
\sigma_{\DC,\rreal} & = &
\sum_{\sigma_1}  {1 \over d}
\int \!  dx_2 \, \bmj_{11'} \! \cdot \! \bmj_{\,22'}
\, 
% \Bigl \langle  
\tilde {J}_{12',21'}^\prime (0)  \; , 
% + \tilde {J}_{12',21'}^{\prime B}(0)  \; , \qqph
\\
 % \nonumber 
\label{eq:sigmaDCimag} 
i   \sigma_{\DC,\iimag}
& = &   \!\!   \lim_{\omega_0 \to 0}
 {1 \over \omega_0 } \sum_{\sigma_1} %\left[ 
\left[ % \left \langle 
{i  e^2 \langle \hat n_{11H} \rangle  \over m}
 + {1 \over d} \! \int \! \!  dx_2 \, \bmj_{11'} \! \cdot \! \bmj_{22'}
\tilde {J}_{12',21'}(0)  \right] . % \right \rangle_\dis . 
\end{eqnarray}
\esubequations
Since we have taken the DC-limit $\omega_0 \to 0$, the imaginary part
$ \sigma_{\DC,\iimag} $ must be strictly equal to 0 (to all orders
in the interaction), which is a useful consistency check.

In App.~\ref{app:Generalcooperon} we show how \Eqs{subeq:sigmaDCreal}
can be massaged into more familiar expressions for $\sigma_\DC$, both
in the absence and presence of interactions [cf.\ 
\Eqs{eq:sigmarrealnonintbeforedisav}, (\ref{eq:sigmaDSnonfinal}),
(\ref{eq:semifinalstep})].

\subsubappendix{Coordinate-Space Path-Integral Representation for $\tilde
  U^a_{\iiijjj}$} 
\label{sec:coordinatePI}

In this subsection we shall derive path integral expressions for the
objects in terms of which the conductivity is expressed in
\Eq{eq:sigmaDCreal}, namely the propagators $\tilde U^a_\iiijjj$
[\Eq{eq:definetildeU}] and the interaction-averaged correlators
$\langle \tilde J^{(\tilde a)} \rangle_V$ [\Eqs{eq:defJ(F/B)}].  We
deviate from GZ's approach, who used a path integral $\int {\cal D}
\bmR \int \bmP$ over both coordinate and momentum space, in that we shall
use coordinate-space-only path integrals $\int \widetilde {\cal D}'
(\bmR)$, because that makes possible a more accurate treatment of the
crucial nonlocal Pauli factors $(\tilde \delta - 2 \tilde \rho)$ in
the effective Hamiltonian $\tilde H^a$ of \Eq{eq:GZlinearresponseall}.

We begin from the power series expansions
(\ref{eq:PIexpandedinfullmaintext}) of the evolution matrix functions
$\tilde U^a_{\iiijjj} (t,t')$ of \Eqs{eq:definetildeU} in powers of
$\tilde h^a_V$, and introduce, as a shorthand for these expansions,
the following coordinate space path integrals:
\begin{eqnarray}
%\nonumber 
\left. \begin{array}{r}
 \tilde U^{F}_{\iiijjj} (t,t') \rule[-2mm]{0mm}{0mm} \\
 \tilde U^{B}_{ji} (t',t)
\end{array} \right\}
\!\!  & =&  \!\! 
\int_{\bmR^a (t')  = \bmr_j}^{\bmR^a (t) =
\bmr_i}  \widetilde {\cal D}' \! \bmR^a
e^{i s_a  \tilde S^a_0 (t,t')/ \hbar }
%\nonumber \\
%& & \times 
\exp \left[ {- i s_a \over \hbar}
 \int_{t'}^{t}    dt_3
 \left\{  \begin{array}{l}
\tilde h^F_{V 3_F \bar 3_F}
\rule[-2mm]{0mm}{0mm}
\\
 \tilde h^B_{V \bar 3_B  3_B} 
\end{array} \right. \right] .   \qqph  \qph
\label{eq:def-path-integral-pureRmaintext}
\end{eqnarray}
Here $s_a$ stands for $s_{F/B} = \pm$,
and the index value $a=F$ or $B$ should be used for the upper or lower
term in the curly bracket, respectively.
The coordinate-space path integral is over
all paths $\bmR^a (t_3)$ that begin at time $t'$ at point $\bmr_j$ and
end at time $t$ at point $\bmr_i$;
the time $t_3$ that is used to parametrize
this path $\bmR^a(t_3)$
is understood to refer to the upper
or lower Keldysh contour for $a = F$ or $B$, respectively
[in this sense, an index $a$ on $t_3$ is implicit,
as in $\bmR^a(t_{3_a})$].
The objects $\tilde S^a_0$ and $\tilde h^a_V$ in
the  exponential factors in \Eq{eq:def-path-integral-pureRmaintext}
are both functionals of the path $\bmR^a (t_3)$:
$S^a_0$ is the standard action for a noninteracting
electron in a disorder potential,
\begin{eqnarray}
  \label{eq:standardfreeactionRonlymaintext}
\tilde S^a_0 (t,t')[ \bmR^a (t_3)] & \equiv &   \int_{t'}^{t}   dt_3
%\tilde L^a_0 \bigl(t_3, \bmR^a (t_3) \bigr) & \equiv &
\left[\toh m \dot \bmR^{a2} (t_3) - V_\imp \bigl( \bmR^a (t_3)
  \bigr)\right]  \;  ,
\end{eqnarray}
whereas in the second exponential, we used
the following shorthand notation:
\bsubequations
  \label{eq:shorthandtildehfunctionR}
\begin{eqnarray}
\tilde h^F_{V 3_F \bar 3_F} &\!\! = \!\! &  \sum_{\alpha = \pm} \tilde w^{F
  \alpha}_{3_F \bar 3_F} \, V_{\alpha \bar 3_F} 
%\\ \nonumber & = &
\; = \; 
\sum_{\alpha = \pm} \tilde w^{F \alpha} \left[t_3, \bmr^F_3 (t_3),
\bmr^F_{\bar 3} (t_3) ] \,
 V_{\alpha } [t_3, \bmr^F_{\bar 3} (t_3) \right] \; , \qqph \qph 
 \\
\tilde h^B_{V \bar 3_B  3_B} & \!\! = \!\! & \sum_{\alpha = \pm}
 V_{\alpha \bar 3_B} \, \tilde w^{B \alpha}_{\bar 3_B  3_B}  
%\\ \nonumber & = & 
\; = \; 
\sum_{\alpha = \pm}
 V_{\alpha } \left[t_3, \bmr^B_{\bar 3} (t_3) \right]
\tilde w^{B \alpha} \left[t_3, \bmr^B_{\bar 3} (t_3),
\bmr^B_{ 3} (t_3) \right]
 \; . \qqph \qph 
\end{eqnarray}
\esubequations
In App.~\ref{app:path-integral} we give an explicit definition of the
path integral \Eq{eq:def-path-integral-pureRmaintext} by time-slicing
the time interval $[t', t]$ [\Sec{sec:expandU}], and a detailed
demonstration that it satisfies the defining \Eqs{eq:tildeUconditions}
[\Sec{app:verifyingdefeqcomprule}].  The explicit derivation given
there shows that, when writing down the path integral
(\ref{eq:def-path-integral-pureRmaintext}), the following points are
to be implicitly understood [see also Fig.~\ref{fig:timeslice} of
App.~D]: (i) The path integral
(\ref{eq:def-path-integral-pureRmaintext}) is simply a short-hand for
the time-ordered power series expansion
(\ref{eq:PIexpandedinfullmaintext}), with $ (-i / \hbar)
\U^{0F}_{\bari_F j_F}$ replaced by $\G^R_{\bari_F j_F}$ and $(i /
\hbar) \U^{0B}_{j_B \bari_B}$ by $\G^A_{j_B \bari_B}$ [cf.\
\Eq{eq:PIexpandedinfullGRGA}].  (ii) For each occurrence of a
``vertex'' $ \tilde h^{F}_{V 3_F \bar 3_F}$ or $ \tilde h^{B}_{V \bar
  3_B 3_B}$, the vertex coordinates $\bmr_3^a (t_3)$ and $\bmr^a_{\bar
  3} (t_3)$ are both associated with the \emph{same} time $t_3$, and
both are assumed to be integrated over in the path integral [as in
\Eq{eq:PIexpandedinfullmaintext}], thereby taking into account the
nonlocal nature of the Hamiltonians $\tilde h^a_{V\iiijjj}$. (iii) The associated
integrations $\int d x_{3, \bar 3}$ are understood to be included in
the measure $ \int \widetilde {\cal D}' \! \bmR^a$ (the prime serves
as reminder of this fact), in addition to the integrations associated
with propagators between vertices. (iv) Vertices are connected by
propagators of the form $ \G^R_{\bari_F j_F}$ or $\G^A_{j_B \bari_B}$
on the forward or backward Keldysh contours, respectively. However,
since these propagators occur under time-ordered integrals anyway,
they can equally well also be written as $ (-i/ \hbar)
\U^{0F}_{\bari_F j_F}$ or $(i /\hbar ) \U^{0B}_{j_B \bari_B}$, as is
convenient [in order to exploit \Eq{subeq:G<GK}] whenever they are
contracted with a density matrix $\tilde \rho^0_{i_F \bari_F}$ or
$\tilde \rho^0_{\bari_B i_B}$.
%\Eq{eq:PIexpandedinfullmaintext}.

{\bbf Now use the path integral representation
  (\ref{eq:def-path-integral-pureRmaintext}) (twice) in
  \Eq{eq:defineJ1231-00} for $\langle \tilde { J}^V \rangle_{V, {\rm
      ns}}^\pdag $, and interchange the order of averages,
  $\left\langle \int \widetilde {\cal D}'\!  (\bmR) \dots \right
  \rangle_V \to \int \widetilde {\cal D}'\!  (\bmR) \left\langle \dots
    \vphantom{\widetilde {\cal D}'\!  (\bmR)} \right \rangle_V$. [The
  latter step could have been postponed until the beginning of
  Section~\ref{sec:integrateoutV}, but is used already here, since it
  simplifies subsequent expressions. Its use, sooner or later, is a
  crucial ingredient in GZ's approach. Its far-reaching consequences
  are discussed in detail in App.~\ref{sec:interchangingaverages}.]
  % which might be worth rereading at this point.]
}  We obtain
\begin{eqnarray}
\langle \tilde {   J}^V_{12',2 \bar 2', {\bar 2}1'} (t_1,t_2; t_0)
\rangle_{V, {\rm ns}}^\pdag  & = &
{1 \over \hbar} \int dx_{0_F,  \bar 0_B}  \, \tilde \rho_{0_F \bar 0_B} 
\nonumber
\\ \label{eq:JrhoRonly} 
& & \phantom{.} \hspace{-0.6cm} \times
\Fint_{2'_F}^{1_F}
\Bint_{\bar 2_B}^{1'_B}
 \widetilde {\cal D}'\!  (\bmR)
\;
\Fint_{0_F}^{2_F}
\Bint_{\bar 0_B}^{\bar 2'_B}
\widetilde {\cal D}'\!  (\bmR) \, 
\tilde {\cal F}_{(t_1,t_0)} [\bmR^a ]  \; , \qqph \qph
\end{eqnarray}
where ${\displaystyle \Fint \Bint \widetilde {\cal D}' \! (\bmR )}$
is used as a shorthand for the
following forward and backward path integral
 between the specified initial and final
coordinates and times:
\begin{eqnarray}
\nonumber 
\Fint_{j_F}^{i_F}
\Bint_{\barj_B}^{\bari_B}
\widetilde {\cal D}' \!  (\bmR) \dots
& \equiv & \int_{\bmR^F (t^F_j)  = \bmr^F_j}^{\bmR^F (t^F_i) =
\bmr^F_i}  \widetilde {\cal D}' \!  \bmR^F (t_3)
\, e^{i \tilde  S_0^F (t^F_i, t^F_j)  / \hbar}
\\
& & \phantom{.} 
\hspace{-0.6cm} \times  \int_{\bmR^B (t^B_j)  = \bmr^B_\barj}^{\bmR^B (t^B_i) =
\bmr^B_\bari} \widetilde {\cal D}' \!  \bmR^B(t_3) \,
e^{- i  \tilde S_0^B (t^B_i, t^B_j)/ \hbar} \dots \; 
\qqph \qqph 
\end{eqnarray}
%$\tilde S_0^a$ is the
%free electron  action of \Eq{eq:standardfreeactionRonlymaintext},
The influence functional
$\tilde {\cal F}_{(t_1,t_0)} [\bmR^a ]$ in \Eq{eq:JrhoRonly}
is defined by the following functional integral
over all configurations of the fields
$V_{\pm 3} =
V_{\pm}(t_3, \bmr_3)$  of \Eqs{eq:newV's}, with $t_3 \in [t_0, t_1]$:
\bsubequations
\label{eq:tildeFtildeB-define}
\begin{eqnarray}
  \label{eq:define-influence-functional-Ronly-nonquadratic}
\tilde {\cal F}_{(t_1,t_0)}
  [\bmR^F (t_3); \bmR^B (t_3)]  & \equiv &
   {\int {\cal D} V_+   \int {\cal D} V_- \;
  e^{{ i \over  \hbar} \left[ S_V^\tot
  - \,  \widetilde {\cal B} \cdot {\cal V} \right](t_1,t_0)}
\over 
  \int {\cal D} V_+   \int {\cal D} V_- \;
  e^{{i\over \hbar} S_V^\tot (t_1,t_0)} } \; ,
\\
%\nonumber
%\widetilde {\cal B} \cdot {\cal V} (t_1,t_0) & \equiv &  \int_{t_0}^{t_1}
%d t_3 \int d \bmr_3 \sum_{\alpha = \pm}
%\Biggl[ s_F \tW^{F \alpha}_{3_F \bar 3_F} \delta
%(\bmr_3 - \bmr_{\bar 3_F}) \Biggr. \\
%&  & % \label{eq:defB-Ronly} 
%\Biggl. 
%+ s_B \tW^{B \alpha}_{\bar 3_B  3_B} \delta
%(\bmr_3 - \bmr_{\bar 3_B})
% \Biggr]  V_\alpha (t_3, \bmr_3) \; ,
%\\
\widetilde {\cal B} \cdot {\cal V} (t_1,t_0) & \equiv &  \int_{t_0}^{t_1}
d t_3 \int d \bmr_3 \sum_{\alpha = \pm}
\tilde {\cal B}_\alpha (t_3, \bmr_3) \,  V_\alpha (t_3, \bmr_3) \; ,
\\
\label{eq:defB-Ronly} 
\tilde {\cal B}_\alpha (t_3, \bmr_3) & \equiv & 
 s_F \tW^{F \alpha}_{3_F \bar 3_F} \delta
(\bmr_3 - \bmr_{\bar 3_F}) + s_B \tW^{B \alpha}_{\bar 3_B  3_B} \delta
(\bmr_3 - \bmr_{\bar 3_B}) \; , \qqph \qph 
\\
\label{eq:Wvertices}
\tW^{a \alpha}_{3 \bar 3} & \equiv &
\theta_{32}  \, \tilde w^{a \alpha}_{3 \bar 3}  +
\theta_{23} \, \tilde w'^{a \alpha}_{3 \bar 3} \; .
\end{eqnarray}
\esubequations
Here $ S_V^\tot (t_1,t_0)$ is given by \Eq{eq:S[V_a]}, and
\Eq{eq:defB-Ronly}, which defines the field $\widetilde {\cal
  B}_{\alpha 3} = {\cal B}_\alpha (t_3, \bmr_3)$, follows from using
\Eqs{eq:shorthandtildehfunctionR} or a primed version thereof, for
$t_3 > t_2$ or $t_3 < t_2$, respectively. The distinction between the
two time orderings, which is reflected in the definition
(\ref{eq:Wvertices}) of the vertices $\tW^{a\alpha}_{3 \bar 3}$ (and
not noted by GZ, since they set $t_0 = t_2$), is necessary, since
\Eq{eq:defineJ1231-00} correspondingly features unprimed or primed
propagators $\U^a_\iiijjj$ or $\U'^a_\iiijjj$, respectively, which have
different vertices [compare \Eqs{eq:wvertices} and
(\ref{eq:definetildewa-prime})].  Note that $\widetilde {\cal
  B}_{\alpha 3}$ is itself a functional of both the paths $\bmR^F
(t_3)$ and $\bmR^B (t_3)$.  The influence functional\footnote{ The
  term ``influence functional'' is used here in precisely the sense in
  which Feynman used it: Our $\tilde {\cal F}_{23} [\bmR^a ]$ is
  analogous to the quantity $F[q(t),q'(t)]$ of Eq.~(12-90) of R. P.
  Feynman and A. R. Hibbs, ``Quantum Mechanics and Path Integrals'',
  McGraw-Hill (1965).}  $\tilde {\cal F}_{(t_1,t_0)} [\bmR^a ]$
describes the effect of all other electrons on a pair (forward and
backward) of singled-out electron trajectories $\bmR^F (t_3)$ and
$\bmR^B (t_3)$ between the initial time $t_0$ and final time $t_1$.
Importantly, this influence functional incorporates the Pauli
principle, via the presence of the
Pauli factor $(\tilde \delta - 2 \tilde \rho)$%_{\bar 3_a 3_a}$ 
in $ \tilde w^{a -}$.% _{\bar 3_a 3_a}$.

\subsubappendix{RPA Approximation \label{sec:RPA}}
\label{sec:RPAapprox}
\label{app:RPA}

To evaluate the influence functional $\tilde {\cal F}_{(t_1,t_0)} $
explicity, our  next task is to perform the functional integrals $\int
\! {\cal D} V_\alpha$ stipulated in
\Eq{eq:define-influence-functional-Ronly-nonquadratic}. As a first
(standard) step toward making these integrals Gaussian, i.e.\ doable,
we apply (following GZ) the RPA approximation: we approximate the
effective action $S_V^\tot$ of \Eq{eq:S[V_a]} by the part quadratic in
the fields $V$, say
\bsubequations
\label{eq:RPA-approx}
\begin{eqnarray}
\label{eq:RPA-approx-a}
  \label{eq:SV2}
i S_V^{(2)} (t_1,t_0)  & = &     
i( S^{0F}_V - S^{0B}_V)
+ \hbar Z^{(2)} \equiv
\label{eq:S[V](2)}
 - \toh \,\bigl[ {\cal V} \cdot \tilde {\cal A} \cdot {\cal V}\bigr]
 (t_1,t_0) \; \\
\label{eq:RPA-approx-b}
%\phantom{.} \hspace{-1cm}
 &=&   - {1 \over 2} \int_{t_0}^{t_1} \! \! d t_3
\int_{t_0}^{t_1} \! \!  d t_4
\! \int \! \! d \bmr_3 \, d \bmr_4 \sum_{\alpha \alpha'}
V_{\alpha 3} \AAA{}^{\alpha \alpha'}_{34} V_{\alpha' 4} \; ,
% \nonumber .
\end{eqnarray}
\esubequations
so that 
\Eq{eq:define-influence-functional-Ronly-nonquadratic} becomes
\begin{eqnarray}
  \label{eq:define-influence-functional-RonlyRPA}
& & \tilde {\cal F}_{(t_1, t_0)}
\stackrel{RPA}{\longrightarrow}
  {\int {\cal D} V_+   \int {\cal D} V_- \;
  e^{-{ 1 \over 2 \hbar} {\cal V} \cdot \tilde {\cal A} \cdot {\cal V} } \;
  e^{-{ i \over \hbar}
  \widetilde {\cal B} \cdot {\cal V} }
\over
  \int {\cal D} V_+   \int {\cal D} V_- \;
  e^{- {1 \over 2 \hbar} {\cal V} \cdot \tilde {\cal A} \cdot {\cal
  V}}} \; .
\qquad \phantom{.}
\end{eqnarray}
To find $\tilde {\cal A}$, we have to find an explicit expression for
the term $ \hbar Z^{(2)} $ in \Eq{eq:RPA-approx-a}, which 
arises from  expanding the factor $\ln Z = 
%\ln[1 + Z^{(1)} +  Z^{(2)} +  {\cal O} (V_a^3)] =  
 Z^{(2)}  + {\cal O} (V_a^3)$ to
second order in $V_a$, using\\
%To find $\AAA$ explicitly, we have to calculate the
%2nd-order contribution $S_V^{(2)}$ [Eq.~(\ref{eq:S[V](2)})] of the
%effective action \Eq{eq:S[V_a]}, by expanding the factor $\ln Z$ to
%second order in $V_a$:
\bsubequations
\begin{eqnarray}
  \label{eq:expandlnZV}
  Z(t_1, t_0)  & =  & 1 + Z^{(1)} + Z^{(2)}  + {\cal O} (V_a^3) \; , 
%\\
%\label{eq:Z1}
% Z^{(1)} & = & - {i\over \hbar}
% \int_{t_0}^{t_1} d t_3 \langle \hat V_{F} (t_3)
% - \hat V_{B} (t_3) \rangle_0 \; = 0 \; ,
\\
\nonumber
 Z^{(2)} & = &
- {1 \over 2 \hbar^2} \int_{t_0}^{t_1} \! d t_3 \!
\int_{t_0}^{t_1} \! d t_4
\left\{  \langle {\cal T} \hat V_{F} (t_3)  \hat V_{F} (t_4)
\rangle \right.
\\
\label{eq:Z2} & & \phantom{.} \hspace{-1cm}
\left. + \langle \overline {\cal T} \hat V_{B}
(t_3)  \hat V_{B} (t_4)
 \rangle
- 2  \langle \hat V_{B} (t_3)  \hat V_{F} (t_4) \rangle_0 \right
\} \; , \qquad \phantom{.}
%\\
%  \ln Z & = &   Z^{(2)}  + {\cal O} (V_a^3) \; .
\end{eqnarray}
\esubequations \noindent
and noting that $Z^{(1)}$ vanishes, since $\hat V_a$ is normal-ordered [cf.\
\Eq{eq:hatV}].
%, hence $ \ln Z =  Z^{(2)}  + {\cal O} (V_a^3) $.
Expressing  Eq.~(\ref{eq:Z2}) through the fields $V_{\alpha
i}$ ($\alpha = \pm 1$) of Eq.~(\ref{eq:newV's}), we find
\begin{eqnarray}
   \hbar  Z^{(2)} & = &  -  \int_{t_0}^{t_1}
t_3 \int_{t_0}^{t_1}  d t_4 \int d \bmr_3 \, d \bmr_4
%\\
 \label{eq:effectiveactionV}
%& & \times 
\Bigl( i V^\pdag_{-3} \tilde \chi_{34} V^\pdag_{+4} +
V^\pdag_{-3} \tilde \eta_{34} V^\pdag_{-4} \Bigr) \; ,
\end{eqnarray}
where $\tilde \chi_\iiijjj$ (the charge susceptibility) and $\tilde
\eta_\iiijjj$ (characterizing charge fluctuations) are defined
as
\bsubequations
\label{eq:defchieta}
\begin{eqnarray}
\label{eq:chieta}
 \tilde  \chi_\iiijjj & \equiv & - i \,
 {2 \, e^2 \over \hbar} \theta (t_\iiijjj)
\langle [ : \! \hat n_{iiI} (t_1) \! :, : \! \hat n_{jjI} (t_j)
\! : ] \rangle_0 \; = \; 
 4 e^2 \hbar \, {\rm Im}  \bigl[ \G^R_\iiijjj \G_{ji}^< \bigr]  ,
\qquad \phantom{.}
 \\
\tilde \eta_\iiijjj &  \equiv &  \frac{e^2 }{2 \hbar}
 \langle \{ : \! \hat n_{iiI} (t_i) \! :,
: \! \hat n_{jjI} (t_j) \! : \} \rangle_0 
\; = \; - {e^2 \hbar } \,  {\rm Re} \bigl[ \G^>_\iiijjj \G^<_{ji}
 \bigr] \; ,
\end{eqnarray}
\esubequations \noindent
with equal spins, $\sigma_i = \sigma_j$ (for $\sigma_i \neq \sigma_j$,
both these quantities vanish).  [\Eqs{eq:defchieta} correspond to
(GZ-II.31) and (GZ-II.32).]  The right-most equalities were obtained
by using Wick's theorem to rewrite the correlators in terms of the
single-particle Green's functions $\tilde G^{<,>}$
[\Eqs{eq:defineG<}].
%App.~\ref{app:spGreensfunctions},
%Eqs.~(\ref{eq:defineelectrongreensfunctions}).  
The  Fourier transforms of $\tilde \chi_\iiijjj$ and
$\tilde \eta_\iiijjj$ satisfy 
$\bar \chi^\ast_{- \bmk} (-\omega) = \bar \chi_\bmk (\omega ) =
\bar \chi_{- \bmk} (\omega)$ and 
$\bar \eta_{- \bmk} (-
\omega) = \bar \eta_\bmk (\omega)$ (thus the latter is real),
and are related by the fluctuation dissipation theorem
[(GZ-II.33)]:
%\bsubequations
 \begin{eqnarray}
    \label{eq:fluc-diss-chi}
% \label{eq:fluc-diss-chi-a}
%  \tilde \eta (\omega ; \bmr_i, \bmr_j ) & = &
% -  \toh   \coth( \hbar \omega/2T)
%\; {\rm Im} \tilde \chi (\omega; \bmr_i,  \bmr_j) \; ,
%\qquad \phantom{.}
%\\
%  \label{eq:fluc-diss-chi-b}
\bar \eta_\bmk (\omega) & = &
 -  \toh  \coth(\hbar \omega/2T) \;
{\rm Im} \, \bar \chi_\bmk (\omega) \; .
\end{eqnarray}
%  \esubequations
%where we have exploited
%the fact that 
%$\bar \chi^\ast_{- \bmk} (-\omega) = \bar \chi_\bmk (\omega ) =
%\bar \chi_{- \bmk} (\omega)$.

Now, if we write the second-order contribution $i S_V^{(2)}(t,t_0)$ in
the form of Eq.~(\ref{eq:RPA-approx-b}), and Fourier
transform,\footnote{Strictly speaking
the Fourier transform (\ref{eq:VAV-Fourier}) is an exact
representation of ${\cal V} \cdot \tilde {\cal A} \cdot {\cal V}$
only if the time integrals in Eq.~(\ref{eq:RPA-approx-b})
  are unbounded, \eg\ for $t_0 =- \infty$ and $t_1 = \infty$. In our
  formalism, this indeed is the case, since we do take the limit $t_0
  \to - \infty$, and may also take $t_1 \to + \infty$ (because the
  $t_1$-dependence drops out, anyway).} we obtain from
Eqs.~(\ref{eq:defineSa}) and (\ref{eq:effectiveactionV}):
\bsubequations
\label{subeq:VAV}
\begin{eqnarray}
  {\cal V} \cdot \tilde {\cal A} \cdot {\cal V} & = &
\int {d \bmk \, d \omega \over (2 \pi )^{d+1}}
\sum_{\alpha \alpha'} % \\
% & & 
\bar V_{\alpha, - \bmk} (- \omega) \bar {\cal A}^{\alpha
\alpha'}_\bmk ( \omega)
 \bar V_{\alpha', \bmk} (\omega) \; ,
 \label{eq:VAV-Fourier}
\\
  \label{eq:S(2)-Fourier}
\bar {\cal A}^{\alpha \alpha'}_\bmk ( \omega) \!\! & = \!\!& -i \left(
\begin{array}{cc} 0 &
 {\displaystyle {\bar \varepsilon_{- \bmk} (- \omega) \over
 \bar V^\inter (\bmk)}}
\\
{\displaystyle {\bar \varepsilon_\bmk (\omega) \over
\bar V^\inter (\bmk)}} &   2 i
\bar \eta_\bmk (\omega) \end{array}
\right)_{\alpha \alpha'} \; , \quad \phantom{.}
%\\
%  \label{eq:def-susceptibility-e}
%\bar \varepsilon_\bmk (\omega) & \equiv & 1 - \bar V^\inter (\bmk) \,
%\bar \chi_\bmk (\omega)  \; ,
\end{eqnarray}
\esubequations
where $\bar \varepsilon_\bmk (\omega) \equiv 1 - \bar V^\inter (\bmk)
\, \bar \chi_\bmk (\omega) $ is the dielectric susceptibility. [The
latter relation is a generalized version of (GZ-II.35); in (GZ-II.36),
GZ added to $\bar \varepsilon_\bmk (\omega)$ an electron-phonon
contribution, which is not important for the present discussion and
neglected by GZ themselves later on, after (GZ-II.75).]  

Having found $\bar {\cal A}$, let us now also find 
and discuss some useful properties of its
inverse, $\bar {\cal A}^{-1}$ [it will be needed
in the next section  after evaluating 
the functional integral \Eq{eq:define-influence-functional-RonlyRPA}].
Using $\bar \varepsilon_{- \bmk} (- \omega) = \bar \varepsilon^\ast_\bmk
(\omega)$, we find
\begin{eqnarray}
  \label{eq:Ainverse-Fourier}
(\bar {\cal A}^{-1})^{\alpha \alpha'}_\bmk ( \omega) &  = & 
\left(\begin{array}{cc}
  \bar I_\bmk (\omega) & i \bar R_\bmk (\omega) \\
  i \bar R_{- \bmk} (-\omega) & 0
\end{array} \right) , 
\end{eqnarray}
\begin{subequations}
with matrix elements given by
\begin{eqnarray}
 \label{eq:defineR-app}
\bar R_\bmk (\omega) & = & 
{\bar V^\inter (\bmk) \over \bar \varepsilon_\bmk ( \omega)}  ,
\\ \label{eq:defineI-app}
\bar I_\bmk (\omega) & = & {2 \bar \eta_\bmk (\omega) 
| \bar V^\inter (\bmk)|^2  \over | \bar \varepsilon_\bmk (\omega) |^2} 
% \; , \\ \label{eq:fluc-dis-RI} & =& 
= - \coth (\hbar \omega /2T ) \, {\rm Im} \, \bar R_\bmk (\omega)
\; , \qqph \qph
\end{eqnarray}
\end{subequations}
where the last equality in \Eq{eq:defineI-app} follows from
\Eq{eq:fluc-diss-chi}.
%\Eqs{eq:defineRI-app} directly imply Eqs.~(\ref{eq:defineIR-wk}).
Note that % the $\bar \chi$-relations stated after \Eq{eq:fluc-diss-chi},
%together with 
the assumptions [stated before \Eq{eq:screenedCoulomb}]
that $\bar V^\inter (\bmk)$ is real and symmetric, imply that
\begin{eqnarray}  \label{eq:Romegakproperties}
\bar R^\ast_{- \bmk}  (- \omega) = \bar R_\bmk (\omega) =
\bar R_{-\bmk} (\omega) \; ,
%\\
%\label{eq:Iomegakproperties} 
\quad \bar I^\ast_{- \bmk} (- \omega) %&=& 
 = \bar I_\bmk (\omega) 
=  \bar I_{- \bmk}  (\omega) = \bar I_\bmk (-\omega) \; ,
\qquad \phantom{.}
\end{eqnarray}
so that the functions $\tilde R_\iiijjj$ and $\tilde I_\iiijjj$ are both
purely real:
%\begin{eqnarray}
%\label{eq:RIxtproperties}
$ \tilde R_{\iiijjj}  =  \tilde R^\ast_{\iiijjj}$ and 
%{\rm and} \qquad
$\tilde I_{\iiijjj}  =  \tilde I^\ast_{\iiijjj} =
\tilde I_{ji} $.
%\label{eq:Ixtproperties}
%\end{eqnarray}
For reference purposes, we note also that their  frequency Fourier
transforms, denoted by $\tilde R_\iiijjj (\omega)$ and $\tilde I_\iiijjj (\omega)$,
satisfy the relations
$  \tilde R_\iiijjj (\omega)  = 
  \tilde R_{ji} (\omega) =   \tilde R^\ast_\iiijjj (-\omega)$ and 
%\bsubequations
 \begin{eqnarray}  \label{eq:RIomegaij}
  \tilde I_\iiijjj (\omega)  =  \tilde I_{ji} (\omega)
 = \tilde I^\ast_\iiijjj (- \omega) =
 \tilde I_\iiijjj (- \omega) 
  \label{eq:RIomegaijcoth}
=  - \coth(\hbar \omega/2T) {\rm Im}
[ \tilde R_\iiijjj  (\omega)]  \; . \qqph 
\end{eqnarray}
%\esubequations
Furthermore, $\tilde R_\iiijjj (\omega)$ is analytic in the upper half
plane, implying  that $\tilde R_\iiijjj$ is proportional to $\theta
(t_\iiijjj)$. In contrast, $\tilde I_\iiijjj$ is symmetric
in its indices and thus nonzero for both $t_\iiijjj > 0$
and $<0$.

The components of  $\AAA^{-1}$ are of course 
related to field correlation functions
of the type
$\langle V_{\alpha i} V_{\alpha' j} \rangle_{V, {\rm ns}}$,
as follows from a simple exercise in Gaussian integration: Introducing 
the generating functional 
\bsubequations
\begin{eqnarray}
  \label{eq:generator-for-V}
Q\mbox{[} \zeta \mbox{]} & \equiv &   \langle e^{-{i\over \hbar} \zeta \cdot
{\cal V}} \rangle_{V, {\rm ns}} \; , 
\\
\label{eq:shortd1d1} [\zeta \cdot {\cal V}] (t_1,t_0) & \equiv &
\int_{t_0}^{t_1}
d t_3 \int d \bmr_3  \sum_{\alpha = \pm}
 \zeta_{\alpha 3} \,  V_{\alpha 3}
\; ,
\end{eqnarray}
\esubequations \noindent
where $\zeta_{\alpha i} = \zeta_\alpha (t_i, \bmr_i)$, with $\alpha =
\pm 1$, are two real source fields, we find
\bsubequations \label{eq:general-QGaussian}
\begin{eqnarray}
  \label{eq:QGaussian}
 Q[\zeta ]  & \; \stackrel{{\rm RPA}}{\longrightarrow} \; &
 {\int {\cal D} V_+   \int {\cal D} V_- \;
  e^{- {1 \over 2 \hbar } {\cal V} \cdot \tilde {\cal A} \cdot {\cal V}} \;
  e^{-{i\over \hbar}  \zeta  \cdot {\cal V} } \over
  \int {\cal D} V_+   \int {\cal D} V_- \;
  e^{- { 1 \over 2 \hbar} {\cal V} \cdot \tilde {\cal A} \cdot {\cal V} } }
\; = \; % \\  & =&   
e^{- {1 \over 2 \hbar} \zeta \cdot \tilde {\cal A}^{-1} \cdot \zeta}
 \; . \qqph \label{eq:QGaussian-result}
\end{eqnarray}
\esubequations
%\end{eqnarray}
%\zeta \cdot {\cal A}^{-1} \cdot \zeta =
%\int_{t_0}^t \! \! d t_1  \, d t_2 \! \int \! \! d \bmr_1 \, d \bmr_2
%\sum_{\alpha \alpha'} V_{\alpha 1}
%(\bcA^{-1})^{\alpha \alpha'}_{12} V_{\alpha' 2}
%\; ,
The field correlators are then easily found to
have the form
\bsubequations
\label{eq:even-odd-correlatorVV}
\begin{eqnarray}
  \label{eq:LL-AA}
{1 \over \hbar} \langle V_{\alpha i} V_{\alpha' j} \rangle_{V, {\rm ns}}
& = & \left. - \hbar {\delta^2 Q\mbox{[}
\zeta \mbox{]} \over \delta \zeta_{\alpha i} \, \delta
\zeta_{\alpha' j}} \right|_{\zeta = 0}
 = \;  (\tilde {\cal A}^{-1})^{\alpha \alpha'}_{\iiijjj} 
% \\
% \label{eq:Ainverserealspace}  \label{eq:int-prop-results}
 =  % & = &
   \left(\begin{array}{cc} \tilde I_{\iiijjj} & i  \tilde R_{\iiijjj} \\
  i  \tilde R_{ji} & 0
\end{array} \right)_{\alpha \alpha'} \qqph \qph
\\
\label{eq:RIvsLRA}
 & = &
- {i \over e^2} \left(\begin{array}{cc} \toh \tilde {\cal L}^K_{\iiijjj} &  
\tilde {\cal L}^R_{\iiijjj} \\
\phantom{\toh}  {\cal L}^A_{\iiijjj} & 0
\end{array} \right)_{\alpha \alpha'} \; . \qquad \phantom{.}
%\\ & = & \label{eq:RIvsLRA}
%- i \left( \begin{array}{cc}
%{1 \over 2 } \, \LL^K_{\iiijjj} &  \LL^R_{\iiijjj} \\
%    \phantom{{1 \over 2} \, }
%     \LL^A_{\iiijjj} & 0  \end{array} \right)_{\alpha \alpha'}
%    \!\!  , \qquad \phantom{.}
\end{eqnarray}
\esubequations
where \Eq{eq:Ainverse-Fourier} has been used, and the functions [cf.\ 
(GZ-II.56) and (GZ-II.57)]
 \begin{eqnarray}
\label{eq:defineRI}
\label{eq:definegreensfunctions}
 \label{eq:defineR-tr}
  (\tilde R/ \tilde I)_{\iiijjj} \! & = & \!
\int \! (d \bmk )(  d \omega ) 
 e^{-i \omega (t_i - t_j) + {i} \bmk \cdot (\bmr_i-\bmr_j)}
 (\bar R/ \bar I)_\bmk (\omega)  
\end{eqnarray}
are defined via their Fourier transforms, given by
\Eqs{eq:defineR-app} and (\ref{eq:defineI-app})
above.  \Eq{eq:RIvsLRA} expresses the general
fact [reviewed in App.~\ref{app:fieldcorrelators}] that the field
correlators can also be written in terms of the standard retarded,
advanced and Keldysh components of the interaction propagator,
$\LL^R_{\iiijjj}$, $\LL^A_{\iiijjj} $ and $\LL^K_{\iiijjj}$, implying that these
are proportional to $\tilde R_\iiijjj$, $\tilde R_{ji}$ and $\tilde I_\iiijjj$
[cf.\ (GZ-III.A14)].  This implies that $ \tilde R_\iiijjj$ is a retarded
propagator and thereby confirms that it is is proportional to $\theta
(t_\iiijjj)$ [as had already been concluded above from the analytic
properties of $\tilde R_\iiijjj (\omega)$].

To obtain explicit expressions for $ \bar R_\bmk (\omega) $,
one needs $\bar \varepsilon_\bmk (\omega)$ and hence 
$\bar \chi_\bmk (\omega)$, for which one has to calculate a
polarization bubble [see App.~F, Fig.~\ref{fig:sigma-cooperon}(e)].  
If $\bar V^\inter (\bmk) = {4 \pi / \bmk^2}$ represents the 
unscreened Coulomb interaction [\Eq{eq:screenedCoulomb} with
$\lambda_0 = 0$] and, 
as is usually the case in the presence of disorder, only
small frequencies and wave numbers are of interest, 
a standard calculation yields [cf.\ \Eq{eq:chi12omegaq-c}
 and (GZ-II.36)]:
\begin{eqnarray}
  \label{eq:Romegak-explicit}
%  \label{eq:dielectricsucs-explicit}
\bar \chi_\bmk (\omega)   =   
- \, { \bmk^2 \, \sigma^\Drude_\DC \over \Dd \bmk^2 - i \omega} 
 , \quad 
  \bar \varepsilon_\bmk  (\omega)  =  1 + {4 \pi \sigma^\Drude_\DC \over
 \Dd \bmk^2 - i \omega } , 
\quad 
 \bar R_\bmk (\omega)  = 
{\Dd \bmk^2 -  i \omega \over    e^2 2 \nud \Dd  \bmk^2} . \qquad 
\end{eqnarray}
%\begin{eqnarray}
% \bar R_\bmk (\omega) \; \simeq \;
%{\Dd \bmk^2 -  i \omega \over    e^2 2 \nud \Dd  \bmk^2} \; .
%\end{eqnarray}

\subsubappendix{Approximating $\tilde \rho^{\ns}_\iiijjj$ by 
$\tilde \rho_\iiijjj^0$} 
\label{sec:rho->rho0}

Even after having made the RPA approximation, the functional integral
in \Eq{eq:define-influence-functional-RonlyRPA} over all field
configurations of $V_\alpha$ is not yet Gaussian. The reason is that
the term ${\cal B'} \cdot   {\cal V} $ in the exponent depends,
via $\tilde w^{a-}$, on the full, interacting density matrix $\tilde
\rho_{\iiijjj}^\ns (t')$, which depends on the fields $V_\alpha$ too, in a
highly nontrivial way.  To make further progress, we shall ultimately
have to neglect the effect of interactions on the single-particle
density matrix, by replacing $\tilde \rho^\ns_{\iiijjj} (t')$ by its
noninteracting (and hence time-independent) version $\tilde
\rho_{\iiijjj}^0$:
\begin{eqnarray}
\label{eq:GZmainapproximation}
\tilde \rho^\ns_{\iiijjj} (t') \; \stackrel{{\rm approx}}{ \longrightarrow} \;
\tilde \rho^0_{\iiijjj} & \equiv & \langle \hat n^\pdag_{\iiijjj S} \rangle_0 \;
%\\
%  \label{eq:rho:t0-t'}
% & = &  \left. \tilde \rho_{\iiijjj}^\ns (t') \right|_{t_0 = t'}
\; .
\end{eqnarray}
GZ use this approximation at two points in their calculation [see the
comment after (GZ-II.43)]: (GZi) to simplify the propagators $\tilde
U^a_\iiijjj$, namely when passing from (GZ-II.40) to (GZ-II.43); and (GZii)
to simplify the thermal weighting factor describing the initial
distribution of electrons, namely to obtain the explicit factor
$\rho_0$ in (GZ-II.49).
 In our formalism, (GZii) would correspond to 
setting $t_0 \to t_2$, \ie\ making the
replacement $\tilde \rho^\ns_{2{\bar 2'}} (t_2, t_0) \to \rho^0_{2{\bar
    2'}}$ in \Eq{eq:defineJ1231} for $\tilde { J}^V_{12', 2 \bar 2' ,
  {\bar 2}1'} (t_1, t_2;t_0)$ and inserting the result into
\Eq{eq:finalGZsigma}, since this would reproduce (GZ-II.49).

We shall use similar but weaker approximations, and proceed in two
separate steps:
\\
(i) We ``linearize'' the exponential factor $\widetilde {\cal B} \cdot
{\cal V}$ in Eq.~(\ref{eq:define-influence-functional-RonlyRPA}) by
making the replacement $\widetilde {\cal B}[\tilde \rho^{\ns}_\iiijjj] \to
\widetilde {\cal B} [\tilde \rho^0_\iiijjj]$, so that the functional
integral (\ref{eq:define-influence-functional-RonlyRPA}) becomes truly
Gaussian in ${\cal V}$ and can readily be performed [see
Sec.~\ref{sec:integrateoutV}]. We thereby neglect the effect of
interactions on all occurences (via $\tilde w^{a -}$ in $\tilde
h^a_V$) of $\tilde \rho^{\ns}$ in the propagators $\tilde U^a_\iiijjj$,
the rationale being that in order to calculate the decoherence rate,
we are interested in how the interaction affects the time-evolved
\emph{propagation} of electrons along time-reversed paths, and not how
it modifies equal-time objects like $\tilde \rho_\iiijjj$.
Diagrammatically, this corresponds to neglecting diagrams which modify
the Keldysh Green's function without affecting the retarded or
advanced ones, \ie\ which modify only the $\tanh$ factor, but not
the propagator $\tilde U_\iiijjj$ in \Eq{eq:GKsummary}. 
%--- Another rationale for this approximation is that
%without it, one would retain in $\widetilde {\cal B} \cdot {\cal V}$
%terms of higher order in $V$, whereas terms of similarly higher order
%were neglected under the RPA approximation for ${\cal V} \cdot \tilde
%{\cal A} \cdot {\cal V}$.

\noindent (ii) 
For the propagator $J^{(\tilde a)}$, which is defined as the sum of
all terms for which the current vertex $j_{22'}$ occurs on contour
$\tilde a$ at time $t_{2_{\tilde a}}$, we neglect all interaction
vertices that occur on the \emph{same} contour $\tilde a$ at earlier
times $t_{3_{\tilde a}}$ or $t_{4_{\tilde a}} \in [t_0 , t_{2_{\tilde
    a}}]$. Thus, in the second lines of
\Eqs{eq:defJ(F)} and (\ref{eq:defJ(B)}), we make the replacements
$\tilde U^{\prime F}_{20} \to \U^0_{20}$ and $\tilde U'^B_{\bar 0 2'}
\to U^0_{\bar 0 2'}$. However, for the opposite contour containing no current
vertex, we include interaction vertices for all times $\in [t_0,
t_1]$.  The rationale for this is that, in diagrammatic language, this
approximation retains only those diagrams for which \emph{both}
current vertices $\hat \bmj_{2 2'}$ and $\hat \bmj_{1 1'}$ are always
sandwiched between a $\GR$- and a $\GA$-function, \ie\ $\tilde G^R
\bmj \, \tilde G^A$.  These are the ones relevant for the Cooperon;
the contributions thereby neglected correspond to the so-called
``interaction corrections'', which feature at least on current vertex
sandwidched between two retarded or advanced functions, \ie\ $\tilde
G^R \hat \bmj \, \tilde G^R$ or $\tilde G^A \hat \bmj \, \tilde G^A$.

Note that this approximation (ii) is much weaker than (GZii): we do
\emph{not} replace $ \tilde \rho^\ns_{2 \bar 2'}$ by $\tilde \rho^0_{2
  \bar 2'}$ in \Eq{eq:defineJ1231} (\ie\ we do not set $t_0 \to t_2$),
but instead use $ \tilde \rho^\ns_{2 \bar 2'} (t_2,t_0) = \tilde
U^F_{20} \tilde \rho^0_{0 \bar 0} \tilde U^B_{\bar 0 \bar 2'} $
[\Eq{eq:tilderhonsUU-ns}] and send $t_0 \to - \infty$.  Also, we
wish to emphasize that ``interaction correction'' terms \emph{can} be
calculated from our formalism if one so chooses, by avoiding our 
approximation (ii) altogether and keeping track of all interaction
insertions on the entire interval $[t_0, t_1]$ of \emph{both} contours
[\Eqs{eq:J1221(F2)FFexpandexplicit-2b} and
(\ref{eq:J1221(F)FFexpandexplicit-3b}) give  examples of such
contributions].  For the sake of greater generality, we shall thus for
the moment use only approximation (i), and postpone the use of (ii) to
\Sec{sec:IFvsPT}.

\subsubappendix{Integrating out the Fields $V_\alpha$ to obtain
$i\tilde S_R +   \tilde S_I$}
\label{sec:integrateoutV}

The approximations discussed in the previous two subsections render
the functional integral
(\ref{eq:define-influence-functional-RonlyRPA}) for $ \tilde {\cal
  F}_{(t_1,t_0)}[\bmR^a ]$ Gaussian. In fact,
\Eq{eq:define-influence-functional-RonlyRPA} is just of the form
(\ref{eq:QGaussian}), with $\zeta \cdot {\cal V}$ replaced by $
\widetilde {\cal B} \cdot {\cal V}$ of Eq.~(\ref{eq:defB-Ronly}),
so that we get 
\begin{eqnarray}
  \label{eq:Gaussian-Integral-Ronly}
\tilde {\cal F}_{(t_1,t_0)} [\bmR^a ] & =&
e^{-{1 \over2 \hbar}
\widetilde {\cal B} \cdot \tilde {\cal A}^{-1} \cdot \widetilde {\cal B}} \; \equiv
 e^{- [i \tilde S_R + \tilde S_I](t_1,t_0)/ \hbar} \; .
\end{eqnarray}
The exponent $(i \tilde S_R + \tilde S_I)[\bmR^a] \equiv \toh
\widetilde {\cal B} \cdot \tilde {\cal A}^{-1} \cdot \widetilde {\cal
  B}$, which is a functional of the paths $\bmR^a$, can be regarded as
an ``effective action'' that describes the effect of interactions on
the ``singled-out'' electron traveling along the paths $\bmR^a$.
The indices ${\scriptstyle R,I}$  are meant to
distinguish terms depending on the interaction propagators $\tilde R$
and $\tilde I$. 
Before working out the effective action explicitly form, however, let
us first collect results to obtain path integral expressions for the
correlators $ \tilde J^{(F/B)}_{12',21'} $ of \Eqs{eq:defJ(F/B)}.
These contain the correlators $\langle \tilde J^V_{12',2 \bar 2',
  {\bar 2}1'} \rangle_{V,\ns}$, for which we use \Eq{eq:JrhoRonly},
with $\tilde {\cal F}_{(t_1,t_0)}$ given by
\Eq{eq:Gaussian-Integral-Ronly}, and $\int dx_{\bar 2}$ integrals,
which we perform in the same way as for the second equalities of
\Eqs{eq:defJ(F/B)}:
\begin{eqnarray}
% \nonumber
\label{eq:intermediateJrho} 
\tilde {   J}^{(F/B)}_{12',2 1'} (t_1,t_2; t_0)
  & = & \pm 
{ 1 \over \hbar}
\int d x_{0_F,\bar 0_B} \, \tilde \rho^0_{0_F \bar  0_B} 
 \\ & & \nonumber \times 
\left\{ \begin{array}{c} 
{\displaystyle
\Fint_{2'_F}^{1_F}
\Fint^{2_F}_{0_F}
\Bint^{1'_B}_{\bar 0_B}
\widetilde {\cal D}' (\bmR) } \rule[-6mm]{0mm}{0mm} \!\! 
\\
{\displaystyle
\Fint_{0_F}^{1_F}
\Bint_{2_B}^{1'_B}
\Bint^{2'_B}_{\bar 0_B} \widetilde {\cal D}' (\bmR) \!\!  }
%\rule[-8mm]{0mm}{0mm}
\end{array} \right\} 
% \\ & & \label{eq:intermediateJrho} \times    
\Bigl.  e^{-[i \tilde  S_R + \tilde S_I](t_1,t_0)/\hbar} 
\Bigr|_{y^{F/B}=0}  \; . \qqph 
\end{eqnarray}
Combined with the current vertex insertions
$\int dx_{2} \, \bmj_{22'} \cdot  \bmj_{11'}  $  of 
\Eq{eq:sigmaDCreal}, we obtain
\begin{eqnarray}
\label{eq:intermediatej22Jrho} 
%\nonumber
\phantom{.} \hspace{-0.8cm}
& &  \int \!\! dx_{2} \, \bmj_{22'} \cdot \bmj_{11'} 
 \sum_{\tilde a = F, B} 
\tilde J^{(\tilde a)}_{12',21'} (t_1, t_2; t_0) \, = \, 
%\\ \nonumber & = &
\int  \!\! d x_{0_F,\bar 0_B} \, \tilde \rho^0_{0_F \bar  0_B} 
\Fint_{0_F}^{1_F}  \!\!\!
\Bint_{\bar 0_B}^{1'_B} \!\!\!
\widetilde {\cal D}' (\bmR ) \quad 
\\ \nonumber 
\phantom{.} \hspace{-0.8cm}
& & \times 
{1 \over \hbar} \Biggl\{
\left[ \hat \bmj (t_{2_F})  \hat \bmj (t_1)
e^{-[i \tilde S_R  + \tilde S_I]
(t_1,t_0)/\hbar}\right]_{y^F=0} 
% \Biggr.
%\\
%& &   
%\Biggl.
- \left[ \hat \bmj (t_{2_B}) \hat \bmj (t_1)
 e^{-[i \tilde S_R  + \tilde S_I](t_1,t_0)/\hbar} \right]_{y^B=0} \Biggr\}
. \quad \phantom{.}
\end{eqnarray}
This expression, which is the first central result of our formalism,
has a simple interpretation: thermal averaging with $\tilde \rho_{0
  \bar 0}^0$ at time $t_0$ ($ \to - \infty$) is followed by
propagation, in the presence of interactions (described by $e^{-[i
  \tilde S_R + \tilde S_I]}$), along all possible paths from time
$t_0$ up to time $t_1$, with insertions of current vertices $\hat \bmj
(t_{2_a})$ at time $t_2$ on either the upper or lower Keldysh contour,
and another current vertex $\hat \bmj (t_1)$ at the final time.
%The insertion
%$\hat \bmj (t_{2_a})$ is defined via its action in a free path
%integral (with $t_j < t_{2_a} < t_i$),
%\begin{eqnarray}
%\Fint_{j_F}^{i_F}
%\Bint_{\barj_B}^{\bari_B}
% \widetilde {\cal D}' (\bmR ) \, 
% \hat \bmj (t_{2_a})
%&\equiv&
%\label{eq:DRDPj2} 
% \left\{ \begin{array}{c} 
%{\displaystyle \int}
% dx_2 \, \tilde U^{0F}_{i2} \, \hat \bmj \, \tilde U^{0F}_{2j}
%\tilde U^{0B}_{\barj \, \bari}
%\rule[-4mm]{0mm}{0mm}
%\\
%{\displaystyle \int} dx_2 \, \tilde U^{0F}_{\iiijjj}  \, \tilde U^{0B}_{\barj 2}
%\, \hat \bmj \, \tilde U^{0B}_{2 \bari}  \end{array}
%\right. \; ,
%\end{eqnarray}
%where  the current operator in the coordinate representation
%is defined, for any two functions $f(\bmr_2) $ and $g(\bmr_2)$,  as
%%\begin{eqnarray}
%  \label{eq:currentoperatorcoordinaterep}
% $ f (\bmr_2) \, \hat \bmj \, g (\bmr_2) = %& = &
%\bigl[ \bmj_{2 2'} f(\bmr_{2'}) g(\bmr_2) \bigr]_{\bmr_2 =
%  \bmr_{2'}} \; .$

Let us now determine the effective action explicitly,
by using \Eq{eq:defB-Ronly}  for $\widetilde {\cal B}$ to evaluate
$\toh \widetilde {\cal B} \cdot \tilde {\cal A}^{-1} \cdot \widetilde {\cal
  B}$:
% Book 7, p. 6  
\begin{eqnarray}
\label{eq:BdotAdotB}
[i \tilde S_R + \tilde S_I](t_1,t_0) [\bmR^a ]
 & =  &
{1 \over 2}
 \sum_{\alpha \alpha'} \int_{t_0}^{t_1} \!\!\! d t_{3}
\int_{t_0}^{t_1} \!\!\! d t_{4}
\, \\
& & \nonumber \phantom{.} \hspace{-4.7cm} 
\times \!\! \Biggl[
2 \theta_{34} \, s_F \tW^{F\alpha}_{3_F \bar 3_F} 
(\tilde {\cal A}^{-1})^{\alpha \alpha'}_{\bar 3_F \bar 4_F} \,
s_F \tW^{F \alpha'}_{4_F \bar 4_F} 
+
s_B \tW^{B\alpha}_{\bar 3_B 3_B} 
(\tilde {\cal A}^{-1})^{\alpha \alpha'}_{\bar 3_B \bar 4_F} \,
s_F \tW^{F \alpha'}_{4_F \bar 4_F} \,
\Biggr. \qph
\\
& & \nonumber
\phantom{.} \hspace{-4.7cm} + \Biggl.
s_F \tW^{F\alpha}_{3_F \bar 3_F} 
(\tilde {\cal A}^{-1})^{\alpha \alpha'}_{\bar 3_F \bar 4_B} \,
s_B \tW^{B \alpha'}_{\bar 4_B  4_B} 
 +
2 \theta_{34} \,
s_B \tW^{B\alpha}_{\bar 3_B  3_B} 
(\tilde {\cal A}^{-1})^{\alpha \alpha'}_{\bar 3_B \bar 4_B} \,
s_B \tW^{B \alpha'}_{\bar 4_B  4_B} 
\Biggr] .
\end{eqnarray}
There are now two somewhat different routes 
to proceed, which lead to two somewhat different 
(but equivalent) representations for the effective action.
The first, followed in the present section, 
exploits symmetries under $3 \leftrightarrow 4$ write
the effective action in terms of as few terms as possible,
leading to expressions [(\ref{eq:defineSiR}), (\ref{eq:Ltildes}), 
or (\ref{eq:defineSiRA}), (\ref{eq:LtildesA})] 
useful for recovering the
Keldysh diagrammatic results for the Cooperon self energy
[(\ref{eq:selfenergies-explicit}) or (\ref{eq:selfenergies-explicit-main})].
The second, summarized in Section~\ref{sec:alteffaction}, does
not combine similar-looking terms, and is useful
for establishing contact with other, more standard
influence-functional approaches. 

Let us proceed with the first route.
Since $(\tilde {\cal A}^{-1})^{\alpha \alpha'}_{34} = ({\cal
  A}^{-1})^{\alpha' \alpha}_{43}$, the integrand in \Eq{eq:BdotAdotB}
for $\widetilde {\cal B} \cdot \tilde A^{-1} \cdot \widetilde {\cal
  B}$ is symmetric under the exchange of variables $\sum_\alpha \int
dt_3 \, d \bmr_3 \leftrightarrow \sum_{\alpha'} \int dt_4 \, d
\bmr_4$.  We have exploited this fact to insert a factor of $2
\theta_{34}$ into the first and last terms of \Eq{eq:BdotAdotB}, which
both \emph{individually} have this symmetry, to obtain time-ordered
integrals for these, which has the advantage of reducing the number of
terms in subsequent expressions.  (We could similarly have inserted $2
\theta_{34}$ into the second and third terms of \Eq{eq:BdotAdotB},
too, but since only their \emph{sum} has the
above-mentioned symmetry, this turns out to be inconvenient.)
  
  More explicit expressions for $\tilde S_{R/I}$ can be
  found with the help of \Eqs{eq:Wvertices} for $ \tW^{a \alpha}_\iiijjj$,
  \Eq{eq:even-odd-correlatorVV} for $\tilde {\cal A}^{-1}$ and
  recalling that $\theta_{34} \tilde R_{\bar 4
    \bar 3} = 0$.  Using the shorthand $(i \tilde S_R/ \tilde S_I)$ to
  present two similar equations in one line, and writing $(i \tilde
  R/\tilde I)_{\bari_a \barj_{a'}} = (i \tilde R/\tilde I) [ t_\iiijjj,
  \bmr^a_{\bari} (t_i) -  \bmr^{a'}_\barj (t_j) ]$, where $t_\iiijjj = t_i -
  t_j$ [and likewise for ${\cal L}^{R,A,K}_{\bari_a \barj_{a'}}$], 
  we find:
% Book 7, p. 177
\begin{eqnarray}
\label{eq:defineSiR}
[i \tilde S_R/ \tilde S_I] (t_1, t_0) \mbox{[}\bmR^a \mbox{]}
& \equiv &
 \sum_{aa'} \int_{t_0}^{t_1} d t_{3_a} \int_{t_0}^{t_1} d t_{4_{a'}}
 \,
(i \tilde L^R / \tilde L^I)_{3_a 4_{a'}}  \mbox{[}\bmR^a \mbox{]} \; ,
\qqph
\end{eqnarray}
\bsubequations
    \label{eq:Ltildes}
\begin{eqnarray}
\label{eq:LtildesFF} 
  - (i \tilde L^R / \tilde L^I)_{3_F 4_F}  & = &
- \theta_{34} \,   s_F s_F \, \tW^{F+}_{3_F \bar 3_F}  \,
\tW^{F\mp }_{4_F \bar 4_F}  ( i \tilde R/ \tilde I)_{\bar 3_F \bar
  4_F}
\\ \nonumber 
& = & 
\phantom{-}  \toh i \:\, \tilde \delta_{3_F \bar 3_F}
\biggl \{  \begin{array}{c}
%2 \tilde \rho^0)_{4_F \bar 4_F}
\!\! [\tilde \delta -  (\theta_{42} + y^F \theta_{24})
2 \tilde \rho^0 ]_{4_F \bar 4_F}
\rule[-2mm]{0mm}{0mm}
\\
\theta_{34} \, \tilde \delta_{4_F \bar 4_F}
\end{array} \!\! \biggr\}
\, \tilde {\cal L}^{R/K}_{\bar 3_F \bar   4_F}
\; , 
\end{eqnarray}
\begin{eqnarray}
%\rule[-7mm]{0mm}{0mm} 
%\\
\label{eq:LtildesBF}
  - (i \tilde L^R / \tilde L^I)_{3_B 4_F}  & = &
- \toh 
%\left\{ \begin{array}{c} 1 \\ \toh \end{array} \right\}
\,   s_B s_F \,
\tW^{F\mp }_{4_F \bar 4_F}  ( 2 i \tilde R/ \tilde I)_{\bar 3_B \bar
  4_F} \, \tW^{B+}_{\bar 3_B 3_B}
\\ & = &     \nonumber
 -  \toh i \:  \,
\biggl \{  \begin{array}{c}
%  (\tilde \delta - 2 \tilde \rho^0)_{4_F \bar 4_F}
\!\!
[\tilde \delta -  (\theta_{42} + y^F \theta_{24})
2 \tilde \rho^0 ]_{4_F \bar 4_F}
\rule[-2mm]{0mm}{0mm}
\\
\toh  \tilde \delta_{4_F \bar 4_F}
\end{array} \!\! \biggr\}
\, \tilde {\cal L}^{R/K}_{\bar 3_B \bar   4_F} \,
\tilde \delta_{\bar 3_B 3_B} \; , 
\end{eqnarray}
\begin{eqnarray}
%  \rule[-7mm]{0mm}{0mm}
%\\
 \label{eq:LtildesFB}
 - (i \tilde L^R / \tilde L^I)_{3_F 4_B}  & = &
- \toh 
%\left\{ \begin{array}{c} 1 \\ \toh \end{array} \right\}
 \,   s_F s_B \, \tW^{F+}_{3_F \bar 3_F}  \,
  ( 2 i \tilde R/  \tilde I)_{\bar 3_F \bar  4_B} 
\, \tW^{B\mp }_{\bar 4_B 4_B}
\\ & = &    \nonumber 
\pm \toh i \:  \, \tilde \delta_{3_F \bar 3_F}
\, \tilde {\cal L}^{A/K}_{ \bar   4_B \bar 3_F}
\,
\biggl \{  \begin{array}{c}
%(\tilde \delta - 2 \tilde \rho^0)_{\bar 4_B 4_B}
\!\!
[\tilde \delta -  (\theta_{42} + y^B \theta_{24})
2 \tilde \rho^0 ]_{\bar 4_B 4_B}
\rule[-2mm]{0mm}{0mm}
\\
\toh \tilde \delta_{\bar 4_B 4_B}
\end{array} \!\! \biggr\}
\; ,  
\end{eqnarray}
%\rule[-7mm]{0mm}{0mm}
%\\
\begin{eqnarray}
\label{eq:LtildesBB}
 - (i \tilde L^R / \tilde L^I)_{3_B 4_B}  & = &
- \theta_{34} \,   s_B s_B \,
 ( i \tilde R/ \tilde I)_{\bar 3_B \bar   4_B}
\, \tW^{B+}_{\bar 3_B  3_B} \,
\tW^{B \mp }_{\bar 4_B 4_B}
\\ & = &     \nonumber 
\mp  \toh i \; \,
 \tilde {\cal L}^{A/K}_{\bar 4_B \bar   3_B} \,
 \tilde \delta_{\bar 3_B 3_B} \,
\biggl \{  \begin{array}{c}
%  (\tilde \delta - 2 \tilde \rho^0)_{\bar 4_B 4_B}
\!\!
[\tilde \delta -  (\theta_{42} + y^B \theta_{24})
2 \tilde \rho^0 ]_{\bar 4_B 4_B}
\rule[-2mm]{0mm}{0mm}
\\
\theta_{34} \, \tilde \delta_{\bar 4_B 4_B}
\end{array} \!\! \biggr\} \; .
\end{eqnarray}
\esubequations
The $\tilde \delta_{\bari i}$ functions in the second lines of
\Eq{eq:Ltildes} will remove one of the coordinate integrations $\int d
x_{\bari, i}$ that are contained in the path integral $\int \widetilde
{\cal D}' \! (\bmR)$.
%Note that the sign
%    for $ - (i \tilde L^R/ \tilde L^I)_{3_a 4_{a'}} $
%is given by $(-1)^{\delta_{aB}} (\pm 1 )^{\delta_{a' B}}$.
The second and third terms of \Eq{eq:BdotAdotB} are equal, as can be
seen by setting $3 \leftrightarrow 4$ and $\alpha \leftrightarrow
\alpha'$ in the third and recalling that $(\tilde {\cal A}^{-1})^{\alpha'
  \alpha}_{43} = (\tilde {\cal A}^{-1})^{\alpha \alpha'}_{34}$; we exploited
this property above to combine those contributions from these terms
that are proportional to $\tilde R_{\bar 3_B \bar 4_F}$ [or $\tilde
R_{\bar 3_F \bar 4_B} $] together into \Eq{eq:LtildesBF} [or
\Eq{eq:LtildesFB}], hence the factors of $2\tilde R$ in these
equations.

Note that if we make approximation (ii) of Sec.~\ref{sec:rho->rho0}, a
useful simplification occurs [which was exploited in App.~A to obtain
\Eqs{eq:LtildesA} from \Eqs{eq:Ltildes}]: all the factors
$(\theta_{4_{a'}2} + y^{a'} \theta_{24_{a'}})$ above then
\label{p:thetay=1}reduce\footnote{\label{f:thetay=1} To see this explicitly, we argue as follows,
  discussing in parallel the cases of $\tilde J^{(\tilde a = F/B)}$,
  having a current vertex on the upper/lower contour and for which we
  have decided to use $y^{\tilde a = F/B} = 0$: if an interaction
  vertex lies on the same contour as the current vertex, \ie\ for
  $\tilde J^{(\tilde a=F/B)}$ on the upper/lower contour at time
  $t_{4_{F/B}}$ (hence $a' = F/B$), approximation (ii) says that it
  must lie at greater times than the current vertex, $t_{4_{F/B}} >
  t_{2_{F/B}}$, implying that $(\theta_{4_{F/B}2} + y^{F/B}
  \theta_{24_{F/B}}) = 1$.  If instead the interaction vertex lies on
  the opposite contour than the current vertex, \ie\ for $\tilde
  J^{(a=F/B)}$ on the lower/upper contour at time $t_{4_{B/F}}$ (hence
  $a' = B/F$), the fact that $y^F + y^B = 1$ (always) and that we
  chose $y^{\tilde a = F/B} = 0$, implying $y^{B/F} = 1$, also gives
  $(\theta_{4_{B/F}2} + y^{B/F} \theta_{24_{B/F}}) = 1$, independent
  of the value of $t_{4_{B/F}}$.  } to 1, because $y^{a'} \neq 1$ was
needed only to deal with interaction vertices occuring at times
$t_{4_{a'}}$ earlier than a current vertex on the same contour $a'$,
and these are precisely the ones that are dropped under approximation
(ii).

\Eqs{eq:Ltildes} for the effective action $(i \tilde S_I + \tilde
S_R)$ constitute the second central result of this section.  It should
be emphasized that in the path integral
(\ref{eq:intermediatej22Jrho}), the Pauli principle is fully accounted
for by the Pauli factors $(\tilde \delta - 2 \tilde \rho)$ in $\tilde
S^R$. The ability to incorporate the Pauli principle into an influence
functional for interacting electrons may be regarded as one of the
main achievements of the formalism developed so far.

This concludes our rederivation of GZ's influence functional.  In the
remaining section B.6, where we show how it is related to diagrammatic
Keldysh perturbation theory, and in the main text, where we use it to
calculated the decoherence rate $\gammaphi$, our analysis differs
significantly from GZ's, since we come to different conclusions.

Let us here just mention one such difference: According to the first
lines of \Eqs{eq:Ltildes}, $i \tilde S_R$ and $\tilde S_I$ are,
respectively, purely imaginary or purely real functionals of the paths
$\bmR^a$, since $\tW^{a \alpha}$, $\tilde R_\iiijjj$ and $\tilde I_\iiijjj$
are all purely real functions. GZ have used this fact to argue that
after averaging $e^{-(i \tilde S_R + \tilde S_I)}$ over all paths [as
required by the path integrals in \Eq{eq:intermediateJrho}], $e^{-
  \tilde S_I/\hbar}$ will produce an exponentially \emph{decaying}
function of time and thereby determine the interaction-induced
decoherence rate, whereas $e^{-i \tilde S_R}$ will just produce an
\emph{oscillating} time dependence, and hence, quite generally, cannot
contribute to decoherence; in particular, they argued that ``$i\tilde
S_R$ can never cancel any contribution from $\tilde S_I$'' [discussion
before (GZ-III.22)].  This general argument would work if the measure
used in the path integral were real; however, it does not apply to the
present case of \Eq{eq:intermediateJrho} where the measure $e^{\pm i
  S_0^{F/B}/ \hbar}$ is \emph{complex}, since the average of a purely
oscillatory function, using a complex measure, can well contain a
decaying component, too. Indeed, it is shown in the main text [end of
\Sec{sec:GZ-classical-paths}] that contributions from $i\tilde S_R$
and $\tilde S_I$ \emph{do} partially cancel each other.

\subappendix{Influence Functional vs. Keldysh  Diagrammatics}
\label{sec:IFvsPT}

\label{sec:perttheory}

To check the general formalism developed above, it is important and
instructive to verify that it can reproduce the standard results of
diagrammatic Keldysh perturbation theory, \emph{before disorder
  averaging}.  We shall do this by expanding the path integrals
(\ref{eq:intermediateJrho}) in powers of the effective action $(i
\tilde S_R + \tilde S_I$):
\begin{eqnarray}
  \label{eq:expandUU}
\Fint_{0_F}^{1_F}
\Bint_{\bar 0_B}^{1'_B}
  \widetilde {\cal D}' (\bmR )
\, e^{-(i \tilde S_R  + \tilde S_I)/\hbar}
& = & \sum_{N=0}^\infty {1 \over N!} \,
\Fint_{0_F}^{1_F}
\Bint_{ \bar 0_B}^{ 1'_B}
 \widetilde {\cal D}' (\bmR )
\\ &  & \nonumber 
\phantom{.} \hspace{-3cm} \times \left[ {- 1  \over \hbar}
\sum_{aa'}
\int_{t_0}^{t_1} d t_{3_a} \int_{t_0}^{t_1} d t_{4_{a'}}
\bigl[ i \tilde L^{R}_{3_a 4_{a'}} + \tilde L^I_{3_a 4_{a'}} \bigr]
 \right]^N \; . 
\end{eqnarray}
Now and henceforth using approximation (ii) of
Sec.~\ref{sec:rho->rho0}, we shall use this expansion to reproduce the
Keldysh expressions for the conductivity in first order perturbation
theory [\Eqs{subeq:Jprime1storder}], and to obtain general expressions
for the first order contributions to the Cooperon before disorder
averaging [\Eqs{eq:selfenergies-explicit}], thereby
reproducing the familiar Keldysh diagrams for the Cooperon self energy
[Fig.~\ref{fig:cooperonselfenergy} of App.~A].

\subsubappendix{First Order Terms and Cooperon Self Energy
$\tilde \Sigma^{R/I}$} 
\label{sec:selfenergy}

\begin{figure}[t]
{\includegraphics[clip,width=\linewidth]{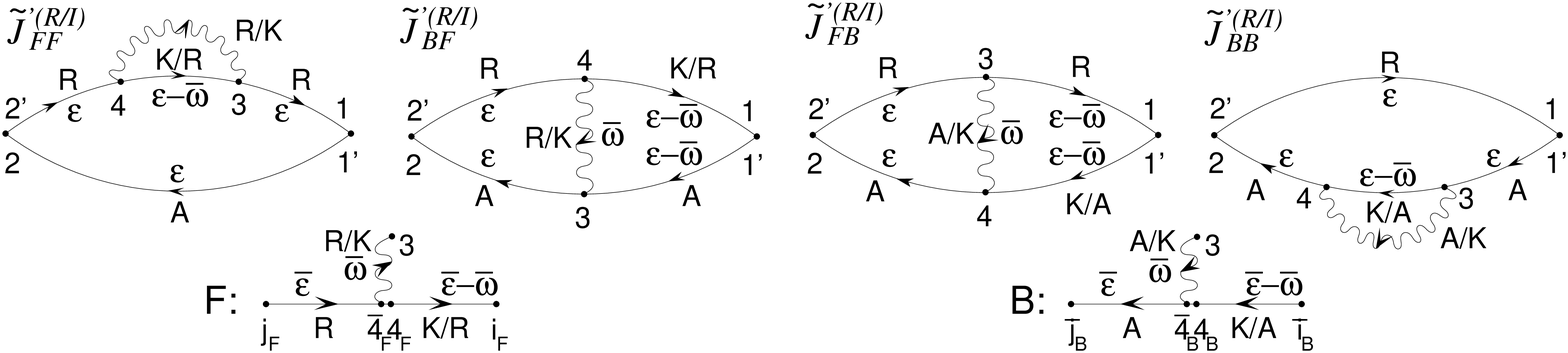}}%
\caption{
  Feynman diagrams for the first-order correlators $\tilde {
    J}^{\prime (R/I)}_{aa'}$ of \Eq{subeq:Jprime1storder}, and for the
  vertices of \Eqs{eq:1-2rho->tanhindicesF}.}
\label{fig:1orderperturbation}
\end{figure}
                            
The $N = 1$ terms of \Eq{eq:expandUU} can be used to obtain the first
order contributions, $\tilde J^{\prime (1)}_{12',21'} (0)$, to the
correlators needed for $\sigma_{\DC, \rreal}$ [\Eq{eq:sigmaDCreal}].
This straightforward, if tedious, excercise is discussed in
App.~\ref{app:1storderperturbation}.  The result can be written as
$\tilde J^{\prime (1)}_{12',21'} (0) = \int (d \ve) \tilde J^{\prime
  (1) \ve}_{12',21'}$, where [see Fig.~\ref{fig:1orderperturbation}]
\bsubequations
 \label{subeq:Jprime1storder}
\begin{eqnarray}
% \label{subeq:Jprime1storder}
%  \label{eq:Jprime1stordergeneral}
% & & \phantom{.} \hspace{-2.5cm}
\tilde J^{\prime (1) \ve}_{12',21'} 
& =  &    [ -  n'_0 ( \hbar  \varepsilon)] \,  ( - \toh  i  \hbar^3 ) \!
 \int   (d \bomega) \, 
\sum_{aa'} \left[\tilde J^{\prime
  (R)}_{aa'} + \tilde J^{\prime (I)}_{aa'} \right] \; , \qquad
\\
% \nonumber % \label{eq:JprimeFF}
 \tilde { J}^{\prime (R/I)}_{FF}  
& = & 
\G^R_{13} (\ve) \,
\G^{K/R}_{3 4} (\ve - \bomega) \, 
 \G^R_{42'} (\ve) \, \G^A_{21'} (\ve)  
\bigl(\LL^R / \LL^K \bigr)_{3 4} (\bomega) \; , 
\rule[-3mm]{0mm}{0mm}
\\
% \nonumber % \label{eq:JprimeBF}
 \tilde { J}^{\prime (R/I)}_{BF}  
& = & 
\G^{K/R}_{14} (\ve - \bomega) \,
\G^{R}_{42'} (\ve ) \, 
 \G^A_{23} (\ve) \, \G^A_{31'} (\ve - \bomega)   
\bigl(\LL^R / \toh \LL^K \bigr)_{3 4} (\bomega) \; , 
\rule[-3mm]{0mm}{0mm} \qqph
\\
% \nonumber %  \label{eq:JprimeFB}
 \tilde { J}^{\prime (R/I)}_{FB} 
& = & 
\G^{R}_{13} (\ve- \bomega) \,
\G^{R}_{32'} (\ve ) \, 
 \G^A_{24} (\ve) \, 
\G^{K/A}_{41'} (\ve - \bomega)   
\bigl(\LL^A / \toh \LL^K \bigr)_{ 43} (\bomega) \; , 
\rule[-3mm]{0mm}{0mm}
\\
% \nonumber %  \label{eq:JprimeBB}
 \tilde { J}^{\prime (R/I)}_{BB}  
& = & 
\G^R_{12'} (\ve) \,
\G^A_{2 4} (\ve ) \, 
 \G^{K/A}_{43} (\ve - \bomega) \, \G^A_{31'} (\ve)  
\bigl(\LL^A / \LL^K \bigr)_{43} (\bomega) \; , 
\end{eqnarray}
\esubequations
where $ \tilde J^{\prime (R/I)}_{aa'} $ denotes a first-order
contribution from $(i \tilde S^R/\tilde S^I)$, with interaction
vertices that lie on contours $a$ and $a'$. 
These expressions agree
with those of standard diagrammatic Keldysh perturbation theory. Thus,
the basic building blocks of the influence functional approach,
including its treatment of the Pauli principle, have survived their
first test.
%\begin{figure}[htbp]
%  \centering
%  \includegraphics[width=\linewidth]{cooperonselfenergy.eps}
%  \caption{First order contributions to the irreducible self energy
%    of the Cooperon, illustrating (a) \Eq{eq:cooperonfirstorderb}, and
%    (b,c) \Eqs{eq:selfenergies-explicit}. The arrows associated with
%    each factor $\tilde G_\iiijjj$ or $\tilde {\cal L}_\iiijjj$ in
%    \Eqs{eq:selfenergies-explicit} are drawn to point from the second
%    index to the first ($j$ to $i$).  Filled double dots denote the
%    occurence of a factor $(\tilde \delta - 2 \tilde \rho)_{4_F \bar
%      4_F}$ on the upper contour or $(\tilde \delta - 2 \tilde
%    \rho)_{\bar 4_B 4_B}$ on the lower contour. Bars on filled dots
%    are used to indicate the barred indices to which the interaction
%    lines is connected.  Both filled and open single dots indicate a
%    delta function $\tilde \delta$; the open dots stand for delta
%    functions that have been inserted to exhaust dummy integrations,
%    as discussed after \Eqs{eq:cooperonfirstorder}.
%    The diagrams in (b) and (c) coincide precisely with those obtained
%    by standard Keldysh diagrammatic perturbation theory for the
%    Cooperon self energy, as depicted, \eg, in Fig.~2 of
%    Ref.~\cite{AVA01}. (There, impurity lines needed for impurity
%    averaging are also depicted; in the present figure, they are
%    suppressed.)  }
%  \label{fig:cooperonselfenergy}
%\end{figure}

Next, we shall derive a general expression for the self energy of the
Cooperon propagator.  Usually, the Cooperon self energy is defined,
after Fourier transforming to momentum space and disorder averaging,
by a Dyson equation of the form $\bcC_\bmq = \bcC{}_\bmq^0 +
\bcC{}_\bmq^0 \overline \Sigma_\bmq \bcC_\bmq$, where $\bcC{}_\bmq^0$,
the free Cooperon in the absence of interactions, is the contribution
to $\langle \G^R \G^A \rangle_\dis$ of time-reversed paths [cf.\ 
\Eq{eq:GGdisorderaverage}].  To identify a similar structure in
position space and before disorder averaging, we need to write the
%Using \Eqs{eq:Ltildes} for $\tilde L^{R/I}_{3_a 4_{a'}}$, it is
%straightforward to verify that the 
first order $(N=1)$ term of \Eq{eq:expandUU} in the form $\U^F_B \cdot
\tilde \Sigma \cdot \U^F_B$, \ie\ a self-energy insertion sandwidched
by two forward-backward propagators $(\tilde U^F_B)^{\bari_F j_F}_{j_B
  \bari_B} \equiv \tilde U^{0F, \bari_F j_F} \tilde U^{0B}_{j_B
  \bari_B} = \hbar^2 \tilde G^{R, \bari_F j_F} \tilde G^{A}_{j_B
  \bari_B}$ (each of which will produce a Cooperon upon disorder
averaging):
\bsubequations
\label{eq:cooperonfirstorder}
\begin{eqnarray}
% \nonumber
%& &  \!\!\!\!\!\!  \!\!\!\!\!\!  \!\!\!\!\!\! 
\label{eq:cooperonfirstordera}
\lefteqn{ % \phantom{.} \hspace{-1.5cm} 
- {\displaystyle  {1 \over \hbar}} \, 
\Fint_{j_F}^{\bari_F}
\Bint_{\barj_B}^{i_B}
 \widetilde {\cal D}' (\bmR )
(i \tilde L^R / \tilde L^I)_{aa'} (t_3, t_4)
%\\ 
%& &  \!\!\!\!\!\!  \!\!\!\!\!\!  \!\!\!\!\!\! 
= %  {s_a s_{a'}} \,
\bigl( \tilde U^F_B \cdot \tilde \Sigma^{R/I}_{aa'} \cdot  \tilde U^F_B
\bigr)^{\bari_F j_F}_{\barj_B i_B}  \;   \qqph \qph }
\\
% & &  % \!\!\!\!\!\!  \!\!\!\!\!\!  \!\!\!\!\!\! 
\label{eq:cooperonfirstorderb}
& =  & % {s_a s_{a'} }
\!\! \int  \! d x_{3_F} d x_{\bar 4_B} \int d x_{\bar 4_F} d x_{3_B}    \,
(\tilde U^F_B)^{\bari_F 3_F}_{\barj_B \bar 4_B}
 \left( \tilde \Sigma^{R/I}_{aa'} \right)^{3_F \bar
  4_F}_{\bar 4_B 3_B} \!\! (\tilde U^F_B)^{\bar 4_F j_F}_{3_B i_B} \; .
\qqph \qqph
\end{eqnarray}
\esubequations
(For ease of recognition, we here and henceforth in this section write
indices associated with the forward $({\scriptstyle F})$ or backward
$({\scriptstyle B})$ paths as superscripts or subscripts,
respectively). As made explicit by \Eq{eq:cooperonfirstorderb}, the
first (or second) dot product on the right-hand side of
\Eq{eq:cooperonfirstordera} indicates integration over the two
coordinates associated with the two ``outgoing'' (or the two
``incoming'') vertices at the corners of the self-energy box [see
Fig.~\ref{fig:cooperonselfenergy}a].  Now, the left-hand side of
\Eq{eq:cooperonfirstordera} contains \emph{two} vertices, associated
with the indices of $(i \tilde L^R / \tilde L^I)_{3_a 4_{a'}}$
[\Eqs{eq:Ltildes}], as insertions into a double path integral, and
therefore contains \emph{four} Green's functions $\tilde G$ [cf.\ the
rule of thumb (\ref{eq:ruleofthumbPI}) of App.~D.3]; however, for
$\U^F_B \cdot \tilde \Sigma \cdot \U^F_B$, we formally need \emph{six}
Green's functions $\tilde G$ and \emph{four} vertices, one for each
corner of the self-energy box.  To achieve this, we proceed as
follows: the two corners to which the interaction lines are connected
[black dots in Fig.~\ref{fig:cooperonselfenergy}] can be naturally
labelled by $a$ and $a'$, which take the values $F/B$, according to
the contour that the corner sits on; let $\bar a$ and $\bar a'$
similarly label the other two, ``free'' corners [empty circles in
Fig.~\ref{fig:cooperonselfenergy}].
% (thus, $s_a s_{a'} = s_{\bar a} s_{\bar a'}$). 
For the free corner $\bar a$ (and similarly for $\bar a'$), we use the identity
($t_k$ is an arbitrary time between $t_i$ and $t_j$)
\begin{eqnarray}
  \label{eq:G=GG}
\tilde G^{R/A}_{i_{\bar a} j_{\bar a}}
= (s_{\bar a} i h ) \int dx_{k_{\bar a}, \bar k_{\bar a}}
 \, \tilde G^{R/A}_{i_{\bar a} k_{\bar a}} \, 
\tilde \delta_{k_{\bar a} \bar k_{\bar a}} \, 
\tilde G^{R/A}_{\bar k_{\bar a} j_{\bar a}} \; , 
  \end{eqnarray}
  taking $R/A$ and $s_{\bar a} = \pm 1$ if $\bar a = F/B$, to write
  one Green's function as the convolution of two, and regard the
  $\tilde \delta_{\bar a}$ function as the ``vertex'' at the
  corresponding free corner of the self energy box.\footnote{
% % Book 6, page 90 and 91
    By using Eq.~(\protect\ref{eq:G=GG}) twice at the two
    free corners, an extra overall phase factor of $( i s_{\bar a}) (
    i s_{\bar a'}) = - s_a s_{a'}$ is generated.  The latter cancels
    the overall phase factor $ (- s_a s_{a'})$ occuring in the first
    lines of Eqs.~(\protect\ref{eq:defineSiR}) for $- (i \tilde L^R
    /\tilde L^I)$, which is why this factor does not occur in the
    first lines of Eqs.~(\protect\ref{eq:selfenergies-explicit}).}  In
  this way, the self-energy contributions $\tilde \Sigma^{R/I}_{a a'}$
  are found to be
%, found by
%inserting \Eqs{eq:Ltildes} for $\tilde L^{R/I}_{aa'}$ into
%\Eq{eq:cooperonfirstordera} and 
given by the first lines of the following equations (summarized
diagrammatically in Fig.~\ref{fig:cooperonselfenergy}):
\bsubequations
  \label{eq:selfenergies-explicit}
  \begin{eqnarray}
\nonumber
\left( \tilde \Sigma^{R/I}_{FF} \right)^{3_F \bar   4_F}_{\bar 4_B 3_B} &=&
%\left \{ \begin{array}{l} 1 \\ \toh \end{array} \right\}
 \, \theta_{34}
\left( \tW^{F+} \tilde \delta_B \cdot \tilde U^F_B \cdot
\tilde \delta_B  \tW^{F \mp} \right)^{3_F \bar   4_F}_{\bar 4_B 3_B}
 {1 \over \hbar} ( i \tilde R / \tilde I)^{3_F \bar 4_F}
\\     \label{eq:SigmaFF} &  = & 
- {i \hbar \over 2} \;
%(-  \toh i \hbar ) \;
(\tilde G^{K/R})^{3_F \bar 4_F} \tilde G^A_{\bar 4_B 3_B}
 ( \tilde {\cal L}^R/ \tilde {\cal L}^K)^{3_F \bar 4_F}
\; , \quad \rule[-6mm]{0mm}{0mm}
\end{eqnarray}
\begin{eqnarray}
\nonumber
\left( \tilde \Sigma^{R/I}_{BF} \right)^{3_F \bar
 4_F}_{\bar 4_B 3_B} &=&
%\left \{ \begin{array}{l} 1 \\ \toh \end{array} \right\}
 \, \phantom{\theta_{34}}
  \left( \tilde \delta^F \tilde \delta_B \cdot \tilde U^F_B \cdot
\tW_{B+} \tW^{F \mp} \right)^{3_F \bar   4_F}_{\bar 4_B 3_B}
{1 \over \hbar}  ( i \tilde R / \tilde I)_{3_B}^{\;\;\;\;  \bar 4_F}
\\ & = &     \label{eq:SigmaBF}
- {i \hbar \over 2} \;
%(-  \toh i \hbar ) \;
  (\tilde G^{K/R})^{3_F \bar 4_F} \tilde G^A_{\bar 4_B 3_B}
 ( \tilde {\cal L}^R/ \toh \tilde {\cal L}^K)_{3_B}^{\;\;\;\;  \bar 4_F}
 ,   \rule[-6mm]{0mm}{0mm} 
\end{eqnarray}
\begin{eqnarray}
\nonumber
 \left( \tilde \Sigma^{R/I}_{FB} \right)^{3_F \bar
  4_F}_{\bar 4_B 3_B} &=&
%\left \{ \begin{array}{l} 1 \\ \toh \end{array} \right\}
 \, \phantom{\theta_{34}}
\left( \tW^{F+} \tW_{B \mp} \cdot \tilde U^F_B \cdot
\tilde \delta_B \, \tilde  \delta^F \right)^{3_F \bar   4_F}_{\bar 4_B 3_B}
{1 \over \hbar}  (i  \tilde R / \tilde I)^{3_F}_{\;\; \;\;  \bar 4_B}
\\     \label{eq:SigmaFB} & = & 
- {i \hbar \over 2} \;
%(-  \toh  i \hbar ) \;
\tilde G^{R, 3_F \bar 4_F} (\tilde G^{K/A})_{\bar 4_B 3_B}
( \tilde {\cal L}^A/ \toh \tilde {\cal L}^K)_{\bar 4_B}^{\; \; \; \,  3_F}
 ,   \rule[-6mm]{0mm}{0mm} 
\end{eqnarray}
\begin{eqnarray}
\nonumber
\left( \tilde \Sigma^{R/I}_{BB} \right)^{3_F \bar
  4_F}_{\bar 4_B 3_B} &=&
%\left \{ \begin{array}{l} 1 \\ \toh \end{array} \right\}
\, \theta_{34}
\left(\tilde \delta^F  \tW_{B \mp}  \cdot \tilde U^F_B \cdot
\tW_{B+}  \tilde \delta^F \right)^{3_F \bar   4_F}_{\bar 4_B 3_B}
{1 \over \hbar}  (i \tilde R / \tilde I)_{3_B \bar 4_F}
\\     \label{eq:SigmaBB} & = &
- {i \hbar \over 2} \;
%(-  \toh  i \hbar ) \;
\tilde G^{R, 3_F \bar 4_F}
(\tilde G^{K/A})_{\bar 4_B 3_B}
( \tilde {\cal L}^A/ \tilde {\cal L}^K)_{\bar 4_B 3_B}
 \; .  \rule[-1mm]{0mm}{0mm}
      \end{eqnarray}
\esubequations
To obtain the second lines of \Eqs{eq:selfenergies-explicit} from the
respective first lines, we proceed similarly as for \Eqs{eq:Ltildes}
[but now with $\theta_{42} + y^{a'} \theta_{24} = 1$, since we use
approximation (ii), as explained in the paragraph after
\Eqs{eq:Ltildes}]. In particular, we exploit the fact that the
time-integrals in a path integral are time-ordered for the upper contour
and anti-time-ordered for the lower contour to replace $\tilde U^{0F}$
by $i \hbar \tilde G^R$ and $\tilde U^{0B}$ by $- i \hbar \tilde G^A$
[cf. \Eq{eq:defineG>}], or, if they are pre- or post-contracted with
$(\tilde \delta - 2 \tilde \rho_0)$, by $i \hbar \tilde G_K$
[\Eqs{eq:GKsummary}].  For example, to obtain \Eqs{eq:SigmaFF} and
(\ref{eq:SigmaBB}) for $\tilde \Sigma^R_{FF/BB}$, we used:
\bsubequations
\begin{eqnarray}
\lefteqn{\left( \tW^{F+} \tilde \delta_B \cdot \tilde U^F_B 
 \cdot \tilde \delta_B \tW^{F-}
\right)^{3_F \bar   4_F}_{\bar 4_B 3_B}}
\\
& = & \nonumber \!\!\!\!  
\int \! d x_{\bar 3_F} d x_{4_B} 
\! \! \int \! d x_{4_F} d x_{\bar 3_B} \, 
\tilde \delta^{3_F \bar 3_F} \, 
\tilde \delta_{\bar 4_B 4_B} \, 
\tilde U^{0F, \bar 3_F 4_F} \, 
\tilde U^{0B}_{4_B \bar 3_B} \, 
\tilde \delta_{\bar 3_B 3_B} \, 
\toh  s_F (1 - 2 \tilde \rho_0)^{4_F \bar 4_F}
\qph 
\\    \label{eq:UWF} & = & 
+ \toh \bigl(i \hbar \, \tilde G^{K, 3_F \bar 4_F} \bigr)
\bigl( - i \hbar \,  \tilde G^A_{\bar 4_B 3_B} \bigr) \; ,
\qquad \phantom{.}
\end{eqnarray}
%\\
\begin{eqnarray}
\lefteqn{\left( \tilde \delta^F \tW_{B-} \cdot \tilde U^F_B 
 \cdot \tW^{B+} \tilde \delta^F 
\right)^{3_F \bar   4_F}_{\bar 4_B 3_B}}
\\
& = & \nonumber \!\!\!\!  
\int \! d x_{\bar 3_F} d x_{4_B} 
\! \! \int \! d x_{4_F} d x_{\bar 3_B} \, 
\tilde \delta^{3_F \bar 3_F} \, 
\toh  s_B (1 - 2 \tilde \rho_0)_{\bar 4_B 4_B} \, 
\tilde U^{0F, \bar 3_F 4_F} \, 
\tilde U^{0B}_{4_B \bar 3_B} \, 
\tilde \delta_{\bar 3_B 3_B} \, 
\tilde \delta^{4_F \bar 4_F}
\qph 
\\ & = &  -
\toh \bigl( i \hbar \, \tilde G^{R, 3_F \bar 4_F} \bigr)
\bigl( i \hbar \, \tilde G^K_{\bar 4_B 3_B} \bigr) \; . \qquad \phantom{.}
\label{eq:UBW}
\end{eqnarray}
\esubequations
Satisfactorily, the second lines of \Eqs{eq:selfenergies-explicit},
summarized diagrammatically in Fig.~\ref{fig:cooperonselfenergy}, are
identical to what one obtains from Keldysh perturbation theory,
as can easily be verified starting from \Eq{eq:selfenergyKeldysh}
of App.~\ref{app:Keldyshpert}. Moreover, they 
are evidently consistent with the first order results listed in
\Eqs{subeq:Jprime1storder} above. (In fact, the latter could have been
used to guess \Eqs{eq:selfenergies-explicit}; the reason for
nevertheless going through the above analysis was to check that the
signs can be organized in a manner that allows for a series to be
summed up.) In Sec.~\ref{sec:cooperonselfenergy-app}, we shall
calculate the Cooperon self-energy explicitly by starting from
\Eqs{eq:cooperonfirstorderb} and (\ref{eq:selfenergies-explicit}) and
performing the disorder averaging diagrammatically.

\subsubappendix{Fate of the Pauli Factor $\Pauli$}
\label{sec:ruleofthumb}

One instructive outcome of the analysis of the previous section is
that we have learnt quite generally how to deal with the Pauli factors
$\Pauli$ occuring in $\tilde S_R$: All Keldysh functions in
\Eqs{subeq:Jprime1storder} and (\ref{eq:selfenergies-explicit}) 
arose from exploiting the identities $\tilde U^0_{i \bari } (\tilde
\delta - 2 \tilde \rho^0)_{\bari j} = (\tilde \delta - 2 \tilde
\rho^0)_{i \bari} \tilde U^0_{\bari j} = i \hbar \tilde G^K_\iiijjj$
[\Eq{eq:GKsummary}].  Since its frequency Fourier transform obeys
[\Eq{eq:GKsummary}] $\tilde G^K_\iiijjj (\ve ) = [1 - 2 n_0 (\hbar \ve )]
[\tilde G^R_\iiijjj - G^A_\iiijjj] (\ve) $, and in \Eqs{subeq:Jprime1storder}
and (\ref{eq:selfenergies-explicit}) all Keldysh functions come in the
combination $\tilde G^K (\ve - \bomega) \tilde {\cal L}^{R/A}
(\bomega)$, we can deduce a rule of thumb: by transforming to the
coordinate-frequency representation, one generates the replacement
\begin{eqnarray}
  \label{eq:1-2rho->tanh}
(\tilde \delta - 2 \tilde \rho^0)
\tilde {\cal L}^{R/A}  \to 
\tanh \Bigl[{ \hbar (\ve - \bomega) \over 2 T } \Bigr]  
\tilde {\cal L}^{R/A}
 (\bomega)    \; .
\end{eqnarray}
Actually, in deriving the general structure of the self-energy above
[\Eq{eq:selfenergies-explicit}], this replacement has, in effect,
already been deduced directly, and to all orders in the interaction,
from the general form of $i \tilde L^R_{a a'}$ in \Eqs{eq:Ltildes}, by
exploiting the fact that in the path integral, each $\tilde L^R_{a
  a'}$ is sandwidched between propagators $\tilde U^0$.  Since this
point is so important, let us spell it out once more: depending on
whether a vertex at time $t_{4_a'}$ sits on the forward (time-ordered)
or backward (anti-time-ordered) contour ($a' = F/B$), the factor
$(\tilde \delta - 2 \tilde \rho^0) \tilde {\cal L}^{R/A}$
occuring in $\tilde L^R_{a a'}$ is sandwidched as follows between two
$\tilde G^R \dots \tilde G^R$ or $\tilde G^A \dots \tilde G^A$
functions [see bottom two diagrams of
Fig.~\ref{fig:1orderperturbation}]:
\bsubequations
 \label{eq:1-2rho->tanhindicesF}
\begin{eqnarray}
%\nonumber 
\phantom{.} \hspace{-.7cm} 
\Big[\tilde G^R_{i_F 4_F} 
 (\tilde \delta - 2 \tilde \rho^0)_{4_F \bar 4_F} \Bigr] 
\tilde {\cal L}^R_{3 \bar 4_F} \, 
\G^R_{\bar 4_F j_F} \! &  \to  &  \! \phantom{-}
\G^K_{i_F \bar 4_F} (\bar \ve - \bomega) \, 
\tilde {\cal L}^R_{3 \bar 4_F} (\bomega) \, 
 %\tanh \Bigl[ {\hbar (\ve - \bomega) \over 2 T} \Bigr]  
\G^R_{\bar 4_F j_F} (\bar \ve) , 
%\tanh[\;] 
\qqph 
% \hspace{-2cm} \phantom{.}
\\
%\nonumber
\phantom{.} \hspace{-.7cm} 
 \G^A_{\barj_B \bar 4_B} \, \tilde {\cal L}^A_{\bar 4_B 3}
\Bigl[(\tilde \delta - 2 \tilde \rho^0)_{\bar 4_B 4_B}
 \G^A_{4_B \bari_B} \Bigr] 
\!  &  \to & \!  -
%\tanh \Bigl[ {\hbar (\bar \ve - \bomega) \over 2 T} \Bigr]  
\G^A_{\barj_B \bar 4_B} (\bar \ve) \, \tilde {\cal L}^A_{\bar 4_B 3} (\bomega) 
\, \G^K_{\bar 4_B \bari_B} (\bar \ve - \bomega) 
% \tanh [\; ] \, 
. \qqph 
% \hspace{-2cm} \phantom{.}\\
 % \label{eq:1-2rho->tanhindicesF}
\end{eqnarray}
\esubequations
Here the left- and right-hand sides are written in the time and
frequency domains, respectively, and the replacement rule
(\ref{eq:1-2rho->tanh}) follows from \Eqs{eq:1-2rho->tanhindicesF}
since $\tilde G^K (\bar \ve - \bomega)$ contains a factor $\tanh [
\hbar (\bar \ve - \bomega) / 2 T ]$. To be very explicit, the arrows
in \Eqs{eq:1-2rho->tanhindicesF} are shorthands for the following
series of manipulations on the above factors of $\G^R_{i_F 4_F}
(\tilde \delta - 2 \tilde \rho^0)_{4_F \bar 4_F}$ or $(\tilde \delta -
2 \tilde \rho^0)_{\bar 4_B 4_B} \tilde G^A_{4_B \bari_B} $ occuring on
the forward or backward contours [indices are now dropped, for
brevity]:
\begin{eqnarray}
\nonumber  
%\label{eq:tanhexplicitly-R}
& &   % \phantom{.} \hspace{-8mm}
 \G^R (\tilde \delta - 2 \tilde \rho^0 ) 
=
   [\G^R - G^A] (\tilde \delta - 2 \tilde \rho^0 ) 
=
 \phantom{-}    \G^K 
\, \stackrel{(1)}{\rightarrow}  \, 
    \phantom{-}  [ \G^R - \G^A ] \, \tanh 
\, \stackrel{(2)}{\rightarrow}  \, 
     \G^R \, \tanh \, ,
\\
\nonumber  
\label{eq:tanhexplicitly-A}
& &  % \phantom{.} \hspace{-8mm}
   (\tilde \delta - 2 \tilde \rho^0 ) \, \G^A
=
    (\tilde \delta - 2 \tilde \rho^0 ) [\G^A - G^R]
=
   - \G^K  
\, \stackrel{(1)}{\rightarrow}  \, 
    -  [ \G^R - \G^A ] \, \tanh
\, \stackrel{(2)}{\rightarrow}  \, 
    \G^A \, \tanh \, . 
\\
\label{eq:tanhexplicitly-R} 
\end{eqnarray}
Beginning in the position-time representation on the left hand side,
we exploit the fact that the upper or lower contours are time- or
anti-time-ordered to add an extra $- \G^{A/R} = 0$ inside the square
brackets, thereby obtaining a $\pm \G^K$. Step (1) indicates Fourier
transforming to the position-frequency domain, in which the $\tanh$
factor becomes explicit. (Step (2) will be discussed later below.)
The expressions obtained after step (1) are the ones used to produce
the right-hand sides of \Eqs{eq:1-2rho->tanhindicesF}; satisfyingly,
the latter are precisely the combinations produced by the Feynman
rules of diagrammatic Keldysh perturbation theory, illustrated in
Fig.~\ref{fig:1orderperturbation}.  [As \Eqs{eq:selfenergies-explicit}
show, the signs work out correctly, too, if the bookkeeping is done
sufficiently carefully].  The above argument is indeed completely
general, and holds for each vertex separately (but with different
$\hbar \bar \ve$'s at each vertex), to all orders in perturbation
theory. Thus, we have succeeded in recovering the Feynman rules from
the influence functional approach.

In \Eqs{eq:1-2rho->tanh} and (\ref{eq:1-2rho->tanhindicesF}), the
variable $\hbar \bar \ve$ represents the energy of the electron line
on the upper (or lower) Keldysh contour before it enters (or after it
leaves) an interaction vertex at which its energy decreases (or
increases) by $\hbar \bomega$ [see lowest two figures in
Fig.~\ref{fig:1orderperturbation}].
%, with $\bomega > 0$ for emission or $\bomega <0$ for absorption. 
The subtraction of $\bomega$ in the argument of $\tanh$ thus reflects
the physics of recoil: emitting or absorbing a photon causes the
electron energy to change by $\hbar \bomega$, and it is this changed
energy $\hbar (\bar \ve - \bomega)$ that enters the Fermi functions
for the accessible final states. (A standard back-of-the-envelope
argument for the origin of the Pauli factor, based on the availability
of initial and final states, is given in 
MDSA-I\cite{MarquardtAmbegaokar04}, Section~V.A.)  
Of course, $\hbar \bar \ve$ will have
different values from one vertex to the next, reflecting the history
of energy changes of an electron line as it proceeds through a Feynman
diagram.

The final step (2) in \Eqs{eq:tanhexplicitly-R} [not contained in
\Eq{eq:1-2rho->tanhindicesF}] indicates an approximation that occurs
\emph{if} one chooses to evaluate the path integral by including only
time-reversed paths [as GZ do, see \Sec{sec:GZ-classical-paths} of
main text]: one thereby drops terms containing interaction vertices at
which $\G^R$ changes to $\G^A$ on the upper contour, or $\G^A$ changes
to $\G^R$ on the lower contour [so-called Hikami box terms], and thus
drops $\G^{A/R} \tanh$ terms on the upper/lower contour.  Of course,
this last step (2) is \emph{optional}; the Hikami terms \emph{can} be
retained, if one so chooses, and we do so in
App.~\ref{sec:cooperonselfenergy-app} when diagrammatically deriving a
Dyson equation for the Cooperon that includes the Hikami box terms.
The result of that analysis is used in the main text
[\Sec{sec:DysonCooperonSelfenergy}] to calculate the decoherence rate;
remarkably and unexpectedly, it turns out that the Hikami-box
contribution to the decoherence rate happens to be zero for the
special form of the interaction propagator used in the main text,
namely the unitary limit of \Eq{eq:recallLCD}.  This fact implies
that, for the specific purpose of deriving the \emph{decoherence rate}
(but not necessarily for other, more general quantities) from an
influence functional, we \emph{may} indeed adopt step (2) and drop
Hikami-box terms. We shall do so henceforth.  For the remaining terms,
comparison of the very left and right-hand sides of
\Eqs{eq:tanhexplicitly-R} clearly shows that one really \emph{can}
simply replace $(\tilde \delta - 2 \tilde \rho^0)$ by $\tanh$, without
worrying about signs, etc., as specified in \Eq{eq:1-2rho->tanh}.

Having adopted step (2) of dropping Hikami-box terms, our rule of
thumb replacement (\ref{eq:1-2rho->tanh}) can quite easily be
implemented ``to all orders'' in the influence
functional approach:  Fourier-transform the kernels $(\tilde i
L^R/ \tilde L^I)_{3_a 4_{a'}}$ [\Eqs{eq:Ltildes}] of the effective
action $(i \tilde S_R + \tilde S_I)$ [\Eq{eq:defineSiR}], and simply
make the replacement (\ref{eq:1-2rho->tanh}) in the
Fourier-transformed version of $i \tilde L^R$, now using the
\emph{same} energy $\hbar \bar \ve \equiv \hbar \ve $ as that which
enters the overall weighting factor $[- n^\prime_0 (\hbar \ve)]$.  The
resulting form of the effective action is summarized in
\Eqs{eq:SIR-LIR-aa-main} to (\ref{subeq:ingredients}) of the main
text, which serve as the starting point of our calculation of the
decoherence rate there.

Diagrammatically speaking, the procedure just proposed amounts to
using the \emph{same} $\ve$ 
%in \emph{all} electron propagators
%$\G^{R/A}(\ve)$ [though still using $\hbar (\ve - \bomega)$ inside
inside each $\tanh[\hbar (\ve - \bomega)/2T] \, \LL^{R/A} (\bomega)$.
\emph{If} one intends to consider \emph{only} self-energy diagrams and
to treat infrared divergent frequency integrals with a
self\-consis\-tently-determined lower cutoff $1/ \tauphi$ (\emph{as GZ in
  fact do themselves in GZ99\cite{GZ2},} and as discussed in detail in
Sec.~\ref{sec:GZ-classical-paths}
and~\ref{sec:DysonCooperonSelfenergy} of the main text) then this
procedure would in fact not introduce any further approximations: the
energy entering and leaving each self-energy insertion then is indeed
the same for all such insertions, so they all \emph{should}
have the same $\tanh[\hbar (\ve - \bomega)/2T] \, \LL^{R/A} (\bomega)$
factors. 
% (and inside a self-energy insertion,
%the replacement $\G^{R/A} (\ve - \bomega) \to \G^{R/A} (\ve)$ is
%harmless and standard as long as $\bomega$ is limited to small values,
%which is the case in the limit of low temperatures, $T \tauel/ \hbar
%\ll 1$, since $\hbar \bomega$ turns out to be limited to be $\lesssim
%T$).
  
Of course, once one includes vertex diagrams too, as needed if one
wants to cure infrared problems ``properly'' (as in GZ99\cite{GZ3})
instead of ``by hand'' (as in GZ99\cite{GZ3}), then the proposed
procedure of using the same $\ve$ everywhere amounts to a further
approximation, since it neglects the accumulation of energy changes
that are generated by vertex terms transferring energy between the
forward and backward contours [as illustrated by the frequencies
$\tomega_1$ and $\tomega_2$ in Fig.~\ref{fig:J122221}].  Nevertheless,
the mistake incurred by this approximation is insignificant,
%, and indeed beyond
%the accuracy of the present approach:
%can be expected (and is indeed found
%\cite{MarquardtAmbegaokar04}) to be insignificant, affecting at most
%the prefactor of $\gammaphi$, but not its generic behavior, since 
since the vertex terms are not ultraviolet divergent, and the frequency
transfers contained therein are limited to the range
$\hbar |\bomega | \ll T$, just as for self-energy terms.
In fact, vertex terms become important only
in the \emph{infrared} limit where $\bomega \simeq 1/t$ (as required, of
course, to cure infrared problems of the self-energy diagrams),
so that we may replace $\bomega$ by 0 wherever else it occurs
in a diagram. More formally, it suffices
to treat the $\bomega$-dependence explicitly only
for that part of a diagram where it occurs as
energy transfer, while Taylor-expanding in $\bomega$ all other
factors of the diagram to which this $\bomega$-dependence
has propagated; only the zeroth-order terms of
this Taylor expansion need to be retained, since the
others contain higher powers of $\bomega \sim 1/t$, and
hence produce contributions with a subleading time dependence.

Note also that the accumulation of energy transfers manisfests itself
only in diagrams of second or higher order in the interaction
propagator. However, the influence functional approach {\bbf proposed
  by GZ and rederived here} features an effective action that is
linear in the interaction propagator, and hence is equivalent to
reexponentiating the \emph{first} order term in the expansion of the
Cooperon in powers of the interaction propagator (as shown explicitly
in DMSA-II\cite{MarquardtAmbegaokar04}).  Hence an accurate
treatment of effects occuring only in second or higher order is beyond
the accuracy of the influence functional approach, {\bbf in both GZ's
  original formulation and the modified version proposed here}. The
accumulation of energy transfers is such an effect.  Fortunately, it
only produces corrections that are subleading in time, as argued
above.

It is shown in the main text that if the replacement
\Eq{eq:1-2rho->tanh} is used in a ``nonperturbative calculation'' of
$\tauphi$ $\grave {{\rm a}}$ la GZ, a result consistent with
conventional wisdom is obtained.  Conversely, the reason why GZ
obtained a different result is that they, in effect, omitted the $-
\bomega$ in the $\tanh$-function in \Eq{eq:1-2rho->tanh}, and hence
lost the physics of recoil, as first suspected 
by Eriksen and Hedegard\cite{EriksenHedegard}.

\subsubappendix{Alternative Representation of Effective Action}
\label{sec:alteffaction}

To facilitate a comparison of the influence functional approach
developed in the present review with that of
MDSA-I\cite{MarquardtAmbegaokar04}, it is convenient to rewrite the
effective action derived in Section~\ref{sec:integrateoutV} and
summarized in \Eqs{eq:defineSiRA}, (\ref{eq:LtildesA}), in the
following form (to be compared to Eqs.~(21) of
MDSA-I\cite{MarquardtAmbegaokar04}):
\begin{eqnarray} 
\phantom{.} \hspace{-6mm}
[i \tilde S_R + \tilde S_I](t_1, t_0)
  &  = &  \toh\int_{t_0}^{t_1} dt_{3}
\int_{t_0}^{t_1}  dt_{4}\,\sum_{aa'=F/B}s_{a}s_{a'}
%  \left\langle     V_{3 {a}}V_{4 {a'}}\right\rangle,
\, (-\toh i) \tilde    {\cal L}^{aa'}_{3_{a} 4_{a'}} \; .
\label{eq:Seff}
\end{eqnarray}
In particular, the integrands are to contain nonzero contributions not
only for $t_{34} > 0$ [as is the case in \Eqs{eq:LtildesA}] but also
for $t_{34}< 0$.  To this end, we follow the second of the routes
mentioned after \Eq{eq:BdotAdotB}. We start from the latter, but
instead of exploiting any $3 \leftrightarrow 4$ symmetries and
inserting any factors of $2 \theta_{34}$, as done in
Section~\ref{sec:integrateoutV} (``route one''), we now write out all
terms explicitly, while still making approximation (ii) of
Sec.~\ref{sec:rho->rho0}, namely to replace all factors of
$(\theta_{4_{a'}2} + y^{a'} \theta_{24_{a'}})$ and $(\theta_{4_{a'}2}
+ y^{a'} \theta_{24_{a'}})$ by 1.  [A perhaps quicker way to obtain
the same results is to start directly from \Eqs{eq:defineSiR},
(\ref{eq:LtildesA}), but to symmetrize the integrands w.r.t.\ $3
\leftrightarrow 4$ by replacing $ \sum_{aa'} {\cal L}_{3_a4_a'}$ by
$\sum_{aa'} \toh \bigl[{\cal L}_{3_a 4_a'} + {\cal L}_{4_a
  3_{a'}}\bigr]$.]  The result can be written in the form of
\Eq{eq:Seff}, with $ {\cal L}^{aa'}_{3_{a} 4_{a'}} $ being a shorthand
for the following expressions:
\begin{eqnarray}
\nonumber %FF
\tilde {\cal L}^{FF}_{3 \bar 3, 4 \bar 4} 
&\!  = \! &
\tilde \delta_{3_F \bar 3_F}
\tilde {\cal L}^{K}_{\bar 3_F \bar   4_F}  
\tilde \delta_{4_F \bar 4_F}
+ 
\tilde \delta_{3_F \bar 3_F}
 \tilde {\cal L}^{R}_{\bar 3_F \bar   4_F}
[\tilde \delta - 2 \tilde \rho^0 ]_{4_F \bar 4_F}
+ 
[\tilde \delta - 2 \tilde \rho^0 ]_{3_F \bar 3_F}
 \tilde {\cal L}^{A}_{\bar 3_F \bar   4_F}
\tilde \delta_{4_F \bar 4_F}  
\\
\nonumber %%BF
\tilde {\cal L}^{BF}_{\bar 3 3, 4 \bar 4} 
&\!  = \! &
\tilde \delta_{\bar 3_B  3_B}
\tilde {\cal L}^{K}_{\bar 3_B \bar   4_F}  
\tilde \delta_{4_F \bar 4_F}
+ 
\tilde \delta_{\bar 3_B 3_B}
 \tilde {\cal L}^{R}_{\bar 3_B \bar   4_F}
[\tilde \delta - 2 \tilde \rho^0 ]_{4_F \bar 4_F}
- 
[\tilde \delta - 2 \tilde \rho^0 ]_{\bar 3_B 3_B}
 \tilde {\cal L}^{A}_{\bar 3_B \bar   4_F}
\tilde \delta_{4_F \bar 4_F} 
\\
\nonumber %FB
\tilde {\cal L}^{FB}_{3 \bar 3, \bar 4 4} 
&\!  = \! &
\tilde \delta_{3_F \bar 3_F}
\tilde {\cal L}^{K}_{\bar 3_F \bar   4_B}  
\tilde \delta_{\bar 4_B 4_B}
- 
\tilde \delta_{3_F \bar 3_F}
 \tilde {\cal L}^{R}_{\bar 3_F \bar   4_B}
[\tilde \delta - 2 \tilde \rho^0 ]_{\bar 4_B 4_B}
+ 
[\tilde \delta - 2 \tilde \rho^0 ]_{3_F \bar 3_F}
 \tilde {\cal L}^{A}_{\bar 3_F \bar   4_B}
\tilde \delta_{\bar 4_B 4_B}  
\\
\nonumber %BB
\tilde {\cal L}^{BB}_{\bar 3 3, \bar 4 4} 
&\!  = \! &
\tilde \delta_{\bar 3_B  3_B}
\tilde {\cal L}^{K}_{\bar 3_B \bar   4_B}  
\tilde \delta_{\bar 4_B 4_B}
- 
\tilde \delta_{\bar 3_B  3_B}
 \tilde {\cal L}^{R}_{\bar 3_B \bar   4_B}
[\tilde \delta - 2 \tilde \rho^0 ]_{\bar 4_B 4_B}
- 
[\tilde \delta - 2 \tilde \rho^0 ]_{\bar 3_B 3_B}
 \tilde {\cal L}^{A}_{\bar 3_B \bar   4_B}
\tilde \delta_{\bar 4_B 4_B} .
\\
\label{eq:seffcompareMarquardt}
\end{eqnarray}
[The double spatial indices, $3 \bar 3$ for the forward
and $\bar 3 3$ for the backward contour, are associated
with the same time $t_3$ and are both integrated over in
the path integral (similarly for $4 \bar 4, \bar 4 4$ and
$t_4$), see point (iii) after
\Eq{eq:shorthandtildehfunctionR}].  As explained in
Sec.~\ref{sec:ruleofthumb}, upon Fourier transforming, the Pauli
factors can be converted via Keldysh Green's functions into $\tanh$
functions.  However, we now need to use a more general replacement
rule (of which the one discussed in Sec.~\ref{sec:ruleofthumb} was a
special case), involving either of the expressions ${\rm th}_{\mp}
\equiv \tanh[{\hbar (\ve \mp \bomega) / 2 T} ] $.  The reason is that
we now have to distinguish two types of vertices: For vertices of
``type one'' [Fig.~\ref{cap:Standard-Keldysh-rules}(a)], the arrows of
the $\tilde L^{R/A}$ and $\tilde G^K$ correlators that get generated
both point in the \emph{same} direction (i.e.\ both away from or both
towards the same vertex), in which case we get the combination $\tilde
L^{R/A}(\bomega) \tilde G^K (\varepsilon - \bomega)$:
\begin{figure}
\begin{center}\includegraphics[%
  width=0.7\columnwidth]{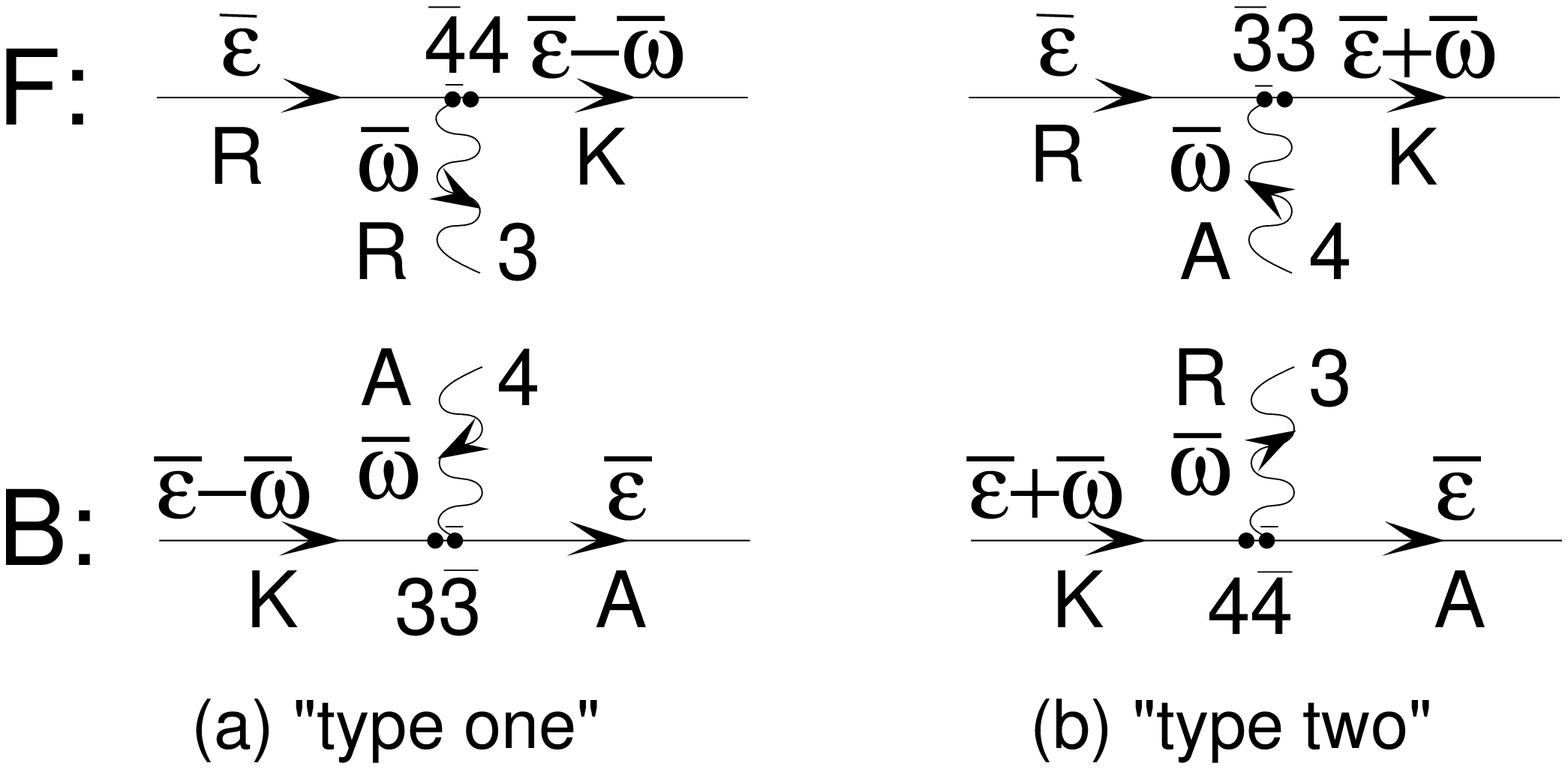}\end{center}
\caption{\label{cap:Standard-Keldysh-rules} (a) Vertices of ``type
  one'' and (b) of ``type two'' arising in Keldysh perturbation
  theory; the accompanying Keldysh Green's functions are $\tilde G^K
  (\ve \mp \bomega)$, respectively, producing Pauli factors $\tanh
  [(\ve \mp \bomega)/2T] $ that dress the associated interaction
  propagators $ \overline {\cal L}^R_\bbmq (\bomega)$ and $\overline
  {\cal L}^A_\bbmq (\bomega)$ [\Eq{subeq:generalizedruleofthumbA}]. }
\end{figure}
\bsubequations
\label{subeq:generalizedruleofthumbA}
\begin{eqnarray}
  \label{eq:generalizedruleofthumbA}
  \tilde L^R_{\bar 3_a \bar 4_F} 
[\tilde \delta - 2 \tilde \rho^0 ]_{4_F \bar 4_F}  
& \!  \!  \to \! \! &  
{\rm th}_- \,  
% \tanh\Bigl[{\hbar (\ve - \bomega) \over 2 T} \Bigr] \,
 \overline {\cal L}^R_\bbmq (\bomega) \; , 
%\\
\quad 
\mbox{[}\tilde \delta - 2 \tilde \rho^0 \mbox{]}_{\bar 3_B 3_B} 
  \tilde L^A_{\bar 3_B \bar 4_{a'}} 
%  & \to &  \tanh\Bigl[{\hbar (\ve - \bomega) \over 2 T} \Bigr] \,
\to {\rm th}_- \, 
 \overline {\cal L}^A_\bbmq (\bomega) \; . \qqph \qqph
\end{eqnarray}
For vertices of ``type two'' (the occurence of which 
was studiously avoided in Section~\ref{sec:integrateoutV}), 
the arrows point in \emph{opposite} directions (one toward, the
other away from the same vertex), 
[Fig.~\ref{cap:Standard-Keldysh-rules}(b)], 
which gives the combination $\tilde
L^{R/A}(\bomega) \tilde G^K (\varepsilon + \bomega)$:
\begin{eqnarray}
  \label{eq:generalizedruleofthumbB}
  \tilde L^R_{\bar 3_a \bar 4_B} 
[\tilde \delta - 2 \tilde \rho^0 ]_{\bar 4_B 4_B} 
% & \to &  \tanh\Bigl[{\hbar (\ve + \bomega) \over 2 T} \Bigr] \,
& \!  \!  \to \! \! &  
{\rm th}_+ \,  
 \overline {\cal L}^R_\bbmq (\bomega) \; , 
\quad
\mbox{[}\tilde \delta - 2 \tilde \rho^0 \mbox{]}_{ 3_F \bar 3_F} 
  \tilde L^A_{\bar 3_F \bar 4_{a'}} 
% & \to &  \tanh\Bigl[{\hbar (\ve + \bomega) \over 2 T} \Bigr] \,
\to {\rm th}_+ \, 
 \overline {\cal L}^A_\bbmq (\bomega) \; . \qqph \qqph 
\end{eqnarray}
\esubequations
Using these replacement rules,
the effective Fourier representations of \Eqs{eq:seffcompareMarquardt}
are readily seen to have the following forms:
\bsubequations
  \label{subeq:LAA'FT}
\begin{eqnarray}
  \label{eq:LAA'FT}
\tilde {\cal L}_{3_a 4_{a'}}
 & = & 
\int (d \bomega) (d \bbmq)
e^{i \bbmq \cdot \left[\bmR^a (t_{3_a}) -
\bmR^{a'} (t_{4_{a'}}) \right]} 
e^{-i  \bomega (t_{3_a} - t_{4_{a'}})}
\overline {\cal L}^{aa'}_\bbmq (\bomega) \, , \qqph
\\
%\overline  {\cal L}^{FF}_\bbmq (\bomega) 
%& = &
%\overline {\cal L}^{K}_\bbmq (\bomega)  
%+ 
%{\rm th}_- \, 
% %\tanh\Bigl[{\hbar (\ve - \bomega) \over 2 T} \Bigr] \,
%\overline {\cal L}^{R}_\bbmq (\bomega) \, 
%+
%{\rm th}_+ \, 
% %\tanh\Bigl[{\hbar (\ve + \bomega) \over 2 T} \Bigr] \,
%\overline {\cal L}^{A}_\bbmq (\bomega) \; ,  \qqph \qqph 
%\\
%\overline  {\cal L}^{BF}_\bbmq (\bomega) 
%& = &
%\overline {\cal L}^{K}_\bbmq (\bomega)  
%+ 
%{\rm th}_- \, 
% % \tanh\Bigl[{\hbar (\ve - \bomega) \over 2 T} \Bigr] \,
%\overline {\cal L}^{R}_\bbmq (\bomega) \, 
%-
%{\rm th}_- \, 
% % \tanh\Bigl[{\hbar (\ve - \bomega) \over 2 T} \Bigr] \,
%\overline {\cal L}^{A}_\bbmq (\bomega) \; ,  \qqph \qqph 
%\\
%\overline  {\cal L}^{FB}_\bbmq (\bomega) 
%& = &
%\overline {\cal L}^{K}_\bbmq (\bomega)  
%- 
%{\rm th}_+ \, 
% % \tanh\Bigl[{\hbar (\ve + \bomega) \over 2 T} \Bigr] \,
%\overline {\cal L}^{R}_\bbmq (\bomega) \, 
%+
%{\rm th}_+ \, 
% % \tanh\Bigl[{\hbar (\ve + \bomega) \over 2 T} \Bigr] \,
%\overline {\cal L}^{A}_\bbmq (\bomega) \; ,  \qqph \qqph 
%\\
%\overline  {\cal L}^{BB}_\bbmq (\bomega) 
%& = &
%\overline {\cal L}^{K}_\bbmq (\bomega)  
%- 
%{\rm th}_+ \, 
% % \tanh\Bigl[{\hbar (\ve + \bomega) \over 2 T} \Bigr] \,
%\overline {\cal L}^{R}_\bbmq (\bomega) \, 
%-
%{\rm th}_- \, 
% % \tanh\Bigl[{\hbar (\ve - \bomega) \over 2 T} \Bigr] \,
%\overline {\cal L}^{A}_\bbmq (\bomega) \; ,  \qqph \qqph 
%\end{eqnarray}
%or, in short,
%\begin{eqnarray}
\overline  {\cal L}^{aa'}_\bbmq (\bomega) 
& = &
\overline {\cal L}^{K}_\bbmq (\bomega)  
+ s_{a'} \, {\rm th}_{-s_{a'}} \, 
% \tanh\Bigl[{\hbar (\ve + \bomega) \over 2 T} \Bigr] \,
\overline {\cal L}^{R}_\bbmq (\bomega) \, 
+ s_a \, {\rm th}_{+ s_a}  \, 
% % \tanh\Bigl[{\hbar (\ve - \bomega) \over 2 T} \Bigr] \,
\overline {\cal L}^{A}_\bbmq (\bomega) \; .  \qqph \qqph 
\end{eqnarray}
\esubequations
\Eqs{eq:Seff} and (\ref{subeq:LAA'FT}) together constitute an
alternative and perhaps more compact expression for the effective
action of \Eqs{eq:SIR-LIR-aa-main} to (\ref{subeq:ingredients}).

%\appeqn
\appendix{Relation between Path Integral and Cooperon}
\label{app:Generalcooperon}

In this appendix we show how the general path integral expression
derived for the conductivity in the main text in terms of 
$\tilde J_{12',21'}$ [\Eqs{eq:sigmaDCreal}
and (\ref{eq:intermediateJrho})], can be rewritten in terms of the
Drude conductivity $\sigma^\Drude_\DC$ and the familiar Cooperon, and
thereby clarify how they are related to the standard relations
familiar from diagrammatic perturbation theory. We begin
[\Sec{sec:conductivity}] by reviewing the noninteracting case before
disorder averaging, then [\Sec{sec:wlnointdisorderave}] recall how
disorder averaging produces the standard result for $\sigma_\DC^\WL$.
Next [\Sec{app:1storderperturbation}] we discuss the first order
interaction contribution and subsequently
[\Sec{sec:definefullCooperon}] generalize the analysis to include
interactions to all orders, before disorder averaging.  
In particular, we elucidate how the average energy $\hbar \epsilon$
of the two counterpropagating trajectories is fixed in this formalism.
Finally, we perform a 
disorder average for the general case 
with interactions [\Sec{sec:appCooperongeneralaveraged}] to
establish a connection to the general Cooperon propagator in the
presence of interactions, and [\Sec{sec:cooperonpostime}] review its
structure in the coordinate space representation.

\subappendix{Noninteracting Limit before Disorder Averaging} 
\label{sec:conductivity}

Let us check that in the noninteracting limit but before disorder
averaging, \Eqs{subeq:sigmaDCreal} for $\sigma_{\DC}$, with $\tilde
J^\prime (0)$ given by \Eqs{sub:finalJ1221omegadefine} and
(\ref{eq:defJ(F/B)}), reproduce familiar expressions for the
conductivity $\sigma^\nonint_\DC$.  If interactions are neglected,
both $\tilde U^a_{\iiijjj}$ and $\tilde U^{\prime a}_\iiijjj$ in
\Eq{eq:defJ(F/B)} reduce to $\tilde U^0_\iiijjj$.  Using
\Eqs{eq:exact-U0-RA} and (\ref{eq:rhoUomegamaintext}) in
\Eqs{eq:defJ(F/B)}, one then readily obtains
\bsubequations
    \label{eq:J01221free}
\begin{eqnarray}
 % & & \phantom{.} \hspace{-1.9cm}
\label{eq:J01221freeF+B}
 \tilde J^{(0)}_{12', 21'} (t_1, t_2; t_0)  \!\! & = & \!\!
%\Bigl[ \tilde J^{0F}_{12', 21'}  +  \tilde J^{0B}_{12', 21'}  \Bigr]
\sum_{\tilde a = F,B}
\tilde J^{(0), \tilde a}_{12', 21'} (t_1, t_2; t_0)  = 
\hbar \, \tilde G^{R}_{12'} \tilde G^{<}_{21'} +
\hbar \, \tilde G^{<}_{12'} \tilde G^{A}_{21'}  . \qqph
\end{eqnarray}
Inserting \Eq{eq:J01221freeF+B} into \Eq{eq:sigmaGZ}, we obtain a
standard expression for $\sigma^\nonint_\DC$, before disorder
averaging. To evaluate its real part $\sigma_{\DC,\rreal}^\nonint$
[\Eq{eq:sigmaDCreal}], we have to Fourier transform $\tilde J$
according to \Eq{sub:finalJ1221omegadefine}.  Writing the result as
$\tilde J^{(0)}_{12', 21'} (\omega_0) = \int (d \ve) \tilde J^{(0),
  \ve}_{12', 21'} (\omega_0)$, we get
\begin{eqnarray}
\label{eq:J01221free-b}
\tilde J^{(0), \ve}_{12', 21'} ( \omega_0)  & =  & 
\hbar \Bigl[ \tilde G^{R}_{12'} (\varepsilon_+) \, 
\tilde G^{<}_{21'} (\varepsilon_- ) \; + 
\;  \tilde G^<_{12'} (\varepsilon_+) 
\, \tilde G^{A}_{21'}  (\varepsilon_-) \Bigr]  , \qqph \qph
\end{eqnarray}
\esubequations
%\begin{eqnarray}
%\label{eq:J01221free-b}
%\tilde J^{(0)}_{12', 21'} (\omega_0)  \!\! & =  & \!\!
% \int  ( d \varepsilon) \, 
%\hbar \Bigl[ \tilde G^{R}_{12'} (\varepsilon_+) \, 
%\tilde G^{<}_{21'} (\varepsilon_- ) \; + 
%\;  \tilde G^<_{12'} (\varepsilon_+) 
%\, \tilde G^{A}_{21'}  (\varepsilon_-) \Bigr]  , \qqph \qph
%\end{eqnarray}
with $\ve_\pm \equiv \ve \pm \toh \omega_0$.  Now expand $\tilde J^{(0),
  \ve}( \omega_0) = \tilde J^{(0), \ve}(0) + \omega_0 \tilde J^{\prime
  \, (0), \ve}(0) $, as needed for \Eq{eq:sigmaDCreal}.  Using $\G^<_\iiijjj
(\ve_\pm) = - n_0 (\hbar \varepsilon_\pm) \bigl[ \tilde G^{R}_\iiijjj -
\tilde G^{A}_\iiijjj \bigr] ( \ve_\pm) $, replacing $\tilde G^{R/A}
(\varepsilon_\pm)$ by $\tilde G^{R/A} (\varepsilon)$, and dropping
terms in $\tilde J^{\prime \, (0), \ve }(0)$ containing
$\partial_\varepsilon \tilde G^{R/A} (\varepsilon)$, since they are
smaller than those kept by a factor $T / \eF$, we obtain
\bsubequations
  \label{sueq:finalAAGsigma-positionspace}
  \begin{eqnarray}
    \label{eq:J(0)}
 \tilde J^{(0), \ve} _{12',21'} (0) & = &     -  
\, n_0 (\hbar \varepsilon ) \,  \hbar \, 
\Bigl[
\tilde G^R_{12'} (\varepsilon)
\tilde G^R_{21'} (\varepsilon)
 - \tilde G^A_{12'} (\varepsilon ) \,  \tilde G^A_{21'} ( \varepsilon)
 \Bigr] 
\; ,  \qqph 
\\
    \label{eq:J'(0)}
 \tilde J^{\prime \,  (0), \ve }_{12',21'} (0) & = &    
 -  n'_0 ( \hbar  \varepsilon) \, \hbar^2 \,
 \tilde G^R_{12'} (\varepsilon) \,
 \tilde G^A_{21'} (\varepsilon) \; . 
  \end{eqnarray}
\esubequations
Here $n'_0 (\xi) \equiv \partial_\xi n_0 (\xi)$, hence, in the $\tilde
J^{\prime \, (0), \ve} (0) $ correlator of \Eq{eq:J'(0)}, the energy
argument $\hbar \varepsilon$ is constrained to be $\lesssim T$.  The
desired result for $\sigma_{\DC,\rreal}^\nonint$ of
\Eq{eq:sigmaDCreal} thus is:
\begin{eqnarray}
  \label{eq:sigmarrealnonintbeforedisav}
  \sigma_{\DC,\rreal}^\nonint & = &
\sum_{\sigma_1}  {1 \over d} 
\int (d \ve) \, \hbar \, 
[ -  n'_0 ( \hbar  \varepsilon)] \, 
\int \! \!  dx_2 \, \bmj_{11'} \! \cdot \! \bmj_{\,22'}
\, \hbar \,  \tilde G^R_{12'} (\varepsilon) \,
 \tilde G^A_{21'} (\varepsilon)  \; . 
\qqph 
\end{eqnarray}
This is a standard result; it still has to be averaged over disorder,
a step that we review in App.~\ref{sec:wlnointdisorderave}.

The $\tilde J^{(0), \ve } (0) $ correlator of \Eq{eq:J(0)}, in which
the energy argument is not constrained, turns out to cancel the
(first) diamagnetic term in \Eq{eq:sigmaDCimag}, implying that
$\sigma^\nonint_{\DC, \iimag} = 0$, as expected.  This cancellation
can be verified, even before disorder averaging, by using an exact
identity,
\begin{eqnarray}
  \label{eq:jGJjG}
 {1 \over d}  \int d x_2 \,
\bmj_{11'} \cdot
\bigl[ \tilde G^{R/A}_{12'} (\ve) \,
\bmj_{22'} \,
\tilde G^{R/A}_{21'} (\ve) \bigr]
&  =  & - {e^2 \over m} \tilde G^{R/A}_{11} (\ve) \, ,
\end{eqnarray}
proven below, to rewrite the contribution from $  \tilde J^{(0)} (0)  $
to \Eq{eq:sigmaDCimag} as follows:
  \begin{eqnarray}
\nonumber %      \label{eq:RR-AAcancelsdia-a}
& & \phantom{.} \hspace{-1cm} \sum_{ \sigma_1} \!
\int \! dx_{2} \,
\bmj_{11'} \! \cdot \! \bmj_{\,22'}
{\tilde J^{(0)}_{12',21'} (0)  \over  \omega_0 d }
 =  {\hbar e^2  \over \omega_0 m }
\sum_{ \sigma_1}
\int  (d \varepsilon) 
n_0 (\hbar \varepsilon )
\,  \Bigl[ \tilde G^R_{11} (\varepsilon) \,
-  \tilde G^A_{11} (\varepsilon) \Bigr]
\\
& & =   - \, {\hbar e^2 \over \omega_0  m} \sum_{ \sigma_1}
\int (d \varepsilon) 
\,  G^<_{11} (\ve) \; = \;
- \sum_{ \sigma_1}
 { i e^2 \langle \hat n_{11H} \rangle \over \omega_0  m}  \; ,
 \qquad \phantom{.}
  \end{eqnarray}
which indeed cancels the first term of \Eq{eq:sigmaDCimag}.
Since the DC conductivity is a real quantity, the latter cancellation
of the two contributions to $\sigma^\DC_\iimag$, namely the
diamagnetic term and a term containing an integral $\int \! d
\varepsilon \, n_0 (\hbar \varepsilon)$ over the entire Fermi sea,
must hold order for order, to all orders, in perturbation theory in
the interaction. Therefore, we shall henceforth not keep track of
these terms, and take $\tilde J$ to represent only those terms that
end up containing a factor $- n'_0 (\hbar \varepsilon)$ that restricts
$\varepsilon$ to the vicinity of the Fermi surface, as in
\Eq{eq:J'(0)} for $\tilde J^{\prime \, (0)}$.

It remains to prove \Eq{eq:jGJjG}. It follows directly from another
exact identity,
% Book 6, p. 142-143
\begin{eqnarray}
  \label{eq:GjG=rG}
  \int \!\! d x_l \,  \bmj_{ll'} \,
\bigl[ \tilde G^{R/A}_{il'} (\ve) \tilde G^{R/A}_{lj} (\ve) \bigr]
& = & - \, {i e \, \bmr_\iiijjj \over \hbar}
\tilde G^{R/A}_\iiijjj (\ve) \; ,
\end{eqnarray}
which can be derived\cite{AAG98} before disorder averaging by evoking
gauge invariance: let $\psi_\lambda (x_i) = \langle x_i | \lambda
\rangle$  and $\xi_\lambda$ be exact eigenfunctions and
eigenvalues of the single-particle Hamiltonian $\hat {H}_0 $ [\ie\ $\hat H_0 |
\lambda \rangle = \xi_\lambda | \lambda \rangle$, cf.\ 
\Eq{eq:definepsifields}], and let $\bmA$ be a spatially uniform vector
potential. Then the gauge-transformed wave-functions $e^{- i e
  \bmA \cdot \bmr_i/\hbar} \psi_\lambda (x_i) \equiv \widetilde
\psi_\lambda (x_i) \equiv \langle x_i | \widetilde \lambda \rangle$
are eigenfunctions of the gauge-transformed Hamiltonian
%\begin{eqnarray}
%  \label{eq:gaugetransformedH0}
$ \hat {  \widetilde H}_0   \equiv 
\hat H_0 (\hat \bmP + e \bmA) = \hat H_0 (\hat \bmP) +
\bmA  \cdot \hat \bmj  + {e^2 \bmA^2 \over 2m} \; ,
$
%\qquad \phantom{.}
%\end{eqnarray}
again with eigenvalue $\xi_\lambda$, \ie\
 $\hat {  \widetilde H}_0 | \widetilde \lambda \rangle =
\xi_\lambda | \widetilde \lambda \rangle $.
Consequently, the gauge-transformed version of
$\tilde G^{R/A}_\iiijjj (\ve)$ %, denoted by a wide tilde,
can be written in two equivalent ways, as follows:
\bsubequations
  \label{subeq:GjG=rG}
  \begin{eqnarray}
\nonumber 
%    \label{eq:gaugetransformedG}
%     \widetilde G^{R/A}_\iiijjj (\ve)
%& \equiv &
e^{- i e \bmA \cdot \bmr_{\iiijjj}/\hbar} \tilde G^{R/A}_\iiijjj (\ve)
%\\
%& = &
= \sum_\lambda \langle x_i | \widetilde \lambda \rangle \frac{1}
{\hbar \ve - \xi_\lambda  \pm i \alpha}
\langle \widetilde \lambda | x_j \rangle 
% \\ 
%\label{x1tildedenominatorx2}
 =  \langle x_i |
{1 \over \hbar \ve - \hat {\widetilde H}_0 \pm i \alpha}
| x_j \rangle \; . \qph
  \end{eqnarray}
\esubequations
Expanding both the left- and right-hand sides 
%\Eqs{eq:gaugetransformedG} and (\ref{x1tildedenominatorx2})
to linear order in $\bmA$, and representing the
latter in terms of the non-gauge transformed wave functions
$\langle x_i | \lambda\rangle = \psi_\lambda (x_i)$,
we obtain
%%%\begin{widetext}
\begin{eqnarray}
\nonumber %  \label{eq:gaugetransfsemifinal}
 - \, {i e \bmA \cdot \bmr_\iiijjj \over \hbar}
\tilde G^{R/A}_\iiijjj (\ve)
%& = & 
%\sum_{\lambda' \lambda}
%\langle x_i | \lambda' \rangle
%\langle \lambda'|
%{1 \over \hbar \varepsilon - \hat H_0 \pm i \alpha}
%\, \bmA \cdot \hat \bmj \,
%{1 \over \hbar \ve - \hat H_0 \pm i \alpha}
%| \lambda \rangle \langle \lambda |
%x_j \rangle  \; \\
%&  & \phantom{.} \hspace{-2cm} = 
= \sum_{\lambda' \lambda}
\psi_{\lambda'} (x_i)
{1 \over \hbar \ve - \xi_{\lambda'} \pm i \alpha}
 \langle  \lambda' | \bmA \cdot \hat \bmj | \lambda \rangle
{1 \over \hbar \ve - \xi_\lambda \pm i \alpha}
\psi_\lambda (x_j) \; .
\end{eqnarray}
This readily yields  \Eq{eq:GjG=rG}, since  the
matrix elements of the current operator are given by
$  \label{eq:jmatrixelements}
 \langle  \lambda' | \hat \bmj | \lambda \rangle
= \int d x_l \, \bmj_{ll'} \bigl[ \psi^\ast_{\lambda'} (x_{l'})  \,
\psi_{\lambda} (x_{l}) \bigr] $.

\subappendix{Disorder Average of Noninteracting Case}
\label{sec:wlnointdisorderave}

Evaluating the disorder average $\langle \G^R \G^A
\rangle_\dis$  needed in \Eq{eq:sigmarrealnonintbeforedisav}  is a
textbook excercise: Introducing an extra dummy integration $\Vol^{-1}
\int d \bmr_1$ into \Eq{eq:sigmaDCreal}, using \Eqs{eq:GijFourier} and
(\ref{subeq:GGdisorderaverageRRorAA}) from
App.~\ref{sec:diagrammatics} and performing the momentum integrals
using \Eq{eq:integrationidentitiesgeneral},
% [with $\bar \bmp' = \bmp$ and
%$\bar \bmp = \bmp' \equiv \bmq - \bmp$ in the second term
%of (\ref{eq:GGdisorderaverage})], 
we find:
\bsubequations
\begin{eqnarray}
\nonumber   
% & & \phantom{.} \hspace{-0.8cm}
  \sigma^\nonint_{\DC}  & =  & 
\int \! (d \varepsilon) 
[-  n'_0 (\hbar \varepsilon)]
\, {2 e^2 \hbar^4  \over d  m^2 \Vol }
\sum_{\bmp \bmp'} \bmp \! \cdot \! \bmp'
 \bcG^R_{\bmp} (\varepsilon)
\bcG^A_{\bmp' } (\varepsilon)
\\ 
& & \times  \left[ \delta_{ \bmp, \bmp'} 
+ 
{\bcG^R_{\bmp'} ( \varepsilon) \,
\bcG^A_{\bmp} (\varepsilon) \,
 \bcC^0_{\bmp + \bmp'}(0) \over \Vol\,  2 \pi \nud \tauel^2 /
  \hbar}
\right] 
\\
%  \sigma^\nonint_{\DC, \rreal} & = & 
%\int (d \varepsilon) 
%[-  n'_0 (\hbar \varepsilon)]
%\, {2 \hbar^2 e^2 \over d \, m^2 }
% \int (d \bmp)
% \bcG^R_{\bmp} (\varepsilon)
%\bcG^A_{\bmp } (\varepsilon)
%\\
%& & \times \left\{ \bmP \cdot \bmP 
%+ 
%\int  (d \bmq) \bmP \cdot (\bmQ - \bmP) \,
%\bcG^R_{\bmq - \bmp} ( \varepsilon) \,
%\bcG^A_{\bmq - \bmp} (\varepsilon) \,
%{ \bcC^0_{\bmq} (0)
%+ \bcD^0_{2 \bmp} (0)
%\over  2 \pi \nud \tauel^2 /
%  \hbar} \; ,
%\qquad \qqph \phantom{.}
%\right\} 
%\\
%&=  &
\label{eq:sigmaDSnonfinal}
% & & % \phantom{.} \hspace{-0.5cm}
&  \simeq  & \; \sigma^\Drude_\DC 
\left[1 - {1 \over \pi \nud
 \hbar} \int d \varepsilon \, \hbar 
\, [-  n'_0 (\hbar \varepsilon)] 
\int (d \bmq) \int_0^\infty \!\!\! d \tau \, 
\tilde  C^0_\bmq (\tau)
\right]  \; . \qqph 
\end{eqnarray}
\esubequations
Here $\sigma^\Drude_\DC = 2 e^2 \nu \Dd$ is the Drude conductivity and
$\Dd = v_F^2 \tauel / d$ is the diffusion constant.  For the second
term of \Eq{eq:sigmaDSnonfinal}, we introduced the variable $\bmq =
\bmp + \bmp'$ and set $\bmq = 0$ everywhere except in
$\bcC^0_\bmq(\omega=0)$, because the latter's infrared singularity as
$\bmq \to 0$ dominates the $\int (d \bmq)$ integral.  [Since
$\bcD^0_{2 \bmp} (0)$ from \Eq{eq:GGdisorderaverage} has no
singularities, its contribution to \Eq{eq:sigmaDSnonfinal} was
dropped.]  The $\int d \varepsilon$ integral in
\Eq{eq:sigmaDSnonfinal}, which trivially equals one, is displayed
here explicitly only for the sake of comparison with later results.

The fact that the weak localization correction is small compared to
the Drude term is often made explicit by expressing the prefactor of
the Cooperon term in terms of the dimensionless conductance $g_\bard
(L)$ [see \Eq{eq:dimconductance}, and the discussion thereafter]:
Using $\int (d \bmq) = a^{\bard - d} \int d^\bard q / (2 \pi)^\bard$
for the momentum integral over the diffusive motion, and introducing,
\eg, the dimensionless variables $u \equiv \tau_{12} /\tauH$ and $z
\equiv q L_H$ (with $L_H = \sqrt{D \tauH}$) [if more convenient, \eg\ 
in the absence of a magnetic field, one could replace $\tauH$ by
$\tauphi$ here) we obtain from \Eq{eq:sigmaDSnonfinal} (times $a^{d-
  \bard}$):
\begin{eqnarray}
\label{eq:sigmaDSdimensionless}
%& & \phantom{.} \hspace{-0.5cm}
\sigma^\nonint_{\bard, \DC} =  
 \sigma_\bard 
\left[1 - {1 \over g_\bard (L_H)} {2 \over \pi}
  \int {d^\bard z \over (2 \pi)^\bard }
\int_{\tauel/\tauH}^\infty \!\!\! d u \, 
\tilde  C^0_{z/L_H} (u \tau_H) 
\right]  \; ,
\end{eqnarray}
where we inserted an ultraviolet cutoff at small times, needed for
$\bar d = 2, 3$.  Appealingly, the prefactor of the Cooperon term
manifestly displays the smallness of $\sigma^\WL_\bard$, via the
largeness of $g_\bard$.

\subappendix{First order Calculation of $\tilde J^\prime$ }
\label{app:1storderperturbation}

\begin{figure}[t]
{\includegraphics[clip,width=\linewidth]{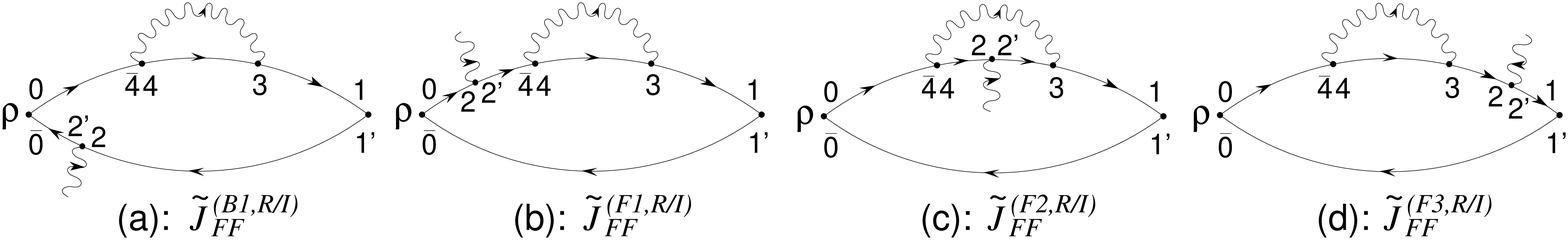}}%
\caption{
 Feynman diagrams for the correlators $\tilde { J}^{(R/I)}_{aa'}$ 
of \Eqs{eq:J1221Faa}.}
\label{fig:realspace1orderperturbation}
\end{figure}

In this section, we illustrate the structure of the perturbation
expansion generated when the influence functional is expanded in
powers of the effective action $(i \tilde S_R + \tilde S_I)$, as in
\Eq{eq:expandUU}: using approximation (ii) of
Sec.~\ref{sec:rho->rho0}, we explicitly calculate the first order
contribution $\tilde J^{\prime (1)}_{12',21'} (0)$ to the correlator
of \Eq{eq:sigmaDCreal}, \ie\ the $\omega_0 = 0$ value of first
derivative $\partial_{\omega_0}$ of the $\omega_0$-Fourier transform
of $\theta_{12} \tilde J^{(1)}_{12',21'} (t_1, t_2; t_0) \equiv
\sum_{aa'} [\tilde J^{( R)}_{aa'} + \tilde J^{(I)}_{aa'} ]$.  Here $
\tilde J^{( R/I)}_{aa'} = \sum_{\tilde a} \tilde J^{(\tilde a,
  R/I)}_{aa'}$ denotes the first-order contribution to $\tilde
J^{(1)}$ that arises from $(i \tilde S^R/\tilde S^I)$ and has
interaction vertices lying on contours $a$ and $a'$, while the index
$\tilde a$ in $ \tilde J^{(\tilde a, R/I)}_{aa'}$ indicates which
contour the current vertex is located on.

Our starting point is \Eq{eq:intermediateJrho}, expanded
to first order in $-(i \tilde S_R / \tilde S_I)/\hbar$, using \Eq{eq:defineSiR}:
%\bsubequations
%  \label{eq:J1221F/Bexpand}
\begin{eqnarray}
\nonumber %  \label{eq:J1221Fexpand}
\tilde {   J}^{(F,R/I)}_{a a'}
  \equiv   - { \theta_{12}  \over \hbar^2}
\int d x_{0_F,\bar 0_B}  \, \tilde \rho^0_{0_F \bar  0_B} \, 
\Fint_{2'_F}^{1_F}
\Fint^{2_F}_{0_F}
\Bint^{1'_B}_{\bar 0_B}
\widetilde {\cal D}' (\bmR) \,
\int_{t_0}^{t_1} d t_{3_a} d t_{4_{a'}}
\left\{ \!\! 
\begin{array}{c} i \tilde L^R \\ 
 \tilde L^I \end{array} \!\! 
\right\}^{y_F = 0}_{3_a 4_{a'}}  \qqph
%(i \tilde L^R/ \tilde L^I)_{3_a 4_{a'}}\Big|_{y_F = 0}  \qph
\\
\nonumber  % \label{eq:J1221Bexpand}
\tilde {   J}^{(B,R/I)}_{aa'} 
  \equiv  \phantom{-} { \theta_{12}  \over \hbar^2}
\int d x_{0_F,\bar 0_B}  \, \tilde \rho^0_{0_F \bar  0_B} \, 
\Fint_{0_F}^{1_F}
\Bint_{2_B}^{1'_B}
\Bint^{2'_B}_{\bar 0_B}
\widetilde {\cal D}' (\bmR) \,
\int_{t_0}^{t_1} d t_{3_a} d t_{4_{a'}}
\left\{ \!\! 
\begin{array}{c} i \tilde L^R \\ 
 \tilde L^I \end{array} \!\! 
\right\}^{y_B = 0}_{3_a 4_{a'}}  \qqph
%(i \tilde L^R/ \tilde L^I)_{3_a 4_{a'}}\Big|_{y_B = 0} \qph
\end{eqnarray}
%\esubequations
If interaction and current vertices occur on the same part (forward or
backward) of the Keldysh contour, then, depending on the relative time
orderings of the vertices, there can be more than one contribution to
each of these quantities, which we shall denote by $\tilde
{J}^{(\tilde a i, R/I)}_{aa'}$, with $i = 1,2,3$, etc.
%$\tilde {J}^{(\tilde a 2)}_{aa'}$, $\tilde
%{J}^{(\tilde a 3)}_{aa'}$, etc.

Consider $\tilde {J}^{(B1,R/I)}_{FF}$ [see
Fig.~\ref{fig:realspace1orderperturbation}(a)], which has two
interaction vertices on the forward contour at times $t_3$ and $t_4$
satisfying $t_0 < t_4 < t_3 < t_1$, and a current vertex on the
backward contour at time $t_2$ satisfying $t_0 < t_2 < t_1$ [in GZ's
approach, who take $t_0 = t_2$, cf.\ Section~\ref{sec:rho->rho0},
these two sets of inequalities are replaced by a single one instead,
namely $t_2 < t_4 < t_3 < t_1$].  Inserting \Eq{eq:LtildesFF} for $(i
\tilde L^R/ \tilde L^I)_{3_F 4_F}$ into the first of the above
equations, we obtain:
% Book 6, p. 152 to 153
\bsubequations
\label{eq:J1221Faa}
   \label{eq:J1221(B)FFexpandexplicit-1}
  \begin{eqnarray}
\nonumber
%    \label{eq:J1221(B)FFexpandexplicit-1a}
 \tilde {   J}^{(B1,R/I)}_{FF}
\!\!\!\! & = & \!\!\!\! 
 -  {i \, \theta_{12} \over 2 \hbar^2} \!\!
\int_{t_0}^{t_1} \!\!\! d t_{3_F} \!\!
\int_{t_0}^{t_{3_F}} \!\!\!\!\! dt_{4_F}
\!\! \int \!\! d x_{0_F,\bar 0_B}
\, \Uz_{1_F 3_F} \,
\delta_{3_F \bar 3_F} \, 
 \Uz_{\bar 3_F 4_F}
\biggl\{ \!\!\!
\begin{array}{c}
(\tilde \delta - 2 \tilde \rho^0)_{4_F \bar 4_F}
\rule[-2mm]{0mm}{0mm} \!\!\!\!
\\
\theta_{34} \, \tilde \delta_{4_F \bar 4_F}
\end{array} \biggr\}
\LL^{R/K}_{\bar 3_F \bar 4_F} 
\\ \nonumber
& & \qquad \times 
\Uz_{\bar 4_F 0_F} \,
\tilde \rho^0_{0_F \bar 0_B} \, 
 \Uz_{\bar 0_B 2'_B} \,
\Uz_{2_B 1_B'}
\\
     \label{eq:J1221(B)FFexpandexplicit-1b}
& = & \!\!\!\! 
{- \toh i  \hbar^2}
 \!\! \int_{- \infty}^{\infty} \!\! d t_{3} \!\!
\int_{-\infty}^{\infty} \!\! dt_{4} \,
\G^R_{13} \,
\G^{K/R}_{3 4} \,
\G^<_{4 2'} \,
\G^A_{2 1'} \,
\LL^{R/K}_{3 4} \; .
\end{eqnarray}
Here integration over repeated spatial indices
such as $0_F$ or $\bar 0_B$ or $3_F$ is implied; those over time are
displayed explicitly, to keep track of the integration boundaries.]
\Eq{eq:J1221(B)FFexpandexplicit-1b} [whose index contractions are
illustrated in Fig.~\ref{fig:1orderperturbation}(a)] follows from the
first line
%(\ref{eq:J1221(B)FFexpandexplicit-1a})
  by relations such as \Eqs{eq:exact-U0-RA} and (\ref{subeq:G<GK})
  (and dropping the subscripts ${\scriptstyle F,B}$ on indices).
  Moreover, taking the limit $t_0 \to - \infty$ [but keeping $t_1$
  fixed], the time integrals were extended to range over $[- \infty,
  \infty]$. This is possible, since $\G^R_\iiijjj$ contains a factor
  $\theta_\iiijjj$, $\G^A_\iiijjj$ a $\theta_{ji}$, and $\LL^R_{34}$ a
  $\theta_{34}$, so that the product of Green's functions under the
  time integrals automatically vanishes for time arguments lying
  outside the integration ranges stipulated by the integration
  boundaries and $\theta$-functions occuring in the first line.
  [However, if $t_0$ had erroneously been replaced by $t_2$ in the
  first line above, as GZ do, the second line would have integration
  limits $\int_{t_2 }^\infty d t_{4_F} \int_{t_4}^\infty dt_{3_F}$,
  since $G^<_{4 2'}$ contains no $\theta_{42}$.]

The case of $\tilde { J}^{(Fi,R/I)}_{FF}$ is similar, but since both the
interaction vertices at times $t_3$, $t_4$ and the current vertex at
time $t_2$ all reside on the forward contour, three separate
have to be considered 
[see Fig.~\ref{fig:realspace1orderperturbation}(b) to (d)], corresponding
to the three possible time orderings, namely (1): $t_0 < t_2 < t_4 <
t_3 < t_1$, or (2): $t_0 < t_4 < t_2 < t_3 < t_1$, or (iii): $t_0 <
t_4 < t_3 < t_2 < t_1$ [since GZ implicitely take $t_0 = t_2$, the
latter two cases do not occur in their approach]:
  \begin{eqnarray}
\nonumber
 \tilde {   J}^{(F1,R/I)}_{FF} \! \! 
\!\!\!\! & = & \!\!\!\!  
{i \, \theta_{12} \over 2 \hbar^2} \!\!
\int_{t_2}^{t_1} \!\!\! d t_{3_F} \!\!
\int_{t_2}^{t_{3_F}} \!\!\!\!\! dt_{4_F}
\!\! \int \!\! d x_{0_F,\bar 0_B}
\, \Uz_{1_F 3_F} \,
\delta_{3_F \bar 3_F} \, 
 \Uz_{\bar 3_F 4_F}
\biggl\{ \!\!
\begin{array}{c}
(\tilde \delta - 2 \tilde \rho^0)_{4_F \bar 4_F}
\rule[-2mm]{0mm}{0mm} \!\!\!\!
\\
\theta_{34} \, \tilde \delta_{4_F \bar 4_F}
\end{array} \biggr\}
\LL^{R/K}_{\bar 3_F \bar 4_F} 
\\
& & \nonumber
\qquad \times 
\Uz_{\bar 4_F 2_F'} \,
 \Uz_{2_F 0_F} \,
\tilde \rho^0_{0_F \bar 0_B} \Uz_{\bar 0_B 1_B'}
\\ & = &     \label{eq:J1221(F)FFexpandexplicit-1b}
 {- \toh i  \hbar^2}
 \!\! \int_{- \infty}^{\infty} \!\! d t_{3} \!\!
\int_{-\infty}^{\infty} \!\! dt_{4} \,
\G^R_{13} \,
\G^{K/R}_{3 4} \,
\G^R_{4 2'} \,
\G^<_{2 1'} \,
\LL^{R/K}_{3 4} \; .
\end{eqnarray}
%\\
\begin{eqnarray}
\nonumber %  
 \tilde { J}^{(F2,R/I)}_{FF}
\!\!\!\! & = & \!\!\!\!  
{i \, \theta_{12} \over 2 \hbar^2} \!\!
\int_{t_2}^{t_1} \!\!\! d t_{3_F} \!\!
\int_{t_0}^{t_2} \!\!\!\!\! dt_{4_F}
\! \! \int \!\! d x_{0_F,\bar 0_B}
\, \Uz_{1_F 3_F} \,
\delta_{3_F \bar 3_F} \, 
\Uz_{\bar 3_F 2_F'} \,
\Uz_{2_F 4_F}
\biggl\{ \!\! \begin{array}{c}
\phantom{\theta_{34}  } \tilde \delta_{4_F \bar 4_F}
\rule[-2mm]{0mm}{0mm} \!\!\!\!
\\
\theta_{34} \, \tilde \delta_{4_F \bar 4_F}
\!\! \end{array} \biggr\}
\LL^{R/K}_{\bar 3_F \bar 4_F} 
\\ \nonumber
& &   
\qquad \times 
\Uz_{\bar 4_F 0_F}
\tilde \rho^0_{0_F \bar 0_B} \Uz_{\bar 0_B 1'_B}
\qqph
\\
    \label{eq:J1221(F2)FFexpandexplicit-2b}
& = &
{- \toh i  \hbar^2}
 \!\! \int_{- \infty}^{\infty} \!\! d t_{3} \!\!
\int_{-\infty}^{\infty} \!\! dt_{4} \,
\G^R_{1 3} \,
\G^{R}_{3 2'} \,
 \G^R_{2 4} \,
 \G^<_{4 1'} \,
\LL^{R/K}_{3 4} \; , 
\end{eqnarray}
%\\
\begin{eqnarray}
\nonumber
 \tilde {J}^{(F3,R/I)}_{FF}
\!\!\!\! & = & \!\!\!\! 
{i \, \theta_{12} \over 2 \hbar^2} \!\!
\int_{t_0}^{t_2} \!\!\! d t_{3_F} \!\!
\int_{t_0}^{t_2} \!\!\!\!\! dt_{4_F}
\!\! \int \!\! d x_{0_F,\bar 0_B} \,
\Uz_{1_F 2'_F} \,
\Uz_{2_F 3_F} \,
\delta_{3_F \bar 3_F} \, 
\Uz_{\bar 3_F 4_F}
\biggl\{ \!\! \begin{array}{c}
\phantom{\theta_{34}  }
\tilde \delta_{4_F \bar 4_F}
\rule[-2mm]{0mm}{0mm} \!\!\!\!
\\
\theta_{34} \,
\tilde \delta_{4_F \bar 4_F}
\end{array} \!\! \biggr\}
\LL^{R/K}_{\bar 3_F \bar 4_F}
\\ \nonumber  
& & \qquad \times 
\Uz_{\bar 4_F 0_F}
\tilde \rho^0_{0_F \bar 0_B} \Uz_{\bar 0_B 1'_B}
\\
    \label{eq:J1221(F)FFexpandexplicit-3b}
& = &
{-  \toh i  \hbar^2}
 \!\! \int_{- \infty}^{\infty} \!\! d t_{3} \!\!
\int_{-\infty}^{\infty} \!\! dt_{4} \,
\G^R_{12'} \,
\G^R_{23}  \,
\G^R_{3 4} \,
\G^<_{4 1'} \,
\LL^{R/K}_{3 4} \; .
  \end{eqnarray}
\esubequations
\Eqs{eq:J1221(F)FFexpandexplicit-1b} corresponds to
Figs.~\ref{fig:1orderperturbation}(b).  The absence of a factor
$(\tilde \delta - 2 \tilde \rho^0)$ in the first lines of
\Eqs{eq:J1221(F2)FFexpandexplicit-2b} and
(\ref{eq:J1221(F)FFexpandexplicit-3b}), and the corresponding absence
of a $\G^K$-function in the respective second lines, reflects the fact
that we took $y^F = 0$ and that $t_4 < t_2$ in these integrals, so
that the factor $(\theta_{42} + y_F \theta_{24} 2 ) 2 \tilde \rho^0$
in \Eq{eq:LtildesFF} for $(i \tilde L^R / \tilde L^I)_{3_F 4_F}$
vanishes. \Eqs{eq:J1221(F2)FFexpandexplicit-2b} and
(\ref{eq:J1221(F)FFexpandexplicit-3b}) are examples of contributions
for which one or more interaction vertices occur on the same contour
as the current vertex, but at \emph{earlier} times.  As discussed in
approximation (ii) of Sec.~\ref{sec:rho->rho0}, such terms contribute
to ``interaction corrections'' but not to decoherence, and thus
will henceforth be be excluded from our considerations.

Adding the two terms [(\ref{eq:J1221(B)FFexpandexplicit-1b}),
(\ref{eq:J1221(F)FFexpandexplicit-1b})] that survive under the said
approximation (ii), we obtain $ \tilde { J}^{(R/I)}_{FF} =
\sum_{\tilde a} \tilde { J}^{(\tilde a 1,R/I)}_{FF} $.  The other
three correlators, $ \tilde { J}^{(R/I)}_{BF}$, $ \tilde {
  J}^{(R/I)}_{FB}$ and $ \tilde { J}^{(R/I)}_{BB}$, can be calculated
in an entirely analogous manner. The results are:
\bsubequations
  \label{subeq:timesaa'}
\begin{eqnarray}
  \label{eq:timesFF}
\tilde { J}^{(R/I)}_{FF} \!\! & =& \!\!\!\!\ 
{- \toh i  \hbar^2}
 \!\! \int_{- \infty}^{\infty} \!\! d t_{3} \, d t_4 \, 
\G^R_{13} \,
\G^{K/R}_{3 4} \,
\Bigl[\G^R_{4 2'} \, \G^<_{2 1'} +
\G^<_{4 2'} \, \G^A_{2 1'} \Bigr]
\bigl(\LL^R / \LL^K \bigr)_{3 4} \; , \qqph \qph
\\
  \label{eq:timesBF}
\tilde { J}^{(R/I)}_{BF} \!\! & = & \!\!\!\!\ 
{- \toh i  \hbar^2}
 \!\! \int_{- \infty}^{\infty} \!\! d t_{3} \, d t_4 \, 
%\biggl\{ \!\! \begin{array}{c}
%1
%\\
%\toh
%\end{array} \!\! \biggr\} \, 
\G^{K/R}_{14} \,
\Bigl[\G^R_{ 42'} \, \G^<_{ 2 3} +
\G^<_{ 42'} \, \G^A_{ 2 3} \Bigr] \,
\G^A_{3 1'}  \,
\bigl(\LL^R / \toh \LL^K \bigr)_{3 4 } \, , \qqph \qph
%\LL^{R/K}_{34} \; , \qqph \qph
\\
  \label{eq:timesFB}
\tilde { J}^{(R/I)}_{FB} \!\! & = & \!\!\!\!\ 
{- \toh i  \hbar^2}
 \!\! \int_{- \infty}^{\infty} \!\! d t_{3} \, d t_4 \, 
%\biggl\{ \!\! \begin{array}{c}
%1
%\\
%\toh
%\end{array} \!\! \biggr\} \, 
\G^R_{13}  \,
\Bigl[\G^{R}_{32'} \,  \G^<_{24 } +
\G^{<}_{32'} \,  \G^A_{24 } \Bigr] \, 
\G^{K/A}_{ 41'} \,
%\LL^{A/K}_{43} 
\bigl(\LL^A / \toh \LL^K \bigr)_{4 3} \, , \qqph \qph
\\ 
 \label{eq:timesBB}
\tilde { J}^{(R/I)}_{BB} \!\! & = & \!\!\!\!\ 
{- \toh i  \hbar^2}
 \!\! \int_{- \infty}^{\infty} \!\! d t_{3} \, d t_4 \,
\Bigl[\G^R_{12'} \, \G^<_{24} \, +
\G^<_{12'} \, \G^A_{24} \Bigr] \, 
\G^{K/A}_{ 4 3} \,
\G^A_{3 1'} \,
\bigl( \LL^A / \LL^K \bigr)_{43} \; .
\end{eqnarray}
\esubequations
Satisfactorily, these expressions agree completely with those
[\Eqs{eq:sumJABC}] obtained in App.~\ref{sec:conductivity-Keldysh}]
using diagrammatic Keldysh perturbation theory.

To obtain $\tilde J^\prime (0)$, we have to Fourier transform
these equations w.r.t.\ $t_{12}$, and then 
calculate  ${J}^{\prime (R/I)}_{aa'} (0) = 
\bigl[ \partial_{\omega_0} {J}^{(R/I)}_{aa'}
(\omega_0)\bigr]_{\omega_0 = 0}$.
For example, $\tilde { J}^{(R/I)}_{FF} (\omega_0)$
is given by
\begin{eqnarray}
\nonumber
 \tilde { J}^{(R/I)}_{FF} (\omega_0)
& = &  {- \toh i  \hbar}
 \int \!  (d \ve) (d \bomega) \, 
\G^R_{13} (\ve_+ ) \,
\G^{K/R}_{3 4} (\ve_+  - \bomega) \, \LL^{R/K}_{3 4} (\bomega) \qquad 
\\   \label{eq:J(F+B)FF}
& & \qquad \times \hbar \Bigl[ \G^R_{42'} (\ve_+) \G^<_{21'} (\ve_-) +
\G^<_{42'} (\ve_+) \G^A_{21'} (\ve_-) \Bigr] \; , 
\end{eqnarray} 
and $ \tilde { J}^{\prime (R/I)}_{FF} (0)$ is easily calculated by
noting that the factor in the second line of \Eq{eq:J(F+B)FF} equals
$\tilde J^{(0)}_{42', 21'} (\omega_0)$ [cf.\ \Eq{eq:J01221free-b}],
whose first derivative is given by \Eq{eq:J'(0)}, namely
$ \tilde J^{\prime \,  (0), \ve }_{42',21'} (0)  =     
 -  n'_0 ( \hbar  \varepsilon) \, \hbar^2 \,
 \tilde G^R_{42'} (\varepsilon) \,
 \tilde G^A_{21'} (\varepsilon) $. Thus, 
the final result for $ \tilde { J}^{\prime (R/I)}_{FF}
(0)$ is
\begin{eqnarray}
\nonumber 
 \tilde { J}^{\prime (R/I)}_{FF} (0) 
=    - \toh  i  \hbar^3 \!\! 
 \int  (d \ve) (d \bomega) \, 
[ -  n'_0 ( \hbar  \varepsilon)] 
\, \G^R_{13} (\ve) \,
\G^{K/R}_{3 4} (\ve \!  - \! \bomega) \, 
 \G^R_{42'} (\ve) \, \G^A_{21'} (\ve)  
\LL^{R/K}_{3 4} (\bomega) . 
\end{eqnarray}
Similar expressions for the other contributions $ \tilde { J}^{\prime
  (R/I)}_{aa'} (0)$ can be derived from \Eqs{subeq:timesaa'} in an
entirely analogous manner, and are given in \Eq{subeq:Jprime1storder}
of App.~\ref{sec:ruleofthumb}.  In each case the combination $\hbar
[\G^R_{i2'} \G^<_{2j} + \G^<_{i2'} \G^A_{2j} ]$ produces a factor
$\hbar^2 [- n_0^\prime (\hbar \ve) ] \G^R_{i2'} (\ve) \G^A_{2j} (\ve)
$.

Actually, it is clear from the above derivation that in \emph{every}
order of perturbation theory in the interaction, such a factor will be
produced for \emph{all} terms that survive the abovementioned
approximation (ii): in analogy to \Eqs{eq:J1221(B)FFexpandexplicit-1b}
and (\ref{eq:J1221(F)FFexpandexplicit-1b}), it will arise from a
factor
\begin{eqnarray}
  \label{eq:originofn'}
 - {1 \over \hbar} \int \!\! d x_{0_F,\bar 0_B} \Bigl[ \Uz_{i_F 0_F} \,
\tilde \rho^0_{0_F \bar 0_B} \, 
 \Uz_{\bar 0_B 2'_B} \,
\Uz_{2_B j_B} -
\Uz_{i_F 2_F'} \,
 \Uz_{2_F 0_F} \,
\tilde \rho^0_{0_F \bar 0_B} \Uz_{\bar 0_B j_B} \Bigr] \; , \qqph
\end{eqnarray}
where $t_{i_F}$ and $t_{j_B}$ are the times
of the \emph{earliest} interaction vertex on the
upper or lower Keldysh contour, respectively.

To conclude this section, we wish to emphasize once more the
significance of the fact, illustrated by \Eqs{subeq:timesaa'} but
valid for all contributions to $\tilde J^{(1)}$ (including the
``interaction corrections''), that all time integrals occuring in
Keldysh perturbation theory can be extended to range over the
\emph{entire} real axis.  Importantly, this implies that the Fourier
transforms that are needed to obtain $\tilde J^{(1)} (\omega_0)$ (and
from there the conductivity) are \emph{always} given by simple
convolution integrals, such as \Eq{eq:J(F+B)FF}. In contrast, in GZ's
calculations all time integrals $\int dt_3 \, dt_4$ have $t_2$ as
lower limit, see \eg\ (GZ-III.A20) and (GZ-III.A23) in
GZ00\cite{GZ3}, whose $t'$ corresponds to our $t_2$.  This means that
instead of obtaining simple convolution integrals, they erroneously
end up with $\sin$ and $\cos$ functions, see (GZ-III.58) and
(GZ-III.61).  This leads to numerous incorrect complications and
conclusions, such as the claimed existence of an ``oscillating
cos-term'' in (GZ-III.70).  Thus, GZ's perturbative analysis in
Sec.~IV of GZ00\cite{GZ3}, in particular their discussion of the
``breakdown of the Fermi golden rule approximation'' in Sec.~IV.B, is
invalid, since its starting point is based on the replacement $t_0 \to
t_2$, which is incorrect (and unnecessary, since the correct limit
$t_0 \to - \infty$ \emph{can} be incorporated into GZ's approach, as
emphasized in Sec.~\ref{sec:rho->rho0} and illustrated explicitly
above).

\subappendix{Thermal Weighting and Path Integral, before Disorder Averaging}
\label{sec:definefullCooperon}

%Book 9, p. 

\begin{figure}[htbp]
\begin{centering}
\includegraphics[width=\linewidth]{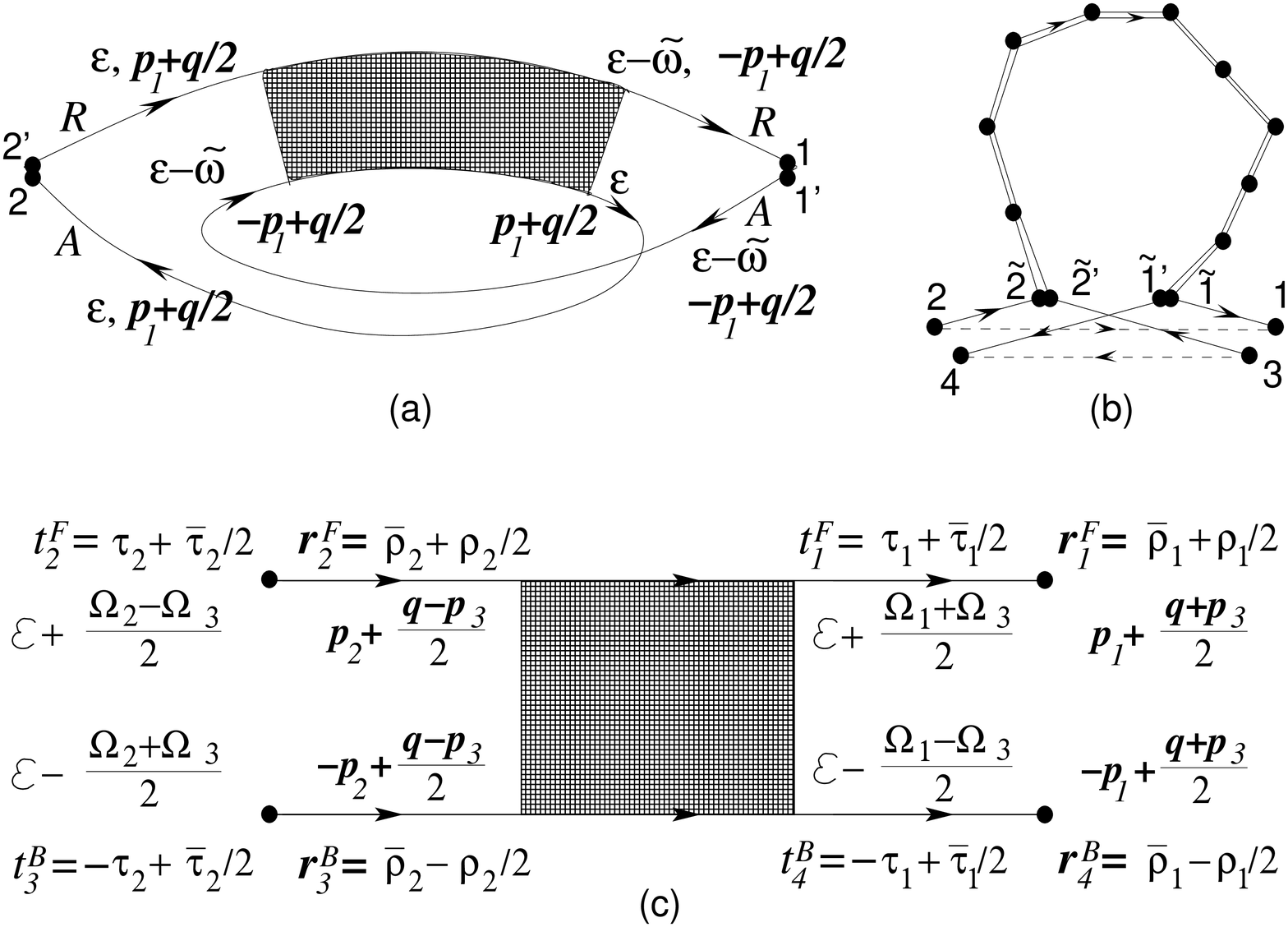}
\end{centering}
\caption{ 
  (a) Diagrammatic depiction of \Eqs{eq:defineJprimevegeneral} 
for $\tilde J^{\prime \, \ve}_{12',21'}$ or
  \Eq{eq:finalsigmawithintmaintext} for $\sigma^\WL_\DC$.  Before
  disorder averaging the black box represents $\tcP^{1 2'}_{21'} ( \ve
  \! - \!  \toh \tomega; - \tomega , \tomega )$, thereafter it
  represents $\tcC^{\ve - {1 \over 2} \tomega}_\bmq (- \tomega ,
  \tomega) / ( 2 \pi \nud \tauel^2/\hbar)$.  (b) Real-space depiction
  of a typical pair of Drude (dashed) and time-reversed (solid)
  trajectories contributing to $\tilde P^{12, \Drude}_{43}$ and
  $\tilde P^{12, \WL}_{43}$, corresponding to
  Eqs.~(\protect\ref{eq:PDruderealspacetime}) and
  (\protect\ref{eq:PWLrealspacetime}), respectively.  (c) Definition
  of variables used for Fourier-transforming the double path integral
  $ \tilde P^{1 2}_{4 3} ({\cal E}, \Omega_1, \Omega_2)$ 
of \Eq{subeq:Cooperonenergytimetime}.
In (c), frequency and momentum variables are chosen such that
$\Omega_1$ and $\Omega_2$ are, respectively, the outgoing and incoming
``Cooperon frequencies'' (\ie\ frequency differences between upper and
lower lines); $\bmq \pm \bmp_3$ are the outgoing and incoming
``Cooperon momenta'' (\ie\ sum of momenta of upper and lower lines);
${\cal E} \pm \Omega_3/2$ are the average (between upper and lower)
frequencies flowing out of our into the Cooperon, respectively.  The
time variables $\tau_{1,2}$ and $\bbtau_{1,2}$ and coordinate
variables $\bmrho_{1,2}$ and $\bbrho_{1,2}$ are purposefully defined
in such a way [\Eqs{eq:defnewtimes}, (\ref{eq:defnewcoordinates})]
that the Fourier exponents in \Eqs{eq:Cooperonenergynewtimetime},
(\ref{eq:Cooperonenergytimetimeinverse}) and
(\ref{eq:finalP=Cooperonrealspace-a}) are free of factors of 2.  (Our
labelling convention differs from that of AAK\protect\cite{AAK82}, 
which has typos involving factors of 2.)}
\label{fig:cooperonrealspace}
\label{fig:cooperondefinition}
\end{figure}

The presence of interactions will, in general, modify the result
(\ref{eq:sigmaDSnonfinal}) for $\sigma_\DC^\nonint$ in two ways:
firstly, it can renormalize the value of $\sigma^\Drude_\DC $, but
this effect is not interesting for present purposes and will be
ignored here. Secondly, it can reduce the life-time of the Cooperon
propagator, thereby contributing to decoherence, which is the effect
we are interested in.  Our goal in this section is to express the
conductivity of \Eq{eq:sigmaDCreal} in terms of double path integral
expressions for $\tilde J^\prime_{12',21'} (0)$, obtained from
\Eq{eq:intermediateJrho}, in a way that is generally valid in the
presence of interactions, \emph{before} disorder averaging, and
properly accounts for thermal weighting via a factor $[- n'_0 (\hbar
\ve)]$, as in \Eq{eq:sigmaDSnonfinal}.  Hence, we will have to find
appropriate Fourier transforms of our path integral expressions that
relate them to the energy $\hbar \ve$.

An important first clue comes from the first order relations
(\ref{subeq:Jprime1storder}) for $\tilde J^\prime_{12',21'} (0)$: each
term contains a factor $\int (d \ve) \hbar^2 [- n_0^\prime (\hbar \ve)
] \G^R_{i2'} (\ve) \G^A_{2j} (\ve)$, thus the current vertex
$\bmj_{22'} $ is always sandwiched between a retarded and advanced
function with energy $\ve$, $\G^R_{i2'} (\ve) \bmj_{22'} \G^A_{2j}$,
and thermal weighting is always governed by a factor $[- n_0^\prime
(\hbar \ve)]$.  As explained after in
App.~\ref{app:1storderperturbation} [just before \Eq{eq:originofn'}]
these properties actually hold in \emph{every} order of perturbation
theory in the interaction, for \emph{all} terms that survive
approximation (ii).  Of course, the other current vertex $\bmj_{11'} $
is similarly sandwiched, too, but in general with a different energy
argument, $\G^A_{\barj 1'} (\ve - \tomega) \bmj_{11'} \G^R_{1 \bari}
(\ve - \tomega)$.  The general expression for that part of the
conductivity containing the Cooperon propagator, relevant for weak
localization, is by definition the sum to all orders of all such terms
containing $[- n'_0] \G^R_{i2'} \bmj_{22'} \G^A_{2j} \dots \G^R_{\barj
  1'} \bmj_{11'} \G^A_{1 \bari}$.  In path integral language, it will
thus have the following form,
\begin{eqnarray}
\label{eq:sigmageneraldefinePI}
\sigma_{\DC} & =  & 
\sum_{\sigma_1}  {1 \over d}
\! \int \! \!  dx_2 \, \bmj_{11'} \! \cdot \! \bmj_{\,22'}
 \int (d \ve) \tilde J^{\prime \, \ve}_{12',21'} \; , 
\end{eqnarray} 
written in analogy to \Eq{eq:sigmaDCreal} for $\sigma_{\DC,\rreal}$,
where the integral equals $ \tilde {J}_{12',21'}^\prime (0) $,
and  $\tilde J^{\prime \, \ve}_{12',21'}$ equals
$[- n_0^\prime (\hbar \ve)]$ times some suitable frequency 
Fourier transform (needed to set the energy to $\ve$)
  of a double path integral
whose forward path connects the points
$\bmr_{2'}$ and $\bmr_1$, while the backward path
connects $\bmr_2$ and $\bmr_{1'}$. 
To find the appropriate expression, we begin by
considering the general double path integral
\begin{eqnarray}
  \label{eq:Cooperonaspathintegral}
  \tilde P^{1 2}_{4 3}
& \equiv & \theta_{12} \, \theta_{34} \, 
\Fint_{\bmR^F (t_2^F) = \bmr^F_2}^{\bmR^F (t_1^F) = \bmr^F_1}
\Bint_{\bmR^B (t_4^B) = \bmr^B_4}^{\bmR^B (t_3^B) = \bmr^B_3}
\Bigl. \widetilde {\cal D}' (\bmR) \,
 e^{-[i \tilde S_R + \tilde S_I]/\hbar} \; , 
\end{eqnarray}
depicted schematically in Figs.~\ref{fig:cooperondefinition}(b) and
\ref{fig:cooperondefinition}(c). It ranges from $\bmr^F_2$ at time
$t_2^{F}$ to $\bmr^F_1$ at time $t_1^{F}$ ($> t_2^F$) on the forward
contour and from $\bmr^B_4$ at time $t_4^{B}$ to $\bmr^B_3$ at time
$t_3^{B}$ ($>t_4^B$) on the backward contour.  These times are
understood to be the limits of the $ \int d t_a$ time integrals in
$\tilde S^a_0$ and $(i \tilde S_R + \tilde S_I)$, and $t_4^{B}$,
$t_3^{B}$ are in general not equal to $t_2^{F}$, $t_1^{F}$, since
they will have to be Fourier transformed independently [as required,
\eg, to properly define the variable $\ve$ in
\Eq{eq:sigmageneraldefinePI}]. For general time arguments, we adopt
the following conventions, depicted in
Fig.~\ref{fig:cooperondefinition}(c), for Fourier transforming from
the time to frequency domain and back:
\bsubequations
\label{subeq:Cooperonenergytimetime}
\begin{eqnarray}
& & \phantom{.} \hspace{-1cm}
  \tilde P^{1 2}_{4 3} 
\equiv  
 \int \! (d {\cal E}) (d \Omega_1)  (d \Omega_2) (d \Omega_3) 
\, 2 \pi \delta (\Omega_3) \,
  \tilde {\cal P}^{1 2}_{4 3} ({\cal E}; \Omega_1, \Omega_2)
\exp i \left\{ 
- t_1^F \! \left[ {\cal E} + {\Omega_1 +  \Omega_3  \over 2} \right]\!
\right. 
\hspace{-1cm} \phantom{.} 
\nonumber 
\\
\label{eq:Cooperonenergytimetime}
& &  \phantom{.} % \hspace{-0.5cm}
%\times \exp \left\{
 \left.
+ \, t_2^F \! \left[ {\cal E} + {\Omega_2 -  \Omega_3 \over 2} \right]\!
-  \, t_4^B \! \left[ {\cal E} - {\Omega_1 -  \Omega_3 \over 2} \right]\!
+ \,  t_3^B \! \left[ {\cal E} - {\Omega_2 +  \Omega_3 \over 2} \right]\!
\right\} , \hspace{-1cm} \phantom{.}
\\
\label{eq:Cooperonenergynewtimetime}
& & \phantom{. \hspace{-1cm}
  \tilde P^{1 2}_{4 3} } = \int (d {\cal E}) \, (d  \omega) \, (d \omega') 
 \, \tilde {\cal P}^{12}_{43} 
({\cal E}; \omega + \toh \omega', \omega - \toh \omega') \, 
e^{-i [\bar \tau_{12} {\cal E} + \ttau_{12} \omega' + \tau_{12} \omega  ]} \; , 
\\
\label{eq:Cooperonenergytimetimeinverse}
& & 
\phantom{.}  \hspace{-1cm} 
  \tilde {\cal P}^{1 2}_{4 3} ({\cal E}; \Omega_1, \Omega_2)
= \int d \tau_1 \,  d \tau_2  \, d \bar \tau_{12} \, 
e^{i[ \tau_1 \Omega_1 - \tau_2 \Omega_2 +  \bar \tau_{12} {\cal E}]}
\tilde P^{12}_{43} (\tau_{12}, \ttau_{12}, \bar \tau_{12}) \; . \qquad 
\end{eqnarray}
  \label{eq:doublefouriertransform}
\esubequations
(For $\tilde P$, the indices ${}^{1 2}_{4 3}$ stand for both
coordinate and time variables, for its frequency Fourier transform
$\tcP$, distinguished from the former by using calligraphic script,
they stand for coordinate variables only; we use a similar convention
for the Cooperon, $\tilde C$ or $\tcC$, defined below.)  For
\Eq{eq:Cooperonenergynewtimetime} we changed frequency variables to
$\omega = \toh (\Omega_1 + \Omega_2)$ and $\omega' = \Omega_1 -
\Omega_2$, and introduced various sum and difference times
[see Fig.~\ref{fig:cooperonrealspace}(c)]: 
\bsubequations
\label{subeq:defnewtimes}
\begin{eqnarray}
  \label{eq:defnewtimes}
 & & \phantom{.} \hspace{-1cm}
 \bbtau_1 \equiv t_1^F + t_4^B  \; ,  \quad
  \tau_1 \equiv {t_1^F - t_4^B \over 2} \; ,  \quad
  \bbtau_2 \equiv {t_2^F + t_3^B } \; , \quad
  \tau_2 \equiv {t_2^F - t_3^B \over 2 } \; , \quad
\\
  \label{eq:defmorenewtimes}
 & & \phantom{.} \hspace{-1cm}
   \tau_{12} = \tau_1 - \tau_2 \; , \quad 
  \bar  \tau_{12} = \bar \tau_1 - \bar \tau_2 \; , \quad 
   \ttau_{12} = {\tau_1 + \tau_2 \over 2} \; .
\end{eqnarray}
\esubequations
On the right-hand side of the back transformation
(\ref{eq:Cooperonenergytimetimeinverse}), $\tilde P^{12}_{43}
(\tau_{12}, \ttau_{12}, \bar \tau_{12})$ by definition is given by
$\tilde P^{12}_{43}$ of \Eq{eq:Cooperonaspathintegral}, with the
understanding that the indices $12,43$ now only specify the path end
points $\bmr^F_1, \bmr^F_2$, $\bmr^B_4, \bmr^B_{3}$, but that the time
arguments $t_1^F$, $t_2^F$, $t_3^B$, $t_4^B$ in
\Eq{eq:Cooperonaspathintegral} are chosen such that
\Eqs{subeq:defnewtimes} hold.

The frequencies introduced in \Eq{subeq:Cooperonenergytimetime} have
evident physical interpretations [see
Fig.~\ref{fig:cooperonrealspace}(c)].  The ``Cooperon frequencies''
$\Omega_1$ and $\Omega_2$ are the outgoing and incoming frequency
differences between upper and lower lines, respectively, while ${\cal
  E} \pm \toh \Omega_3$ are the average (between upper and lower)
frequencies flowing out of or into the Cooperon.  In general, the
presence of external time-dependent fields would require $\Omega_3$,
the total frequency difference between outgoing and incoming lines, to
be nonzero. However, for the present purpose of calculating the
conductivity in linear response, such external fields can be set to
zero; hence in \Eq{eq:Cooperonenergytimetime}
we use a delta-function to set $\Omega_3$ equal to zero,
thus recovering translational invariance in time for $ \tilde P^{1
  2}_{4 3} $.

Having identified the meaning of the frequency arguments ${\cal E}$,
$\Omega_1$ and $\Omega_2$ [Fig.~\ref{fig:cooperonrealspace}(c)], and
inspecting the frequency labels of the standard diagrammatic depiction
[Fig.~\ref{fig:cooperonrealspace}(a), where an integral over the
``internal'' frequency $\tomega$ is implied] of the current-current
correlator needed for the conductivity, it becomes evident that the
average frequency is ${\cal E} = \ve - \toh \tomega$, while the
outgoing and incoming Cooperon frequencies are $\Omega_{1} = -
\tomega$ and $\Omega_2 = \tomega$, respectively (\ie\ $\omega = 0$ and
$ \omega' = - 2 \tomega$).  Moreover, the upper line runs from
$\bmr_{2'}$ to $\bmr_1$, while the lower line runs backwards in time
from $\bmr_{1'}$ to $\bmr_2$.  Thus, the particular Fourier
transformed version of $\tilde P$ needed for $\tilde J^{\prime \,
  \ve}_{12',21'}$ in \Eq{eq:sigmageneraldefinePI} is
\begin{eqnarray}
\label{eq:defineJprimevegeneral}
\tilde J^{\prime \, \ve}_{12',21'}
& = & [ - n' (\hbar \ve )] 
\int (d \, 2 \tomega) \, 
\tcP^{1 2'}_{21'} \bigl( \ve - \toh \tomega  ; 
- \tomega, \tomega \bigr)  \; .
\end{eqnarray}
To check that that the normalization factors and frequency assignments
are correct, let us expand $\tilde P^{12'}_{21'}$ in
\Eq{eq:defineJprimevegeneral} to zeroth and first order in the
interaction in order to calculate $\tilde J^{\prime \, \ve}_{12',21'}
= [\tilde J^{\prime (0), \, \ve} + \tilde J^{\prime (1)\,
  \ve}]_{12',21'}$, and compare the results to our previously-obtained
expressions for these [\Eqs{eq:J'(0)} and
(\ref{subeq:Jprime1storder})].  To this end, we begin with $\tilde
P^{12}_{65}$, as given in \Eq{eq:Cooperonaspathintegral} and with
general indices, expand it to first order in ${-[i \tilde S_R + \tilde
  S_I]/\hbar}$, and express the resulting terms in terms of $\tilde
G^{K/R/A}_\iiijjj$ functions.  The details are analogous to those
presented in Sections~\ref{sec:conductivity}
and~\ref{app:1storderperturbation} to derive $\tilde
J^{(0)}_{12',21'}$ and $\tilde J^{(1)}_{12',21'}$ from $\tilde
J^{F/B}_{12',21'}$ of \Eq{eq:intermediateJrho} (except that the
latter's first line is not needed for $\tilde P^{12}_{65}$, and the
limits of the path integrals are different). The result can be written
as $\tilde P^{12}_{65} = \tilde P^{(0)}_{12,65} + \sum_{aa'}
\left[\tilde P^{(R)}_{aa'} + \tilde P^{(I)}_{aa'} \right]_{12,65}$,
where $\tilde P^{(0)}_{12,65}$ and $\tilde P^{(R/I)}_{aa'}$ are given
by \Eqs{eq:J01221freeF+B} and (\ref{subeq:timesaa'}), respectively
[with $(12',21') \to (12,65)$], except that all occurences of the
combination $ \hbar [ \tilde G^{R}_{i2'} \tilde G^{<}_{2j} + \tilde
G^{<}_{i2'} \tilde G^{A}_{2j}] $ have to be replaced by $\hbar^2
\tilde G^{R}_{i2} \tilde G^A_{6j} $.  Fourier transforming the result
for $\tilde P^{12}_{65}$ [via \Eq{eq:Cooperonenergytimetimeinverse}]
to obtain $ \tilde {\cal P}^{1 2}_{65 } ({\cal E}; \Omega_1,
\Omega_2)$, specifying the spatial indices as $(12,65) \to (12',21')$ and
then integrating as stipulated in \Eq{eq:defineJprimevegeneral}, one
recovers $\tilde J^{\prime \, \ve}_{12',21'} = [\tilde J^{\prime (0),
  \, \ve} + \tilde J^{\prime (1)\, \ve}]_{12',21'}$, with the first
and second terms given by \Eqs{eq:J'(0)} and
(\ref{subeq:Jprime1storder}), respectively, as expected.  Thus, our
check worked.  [Also, the reason for the 2 in $\int (d \, 2 \tomega)$
in \Eq{eq:defineJprimevegeneral} becomes clear: $ \tilde {\cal P}^{1
  2}_{65 } ({\cal E}; \Omega_1, \Omega_2)$ turns out to contain
factors of $2 \pi \delta (\Omega_1 - \Omega_2)$ or $2 \pi \delta
(\Omega_1 - \Omega_2 + \dots )$ for self-energy and vertex terms,
respectively, which under the integral $\int (d \, 2 \tomega) \tilde
{\cal P}^{1 2'}_{2 1'} (\ve - \toh \tomega; - \tomega, \tomega)$ of
\Eq{eq:defineJprimevegeneral} have to collapse to unity, $\int (d \, 2
\tomega) 2 \pi \, \delta (\dots - 2 \tomega) = 1$.]

Finally, let us rewrite \Eq{eq:defineJprimevegeneral} in a more suggestive
form. Transforming back to the time domain using
\Eq{eq:Cooperonenergytimetimeinverse} and writing
the result in terms of the time variables of \Eq{eq:defmorenewtimes}, 
we find
\bsubequations
\label{subeq:defineJprimetime}
\begin{eqnarray}
\label{eq:defineJprimetimes-a}
\tilde J^{\prime \, \ve}_{12',21'}
&  =  &  [ - n' (\hbar \ve )] \, 
\int_0^\infty d \tau_{12}  \, 
\tilde P^{1 2', \ve}_{21'} (\tau_{12}) \, , 
\\
\label{eq:defineJprimetimes-b}
\tilde P^{1 2', \ve}_{21'} (\tau_{12}) 
 &  = & 
\int_{-\infty}^\infty \! d \ttau_{12} 
\int (d \, 2\tomega) \,  e^{- i 2 \tomega \ttau_{12}}
 \int_{-\infty}^\infty \! \!   d \bar \tau_{12} \, 
 e^{i \bar \tau_{12} (\ve - {1 \over 2} \tomega)}
\\
& & \nonumber \hspace{1cm} \times 
 \tilde P^{12'}_{21'} (\tau_{12}, \ttau_{12}, \bar \tau_{12}) \; , 
\qqph \qph 
\\
\label{eq:defineJprimetimes-c}
& \! = \! & 
\int_{-\infty}^\infty \! \! \!  d \bar \tau_{12} \, 
e^{i \bar \tau_{12} \ve }
\tilde P^{12'}_{21'} (\tau_{12}, - {\textstyle {1 \over 4}} 
\bar \tau_{12}, \bar \tau_{12}) \; .
\end{eqnarray}
\esubequations

{\bbf We need to consider $\tilde P^{12'}_{21'} $ only in the limit $\bmr_2
\to \bmr_1$, since the Cooperon contribution to it is negligible for $
|\bmr_1 - \bmr_2| \gtrsim \lambda_{\rm F} $, where $\lambda_{\rm F}$ is
the Fermi wavevector (assumed to be much smaller than the mean free
path, $\lambda_{\rm F} \ll \lel$).
%[The notation for $\tilde P^{1 2', \ve}_{21'} (\tau_{12})$ takes it to
%be understood that the indices $(12',21')$ refer only to position
%coordinates, whereas the time arguments $t_1^F$, $t_2^F$, $t_3^B$,
%$t_4^B$ are as indicated in \Eq{eq:Cooperonaspathintegral-maintext},
%and are converted to $\tau_{12}$, $\ttau_{12}$ and $\bar \tau_{12}$
%via \Eq{eq:tau12tildetau12maintext}.]
The purpose of the time integrals in \Eq{eq:defineJprimetimes-b} is to
project out from the general path integral $\tilde P^{12'}_{21'}$ of
\Eq{eq:Cooperonaspathintegral}, defined in the position-time domain,
an object depending in an appropriate way on both the average
propagation time $\tau_{12}$ of the forward and backward paths and the
energy $\ve$ occuring in the thermal weighting factor.  (The
simultaneous specification of both a time and an energy does not
violate the time-energy uncertainty relation, as incorrectly argued by
GZ\cite{GZ05}, because $\tilde P^{1 2', \ve}_{21'} (\tau_{12})$ is
constructed from \emph{two} electron propagators, not one). To see how
this projection works in detail, we use \Eqs{eq:defmorenewtimes} to
write the time differences $\tau_{12}$, $\ttau_{12}$ and $\bar
\tau_{12}$ as follows:
\begin{eqnarray}
\nonumber
  \tau_{12} & = &
\toh \bigl[ (t_1^F - t_2^F)  + ( t_3^B - t_4^B ) \bigr] , \; \quad
\ttau_{12} =  \toq \bigl[ (t_1^F + t_2^F )
- ( t_3^B + t_4^B )\bigr] , \qph
  \\   \label{eq:tau12tildetau12maintext}
\bar \tau_{12} & = &
(t_1^F - t_2^F)  - ( t_3^B - t_4^B ) \; .
\end{eqnarray}
The $\int d \bar \tau_{12}$ integral in \Eq{eq:defineJprimetimes-b}
fixes the average energy of the upper and lower electron lines (in
diagrammatic language) to be $\ve - \tomega/2$ [where $\bar \tau_{12}$
is the length difference between the forward and backward pieces of
the contour].  The $\int (d \tomega)$ integral averages over all
possible frequency differences $\tomega$ between the upper and lower
electron lines, as is necessary when vertex terms are present that
transfer energy between them. And finally, the $ \int d \ttau_{12}$
integral projects out the $\ttau_{12}$-dependence of $P^{12'}_{21'}$
[where $\ttau_{12}$ is half the difference between the midpoints of
the forward and backward pieces of the contour].  The only remaining
time variable, $\tau_{12}$, is the average of the lengths of the
forward and backward pieces, and can be viewed as the ``observation
time'' as a function of which $\tilde P^{1 2', \ve}_{21'} (\tau_{12})
$ will decay.  $\tilde P^{1 2', \ve}_{21'} (\tau_{12})$ will contain a
contribution resulting from time-reversed paths that corresponds to
the full Cooperon in the position-time representation, $\tilde
C_{\bmrho = 0} (\tau_{12})$. The time scale on which it decays is the
desired decoherence time $\tauphi$.

Now, the $\int (d \tomega)$ integral in \Eq{eq:defineJprimetimes-b}
yields $\delta (\ttau + {\textstyle {1 \over 4}} \bar \tau)$ [here and
henceforth we drop the subscripts on $\tau$, $\bar \tau$ and $\ttau$],
leaving us to consider a path integral with time arguments $ \tilde
P^{1 2'}_{21'} (\tau, - {\textstyle {1 \over 4}} \bar \tau, \bar \tau)
$, as indicated in \Eq{eq:defineJprimetimes-c}.  These time arguments
can be obtained by choosing, \eg, $t_1 = t_3 = \toh \tau$ and $t_{2,4}
= - \toh (\tau \pm \bar \tau)$, resulting
in:% [cf.\ \Eq{eq:defineJprimetimes-c}]:
\begin{eqnarray}
\label{eq:Pfixedenergy-a}
\tilde P^{1 2', \ve}_{21'} (\tau)
& \!\! = \!\! & \!\!  
\int_{-\infty}^\infty \!   d \bar \tau \, 
e^{i \bar \tau \ve } \;
\Fint_{\bmR^F (-{\tau \over 2}  - {\bar \tau \over 2})
 = \bmr_{2'}}^{\bmR^F ({\tau \over 2} ) = \bmr_1}
\Bint_{\bmR^B (-{\tau \over 2} +  {\bar \tau \over 2}) 
= \bmr_2}^{\bmR^B ({\tau \over 2}) = \bmr_{1'}}
\Bigl. \widetilde {\cal D}' (\bmR) \,
 e^{-[i \tilde S_R + \tilde S_I]/\hbar} \; .
\qqph \qph
\end{eqnarray}
\Eq{eq:sigmageneraldefinePI}, together with
(\ref{eq:defineJprimetimes-a}) and (\ref{eq:Pfixedenergy-a}), are the
central results of this section, because they express the conductivity
in terms of a general path integral influence functional, with thermal
weighting taken properly into account.  The main difference to the
path integral (\ref{eq:sigmageneraldefinePI-MAINtext-b}) used 
in the main text (and by GZ) is that
in \Eq{eq:Pfixedenergy-a} the duration of the forward and backward
paths differs by a time $\bar \tau$ that is being integrated over in
$\int d \bar \tau \, e^{i \bar \tau \ve }$. The remainder of this
section is devoted to justifying the replacement of
\Eq{eq:Pfixedenergy-a} by the simpler
\Eq{eq:sigmageneraldefinePI-MAINtext-b}.

%An interpretation
%of the meaning of the various integrals occuring in
%\Eqs{subeq:defineJprimetime} may be found in the last paragraphs of
%App.~\ref{sec:thermalaveragingAppA}, which we suggest should be reread
%at this point [\Eqs{eq:defineJprimetimes-bmaintext} and
%(\ref{eq:tau12tildetau12maintext}) there correspond to
%\Eqs{eq:defineJprimetimes-b} and (\ref{subeq:defnewtimes}) here].
%There we also describe an approximation scheme for simplifying these
%integrals significantly, leading to
%\Eq{eq:sigmageneraldefinePI-MAINtext} of the main text.

The combination $\int d\ve \int d \bar \tau$ of integrals from
\Eqs{eq:sigmageneraldefinePI} and (\ref{eq:Pfixedenergy-a})
have the effect of fixing the average energy of the forward and
backward trajectories to be close to the Fermi energy, with energy
spread of roughly $\pm T$ (in a way reminiscent of App.~B of the
review\cite{ChakravartySchmid86} by Chakravarty and Schmid).
To see this, consider first the noninteracting limit (\ie\ ignore $i
\tilde S_R + \tilde S_I$) in the semiclassical approximation, where
the path integrals in \Eq{eq:Pfixedenergy-a} are restricted to all
possible classical forward and backward paths $\bmr_\class^{F/B}(t_3)$
having the specified boundary conditions, with corresponding classical
actions $S_{0,\class}^{F/B} ({\textstyle {\tau \over 2}}, -{\textstyle
  {\tau \over 2}} \mp {\textstyle {\bar \tau \over 2}}) $.  Since
these paths follow diffusive trajectories through a disordered
potential landscape, for any given $\tau$ and $\bar \tau$ the path
integral still includes many such classical paths, with a range of
different classical energies (and correspondingly different diffusion
constants). Now, the energy integral in
\Eq{eq:sigmageneraldefinePI-MAINtext-a} restricts the $\int d \bar
\tau$ integral in \Eq{eq:Pfixedenergy-a} to the range $|\bar \tau|
\lesssim \hbar /T$, since
\begin{eqnarray}
  \label{eq:deltaTfixesttau}
  \int d \ve [-n'(\hbar \ve)] e^{i \ve \bar \tau}
  = {\pi \hbar \bar \tau T \over \sinh( \pi \hbar \bar \tau T)} \; . 
\end{eqnarray}
The relevant values of $\bar \tau$ are thus much smaller than the
typical propagation times $\tau$ relevant for determining the
decoherence time [$\tau \simeq \tauphi \sim \hbar g(L_\varphi)/T \gg
\hbar /T$, see \Eq{eq:definetauphig}], so that the classical actions
can be expanded\cite{ChakravartySchmid86} to first order in $\bar
\tau$,
\begin{eqnarray}
\label{eq:Sttauexpand}
S_{0,\class}^{F/B} % [\bmr_\class^{F/B} (t_3)]
({\textstyle {\tau \over 2}}, 
-{\textstyle {\tau \over 2}} \mp {\textstyle {\bar \tau \over 2}}) \simeq  
S_{0,\class}^{F/B} %[\bmr_\class^{F/B} (t_3)]
({\textstyle {\tau \over 2}}, 
-{\textstyle {\tau \over 2}}) \mp 
\toh  \bar \tau  {\cal E}_\class^{F/B} %[\bmr_\class^{F/B} (t_3)] 
\; , 
\qqph
\end{eqnarray}
where ${\cal E}_\class^{F/B}$ is the classical energy at the endpoint
of the corresponding classical path $\bmr_\class^{F/B} (t_3)$.  
Using this in \Eq{eq:Pfixedenergy-a},
the $\int d \bar \tau$ integral is seen to  fix the
average classical energy of the forward and backward classical paths
to be close to the Fermi energy $\ve_F = 0$, with an energy
spread of order $T$:
\begin{eqnarray}
\label{eq:fixenergytoEFmoduloT}
\int_{-\infty}^\infty \!   d \bar \tau \, 
{\pi \hbar \bar \tau T \over \sinh( \pi \hbar \bar \tau T)} 
\, e^{i  \bar \tau  \toh (
    {\cal E}_\class^F + {\cal E}_\class^B )}
  = \int d \ve [-n'(\hbar \ve)] \delta \Bigl( \ve - 
  \toh \left(
    {\cal E}_\class^F  + {\cal E}_\class^B \right) \Bigr)
  \; . \qqph
\end{eqnarray}
(The right-hand side follows from using the integral representation
(\ref{eq:deltaTfixesttau}) for the $\sinh$-function.) Note that the
energy spread is consistent with the time-energy uncertainty relation
in the limit of present interest, $\tau T \gg \hbar $.

Now, in the absence of interactions, the only effect of fixing this
average energy $\ve$ to be roughly $\ve_F$ is that the velocity
appearing in the diffusion constant is the Fermi velocity, $D = v_F^2
\tauel / d$. However, in the presence of interactions, the energy
$\ve$ also plays a role in determining the phase space available for
electrons to get scattered upon absorbing or emitting a noise quantum.
In particular, in perturbative calculations it shows up in the $\tanh
[ \hbar (\ve \mp \bomega)/2T]$-factors of the Keldysh electron Green's
functions $\tilde G^K (\ve \mp \bomega)$.  In our influence functional
approach this can be kept track of by replacing \Eq{eq:Pfixedenergy-a}
by \Eq{eq:sigmageneraldefinePI-MAINtext-b}, which mimicks the effect
of the former's integral $\int d \bar \tau e^{i \ve \bar \tau}$ by
using (i) forward and backward paths of equal duration $\tau$ and (ii)
an effective action whose time integration boundaries are fixed at
$\pm \tau/2$, but which \emph{depends explicitly on the average
  propagation energy $\ve$}.  Note that GZ's approach in effect
employs the same simplification, since they likewise have no $\int d
\bar \tau e^{i \ve \bar \tau}$ integral and use forward and backward
paths of equal duration $\tau$.

The $\ve$-dependence of the effective action enters through the Pauli
factor $(\tilde \delta - 2 \tilde \rho)$ occuring in $\tilde S_R$
[\Eqs{eq:LtildesA} or (\ref{eq:seffcompareMarquardt})], which we treat
differently from GZ. In our approach, it produces factors of $\tanh [
\hbar (\ve \mp \bomega)/2T]$ in the frequency representation of
$\tilde S_R$ [cf.\ \Eqs{eq:modifiedRAtanh} or (\ref{subeq:LAA'FT})],
chosen in such a way as to be consistent with Keldysh perturbation
theory, as discussed in Sec~\ref{sec:Paulifactorshort} and (more
extensively) \ref{sec:ruleofthumb}, \ref{sec:alteffaction}.  In GZ's
approach, the $\tanh$-arguments contain $\ve$ instead of $\ve \mp
\bomega$ (\ie\ their effective action depends on the average energy
too).  However, lacking the $\mp \bomega$ recoil shifts, the
$\tanh$-terms turn out to yield zero after averaging over random
walks, so that $\langle i \tilde S_R^\GZ \rangle_\rw \simeq 0$.

The strategy just described for arriving at forward and backward paths
of equal duration is of course not exact; but it is sufficiently
accurate for our purposes: the errors incurred by it are of order
$\hbar /(T \tau)$ ($\ll 1$ for $\tau \sim \tauphi$), as can be shown
by a detailed comparison with Keldysh diagrammatic perturbation theory
(App.~B.6.2 of this review, and App.~A.3 of
DMSA-II\cite{MarquardtAmbegaokar04}).

}

\subappendix{General Cooperon, after Disorder Averaging} 
\label{sec:appCooperongeneralaveraged}

Let us now disorder average \Eqs{eq:defineJprimevegeneral}, in order
to arrive at an expression for $\sigma_\DC^\WL$ in terms of a general
Cooperon propagator, in the presence of interactions.  To this end, we
have to Fourier transform from position to wave-number variables,
\begin{eqnarray}
& & \phantom{.} \hspace{-1cm}
  \tilde {\cal P}^{1 2}_{4 3} ({\cal E}; \Omega_1, \Omega_2)  = 
\, {1\over  \Vol^2} \sum_{\bmp_1, \bmp_2, \bmp_3 , \bmq} 
\, \delta_{\bmp_3,0} \, 
\bcP^{{\cal E}, \Omega_1, \Omega_2}_{\bmq , \bmp_1 , \bmp_2}
\exp i \left\{
 \bmr_1^F \! \cdot \! \left[   \bmp_1 +  {\bmq + \bmp_3  \over 2}
 \right]
\right. \nonumber 
\\
\label{eq:Cooperonmomentumspace}
& &  \phantom{.} \hspace{-0.5cm}
\left. 
-  \bmr_2^F\! \cdot \!  \left[\bmp_2 +  {\bmq - \bmp_3  \over 2}  \right]\!
+  \bmr_4^B \! \cdot \! \left[  - \bmp_1  + {\bmq + \bmp_3  \over 2} \right]\!
-  \bmr_3^B \! \cdot \! \left[ - \bmp_2  + {\bmq - \bmp_3  \over 2}  
\right] \! \right\} \; . \qqph 
\end{eqnarray}
as depicted in Fig.~\ref{fig:cooperonrealspace}(c).  (Again, the
$\delta_{\bmp_3, 0}$ guarantees translational invariance.)  According
to the standard diagrammatic approach for disorder averaging [cf.\ 
Fig.~\ref{fig:sigma-cooperon} in App.~\ref{sec:diagrammatics}], the
disorder average of $\bcP^{{\cal E}, \Omega_1, \Omega_2}_{\bmq ,
  \bmp_1 , \bmp_2}$ can be separated into a ``Drude'' and a
``weak-localization'' contribution,
\begin{eqnarray}
\nonumber
\bigl \langle
 \bcP^{{\cal E}, \Omega_1, \Omega_2}_{\bmq , \bmp_1 , \bmp_2}
\bigr \rangle_\dis
& = & 
\hbar^2 \, 
\bcG^R_{{1\over 2} \bmq +\bmp_1 } ({\cal E} + \! \toh \Omega_1) \, 
\bcG^A_{{1\over 2} \bmq -\bmp_1} ({\cal E} \! - \! \toh \Omega_1) \, 
\left\{ 2 \pi (\Omega_1 - \Omega_2)  \, \delta_{\bmp_1, \bmp_2} 
\phantom{{\bcC_\bmq^{\cal E} \over  \tauel^2} }\right.
\qph 
\\ 
  \label{eq:barPqQppexplicit}
& & \left. +  \; 
\bcG^R_{{1\over 2} \bmq + \bmp_2} ({\cal E} + \! \toh \Omega_2) \, 
\bcG^A_{{1\over 2} \bmq -\bmp_2} ({\cal E} \! - \! \toh \Omega_2) \, 
{\bcC{}_\bmq^{\cal E} (\Omega_1, \Omega_2) 
\over \Vol \, 2 \pi \nud \tauel^2 / \hbar}  \right\} , \qqph
\end{eqnarray}
where in the second term the
contributions from the four external electron lines 
were separated and a conventional prefactor 
 $(2 \pi \nud \tauel^2 / \hbar)^{-1}$ was split off. 
The normalization of the general Cooperon in the presence of
interactions, $\bcC{}^{\cal E}_\bmq ( \Omega_1, \Omega_2)$, is fixed by
requiring that when interactions are switched off, it reduces to its
free version, $\bcC{}^0_\bmq (\Omega_1)$,
%= [\Dd \bmq^2 - i \Omega_1 + \gammaH]^{-1}$, 
according to
%\bsubequations
\begin{eqnarray}
  \label{eq:freefromfullCooperon}
\bcC{}^{\cal E}_\bmq (\Omega_1, \Omega_2) 
& \stackrel{{\rm no \; int}}{\longrightarrow} &
2 \pi \, \delta (\Omega_1 - \Omega_2) \, \bcC{}^0_\bmq
( \Omega_1) \; , 
\quad 
\bcC{}^0_\bmq (\Omega_1) = { 1 \over \Dd \bmq^2 - i \Omega_1 + \gammaH} \;
. \qqph  \qph 
%\bigl(\toh \Omega_1 + \toh \Omega_2  \bigr) \; , 
\end{eqnarray}
Just as $\bcC{}^0_\bmq
(\Omega_1)$, the full Cooperon $\bcC{}^{\cal E}_\bmq (\Omega_1, \Omega_2) $
does not depend on the external momenta $\bmp_{1,2}$, because, in
diagrammtic terms, it is separated from external lines by impurity
lines.

In a purely diagrammatic approach, where one typically
works exclusively in the wavenumber-frequency domain,
\Eq{eq:barPqQppexplicit} would be the standard starting point
for further calculations.
Since the dominant contribution to $\bcC{}_\bmq^{\cal E} (\Omega_1,
\Omega_2)$ typically  comes from small $\bmq$ (with $q \lel 
\ll 1$) and small $\Omega_{1,2}$ (with $\Omega_{1,2} \tauel \ll 1$),
while ${\cal E}$ is likewise small ($\lesssim T$), it is customary
to neglect the terms $\pm \toh \bmq$ and $ {\cal E} \pm \toh \Omega_{1,2}$ 
in the arguments of the external 
$\bcG^{R/A}$ functions, which simplifies the $\int d\bmp_{1,2}$ 
integrals. To explore the effects of interactions, one 
would proceed to expand $\bcC{}_\bmq^{\cal E} (\Omega_1, \Omega_2) $ 
in powers of the interaction propagator, etc.

Instead, here we shall use the general \Eqs{eq:Cooperonmomentumspace}
and (\ref{eq:barPqQppexplicit}) for $ \tilde {\cal P}^{1 2}_{43}
({\cal E}; \Omega_1, \Omega_2) $ to analyse the general structure of
the disorder-averaged version of \Eq{eq:sigmageneraldefinePI}, as
needed for $\bigl\langle \sigma_{\DC} \bigr \rangle_\dis$. As
intermediate result, we obtain
 \bsubequations
 \begin{eqnarray}
   \label{eq:newintermediatestep}
  & & \phantom{.} \hspace{-0.7cm}
\int \! dx_{2} \,
\bmj_{11'} \cdot \bmj_{\,22'}
\bigl \langle \tilde {\cal P}^{1 2'}_{21'} ({\cal E}; \Omega_1, \Omega_2) 
\bigr \rangle_\dis
\\ \nonumber 
& & \phantom{.} \hspace{-0.5cm} = 
{e^2 \hbar^4 \over 4 m^2 \Vol^2} \sum_{\bmp_1 \bmp_2
  \bmq}
 (\bmq + \bmp_1 - \bmp_2) \! \cdot \!   (\bmq - \bmp_1 + \bmp_2) 
\bigl \langle
\bcP^{{\cal E}, \Omega_1, \Omega_2}_{\bmq , \bmp_1 , \bmp_2}
\bigr \rangle_\dis
\int d \bmr_2  
e^{i ( \bmr_1 - \bmr_2)\cdot (\bmp_1 - \bmp_2)}
\\    \label{eq:semifinalstep}
& & \phantom{.} \hspace{-0.5cm}
= \sigma^\Drude_\DC \, 2 \pi \hbar 
\left[2 \pi (\Omega_1 - \Omega_2)  
 - {1 \over \pi \nud
 \hbar} \int (d \bmq)  \, 
\bcC{}^{{\cal E}}_\bmq (\Omega_1, \Omega_2)
\right] \; ,
 \end{eqnarray}
 \esubequations
 where \Eq{eq:integrationidentitiesgeneral} was used (under neglect of
 ${\cal E} \pm \toh \Omega_{1/2}$ in the frequency arguments of all
 electron Green's functions) to perform the momentum integrals, \ie\ 
 the $\int (d\bmq)$ integral for the Drude contribution to
 $\bcP^{{\cal E}, \Omega_1, \Omega_2}_{\bmq , \bmp_1 , \bmp_2}$,
 and the $\int (d \bmp_1)$ integral for the Cooperon contribution (for
 the latter, the $\toh \bmq$ arguments in the external electron leg
 Green's functions were neglected).  Inserting 
 \Eqs{eq:defineJprimevegeneral} and (\ref{eq:semifinalstep})
 into \Eq{eq:sigmageneraldefinePI}, we readily find:
\begin{eqnarray}
  \label{eq:finalsigmawithintmaintext}
  \sigma_\DC =  \sigma^\Drude_\DC 
\left[ 1 - {1 \over \pi \nud  \hbar} 
\int d \ve  \, \hbar 
\, [-  n'_0 (\hbar \ve)] 
\int (d \bmq)  \int (d \, 2  \tomega) \, 
\bcC{}^{\ve - {1 \over 2} \tomega}_\bmq
(- \tomega , \tomega) \right]  . \qqph
\end{eqnarray}
\Eq{eq:finalsigmawithintmaintext} is the desired generalization of
\Eq{eq:sigmaDSnonfinal} [and in the absence of interactions, duly
reduces to the latter, via \Eq{eq:freefromfullCooperon}].
%If so desired, the
%factor $\int (d \, 2 \tomega) \, \tcC^{\ve - {1 \over 2} \tomega}_\bmq
%(- \tomega , \tomega) $ can be written as $\int d \tau_{12} \, \tilde
%C^{\ve}_\bmq (\tau_{12})$, where $\tilde C^{\ve}_\bmq (\tau_{12})$ is
%defined analogously to \Eq{eq:defineJprimetimes-b}.

\subappendix{Cooperon in Position-Time Domain}
\label{sec:cooperonpostime}

For our present purpose of relating the diagrammatic and path integral
aproaches to each other, it is instructive to understand the
consequences of \Eq{eq:barPqQppexplicit} also in path integral
language.  To this end, let us transcribe \Eq{eq:barPqQppexplicit}
back into the position-time domain, in which the Cooperon is defined
as:
\begin{eqnarray}
  \label{eq:Cpositiontimenew}
  \tilde C^{\cal E}_\bmrho (\tau_1, \tau_2) & \equiv &
\int (d \bmq) (d \Omega_1 ) (d \Omega_2 ) \, 
e^{i (\bmrho \cdot \bmq - \Omega_1 \tau_1 + \Omega_2 \tau_2)}
\, \bcC{}_\bmq^{\cal E} (\Omega_1, \Omega_2) \; . 
\end{eqnarray}
Inserting \Eqs{eq:Cooperonmomentumspace} and
(\ref{eq:barPqQppexplicit}) into \Eq{eq:Cooperonenergytimetime} yields
$ \bigl \langle \tilde P \bigr \rangle_\dis = \tilde P^\Drude + \tilde
P^\WL$, with
\bsubequations
  \begin{eqnarray}
    \label{eq:PDruderealspacetime}
&&  \phantom{.} \hspace{-0.8cm}
   \tilde P^{12, \Drude}_{43}  =  \hbar^2 
    \tcG^R_{\bmr_{12}} (t_{12}) \tcG^A_{\bmr_{43}} (t_{43}) \; ,
\\ \nonumber
& &  \phantom{.} \hspace{-0.8cm}
    \tilde P^{12, \WL}_{43}  =  
\int {d \tilde \bmr_1 \,  d \tilde \bmr_2 \,  
d \tilde t_1 \,  d \tilde t'_1 \,  d \tilde t_2 \, d \tilde t'_2 
 (d{\cal E})  \over 2 \pi \nu \tauel^2 / \hbar}
e^{- i {\cal E} (\tilde t_1 + \tilde t'_1 - \tilde t_2 - \tilde t'_2)}
\, \tilde C^{\cal E}_{\tilde \bmr_1 - \tilde \bmr_2} 
\bigl( \toh (\tilde t_1 - \tilde t'_1), 
       \toh (\tilde t_2 - \tilde t'_2) \bigr) 
\\
    \label{eq:PWLrealspacetime}
& & \times \hbar^2 
\tcG^R_{\bmr_1 - \tilde \bmr_1} (t_1 - \tilde t_1) \, 
\tcG^A_{\bmr_4 - \tilde \bmr_1} (t_4 - \tilde t'_1) \, 
\tcG^R_{\tilde \bmr_2 -  \bmr_2} (\tilde t_2 - t_2) \, 
\tcG^A_{\tilde \bmr_2 -  \bmr_3} (\tilde t'_2 - t_3) \; .
  \end{eqnarray}
\esubequations
  Fig.~\ref{fig:cooperonrealspace}(b) offers an intuitive interpretation
  of these expressions: $ \tilde P^{12, \Drude}_{43}$ gives the
  amplitude for propagation from $(\bmr_2, t_2) \to (\bmr_1 , t_1)$
  (forward in time) times that for $(\bmr_3, t_3) \to (\bmr_4 , t_4)$
  (backward in time).  And $ \tilde P^{12, \WL}_{43}$ gives the
  amplitude for forward propagation from $(\bmr_2, t_2) \to (\tilde
  \bmr_2, \tilde t_2) \to (\tilde \bmr_1, \tilde t_1) \to (\bmr_1,
  t_1)$, times that for backward propagation from $(\bmr_3, t_3) \to
  (\tilde \bmr_2, \tilde t'_2) \to (\tilde \bmr_1, \tilde t'_1) \to
  (\bmr_4, t_4)$. The middle part of the forward and backward paths
  have the same beginning and end points in space, albeit not in time,
  and hence can interfere constructively if the paths connecting them
  are time-reversed partners.
  
  The approximation mentioned above of neglecting $\pm \toh \bmq$ and
  $ {\cal E} \pm \toh \Omega_{1,2}$ in the arguments of external
  $\bcG^{R/A}$ functions has a counterpart in the position-time
  domain: when performing the integrals in \Eq{eq:PWLrealspacetime},
  it corresponds to exploiting the fact that $\tcG_\bmr (\tilde t)$
  has a short range in space $(| \bmr| \lesssim \tauel)$ and time
  $(|\tilde t| \lesssim \lel$) [cf.\ \Eq{eq:shortrangeGtimeresult-b}].
  To be explicit, the latter fact means that the disordered Green's
  functions occuring in the second line of \Eq{eq:PWLrealspacetime}
  act effectively as delta-functions in time as far as the factor
  $e^{-i{\cal E} ( \; )} \tilde C^{\cal E}_{\tilde \bmr_1 - \tilde
    \bmr_2} (\; )$ is concerned. Thus, in the latter we may make the
  replacements $\tilde t_1 \to t_1$, $\tilde t'_1 \to t_4$, $\tilde
  t_2 \to t_2$, $\tilde t'_2 \to t_3$, after which the four
  time-integrals each yield a zero-frequency Green's function, $\int d
  \tilde t \, \tcG_\bmr (\tilde t) = \tcG_\bmr (\ve = 0)$.
  Introducing the sum and difference coordinates
\begin{eqnarray}
  \label{eq:defnewcoordinates}
& &  \phantom{.} \hspace{-1cm}
 \bbrho_1 \equiv {\bmr_1^F + \bmr_4^B \over 2} \; ,  \quad
  \bmrho_1 \equiv {\bmr_1^F - \bmr_4^B } \; ,  \qquad
  \bbrho_2 \equiv {\bmr_2^F + \bmr_3^B \over 2} \; , \quad
  \bmrho_2 \equiv {\bmr_2^F - \bmr_3^B } \; , \qqph 
\end{eqnarray}
recalling similar definitions (\ref{subeq:defnewtimes}) for
the time variables, and  shifting the space integrations according
to $\tilde \bmr_i \to \tilde \bmr_i + \bbrho_i$ for $i = 1,2$,
\Eq{eq:PWLrealspacetime} gives: 
\begin{eqnarray}
 \nonumber
      \tilde P^{12, \WL}_{43} & = & 
\int \!\! d \tilde \bmr_1 \,  d \tilde \bmr_2 \int  (d{\cal E}) \, 
e^{- i {\cal E} \bbtau_{12}} \, 
{\tilde C^{\cal E}_{\bbrho_1 - \bbrho_2 + \tilde \bmr_1 - \tilde \bmr_2} 
\bigl( \tau_1, \tau_2) 
\over 2 \pi \nu \tauel^2 / \hbar} \, 
\\
    \label{eq:PWLrealspacetimec}
& & \times \hbar^2 
\tcG^R_{{1 \over 2} \bmrho_1 - \tilde \bmr_1} (0) \, 
\tcG^A_{-{1 \over 2} \bmrho_1 - \tilde \bmr_1} (0) \, 
\tcG^R_{\tilde \bmr_2 -  {1 \over 2} \bmrho_2} (0) \, 
\tcG^A_{\tilde \bmr_2 +  {1 \over 2} \bmrho_2} (0) . \qqph
\end{eqnarray}
Since the zero-frequency Green's functions $\tcG^R_\bmr (0) $
decay with distance as $e^{- |\bmr|/2 \lel}$, we note that
$\tilde \bmr_i \simeq \toh \bmrho_i \simeq - \tilde \bmr_i$,
which implies that $|\tilde \bmr_i| \lesssim \lel$ and $|\bmrho_i| \lesssim
\lel$. Thus, we may drop the terms $\tilde \bmr_1 - \tilde \bmr_2$
from the argument of  
$\tilde C^{\cal E}$ in \Eq{eq:PWLrealspacetimec}, whereupon
the spatial integrations can be performed 
explicitly,  using 
\bsubequations
    \label{eq:smeared-Delta}
\begin{eqnarray}
& &  \int_{- \infty}^\infty d \tilde t_i 
 \int_{- \infty}^\infty d \tilde t_j 
 \int d \tilde \bmr_l \;
\tcG^{R/A}_{il} (\tilde t_i) \tcG^{A/R}_{lj} (\tilde t_j)  =  
{ \delta_{\lel} (r_\iiijjj) \over \eF^2} \; ,
\\
& & \tilde \delta_\lel (r_\iiijjj)   \equiv 
\left( {  \lel \kF^3 \over 4 \pi } \right) 
e^{- r_\iiijjj / 2 \lel} {\sin(\kF
  r_\iiijjj) \over  r_\iiijjj } \; , \qqph \qph
\end{eqnarray}
\esubequations
where $\tilde \delta_{\lel} (r_\iiijjj) $ is a ``smeared-out delta
function'' of normalization $\int d \bmr_\iiijjj \, \tilde \delta_\lel
(r_\iiijjj) = 1$ and width $\simeq 1/\kF$, the Fermi wavelength (since
$1/\kF \ll \lel $, the width is set by the oscillating factor $\sin
(\kF r)/ r$, not by the exponential $e^{- r / 2\lel}$).  Thus, 
\Eq{eq:PWLrealspacetimec} becomes:
\begin{eqnarray}
    \label{eq:finalP=Cooperonrealspace-a}
\tilde P^{12, \WL}_{43} & = & 
{\hbar^2 \over \eF^4}
\tilde \delta_\lel (\bmrho_1) \tilde \delta_\lel   (\bmrho_2) \,
\int (d {\cal E}) \, e^{- i {\cal E} (\bbtau_1 - \bbtau_2)} \, 
  {\tilde C^{\cal E}_{\bbrho_1 - \bbrho_2} (\tau_1, \tau_2) 
\over 2 \pi \nu \tauel^2 / \hbar} \, .
\end{eqnarray}
This useful result clarifies the relation between the coordinates
1,2,3,4 of $\tilde P^{12, \WL}_{43}$, and the times and spatial
coordinates relevant for the Cooperon. In particular, we see that $
\tilde P^{12, \WL}_{43} $ is nonzero only for $|\bmrho_1| = |\bmr_1 -
\bmr_4| \lesssim 1 / \kF$ and $|\bmrho_2| = |\bmr_2 - \bmr_3| \lesssim
1/\kF$.  Moreover, if we want to describe a Cooperon with a specified
average energy ${\cal E}$, we need to Fourier-transform $ \tilde P^{12,
  \WL}_{43} $ with $\int \! d \bar \tau_{12} e^{i {\cal E}
 \bar \tau_{12}}$. Note that for $\tilde P^{1 2', \WL}_{21'}$, 
as needed in \Eqs{subeq:defineJprimetime},
the Cooperon position argument is identically zero,
${\bbrho_1 - \bbrho_2} = 0$, while 
$|\bmrho_1| = |\bmr_1 - \bmr_2|$
ensures that $\bmr_1$ and $\bmr_2$ lie close together.

%\appeqn
\appendix{Time-Slicing of Path Integral for $\tilde U^a_{\iiijjj}$}
\label{app:path-integral}

In this appendix, we give an exlicit time-slicing definition for the
path integral representation
(\ref{eq:def-path-integral-pureRmaintext}) of the propagators $\tilde
U^a_{\iiijjj}$ used the main text, and derive various properties thereof.
Our discussion is very (perhaps overly) detailed, since the object of
interest is somewhat unconventional, namely a path integral for a
non-local hamiltonian. We begin [Secs.~D.1 to D.3] by defining it in
terms of a path integral $\int {\cal D} \bmR \int {\cal D} \bmP$ over
paths in both coordinate and momentum space, which is the form used by
GZ; then [Sec.~\ref{sec:expandU}] we explicitly perform the $\int
{\cal D} \bmP$ integral to arrive at a ``coordinate-space-only'' path
integral $\int \widetilde {\cal D}' \bmR$, which is the form used in
App.~B.5 to B.8.  Finally [Sec.~\ref{sec:b=0results}] we present
explicit expressions for the effective Hamiltonian $\bar H^a_n$ in the
position-momentum representation used by GZ, and
[Sec.~\ref{sec:SRSI-RP}] recover from this GZ's expressions for the
effective action $(i\bar S_R + \bar S_I)[\bmR^a, \bmP^a]$.

\begin{figure}[t]
{\includegraphics[clip,width=0.98\linewidth]{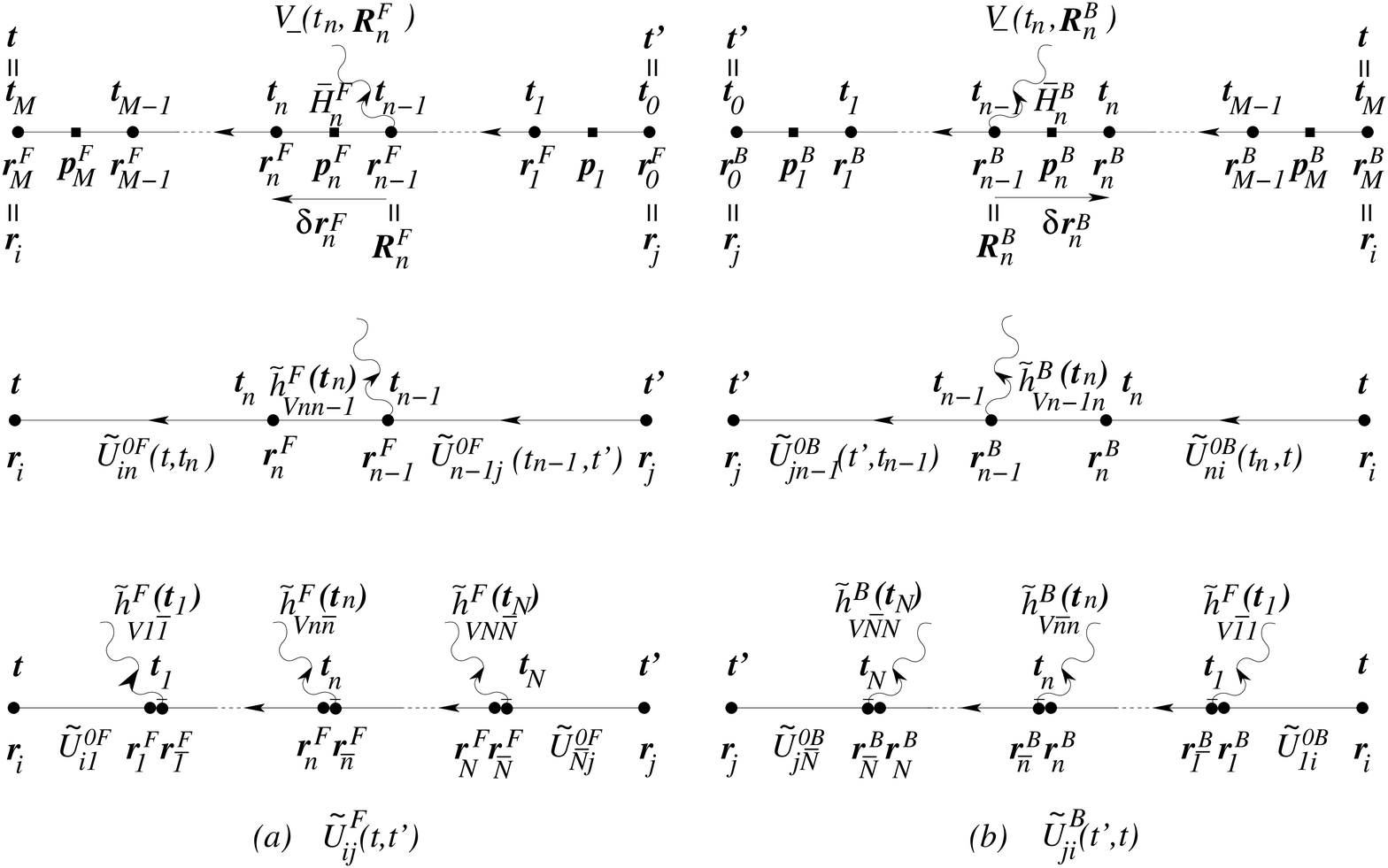}}%
\caption{
  Three representations of the propagators (a) $\U^F_\iiijjj (t, t')$ and
  (b) $\U^B_{ji} (t', t) $, with $t> t'$. Arrows point from the second
  to the first index of propagators. The first row illustrates the
  position-momentum (dots-squares) time-sliced path integral
  representation of \Eq{eq:def-path-integral} (with the choice $b_a =
  \delta_{a B}$ in cf.\ \Eq{eq:deltarR}, so that $\bmR^a_n =
  \bmr^a_{n-1}$); the wavy line indicates which end of the $n$-th time
  slice the interaction field $V_- (t_n, \bmr_{n-1}^a)$ is attached
  to.  The second row depicts the first order perturbation expansion
  of \Eq{eq:PIexpandUa-2}, obtained after performing the momentum path
  integral, using \Eq{eq:identifyhV} to convert $\bar h^a_V$ to
  $\tilde h_V^a$. The third row shows the $N$-th order perturbation
  term of \Eq{eq:PIexpandedinfull}. The double dots remind us that the
  vertices $\tilde h^F_{V n \bar n}$ and $\tilde h^B_{V \bar n n}$ are
  nonlocal (since they contain factors of $\tilde \rho_{n \bar n}$ or
  $\tilde \rho_{\bar n n}$): they arise from ``pulling together'' the
  two local vertices at times $t_n$ and $t_{n-1}$ of the second row of
  this figure into a single nonlocal vertex at time $t_n$, with which
  we hence associate a double integration $\int d x^F_{n \bar n}$ or
  $\int d x^B_{\bar n n}$.  The dot carrying a bar indicates which of
  these two integration variables occurs in the argument of $V_-
  (\bmr_{\bar n})$, namely the one drawn on the side of earlier
  times.}
\label{fig:timeslice}
\end{figure}

\subappendix{Time-Slicing Definition}

The propagators $\tilde U^a_{\iiijjj}$ are defined by the requirement that
they have to satisfy both the conditions \Eqs{eq:tildeUconditions}.
This fact can be used to give meaning to the formal path integral of
\Eq{eq:def-path-integral-pureRmaintext},  by using the standard
time-slicing procedure to construct an object that satisfies this
requirement.  To this end, we divide the interval
$[t', t]$ into $M = (t - t')/\epsilon$ time intervals, with $t_n = t'
+ n \epsilon $ for $n = 0, \dots M$, and write $\bmr^a_n = \bmr^a(t_n)
$ [$\bmr^a_0 = \bmr_j$, $\bmr^a_M = \bmr_i$] and $ \bmp^a_n =
\bmp^a(t_n) $. Then the following construction, illustrated in the
first row of Fig.~\ref{fig:timeslice}, has the desired properties:
\bsubequations
\begin{eqnarray}
\label{eq:def-path-integral}
\left. \begin{array}{r}
 \tilde U^F_{\iiijjj} (t,t') \\
 \tilde U^B_{ji} (t',t)
\end{array} \right\} \!\!
 & \equiv  & \!\!  \delta_{\sigma_i \sigma_j}
\lim_{M \to \infty} \prod_{n=1}^{M-1} \Bigl( \int d \bmr^a_{n} \Bigr)
\prod_{n=1}^M \Bigl( \int {d \bmp^a_n \over (2 \pi )^d} \Bigr)
e^{{  (i s_a \epsilon/  \hbar) } \sum_{n=1}^M \bar L^a_n } \; \qqph
\qph \\
& \equiv & 
\int \! {\cal D} \bmR \! \! \int \! {\cal D} \bmP \, e^{(i s_a / \hbar) \bar 
S^a [\bmR^a, \bmP^a] } \; 
. \qqph
\qph 
\end{eqnarray}
\esubequations
The second line, with action $\bar S^a = \epsilon \sum_n \bar L^a_n$,
is a formal shorthand for the detailed time-slicing construction of
the first line.  Here and below, $t > t'$, the index value $a=F$ or
$B$ should be used for the upper or lower term in the curly bracket,
and $s_a$ stands for $s_{F/B} = \pm$.  The multiple products in
\Eq{eq:def-path-integral} contain one momentum integral ($M$ in total)
for each interval, and one position integral ($M-1$ in total) for each
boundary between intervals (see Fig.~\ref{fig:timeslice}). The
Lagrangian $ \bar L^a_n $ and Hamiltonian $\bar H^a_n \equiv \bar H^a
\bigl( t_n , \bmR^a_n, \bmP^a_n \bigr)$ associated with the $n$-th
interval are given by (here $\bmP^a_n \equiv \hbar \bmp^a_n$):
\bsubequations
\label{eq:defineHn}
\begin{eqnarray}
\bar L^a_n & \equiv &
\bmP^a_n \cdot {\delta \bm{r}^a_n
 \over \epsilon} -  \bar  H^a_n \; ,
\\ \label{eq:defineH^aslicen}
\bar H_n^a &  \equiv & 
\int d (\delta \bmr^a_n) e^{ - i s_a \bmp_n^a \cdot \delta \bmr^a_n }
\tilde H^a (t_n, \bmR_n^a + s_a (1-b_a)
\delta \bmr^a_n, \bmR_n^a - s_a b_a \delta \bmr^a_n) \; . \qqph \qph 
\end{eqnarray}
\esubequations
Here we  introduced relative and ``asymmetric
center-of-mass'' coordinates for the $n$-th interval,
\begin{eqnarray}
\label{eq:deltarR}
\delta \bmr^a_n = \bmr^a_n - \bmr^a_{n-1} \,  , \quad
\bmR^a_n =
\left\{ \begin{array}{l}
\bmr^F_{n-1} \\
\bmr^B_n \end{array} \right. \!\!\! + s_a b_a \delta \bmr^a_n
\, = \,
\left\{ \begin{array}{l}
\bmr^F_n \\
\bmr^B_{n-1} \end{array} \right. \!\!\! - s_a(1- b_a) \delta \bmr^a_n \, ,
\qph 
\end{eqnarray}
where the ``asymmetry parameter'' $b_a$ is a real number with $0 \le
b_a \le 1$, which in general can be different for $a=F$ or $B$.  The
actual values chosen for $b_a$ do not affect any of the final
results, hence they can be chosen according to taste or convenience, or
left unspecified, as we shall do for now. 
%(According to GZ's notation
%in the paragraph before (GZ-II.58), they apparently take $b_F = b_B =
%\toh$.)  
It is to be understood that under the path integral, the
notation $ \bmR^a (t_n)$ and $ \bmP^a (t_n)$ [\eg\ as arguments of the
fields $V_\pm (t_n, \bmR^a (t_n)$], should be interpreted as
$\bmR^a_n$ and $\bmP^a_n$, respectively.

The arguments of $\tilde H^a$ in \Eq{eq:defineH^aslicen} were
purposefully constructed such that the inverse Fourier transform of
\Eq{eq:defineH^aslicen} yields
\begin{eqnarray}
  \label{eq:HaRPrdr}
  \int {d \bmp_n^a \over (2 \pi )^d} \,
e^{ i s_a \bmp^a_n \cdot \delta \bmr^a_n}
 \bar H^{a} (t_n, \bmR_n^a, \bmP^a_n )  
%\; = \;
%  \tilde H^a (t_n, \bmR_n^a + s_a  (1-b_a) \delta \bmr^a_n, \bmR_n^a - s_a b_a
%  \delta \bmr^a_n) 
&=&
\left\{ \begin{array}{l}
\tilde H^F (t_n, \bmr^F_n, \bmr^F_{n-1}) \; ,
\\
\tilde H^B (t_n, \bmr^B_{n-1}, \bmr^B_n) \; .
\end{array} \right.
\end{eqnarray}
This equation can be regarded as the defining relation for $\bar
H^a_n$ (and \Eq{eq:defineH^aslicen} as its consequence): $\bar H^a_n$
is the (generally asymmetric) Fourier transform, with respect to the
relative coordinate $\delta \bmr^a_n $, of $\tilde H^a (t_n)$, in
which the position arguments $\bmr^a_n$ and $\bmr^a_{n-1}$ occur in a
time-ordered or anti-time-ordered fashion for $a=F$ or $B$,
respectively (\ie\ the coordinate associated with the later time,
$t_n$, appears to the left or right of the earlier time, $t_{n-1}$,
respectively). This, of course, is required to ensure that the path
integral representation for $\tilde U^F_\iiijjj (t,t')$ and $\tilde
U^B_{ji}(t', t)$ produces time-ordered and anti-time-ordered
expressions, respectively, as illustrated in Fig.~\ref{fig:timeslice}.
The reason for using a factor $s_a$ in the Fourier transform
exponentials $e^{- i s_a \bmr^a \cdot \bmp^a}$ in the definition
(\ref{eq:defineH^aslicen}) of $\bar H^a_n$ and its inverse,
\Eq{eq:HaRPrdr}, is simply that the factor $e^{ i s_a \bmr^a \cdot
  \bmp^a}$ occuring in the latter is generated by the combination $i
s_a \bar L^a_n$ in the action of \Eq{eq:def-path-integral}.  Finally,
note also that $\bar H^a(\bmR^a, \bmP^a)$ is independent of $\bmP^a$
if and only if $\tilde H^a( \bmr^a_i, \bmr^a_\bari)$ is proportional
to $\tilde \delta (\bmr^a_i - \bmr^a_\bari)$.

\subappendix{Verifying the Defining Equations and Composition Rule}
\label{app:verifyingdefeqcomprule} \nopagebreak
It is straightfoward to verify that \Eq{eq:def-path-integral}
satisfies all the requirements expected of a propagator.  We shall now
first show that it fulfills the defining conditions for $\tilde
U^a_{\iiijjj}$, namely \Eqs{eq:tildeUconditions}, and then check
that it satisfies the usual composition rule.
Since the manipulations for $a=F$ and
$a=B$ are very similar, but differ in numerous minor details, we shall
mostly consider the former case only. Hence, $a$ will be understood to
stand for $F$ below, except when explicitely noted otherwise.

\emph{Normalization:} To recover the normalization condition
\Eq{eq:tildeU-bound}, take the limit $t \to t'$ by taking $M=1$ and
$\epsilon \to 0$. Then the entire path integral reduces simply to
\begin{eqnarray}
 \lim_{t \to t'}  \tilde U^a_{\iiijjj} (t, t') =  \delta_{\sigma_i \sigma_j}
 \int \! {d \bm{p}^a_1 \over (2 \pi )^d} \, \,
e^{i \bm{p}^a_1 \cdot (\bm{r}^a_i - \bm{r}^a_j)}  = \tilde \delta_{\iiijjj} \; .
%\delta^{(d)} ( \bmr_i - \bmr_j ) \; . \qqph 
\end{eqnarray}

\emph{Equation of motion:}
To recover  the equation of motions for  $\tilde U^F_\iiijjj$ and
 $\tilde U^B_{ji}$,
namely \Eqs{eq:tildeU-diff-F} and (\ref{eq:tildeU-diff-B}),
add one time slice in \Eq{eq:def-path-integral} $(M \to M+1$,
so that now  $\bmr^a_i = \bmr^a_{M+1}$),
and expand the corresponding exponential
$e^{(i s_a \epsilon / \hbar)L^a_{M+1} }$ to first order in $\epsilon$:
\bsubequations
\label{eq:EOMforU}
 \begin{eqnarray}
\nonumber
\label{eq:EOMforU-a}
  \tilde U^F_{\iiijjj} (t + \epsilon, t') \!\!
 & = & \!\! \sum_{\sigma_M} \delta_{\sigma_i  \sigma_M} \!
\int \! \! d \bmr^F_M \!\! \int \! {d \bmp^F_{M+1} \over (2 \pi)^d} \:
  e^{i
 \bm{p}^F_{M+1} \cdot \delta \bmr^F_{M+1}} \! \left[ 1 -
{i \epsilon \over \hbar} \bar H^F_{M+1} \right]  \! U^{F\pdag}_{M j} (t, t')
\\ \label{eq:EOMforU-b}
& = & U^{F\pdag}_{\iiijjj} (t, t')  - {i \epsilon \over \hbar}
\int \! d x^F_M \,
 \tilde H^F_{iM} (t)\,  \tilde U^F_{Mj} (t, t') \; ,
\\
\nonumber \label{eq:EOMforU-c}
  \tilde U^B_{ji} (t', t + \epsilon) \!\!
 & = &  \!\! \sum_{\sigma_M}  \delta_{\sigma_i  \sigma_M} \!
\int \! \! d \bmr^B_M \!\! \int \! {d \bmp^B_{M+1} \over (2 \pi)^d} \,
U^{B\pdag}_{jM} (t', t)\,   e^{- i
 \bm{p}^B_{M+1} \cdot \delta \bmr^B_{M+1}} \! \left[ 1 +
{i \epsilon \over \hbar} \bar H^B_{M+1} \right]
\\
\label{eq:EOMforU-d}
& = & U^{B\pdag}_{ji} (t', t)  + {i \epsilon \over \hbar}
\int \! d x^B_M \,  \tilde U^B_{jM} (t', t) \,
 \tilde H^B_{Mi} (t)  \; .
\end{eqnarray}
\esubequations
Here \Eqs{eq:HaRPrdr} and  $\bmr^a_{M+1} = \bmr^a_i$ were
used to obtain \Eqs{eq:EOMforU-b} and (\ref{eq:EOMforU-d}), which, in
the limit $\epsilon \to 0$, reproduce \Eqs{eq:tildeU-diff-F} and
(\ref{eq:tildeU-diff-B}).

\emph{Composition rule:} 
\label{app:composition}
Next we check that \Eq{eq:def-path-integral} also satisfies the
 usual composition rules for propagators, namely
\begin{eqnarray}
\nonumber %  \label{eq:composition}
\int d x_1 \, U^{F\pdag}_{i1} (t, t_1)  \, U^{F\pdag}_{1j} (t_1, t')  =
U^{F\pdag}_{\iiijjj} (t, t')    \; ,
\quad
\int d x_1 \, U^{B\pdag}_{j1} (t', t_1)  \, U^{B\pdag}_{1i} (t_1, t)  =
U^{B\pdag}_{ji} (t', t)   . 
\end{eqnarray}
To this end, let $M_1$ be the number of intervals between $t_1$ and $t'$,
\ie\ write $t_1 = t' + \epsilon M_1$ and $\bmr_1 = \bmr_{M_1}$.
Then, by concatenating two  expressions
of the form \Eq{eq:def-path-integral} for
$ U^{F\pdag}_{i1}$ % (t, t_1)$ 
and $  U^{F\pdag}_{1j}$, % (t_1, t')$,
we find that the left-hand side of the above equation %\Eq{eq:composition}
can be written, up to a factor
$\delta_{\sigma_i \sigma_1} \delta_{\sigma_1 \sigma_j}$, as
\begin{eqnarray}
\nonumber 
& & \phantom{.} \hspace{-0.8mm}
\int \! d \bmr^F_1
\lim_{M \to \infty}
\prod_{n=M_1 + 1}^{M-1}
 \Bigl( \int \! d \bmr^F_{n} \Bigr)
\prod_{n=M_1+1}^M \Bigl( \int \! {d \bmp^F_n \over (2 \pi )^d} \Bigr)
e^{(i \epsilon / \hbar ) \sum_{n=M_1+1}^{M} \bar L^F_n }
%\\
%& = & \times
%\phantom{\int \! d \bmx_1 \delta_{\sigma_i \sigma_1} \delta_{\sigma_1 \sigma_j}
%\lim_{M \to \infty} }
\\ \label{eq:concatenate}
& & \times \prod_{n=1}^{M_1-1}
 \Bigl( \int \! d \bmr^F_{n} \Bigr) \!\!
\prod_{n=1}^{M_1} \Bigl( \int \! {d \bmp^F_n \over (2 \pi )^d} \Bigr)
e^{(i \epsilon / \hbar ) \sum_{n=1}^{M_1} \bar L^F_n }
\; . \qquad  \phantom{.}
\end{eqnarray}
This is equal to $\tilde U^F_\iiijjj (t,t')$ as given by
\Eq{eq:def-path-integral}, since
$\int d \bmr^F_1 = \int d \bmr^F_{M_1}$.
The derivation for $\tilde U^B_{ji}$ is entirely analogous.

\subappendix{Power Series Expansion in $\tilde h^a$:}

The power series
expansions of $\tilde U^F_\iiijjj (t,t')$ and $\tilde U^B_{ji}(t',t)$
in powers of $\tilde h^F_V$ and  $\tilde h^B_V$ are given by
\Eq{eq:PIexpandedinfullmaintext}. To illustrate how they come about from the
time-slicing definition (\ref{eq:def-path-integral}) of the path
integral, we begin by considering only the first order terms 
(the higher order terms will be discussed subsequently).
To this end, we expand each factor
$e^{(i s_a \epsilon / \hbar) \bar L^a_{n_1}}$ in
\Eq{eq:def-path-integral} to linear order in $\bar h^a_{V n_1}$, to
obtain $e^{(i s_a \epsilon / \hbar) \bar L^{0a}_{n_1}} + e^{ i s_a
  \bmp^a_{n_1} \cdot \delta \bmr_{n_1}^a} (- s_a i
\epsilon/\hbar) \bar h^a_{Vn_1}$. Here $\bar L^{0a}_{n_1}$ is the
$V$-independent part of $\bar L^a_{n_1}$, and for the second term, all
contributions of order $\epsilon^2$ or higher were dropped (in
particular, we replaced $e^{-(i s_a \epsilon/\hbar)\bar h^a_{0 n_1}}$ by 1).
Then, to leading order in $\epsilon$, \Eq{eq:def-path-integral}
readily yields the following expression:
\bsubequations
  \label{eq:PIexpandUa}
\begin{eqnarray}
\lefteqn{
\left. \begin{array}{l}
\tilde U^F_{\iiijjj} (t,t') - \tilde U^0_\iiijjj (t,t') \rule[-4mm]{0mm}{0mm}\\
\tilde U^B_{ji} (t',t) - \tilde U^0_{ji} (t',t)
\end{array} \right \} \; = \;  
\delta_{\sigma_i \sigma_j} \lim_{M \to \infty}
{(-s_a i \epsilon) \over \hbar}}
\\ \nonumber & & 
\times \sum_{n_1=1}^M \left\{\left[
\prod_{n=n_1+1}^{M-1} \Bigl( \int d \bmr^a_{n} \Bigr)
\prod_{n=n_1+1}^{M} \Bigl( \int {d \bmp^a_n \over (2 \pi )^d} \Bigr)
e^{{ (i s_a \epsilon/  \hbar) } \sum_{n=n_1+1}^{M} \bar L^{0a}_n }\right]
\right.  \qquad \phantom{.}
\\ \nonumber
& & \left. \times \left[
\int \! \! d \bmr^a_{n_1} \int \! \! d \bmr^a_{{n_1}-1}
\int {d \bmp_{n_1}^a \over (2 \pi)^d}
\,  e^{ i s_a \bmp^a_{n_1} \cdot \delta \bmr_{n_1}^a }
\bar h^a_{V{n_1}} \right] \right. 
\\
& & \nonumber
 \left. \times 
\left[ \prod_{n=1}^{n_1-2} \Bigl( \int d \bmr^a_{n} \Bigr)
\prod_{n=1}^{n_1 -1} \Bigl( \int {d \bmp^a_n \over (2 \pi )^d} \Bigr)
e^{{(i s_a \epsilon/  \hbar) } \sum_{n=1}^{n_1 -1} \bar L^{0a}_n }
\right] \right\} + \dots
\qquad \phantom{.}
\end{eqnarray}
%\\ 
\begin{eqnarray} 
\nonumber
& = &   \delta_{\sigma_i \sigma_j}
 \lim_{\epsilon \to 0}  {(- s_a  i \epsilon)
 \over \hbar}
\sum_{{n_1}=1}^M
\int \! \! d \bmr^a_{n_1} \int \! \! d \bmr^a_{{n_1}-1}
\\ & &   \label{eq:PIexpandUa-2} \times 
\left\{ \begin{array}{l}
\tilde U^{0F}_{i{n_1}} (t,t_{n_1}) \, \tilde h^F_{V {n_1} \, {n_1}-1} (t_{n_1})
\, \tilde U^{0F}_{{n_1}-1 \, j} (t_{{n_1}-1},t') \rule[-3mm]{0mm}{0mm}
\\
 \tilde U^{0B}_{j \, {n_1}-1 j} (t',t_{{n_1}-1}) \,
\tilde h^B_{V {n_1 -1 } \, {n_1}} (t_{n_1}) \,
\tilde U^{0B}_{{n_1}\, i } (t_{n_1},t) \,
\end{array} \right.
  + \dots
\\
  \label{eq:PIexpandUa-3}
& = & -   {i s_a \over \hbar } \int_{t'}^t d t_1 \int dx_{1, \bar 1}
\left\{ \begin{array}{l}
 \tilde U^{0F}_{i1} \tilde h^F_{V 1 \bar 1} \tilde U^{0F}_{\bar 1j}
\rule[-3mm]{0mm}{0mm} \\
 \tilde U^{0B}_{j \bar 1 } \tilde h^B_{V \bar 1  1}  \tilde U^{0B}_{1i}
\end{array} \right.
+ \dots\; ,
\end{eqnarray}
\esubequations
in agreement with the $N=1$ terms of \Eq{eq:PIexpandedinfullmaintext}.
For \Eq{eq:PIexpandUa-2}, which is illustrated in
the second row of Fig.~\ref{fig:timeslice}, 
we have evoked \Eq{eq:HaRPrdr}
to make the identification
\begin{eqnarray}
  \label{eq:identifyhV}
\int {d \bmp^a_{n_1} \over (2 \pi )^d}
e^{ i s_a \bmp^a_{n_1} \cdot \delta \bmr_{n_1}^a}
\bar h^a_{V{n_1}} =
\left\{ \begin{array}{l}
\tilde h^F_{V {n_1} \, {n_1}-1} (t_{n_1}) \rule[-4mm]{0mm}{0mm}
\\
\tilde h^B_{V {n_1 -1} \, {n_1}} (t_{n_1})
\end{array} \right.
\; .
\end{eqnarray}
From the above excercise, we extract the following rule of thumb: when
a function $\bar f^a (t_1) \equiv \bar f^a \bigl( t_1, \bmR^a (t_1),
\bmP^a (t_1) \bigr) $ [\eg\ $\bar h_V^a$ above] occurs at time $t_1$
along the forward or backward parts of the Keldysh path integral $\int
{\cal D} \bmR^F {\cal D} \bmR^B $, the $\int d \bmp^a_{n_1} e^{ i s_a
  \bmp^a_{n_1} \cdot \delta \bmr^a_{n_1}} $ momentum integral at that
the corresponding time slice $t_{n_1} = t_1$ converts it into $\tilde
f^F_{ n_1 n_1-1} (t_{n_1})$ or $\tilde f^B_{ n_1 -1 n_1} (t_{n_1})$.
Combining this with the propagators implicit in $e^{i s_a \bar
  S^a_0}$, generates terms of the form $\tilde U^{0F}_{i1} \tilde
f^F_{1 \bar 1} \tilde U^{0F}_{\bar 1 j}$ or $\tilde U^{0B}_{j \bar 1}
\tilde f^B_{\bar 1 1} \tilde U^{0B}_{1 i}$, respectively [where $\bar
f^a$ and $\tilde f^a$ are Fourier transform pairs,
in analogy to $\bar H^a$ and $\tilde H^a$ of 
\Eqs{eq:defineH^aslicen} and (\ref{eq:HaRPrdr})].  To be explicit, we
have
\begin{eqnarray}
  \label{eq:ruleofthumbPI}
 \delta_{\sigma_i \sigma_j} \int_{\bmR^a (t_j)  = \bmr_j}^{\bmR^a (t_i) =
\bmr_i} \! \! {\cal D} \bmR^a
\! \int  \! \! {\cal D} \bmP^a  \,
e^{i s_a \bar S^a_0 (t,t')} \, \bar f^a (t_1)
& = &
\int d x_{1, \bar 1}
\left\{ \begin{array}{l}
 \tilde U^0_{i1} \tilde f^F_{ 1 \bar 1} \tilde U^0_{\bar 1j}
\rule[-4mm]{0mm}{0mm} \\
\tilde U^0_{j \bar 1 } \tilde f^B_{ \bar 1  1}  \tilde U^0_{1i}
\end{array} \right. \; . \qqph \qph 
\end{eqnarray}
Having found the rule (\ref{eq:ruleofthumbPI}), it is straightforward
to go beyond the first order and to recover the full perturbation
expansion from the path integral:
\bsubequations
  \label{eq:PIexpandedinfull-all}
\begin{eqnarray}
  \label{eq:PIexpandedinfull-1}
\lefteqn{\phantom{.} \hspace{-0.7cm}
\left. \begin{array}{l}
\rule[-4mm]{0mm}{0mm}
   \tilde U^F_{\iiijjj}(t_i,t_j) \\
   \tilde U^B_{ji}(t_j,t_i)
\end{array} \right\}
 \,  = \;   \delta_{\sigma_i \sigma_j} \int_{\bmr_j}^{
\bmr_i} \! \! {\cal D} \bmR^a(t_1)
\! \int  \! \! {\cal D} \bmP^a (t_1) \,
e^{\pm \left[ i \bar S^a_0 (t,t') - \int_{t'}^t d t_1 \bar h^a_V (t_1)
 \right]}
}
\\ \nonumber
\phantom{.} \hspace{-2cm}
& =&
 \delta_{\sigma_i \sigma_j} \int_{\bmr_j}^{\bmr_i} \! \! {\cal D} \bmR^a(t_1)
\! \int  \! \! {\cal D} \bmP^a (t_1) \,
e^{\pm  i \bar S^a_0 (t,t')}
\\
& & \nonumber 
\times \sum_{N=0}^\infty {(\mp i)^N \over \hbar^N}
\int_{t_j}^{t_i} \!\! d t_1
\int_{t_j}^{t_1} \! \!d t_2
 \dots
\int_{t_j}^{t_{N-1}} \! \! d t_N
 \bar h^a_V (t_1)  \bar h^a_V (t_2) \dots  \bar h^a_V (t_N)
\\ & =&
\nonumber 
  \sum_{N=0}^\infty % { 1 \over \hbar^N}
\int_{t_j}^{t_i} \!\! d t_1
\int_{t_j}^{t_1} \! \!d t_2
 \dots
\int_{t_j}^{t_{N-1}} \! \! d t_N
\int \! dx_{1,\bar1} dx_{2,\bar 2} \dots dx_{N, \bar N} \,
\\ & & \label{eq:PIexpandedinfull} % \nonumber 
\times \left\{ \begin{array}{l} (-i/ \hbar)^N
\tilde U^{0F}_{i1} \tilde h^F_{V 1 \bar 1}  \tilde U^{0F}_{\bar 12}
\dots \tilde h^F_{V N \bar N}  \tilde U^{0F}_{\bar Nj} \rule[-4mm]{0mm}{0mm}
\\
(+i/ \hbar)^N
\tilde U^{0B}_{j \bar N} \tilde h^B_{V \bar N N}
\dots \tilde U^{0B}_{2 \bar 1} \tilde h^B_{V \bar 1 1} \tilde U^{0B}_{1i}
\end{array} \right. \; . \qquad \phantom{.}
\\ \label{eq:PIexpandedinfullGRGA}
& = &   \sum_{N=0}^\infty % { 1 \over \hbar^N}
\int_{t_j}^{t_i} \!\! d t_1  \dots d t_N
\int \! dx_{1,\bar1} \dots dx_{N, \bar N} \,
 \left\{ \begin{array}{l} 
\tilde G^R_{i1} \tilde h^F_{V 1 \bar 1}  \tilde G^R_{\bar 12}
\dots \tilde h^F_{V N \bar N}  \tilde G^R_{\bar Nj} \rule[-4mm]{0mm}{0mm}
\\
\tilde G^A_{j \bar N} \tilde h^B_{V \bar N N}
\dots \tilde G^A_{2 \bar 1} \tilde h^B_{V \bar 1 1} \tilde G^A_{1i}
\end{array} \right. .  \qqph \qph
\end{eqnarray}
 \esubequations
 \Eq{eq:PIexpandedinfull}, which is illustrated in the third row of
 Fig.\ref{fig:timeslice}, was obtained from the line preceding it by
 multiple applications of the rule of thumb (\ref{eq:ruleofthumbPI}),
 and reproduces the expansions of \Eqs{eq:PIexpandedinfullmaintext}.
 For \Eq{eq:PIexpandedinfullGRGA} we recalled \Eq{eq:exact-U0-RA} to
 set $\tilde U^{0F/B} = \pm i \, \hbar \, \tilde G^{R/A}$ along the
 forward or backward contours, respectively.

\subappendix{Coordinate-Space-Only Path Integral}
\label{sec:expandU}

Since the power series expansions (\ref{eq:PIexpandedinfull}) for
$\tilde U^a_\iiijjj$ do not contain any explicit momentum integrals, they
may be used as starting points for deriving 
coordinates-only path integral expressions
containing no $\int {\cal D} \bmP^a$ integrations at all, 
so that only the coordinate integrations $\int {\cal
  D} \bmR^a$ remain. To this end, we simply perform the $\int {\cal
  D}\bmP^a$ integrals in the definition of the free propagators
$\tilde U^{0a}_\iiijjj$ explicitly, with the well-known result:
\bsubequations
  \label{eq:def-Ronly-PI}
\begin{eqnarray}
\nonumber %\label{eq:def-path-integral-0}
\lefteqn{
\left. \begin{array}{r}
 \tilde U^{0F}_{\iiijjj} (t,t') \\
 \tilde U^{0B}_{ji} (t',t)
\end{array} \right\}
  \equiv    \delta_{\sigma_i \sigma_j}
\lim_{M \to \infty} \prod_{n=1}^{M-1} \Bigl( \int d \bmr^a_{n} \Bigr)
\prod_{n=1}^M \Bigl( \int {d \bmp^a_n \over (2 \pi )^d} \Bigr)}
\\
& & \times \exp \left[   {i s_a \epsilon \over \hbar}  \sum_{n=1}^M
\left(\hbar \bm{p}^a_n \cdot {\delta \bm{r}^a_n
 \over \epsilon} -  {\hbar^2 \bmp^{a 2} \over 2 m} - V_\imp (\bmR^a_n)
\right) \right]  \qquad \phantom{.}
\\ \nonumber
 & = & \!\!
\delta_{\sigma_i \sigma_j}
\left[{ m \over 2 \pi i s_a \epsilon \hbar  }
\right]^{Md/2} \prod^{M-1}_{n=1}
\int d \bmr^a_n \exp
 \left[   {i s_a \epsilon \over \hbar}  \sum_{n=1}^M
\left( {m \over 2} \left[\delta \bmr^a_n \over \epsilon \right]^2  
-  V_\imp (\bmR^a_n) \right)  \right] \qph
\\ \label{eq:def-path-integral-0-measure}
& \equiv & \!\!
\int_{\bmR^a (t')  = \bmr_j}^{\bmR^a (t) =
\bmr_i} \widetilde {\cal D} \bmR^a e^{(i s_a / \hbar)
\tilde S^a_0 (t,t') } \, , 
\\
\label{eq:standardfreeactionRonly}
& & \tilde S^a_0 (t,t')[ \bmR^a (t_3)]  \equiv    \int_{t'}^{t}   dt_3
%\tilde L^a_0 \bigl(t_3, \bmR^a (t_3) \bigr) & \equiv &
\left[\toh m \dot \bmR^{a2} (t_3) - V_\imp \bigl( \bmR^a (t_3)
  \bigr)\right]  \; .
\end{eqnarray}
\esubequations
Here $ \tilde S^a_0$ is the standard action for a noninteracting
electron in a disorder potential, and the tilde indicates that [in
contrast to $\bar H^a$ of \Eq{eq:defineH^aslicen}] it is a functional
of $\bmR^a (t_3)$ only, not of $\bmP^a (t_3)$ too.  The tilde on $\int
\widetilde {\cal D} \bmR$ in \Eq{eq:def-path-integral-0-measure}
indicates that the measure includes the prefactor in the line above
it.  Now, if we take the power series expansion
(\ref{eq:PIexpandedinfull}) for $\tilde U^{a}_\iiijjj$ and insert
\Eq{eq:def-path-integral-0-measure} for each occurrence of $\tilde
U^{0a}_\iiijjj$, we obtain for $\tilde U^{a}_\iiijjj$ a coordinate-only path
integral expression with a precise (though cumbersome) time-slicing
definition.  In the main text, we have used for the path integral so
obtained the formal path integral notation
(\ref{eq:def-path-integral-pureRmaintext}), with actions defined by
\Eqs{eq:standardfreeactionRonlymaintext} and
(\ref{eq:shorthandtildehfunctionR}), and measure $\int \widetilde
{\cal D}' \bmR$, where the prime reminds us of the double position
integrals $\int dx_{i, \bari}$ occuring in \Eq{eq:PIexpandedinfull}.
The points discussed after \Eq{eq:shorthandtildehfunctionR} in the
main text all follow directly from the explicit construction given
above.

\subappendix{Explicit Expressions for $\bar H^a_n$}
\label{sec:b=0results}

The material presented up to now in this appendix was general,
applicable to \emph{any} nonlocal Hamiltonian of the form $\tilde
H^a_{\iiijjj} = \tilde \delta_\iiijjj h_{0j} + \tilde h^a_{V \iiijjj}$.  Let us now
be more concrete and specialize to the Hamiltonian defined by
\Eqs{eq:GZlinearresponseall}, in order to verify GZ's expression for
the effective action derived for their $\int {\cal D} \bmR \int {\cal
  D} \bmP$ path integral.

Inserting \Eqs{eq:GZlinearresponseall}
into \Eq{eq:defineHn}, we readily find that
\bsubequations
 \label{eq:barHprn}
\begin{eqnarray}
 \label{eq:barH3termspr}
      \bar H^a_n
  &= &
\bar h_0 (\bmR_n^a, \bmP_n^a) \, + \,
\bar h^a_V (t_n, t_0; \bmR_n^a, \bmP_n^a)  \; = \;
 \bar h_{0n} + \bar h^a_{Vn} \; ,
\\
 \label{eq:barwpr-app}
\bar h^a_{Vn} &=&
\sum_{\alpha = \pm} \bar w^{a \alpha } (t_n, t_0; \bmR_n^{a}, \bmP_n^a) 
\,   V_\alpha (t_n, \bmR_n^a) \; ,
\\
  \label{eq:define-bar-h0}
\bar h_0 (\bmR^a, \bmP^a) & \equiv & 
{ \bmP^{a2} \over 2m} +
V_{\rm imp}(\bmR^a) - \mu \;  ,
\\
  \label{eq:definewoperator}
 \bar  w^{a \pm} \bigl(t, t_0; \bmR^a, \bmP^a ) & \equiv & \!\! 
\left\{\!\!  \begin{array}{l} e \\
e s_a \toh e^{ - i (b_a - \delta_{a B})  \bnabla_{\bmp^a}
  \cdot \bnabla_{\bmR^a} }
[1 - 2 \bar \rho^a \bigl( t,t_0;
{\bmR^{\prime a}}, \bmP^a \bigl) ]\Bigr. 
\end{array} \right. \!\! ,
\\
 \label{eq:define-bar-rho}
\bar \rho^a (t,t_0; \bmR^a, \bmP^a) & =  & \!\! \! \int \!\! d \bmr^a  e^{- i
s_a  \bmp^a
\cdot \bmr^a}   \tilde \rho^\ns (\bmR^a \! + \! s_a (1-b_a) \bmr^a,
\bmR^a \! - \! s_a b_a \bmr^a)  . \qqph \qph %
% \; = \; \bar \rho^\ast (t,t_0; \bmR^a, \bmP^a) . \qquad \phantom{.}
\end{eqnarray}
\esubequations
\noindent
Here $\bar h_0 (\bmR^a, \bmP^a)$ and $\bar \rho^a (t,t_0;\bmR^a,
\bmP^a)$ are, respectively, the free Hamiltonian and the
single-particle density matrix (in the presence of interactions but
without source terms) in the mixed representation.  In the definition
(\ref{eq:definewoperator}) of $\bar w^{a-}$, it is to be understood
that $\bmR^{\prime a}$ should be equated to $\bmR^a$ after evaluating
the action of the exponential differential operator on the
function $ V_-(t_n, \bmR_n^a)$ to the right of $\bar w^{a-} (t, t_0;
{\bmR^a}, \bmP^a)$ in \Eq{eq:barwpr-app}, and all equations derived
therefrom.

For general choices of $b_a$, the shift operator $e^{- i (b_a -
  \delta_{a B}) \bnabla_{\bmp^a_n} \cdot \bnabla_{\bmR^a_n} } $ in
Eq.~(\ref{eq:definewoperator}) 
%for $ \bar w^{a-} \bigl(t, t_0; \bmR^a, \bmP^a )$ 
is needed for the following reason: In the defining
Eqs.~(\ref{eq:tildeHa}) for $\tilde H^F_{i \bari}$ and $\tilde
H^B_{\bari i}$, the arguments of the field $V_{- \bar i}$ are
evaluated at $\bmr^F_\bari$ and $\bmr^B_\bari$, respectively.  When
considering the $n$-th interval (for which $\bmr_i^a = \bmr_n^a$,
$\bmr_\bari^a = \bmr_{n-1}^a$), these arguments of $V_{- \bari }$
become $\bmr^F_\bari = \bmr_{n-1}^F = \bmR^F_N - b_F \delta \bmr^F_n$
and $\bmr^B_\bari = \bmr^B_{n-1} = \bmR^B_n - (1- b_B) \delta
\bmr_n^B$ [cf.\ \Eq{eq:deltarR}], which are evidently shifted relative
to the argument at which the field $V_- (t_n , \bmR^a_n)$ is evaluated
in \Eq{eq:barwpr-app}, namely $\bmR^a_n$, by an amount
%$ - s_a b_a \delta \bmr^a_n$.  
$ - s_a (b_a - \delta_{aB})\delta \bmr^a_n$.  The exponential shift
operator implements this shift [as can be verified by inserting
\Eqs{eq:barHprn} into \Eq{eq:HaRPrdr} to recover $H^F_{i \bari }$ and
$H^B_{\bari i}$].  Evidently, though, one \emph{can} achieve $\bmR^F_n =
\bmr_{n-1}^F (= \bmr^F_j)$ and $\bmR^B_n = \bmr_n^B (= \bmr^B_i)$ and
hence avoid the need for shifts, by making the special, ``maximally
asymmetric'' choice $b_a = \delta_{aB}$.  Indeed, for this choice,
which we shall adopt henceforth, the exponential shift operators $e^{-
  i (b_a - \delta_{a B}) \bnabla_{\bmp} \cdot \bnabla_{\bmR} } $
reduce to unity.  Moreover, since $\bmR^a_n$ then depends on only one
of the position coordinates $\bmr^a_n$ and $\bmr^a_{n-1}$ associated
with the $n$-th time interval, namely the second, 
%namely
%\begin{eqnarray}
%  \label{eq:Rnmaxasym}
%  \bmR^F_n = \bmr^F_{n-1}, \qquad
%  \bmR^B_n = \bmr^B_n,
%\end{eqnarray}
so does $\bar H^a_{n} = \bar H^a (t_n, \bmr^a_{n-1}, \bmP^a_n)$, 
%\begin{eqnarray}  \label{eq:alternativeHaslice}
%\bar H^F_n & = & \bar H^F(t_n, \bmr^F_{n-1}, \bmP^F_n)
%\; , \qquad \bar H^B_n = \bar H^B(t_n, \bmr^B_{n},
%\bmP^B_n) \; ,
%\end{eqnarray}
which greatly simplifies subsequent manipulations.  The ``price''
to be paid for this simplification is not high -- one merely has to
remember that the definitions (\ref{eq:defineH^aslicen}) of $\bar
H^a_n$ in terms of the Fourier transforms of $\tilde H^a$ and $\tilde
\rho^a $ with respect to the relative coordinate
 become fully asymmetric:
\bsubequations
\label{eq:barHpr-APP}
\begin{eqnarray}
      \bar H^F_n
 & \! \!  \! \equiv & \!\! \! \! \int \! d (\delta \bmr_n^F) 
e^{- i \bmp^F_n \cdot \delta \bmr^F_n}
     \tilde H^F ( t_n, \bmr^F_{n-1} + \delta \bmr^F_n, \bmr^F_{n-1} )
\, = \bar h_0 (\bmr^F_{n-1}, \bmp^F_n) \, + \, \bar h^F_{Vn} \; ,
\qqph \qph %\nonumber 
\\
      \bar H^B_n
 & \!\! \equiv & \!\! \! \! \int \! d (\delta \bmr_n^B) 
e^{ {i}  \bmp_n^B \cdot \delta \bmr^B_n}
     \tilde H^B ( t_n, \bmr^B_{n-1}, \bmr^B_{n-1}  - \delta \bmr^B_n)
\, = \,  \bar h_0 (\bmr^B_{n-1}, \bmp^B_n) \, + \, \bar h^B_{Vn} ,
 \qqph \qph   \end{eqnarray}
\esubequations
% \begin{eqnarray}
%\label{eq:hVbarexplicit}
%\label{eq:hVFbarexplicit}
%\bar h^F_{Vn}
%& = & \sum_{\alpha = \pm} \bar w^{F \alpha}_n \, 
% V_\alpha (t_n,\bmr^F_{n-1})
%\; \qquad \bar h^B_{Vn} =
% \sum_{\alpha = \pm}  V_\alpha (t_n,\bmr^B_{n-1}) \, 
%\bar w^{B \alpha}_n
%\; ,
%\qquad \phantom{.}
%\end{eqnarray}
where $\bar h^a_{Vn} = \sum_{\alpha = \pm} \bar w^{a \alpha}_n \, 
 V_\alpha (t_n,\bmr^a_{n-1})$, with  $\bar w^{a+}_n = e$ and
$\bar w^{a-}_n = e \toh s_a [1 - 2 \bar \rho^a_n]$,
and $\bar \rho^a_n$ is defined in terms of $\tilde \rho^\ns_\iiijjj (t_n, t_0)$
by Fourier transform relations [\Eq{eq:define-bar-rho}] 
that are analogous to those [\Eqs{eq:barHpr-APP}]
for $\bar H^a_n$ in terms of $\tilde H^a_\iiijjj (t_n)$. 
%\bsubequations
%\label{eq:define-bar-rho-app}
%\begin{eqnarray}
%\bar \rho^F_n & \!\! \equiv \!\! & \!\!
%\bar \rho^F (t_n, t_0;\bmr^F_{n-1}, \bmp^F_n)
%\equiv \int  \!\! d (\delta \bmr^F_n) \, \,
%e^{- i \bmp^F_n \cdot \delta \bmr^F_n}
%     \tilde \rho   (t_n,t_0;  \bmr^F_{n-1} + \delta \bmr^F_n,
%     \bmr^F_{n-1}) , \qqph \qph 
%\\
%\bar \rho^B_n & \!\! \equiv \!\! & \!\! 
%\bar \rho^B (t_n, t_0; \bmr^B_{n-1}, \bmp^B_n) \equiv \int \!\! d (\delta
%\bmr^B_n)  \, \,
%e^{i \bmp^B_n \cdot \delta \bmr^B_n}
%     \tilde \rho (t_n,t_0; \bmr^B_{n-1}, \bmr^B_{n-1}  - \delta \bmr^B_n) .
%     \qquad \phantom{.}
%\end{eqnarray}
%\esubequations 

\subappendix{GZ's Effective Action in Position-Momentum Representation}
\label{sec:SRSI-RP}

Having specified the time-sliced versions of $\bar H^a$ in the
position-momentum representation, it is straightforward to also derive
the effective action $(i \bar S_R + \bar S_I)[\bmR^a, \bmP^a]$ for this
representation: simply repeat the strategy followed in App.~B.5.5 to
5.8, but use the position-momentum representation (denoted by bars
instead of tildes) throughout.  Since the details are very analogous,
we shall be very brief, and indicate only the main differences.

The starting point is again \Eq{eq:JrhoRonly} for 
$\langle \tilde {   J}^V_{12',2 \bar 2', {\bar 2}1'} (t_1,t_2; t_0)
\rangle_{V, {\rm ns}}^\pdag$, but with the coordinates-only path
integral measure
${\displaystyle \Fint \Bint \widetilde {\cal D}' \! (\bmR )}$
replaced by a position-momentum
path integral measure,
${\displaystyle \Fint \Bint {\cal D}(\bmR \bmP)}$, which is
a shorthand for
\begin{eqnarray}
\Fint_{j_F}^{i_F}
\Bint_{\barj_B}^{\bari_B}
{\cal D} (\bmR \bmP) \dots \! \! 
&  \equiv & \!\! \phantom{\times}
 \int_{\bmR^F (t_j)  = \bmr_{j}}^{\bmR^F (t_i) =
\bm{r}_i}  \!\!\! {\cal D} \bmR^F (t_3)  \int {\cal D} \bmP^F (t_3)
e^{i \bar S_0^F  (t_i,t_j)/ \hbar} \\
& & \!\! \times 
\int_{\bmR^B (t_{j})  = \bmr_{\barj}}^{\bmR^B (t_{i}) =
\bm{r}_{\bari}} \!\!\! {\cal D} \bmR^B(t_3)  \int {\cal D} \bmP^B (t_3)
e^{-i \bar S_0^B (t_i,t_j)/ \hbar} \dots \; .
\qquad \phantom{.} \qph 
\end{eqnarray}
$\bar S_0^a [\bmR^a (t_3), \bmP^a(t_3)]$ in the weighting factor is
the action for a single, free electron,
\begin{eqnarray}
  \label{eq:defineS0-electrons}
  \bar S_0^a (t_i,t_j) = \int_{t_j}^{t_i} d t_3
\left[ \bmP^a (t_3) \cdot \partial_{t_3} \bmR^a (t_3) -
\bar h_0 \bigl(\bmR^a (t_3), \bmP^a (t_3) \bigr)
\right] \; , \qqph 
\end{eqnarray}
and the bar on $\bar S_0$ (and $\overline {\cal B}$, $\bar S_{R/I}$
below) indicates that [in contrast to $\tilde S^a_0$, $\tilde {\cal
  B}$, $\tilde S_{R/I}$ of App.~B] they are  functionals of $\bmR^a
(t_3)$ \emph{and} $\bmP^a (t_3)$, not of $\bmR^a (t_3)$ only. 
In \Eqs{eq:tildeFtildeB-define}, $\tilde {\cal B}_{\alpha
  3}$ is replaced by 
\bsubequations
\label{eq:defbarB-Ronly} 
\begin{eqnarray}
\overline {\cal B}_\alpha (t_3, \bmr_3) & \!  \equiv \! & 
\sum_a  s_a \bW^{a \alpha}_{3_a} \delta
\bigl(\bmr_3 - \bmR^a (t_3) \bigr)  \; , \qqph 
\bW^{a +}_{3_a} \; = \;  e \; , 
\label{eq:barWvertices}
%\bW^{a \alpha}_{3}  \; \equiv  \; 
%\theta_{32}  \, \bar w^{a \alpha}_{3 }  +
%\theta_{23} \, \bar w'^{a \alpha}_{3 } \; , \qquad \qph
\\
\bW^{a -}_{3_a} & = & e \toh s_a \bigl[1 - 
(\theta_{32} + y^a \theta_{23}) 
\bar \rho^a (t_{3}, t_0; \bmR^a (t_3), \bmP^a (t_3)\bigr) \bigr] \; . 
\end{eqnarray}
\esubequations
Now use precisely the same set of approximations and arguments as in
App.~B.5.6 to B.5.8 to derive the effective action $i \bar S_R + \bar
S_I$. One readily arrives at an equation just like
(\ref{eq:defineSiR}), but with $(i \tilde L^R / \tilde L^I)$ of
\Eqs{eq:Ltildes} replaced by\footnote{The $\theta_{34} (i \tilde R/
  \tilde I)_{34}$ occuring in \Eqs{eq:Ltildes} was written as $\toh (2
  i \tilde R/ \tilde I)_{34}$ here, exploiting the symmetry $\tilde
  I_{34} = \tilde I_{43}$.}
\begin{eqnarray}
  \label{eq:barLfinal}
  (i \bar L^R/ \bar L^I)_{3_a 4_{a'}}  &  = &
\toh  s_a s_{a'} \bW^{a +}_{3_a} \bW^{a'\mp}_{4_{a'}} (2 \tilde R / \tilde
  I)_{3_a 4_{a'}}
\; ,     \label{eq:barLwaa}
\end{eqnarray}
where the density matrix occuring in $\bW^{a -}$ now is the free one, 
$\bar \rho^a_0$.  Multiplying out the terms in \Eq{eq:barLwaa}
explicitly (and setting $(\theta_{4_{a'} 2} + y^{a'} \theta_{2
  4_{a'}}) = 1$ for reasons explained in footnote~\ref{f:thetay=1}  
on page~\pageref{p:thetay=1}), we find
\bsubequations
  \label{eq:multiplyoutSRSI}
  \begin{eqnarray}
\label{eq:multiplyoutSR}
\bar   S_R  (t_1,t_0) & = & {e^2 \over 2} 
\int_{t_0}^{t_1} d t_3 \int_{t_0}^{t_1} d t_4
\Biggl\{
%\\
%& & \nonumber
%%\phantom{.} \hspace{-1cm} 
%\qquad \phantom{+} \Bigl[1 - 2  \bar \rho^{F}_0 
%\bigl(\bmR^F (t_4), \bmP^F (t_4)\bigr) \Bigr]  
% \sum_a s_a \tilde R [t_{34}, \bmR^a(t_3) - \bmR^F (t_4)]
%\\
%& & \nonumber
%%\phantom{.} \hspace{-1cm} 
%\qquad   + \Bigl[1 - 2  \bar \rho^{B}_0 \bigl(\bmR^B (t_4), \bmP^B
% (t_4)\bigr) \Bigr]
%\sum_a s_a  \tilde R [t_{34}, \bmR^a(t_3) - \bmR^B (t_4)]
%\Biggr\} \; , 
%\Bigl[1 - 2  \bar \rho^{F}_0 
%\bigl(\bmR^F (t_4), \bmP^F (t_4)\bigr) \Bigr]  
%\\ \nonumber & & 
%\times 
% \Bigl( \tilde R [t_{34}, \bmR^F(t_3) - \bmR^F (t_4)]
%- \tilde R [t_{34}, \bmR^B(t_3) - \bmR^F (t_4)] \Bigl)
\Bigr. \\
& &  \nonumber 
\phantom{.} \hspace{-2.4cm} \Bigl[1 - 2  \bar \rho^{F}_0 
\bigl(\bmR^F (t_4), \bmP^F (t_4)\bigr) \Bigr]  
 \Bigl( \tilde R [t_{34}, \bmR^F(t_3) - \bmR^F (t_4)]
- \tilde R [t_{34}, \bmR^B(t_3) - \bmR^F (t_4)] \Bigl) + 
\\
& & \nonumber
\phantom{.} \hspace{-2.4cm}
   \Bigl[1 - 2  \bar \rho^{B}_0 \bigl(\bmR^B (t_4), \bmP^B
 (t_4)\bigr) \Bigr]
\Bigl(  \tilde R [t_{34}, \bmR^F(t_3) - \bmR^B (t_4)]
- \tilde R [t_{34}, \bmR^B(t_3) - \bmR^B (t_4)] \Bigr) \Biggr\} \; , 
%\\
%& & \nonumber 
% \phantom{{e^2 \over 2} }
%\Bigl. 
% +  \Bigl[1 - 2  \bar \rho^{B}_0 \bigl(\bmR^B (t_4), \bmP^B
% (t_4)\bigr) \Bigr]
%\\ & & \nonumber \times 
%\Bigl(  \tilde R [t_{34}, \bmR^F(t_3) - \bmR^B (t_4)]
%- \tilde R [t_{34}, \bmR^B(t_3) - \bmR^B (t_4)] \Bigr) \Biggr\}  , 
\\
\nonumber
 \bar  S_I  (t_1,t_0) & = & {e^2 \over 2}
\int_{t_0}^{t_1} d t_3 \int_{t_0}^{t_1} d t_4
\Bigl\{ \tilde I [t_{34}, \bmR^F(t_3) - \bmR^F (t_4)]
- \tilde I [t_{34}, \bmR^B(t_3) - \bmR^F (t_4)] \Bigr. \\
& &     \label{eq:multiplyoutSI}
\phantom{{e^2 \over 2} }
\Bigl. - \tilde I [t_{34}, \bmR^F(t_3) - \bmR^B (t_4)]
+ \tilde I [t_{34}, \bmR^B(t_3) - \bmR^B (t_4)] \Bigr\} \; . 
\qquad \phantom{.}
  \end{eqnarray}
\esubequations
This reproduces GZ's expressions for the effective action, since
Eqs.~(\ref{eq:multiplyoutSRSI}) are the analogues of (GZ-II.54) and
(GZ-II.55) [our 1st, 2nd, 3rd and 4th terms, having $aa' = FF$, $BF$,
$FB$, $BB$, correspond to GZ's 1st, 4th, 3rd and 2nd terms,
respectively]. The only difference is that in their Pauli factor, GZ
have evidently replaced our $ \bar \rho^{a}_0 \bigl(\bmR^a (t), \bmP^a
(t)\bigr)$ by $ n \bigl(\bmR^a (t), \bmP^a (t)\bigr)$, which they
define as the Fermi function $n (\bar h_0)$, evaluated at energy
$\bar h_0 (\bmR^a (t), \bmP^a (t)\bigr)$.

GZ offered no justification for the latter replacement in 
GZ99\cite{GZ2}, but have defended it
in subsequent papers\cite{GZS} by arguing that it amounts to a
quasiclassical approximation that neglects terms of order $\hbar$.  We
have argued in a previous publication\cite{vonDelftJapan02} 
that the ``small parameter''
that would protects this approximation is actually $\tauel \hbar / T$,
which evidently is \emph{not} small in the $T \to 0$ limit of present
interest.  Much more alarming, though, is that when averaging over all
self-returning random walk paths, GZ proceed to make the assumption
that ``$n_0$ depends only on the energy \emph{and not on time} (our
emphasis), because the energy is conserved along the clasical path''
[see discussion after Eq.~(GZ-II.68)]. As argued in Sec.~4 of the main
text, however, this neglects recoil, and produces incorrect results. A
more accurate way of treating the Pauli factor, that properly includes
recoil, is discussed in Sec.~3 of the main text.

%\appeqn
\appendix{Diagrammatic Keldysh Approach}
\label{app:Keldysh}

In order to facilitate comparison between GZ's notation and our's,
this appendix collects some standard definitions (following Rammer and
Smith\footnote{\label{RS86}
J. Rammer and H. Smith, Rev.\ Mod.\ Phys.\ {\bf 58}, 323 (1986).
}) and results for electron and field correlators used
in the Keldysh approach. [Where relevant, GZ's notation is given in
brackets.]  Below, subscripts $i$ are abbreviations for $(t_i,x_i)$
when used for fermion fields or for $(t_i,\bmr_i)$ when used for
interaction fields.  $\tilde G_{\iiijjj}$ is a shorthand for $ \tilde
G_{\iiijjj} (t_{\iiijjj}) \equiv G (t_{\iiijjj}; x_i, x_j)$ [and similarly for
$\LL_{\iiijjj}$], i.e.\ the time-argument, when not displayed explicitly,
will be understood to be $t_{\iiijjj} = t_i - t_j$.  As elsewhere, the
tilde signifies the matrix structure in coordinate space, while bold
symbols are used for matrices in Keldysh space, e.g.\ $\bG_{\iiijjj}$.

\subappendix{Electron Correlators}

\label{app:spGreensfunctions}
\label{app:gen-def}
\label{app:def-GF}

We  begin with the electronic Green's functions $\G_{\iiijjj}$,
and consider for the moment only those
for free, noninterating electrons (i.e.\ evaluated
for $V_a = 0$):  the basic correlators
\bsubequations
\label{eq:defineG><}
\begin{eqnarray}
\label{eq:defineG<-E}
  \G_{\iiijjj}^<  & \equiv  & \phantom{-} {i\over \hbar}
 \langle \hat \psi_I^\dag
(t_j,x_j)  \hat \psi^\pdag_I (t_i,x_i) \rangle_0 \quad  \; \; [=
G_{12}^\GZ (\iiijjj)]  , \quad \phantom{.}
\\
\label{eq:defineG>-E}
   \G_{\iiijjj}^> & \equiv &
-{i\over \hbar}
\langle \hat \psi^\pdag_I (t_i,x_i) \hat \psi_I^\dag (t_j,x_j) \rangle_0
\, \quad  \, \; [= G_{21}^\GZ (\iiijjj)] , \quad \phantom{.}
\end{eqnarray}
\esubequations\noindent
are used as follows to construct the
time-ordered, anti-time-ordered, retarded, advanced,
Keldysh and contour-ordered Green's functions,
respectively:
\bsubequations
\label{eq:defineelectrongreensfunctions}
\begin{eqnarray}
  \G^T_{\iiijjj} & \equiv & \phantom{-}
\theta (t_{\iiijjj})  \G^>_{\iiijjj} + \theta (t_{ji})  \G^<_{\iiijjj}
  \quad [= G_{11}^\GZ (\iiijjj)] , \quad \phantom{.}
%& = & \theta_{\iiijjj}  \G^>_{\iiijjj} + \theta_{ji}  \G^<_{\iiijjj}
%\qquad \qquad \qquad [= \G_{11}^\GZ (\iiijjj)] \; ,
\\
 \G^{\overline T}_{\iiijjj} & \equiv & \phantom{-}
\theta (t_{ji})  \G^>_{\iiijjj} + \theta (t_{\iiijjj})  \G^<_{\iiijjj}
%\theta_{ji}  \G^>_{\iiijjj} + \theta_{\iiijjj}  \G^<_{\iiijjj}
 \quad [= G_{22}^\GZ (\iiijjj)] , \quad \phantom{.}
\\
  \label{eq:GRGA-1}
  \G^R_{\iiijjj} & \equiv & \phantom{-} \theta (t_{\iiijjj})
 ( \G^>_{\iiijjj} - \G^<_{\iiijjj} ) \; \,
\qquad [= \G^{R, \GZ} (\iiijjj)] , \quad \phantom{.}
\\
  \G^A_{\iiijjj} & \equiv & - \theta (t_{ji}) ( \G^>_{\iiijjj} - \G^<_{\iiijjj} ) \; \,
 \qquad [= G^{A, \GZ} (\iiijjj)] , \quad \phantom{.}
\\
  \G^T_{\iiijjj} & =  &  \G^<_{\iiijjj} + \G^R_{\iiijjj} = \G^>_{\iiijjj} + \G^A_{\iiijjj} \; ,
\\
  \G^{\overline T}_{\iiijjj} & =  &
 \G^<_{\iiijjj} - \G^A_{\iiijjj} = \G^>_{\iiijjj} - \G^R_{\iiijjj} \; ,
\\
\label{eq:GKeldyshdef}
  \G^{K}_{\iiijjj} & \equiv  &  \G^>_{\iiijjj} + \G^<_{\iiijjj}
= \G^R_{\iiijjj} -  \G^A_{\iiijjj} + 2 \G^<_{\iiijjj}  \; ,
\\
  \label{eq:defineU0}
  \tilde U^{0}_{\iiijjj} & \equiv  &  i (\G^R_\iiijjj - \G^A_\iiijjj )
= i(  \G^>_{\iiijjj} - \G^<_{\iiijjj})  \; ,
  \\
  \label{eq:contour-G}
  \G^{c}_{\iiijjj} & \equiv &
\label{eq:contour-G<>}
\left\{ \begin{array}{cc}
\G^>_{\iiijjj} & \quad \mbox{for} \quad t_i \, >_c \, t_j \; ,
\\
\G^<_{\iiijjj}
& \quad \mbox{for} \quad t_i \, <_c \, t_j \; ,
\end{array} \right.
\end{eqnarray}
\esubequations \noindent
where  $t_i >_c t_j$ means that $t_i$ is further along the
Keldysh contour than $t_j$, and ${\cal T}_{c} $ denotes
contour-ordering along this contour.
(The Keldysh contour runs from the initial time $t_0$ to $+ \infty$
and back.) Under complex conjugation, the following relations hold:
\begin{eqnarray}
 (\tilde G^{R/A}_\iiijjj)^\ast  =  \tilde G^{A/R}_{ji}  \, , \qquad
 (\tilde G^{K}_\iiijjj)^\ast  =   - \tilde G^{K}_{ji}  \, , \qquad 
 (\tilde G^{</>}_\iiijjj)^\ast   =  - \tilde G^{</>}_{ji}  \, .
    \label{eq:complexconjugation}
\end{eqnarray}
It is customary to represent the contour-ordered Green's function
$\G_{\iiijjj}^c$ by a $2\times 2$ matrix $ \bG{}^0_{\iiijjj}$ in Keldysh space,
whose components are the quantum-statistical averages
of contour-ordered operator products,
\begin{eqnarray}\label{eq:contour-ordered-aa}
 (\bG{}^0_{\iiijjj})^{aa'} & \equiv & \langle \hbG{}_{\iiijjj}^{aa'}
 \rangle_0 \; , 
\qquad \hbG{}_{\iiijjj}^{aa'} \equiv 
 - {i\over \hbar} {\cal T}_{c}\,
\left[ \hat \psi_I^a (t_i, x_i) \, \hat \psi_I^{a' \dagger}
  (t_j, x_j)  \right] \; ,
  \end{eqnarray}
and are  labeled
by indices $a,a'$ that take the values $F$ and $B$,
with the convention that if $a = F$ (or $B$), then $t_i$
resides on the forward (or backward) part of the Keldysh contour,
and similarly for $a'$ and $t_j$. In matrix notation, we have
\begin{eqnarray}
  \label{eq:G-Keldysh-Matrix}
  \bG{}^0_{\iiijjj} & = & \langle \hbG{}_{\iiijjj} \rangle_0 \; = \; 
- {i \over \hbar }
\, \langle {\cal T}_c \bmpsi_i \bmpsi^\dag_j
\rangle_0 \; = \; 
\left( \begin{array}{cc}
  \G^{T}_{\iiijjj}  &   \G^{<}_{\iiijjj}
\\
  \G^{>}_{\iiijjj}  &   \G^{\overline T}_{\iiijjj}
\end{array}\right) \; ,   
\end{eqnarray}
where we used a boldface notation for the fermion fields to indicate
that it has two components in Keldysh space, $\bmpsi_i \equiv {\hat
  \psi_{I}^F (t_i, x_i) \choose \hat \psi_{I}^B (t_i, x_i)}$.  [Note
that $(\bG^0_{\iiijjj})^{aa'}$ corresponds to GZ's $G^\GZ_{aa'}(\iiijjj)$,
with $F\to 1$ and $B \to 2$]. A more convenient, since tridiagonal,
form is obtained using the representation\footnote{ A. I. Larkin and
  Yu. N. Ovchinnikov, Zh.\ Eksp.\ Theor.\ Fiz.\ {\bf 68}, 1915 (1975)
  [Sov.\ Phys.\ -- JETP {\bf 41}, 960 (1975)].  This is also the form
  used by Rammer and Smith (see footnote~\ref{RS86}). }
\bsubequations  \label{eq:Keldyshrep}
\begin{eqnarray}
\label{eq:transformpsi}
\unbmpsi_i & \equiv &  \bL \btau^3 \bmpsi_i \; , \qquad
\unbmpsi^\dag_i \equiv  \bmpsi_i^\dag \bL^\dag \; ,
\qquad 
 \chbG{}_{\iiijjj} \; \equiv \; 
- {i \over \hbar }
\left[ {\cal T}_c \unbmpsi_i \unbmpsi^\dag_j \right] \; ,
\\
  \label{eq:Larkin-Ovchinnikov}
   \label{eq:tridiagonalKeldysh}
 \cbG{}^0_{\iiijjj} & \equiv &
    \langle  \chbG{}_{\iiijjj} \rangle_0 \;  = \;   
\bL \btau^3   \bG{}^0_{\iiijjj}  \bL^\dagger 
 =  \left( \begin{array}{cc} \G^R_{\iiijjj} & \G^K_{\iiijjj} \\ 0 &
    \G^A_{\iiijjj}  \end{array} \right) \; ,
\end{eqnarray}
\esubequations
where $\btau^{1,2,3}$ denote the Pauli matrices
acting in Keldysh space,
$\bL  = {1 \over \sqrt 2} {1 \; -1 \choose 1 \; \phantom{-}1} $,
and Eq.~(\ref{eq:tridiagonalKeldysh}) follows
from the definitions (\ref{eq:defineelectrongreensfunctions}).

For future reference, note also that density operators
$\hat n_{ij I}^{a} (t_1)$ located on the forward or backward
branches of the Keldysh contour have the following
representations (suppressing the time argument), for $a = F,B$:
\bsubequations  \label{eq:densities-K}
\begin{eqnarray}  \label{eq:definen-K}
\hat n_{ij I}^{a}  & = & \psi^{a\dag}_{j} \psi^a_i =
\bmpsi^\dag_j \bmP_a \bmpsi_i^\pdag =
\unbmpsi_{jI}^\dag \underline{\bmP_a} \,\unbmpsi_{iI} \; ,
\\
\bmP_{F/B} & = & {1 \over 2} ( \bmone \pm \btau^3 ) \; ,
\qquad
\unbmP_{F/B}  = \bL \btau^3   \bmP_{F/B} \bL^\dagger
\, = \, {1 \over 2} ( \btau_1 \pm \bmone) \; .
\end{eqnarray}
\esubequations

\subappendix{\label{app:fieldcorrelators} Field Correlators}

Next we consider the ``interaction propagators'' $\LL_{\iiijjj}$,
i.e.\ correlators involving the real, bosonic fields $V_i$
that were introduced via the Hubbard-Stratonovich
transformation (\ref{eq:Hubbard-Stratonovitch}).
Below we shall use $V_i$ as a shorthand
for $V_a (t_i,\bmr_i)$, taking it to be understood
that if $a = F$ (or $B$), then $t_i$ resides  on the forward
(or backward) parts of the Keldysh contour. The basic correlators
\begin{eqnarray}
  \label{eq:L><} \label{eq:defineV><}
   \LL^{<}_{\iiijjj}  & \equiv  &
 {i e^2 \over \hbar} \langle V_{j }  V_{i} \rangle_{V} \; \equiv \; 
   \LL^{>}_{ji}  \; ,
\end{eqnarray}
are averaged over all field configurations according to
\Eq{eq:V-average}.  The definitions of the correlators $\LL^T_{\iiijjj}$,
$\LL^{\overline T}_{\iiijjj}$, $\LL^{R}_{\iiijjj}$, $\LL^{A}_{\iiijjj}$,
$\LL^K_{\iiijjj}$ and $\LL^c_{\iiijjj}$ in terms of $\LL^{<}_{\iiijjj}$ and
$\LL^{>}_{\iiijjj}$ are identical to those of the corresponding electronic
$\G_{\iiijjj}$'s in terms of $\G^<_{\iiijjj}$ and $\G^>_{\iiijjj}$ in
\Eqs{eq:defineelectrongreensfunctions}.  The matrix representation
$\bLL_{\iiijjj}$ of the contour-ordered interaction propagator
$\LL^c_{\iiijjj}$, with matrix elements
\bsubequations  \label{eq:define-LL-correlators}
\begin{eqnarray}  \label{eq:LL-matrix-elements}
\bLL{}^{aa'}_{\iiijjj} \equiv
 { i  e^2 \over \hbar}  \langle {\cal T}_c
\, V_{ai} V_{a'j} \rangle_{V} \, = \,
 \bLL{}^{a'\! a}_{ji} \; ,
\end{eqnarray}
takes a form analogous to Eq.~(\ref{eq:G-Keldysh-Matrix}),
namely:
\begin{eqnarray}
  \label{eq:def-interaction-Keldysh-matrix}
  \bLL_{\iiijjj} &   \equiv &
\label{eq:def-interaction-Keldysh-matrix-explicit}
  { i  e^2 \over \hbar}
{\displaystyle \left( \begin{array}{cc}
  \langle V_{Fi} V_{Fj} \rangle_{V}   &
  \langle V_{Fi} V_{Bj} \rangle_{V}   \\
  \langle V_{Bi} V_{Fj} \rangle_{V}   &
  \langle V_{Bi} V_{Bj} \rangle_{V}
\end{array}
\right) }
\; = \; %\\ & = &
\left( \begin{array}{cc}
  \LL^{T}_{\iiijjj}  &   \LL^{<}_{\iiijjj}
\\
  \LL^{>}_{\iiijjj}  &   \LL^{\overline T}_{\iiijjj}
\end{array}
\right)  \;  .
\end{eqnarray}
\esubequations
Following AAG\cite{AAG98}, we shall use the transformation
$\btL = {1 \over \sqrt 2} {1 \; \phantom{-} 1 \choose 1 \; -1} $
 to obtain a tridiagonal representation, reminiscent of
\Eq{eq:tridiagonalKeldysh},
\bsubequations
   \label{eq:tridiagonalKeldysh-LL}
 \begin{eqnarray}
   \label{eq:tridiagonalKeldysh-LLa}
\bcL_{\iiijjj} & \equiv &
  \btL \bLL_{\iiijjj}  \btL^\dagger \;
 =
\left( \begin{array}{cc} \LL^K_{\iiijjj} & \LL^R_{\iiijjj} \\
    \LL^A_{\iiijjj} & 0  \end{array} \right)
=
   \label{eq:tridiagonalKeldysh-LLb}
  { i  e^2 \over \hbar}
{\displaystyle \left( \begin{array}{cc}
 2 \langle V_{+i} V_{+j} \rangle_{V}   &
 \phantom{\toh} \langle V_{+i} V_{-j} \rangle_{V}   \\
  \phantom{2} \langle V_{-i} V_{+j} \rangle_{V}   &
  \toh \langle V_{-i} V_{-j} \rangle_{V}
\end{array} \right) } \qquad \phantom{.} \qqph
\\
& = &
%   \label{eq:tridiagonalKeldysh-LLb}
%  { i  e^2 \over \hbar}
%{\displaystyle \left( \begin{array}{cc}
% 2 \langle V_{+i} V_{+j} \rangle_{V}   &
% \phantom{\toh} \langle V_{+i} V_{-j} \rangle_{V}   \\
%  \phantom{2} \langle V_{-i} V_{+j} \rangle_{V}   &
%  \toh \langle V_{-i} V_{-j} \rangle_{V}
%\end{array} \right) } %\qquad \phantom{.}
%\; = \; %\\ & = &
   \label{eq:tridiagonalKeldysh-LLc}
     e^2  \left( \begin{array}{cc}
       2 i  \, \tilde I_{\iiijjj} & - \tilde R_{\iiijjj} \\
      - \tilde R_{ji} & 0 \
    \end{array}\right), \qquad \phantom{.}
\end{eqnarray}
\esubequations
with matrix elements to be denoted by $\bcL^{\alpha \alpha'}_{\iiijjj}$,
where $\alpha, \alpha' $ take the values $\pm$.  The last equality
of \Eq{eq:tridiagonalKeldysh-LLa}
was obtained by using
\begin{eqnarray}  \label{eq:newwayforVpm}
 \left(\begin{array}{cc}
  \sqrt{2}    & 0 \\
  0 & {1 \over \sqrt{2}}
\end{array} \right)
{ V_{+i} \choose V_{-i} }
 = \btL
{ V_{Fi} \choose V_{Bi} }\; ,
\end{eqnarray}
[cf. Eq.~(\ref{eq:newV's})] to rewrite $\btL \bLL_{\iiijjj} \btL^\dagger$
in terms of the correlators $e^2 \langle V_{\alpha i} V_{\alpha' j}
\rangle_{V}$. The relations (\ref{eq:tridiagonalKeldysh-LLa}) are
general. The explicit expressions for these correlators given by
\Eq{eq:tridiagonalKeldysh-LLc}, which are specific for the present
model, follow from \Eq{eq:LL-AA}.  [Incidentally, comparing
\Eqs{eq:tridiagonalKeldysh-LLa} and (\ref{eq:tridiagonalKeldysh-LLc})
proves Eq.~(\ref{eq:RIvsLRA}].  Using the explicit forms for $\tilde
R_\iiijjj$ and $\tilde I_\iiijjj$ of Eqs.~(\ref{eq:defineRI}), it can easily
be checked that
   \begin{eqnarray}
    \label{eq:realitypropertiesLRAK}
\bigl( \LL^{R/A}_\iiijjj \bigr)^\ast  & = &    \LL^{R/A}_\iiijjj  =
    \LL^{A/R}_\ji  \; , \qquad
%\\
\bigl( \LL^{K}_\iiijjj \bigr)^\ast \; = \; %& = &  
-  \LL^{K}_\iiijjj = -    \LL^K_\ji  \; ,
  \end{eqnarray}
  and that their  Fourier transforms w.r.t.\ $t_\iiijjj$ 
satisfy the relations 
\begin{eqnarray}
  \label{eq:LRLALKomegaij}  \label{eq:LRLAomegaij}
\tilde {\cal L}^R_\iiijjj (\omega)   & \! \! = \! \! &  
\tilde {\cal L}^R_{ji} (\omega)
= \tilde {\cal L}^{R \ast}_\iiijjj (- \omega) =
\tilde {\cal L}^{A \ast}_\iiijjj (\omega)
\; , \qph
% \\
  \label{eq:LKomegaij}
\tilde {\cal L}^K_\iiijjj (\omega)  =  % & \! = \! & 
\tilde {\cal L}^K_{ji} (\omega) = - \tilde {\cal L}^{K \ast}_\iiijjj ( \omega)
\; ,  \qqph \;\; 
%\end{eqnarray}
%\begin{eqnarray}
\\  \label{eq:fluc-diss-LL}
  \LL^K_\iiijjj (\omega)  &  = &
  \coth ( \hbar \omega /2 T) \bigl[
  \LL^R_\iiijjj (\omega)  -   \LL^A_\iiijjj ( \omega) \bigr]
\; . \quad \phantom{.}
\end{eqnarray}
\Eq{eq:fluc-diss-LL} [cf.  \Eq{eq:RIomegaijcoth}] has the form
required by the fluctuation-dissipation theorem.

Explicit expressions for the interaction propagators are most readily
written down in the Fourier representation.  For disordered metals,
where small frequencies and wave numbers dominate, we obtain from
$\bar {\cal L}^R = - e^2 \bar R$ and (\ref{eq:Romegak-explicit}) the
following relations (in agreement with Eq.~(5.8) of AAG\cite{AAG98}):
\begin{eqnarray}
  \label{eq:LR-explicit}
 \bar {\cal L}_\bmq^R(\omega) \; \simeq \;
- {\Dd \bmq^2 - i \omega\over   2 \nud \Dd \, \bmq^2} \; , \qquad
 \bar {\cal L}_\bmq^K(\omega)
 =  2 \, i \coth (\hbar \omega / 2T) \, {\rm Im} 
 \bar {\cal L}_\bmq^R(\omega) \; . 
\end{eqnarray}

\subappendix{Keldysh Perturbation Theory}
\label{app:Keldyshpert} 

In this section, we recall how the Feynman rules for 
 Keldysh perturbation theory are derived, and use
them to obtain an expression for the self energy $\cbSigma$ of the Keldysh 
electron Green's function [\Eq{eq:selfenergyKeldysh}]. 

In the Keldysh approach, expectation values of the form
occuring in Eq.~(\ref{eq:rho-interaction}) are written as follows 
(following Rammer and Smith, see footnote~\ref{RS86}):
\bsubequations
\label{eq:defKeldyshexp}
\begin{eqnarray}
\label{eq:defKeldyshexp-1}
\langle \hat O (t) \rangle_K & \equiv &
{ \langle \hat U_{IB} (t, t_0) \hat O_I (t) \hat U_{IF} (t,t_0)
\rangle_0 \over
 \langle \hat U_{IB} (t, t_0) \hat U_{IF} (t,t_0)
\rangle_0 } % \qquad \phantom{.}
%\\  & = &
\label{eq:defKeldyshexp-2}
\;  = \; { \langle{\cal T}_c \, \hat S_{c i} \, \hat S_{c \tilde v}
\,  \hat O_I (t) \rangle_0
\over  \langle{\cal T}_c \, \hat S_{c i} \, \hat S_{c \tilde v}
\rangle_0 } \; , \qqph 
\\ \label{eq:defKeldyshexp-3}
\hat S_{c i} & \equiv & {\cal T}_c \, e^{-{i\over \hbar}
 \int_c dt_3 \, \hat
  H_{iI} (t_3) } \, , \qquad \;
% \\ 
\label{eq:defKeldyshexp-4}
\hat S_{c \tilde v}
\; \equiv \; {\cal T}_c \, e^{- {i\over \hbar} \int_c dt_3 \, \hat
  v_I (t_3) } \, ,
\end{eqnarray}
\esubequations
where $\int_c dt_1$ and ${\cal T}_c$ indicate integration and time
ordering along the familiar Keldysh contour [$\hat H_{iI}$ and $\hat
v$ are defined in Eqs.~(\ref{eq:defHint}) and (\ref{eq:defvhat})].  In
\Eq{eq:defKeldyshexp-2}, the operator $\hat O_I (t)$ can be written as
either $\hat O_I^F (t)$ or $\hat O_I^B (t)$, where the superscripts
indicate that the operator resides on the upper or lower branch of the
Keldysh contour, since the contribution from the portion of the
Keldysh contour from $t$ to $\infty$ cancels that from $\infty$ back
to $t$. Consequently, we can also represent $\hat O_I (t)$ as ${1
  \over 2 } \mbox{[} \hat O_I^F + \hat O_I^B\mbox{]}(t)$, which turns
out to be most convenient and will be used henceforth. For examples,
the reduced single-particle matrix $\tilde {\bm{\rho}}^\pdag_{11'}
(t,t_0) $ of Eq.~(\ref{eq:rho-interaction}) can be written
as\footnote{An alternative but equivalent form to \Eq{eq:reddenmat-K}
  is often used (\eg\ by AAG\cite{AAG98}, Eq.~(5.1), where
  the factor 2 in front of $\hat \tau_1$ is a typo), namely $ \tilde
  {\bm{\rho}}^\pdag_{11'} = \langle \unbmpsi_{1'}^\dag \toh (\btau^1
  \!  - \! \bmone) \unbmpsi_1^\pdag \rangle_K = - i \hbar [ \toh
  (\btau^1 \! - \! \bmone) \cbG{}^\full_{11'} ] $, where it is to be
  understood that $t_{1'} = t_1 + 0^+$.}
%\bsubequations
\begin{eqnarray}
\label{eq:reddenmat-a}
 \tilde {\bm{\rho}}^\pdag_{11'} (t_1,t_0) & \! = \! &
\toh \langle \hat n_{11' I}^F
 + \hat n_{11' I}^B  \rangle_K =
 \langle  \unbmpsi_{1'}^\dag  \toh \btau^1
 \unbmpsi_1^\pdag \rangle_K  % \qquad \phantom{.} \qph
% \\ 
\label{eq:reddenmat-K}
= - i \hbar  \Tr_K \bigl[ \toh \btau^1   \cbG{}^\full_{11'} \bigr] 
\; , \qqph \qph
\end{eqnarray}
%  \esubequations
Here $\Tr_K$ denotes a trace over Keldysh indices, $
\cbG{}^\full_{11'} = \langle \chbG{}_{\iiijjj} \rangle_K$ (and likewise $
\chbG{}^\nns_{11'} = \langle \chbG{}_{11'} \rangle_{K, \nns}$, which
will occur below, too), $\chbG{}_\iiijjj$ has the<same matrix structure as
in \Eq{eq:transformpsi}, and the superscript ``full'' (or ``ns'')
indicates that the average is to be evaluated in the presence of the
full interaction and including (or excluding) all external
perturbations, \ie\ with $\langle \; \rangle_K$ (or $\langle \;
\rangle_{K,\nns}$) instead of $\langle \; \rangle_0$.  As a check, we
note that in the absence of interactions, \Eq{eq:reddenmat-K} reduces
to $ - i \hbar \toh \tilde G^K_{11'} = \toh \langle \psi_{1'}^\dag
\psi_1^\pdag - \psi_{1}^\pdag \psi_{1'}^\dag \rangle_0 \; , $ which is
equal to the desired result of $ \langle \psi_{1'}^\dag \psi_1^\pdag
\rangle_0 $ (recall that $ \psi_{1'}^\dag $ and $ \psi_1^\pdag $
anticommute, since $x_{1'} $ is equated to $x_{1}$ only at the very
end of the calculation).

By writing $\int_c dt_3 \hat v_I (t_3) =
\int_{t_0}^\infty dt_3 [\hat v^F_{3I} - \hat v_{3I}^B]$,
and switching to the Keldysh representation of
Eq.~(\ref{eq:transformpsi}), $S_{c \tilde v}$ takes
the form
\bsubequations
  \label{eq:SctildeV-K}
  \begin{eqnarray}
    \label{eq:eq:SctildeV-K-1}
\hat S_{c\tilde v}  & = &
{\cal T}_c \, e^{-{i \over \hbar}
 \int_{t_0}^\infty dt_3 \int d x_{3, \bar 3} \,
 \hat \unv_{3 \bar 3} (t_3) } \qquad \phantom{.}
\\
 \hat \unv_{3 \bar 3} (t_3) & \equiv & \tilde v_{\bar 3 3} (t_3)
 \bigl[\hat n^F_{3 \bar 3 I}  - \hat n^B_{3 \bar 3I}  \bigr]
 \rule[0.6cm]{0cm}{0cm}
% \\ & = &
\; = \;    \label{eq:eq:SctildeV-K-3}
\tilde v_{\bar 3 3} (t_3) \,  \unbmpsi^\dag_{\bar 3} \bmone \unbmpsi_3
\; .
  \end{eqnarray}
\esubequations
As a special case of \Eq{eq:eq:SctildeV-K-3}, we note that the
external perturbation, $\hat H_\ext$ of \Eq{eq:externalperturbation},
generates vertices of the form
\begin{eqnarray}
  \label{eq:externalvertex}  ( -i / \hbar) \,
 \underline{\hat h}^\ext_{22'} =
 ( -i / \hbar)  \, h^\ext_{22'} \, \unbmpsi^\dag_{2'} \bmone
 \unbmpsi_2 \; .
\end{eqnarray}
To linear order in $\hat h^\ext$, where each fermion line
is simply decorated by the insertion of a single external vertex,
we thus have the Feynman rule that
each full $ \cbG{}^\full_\iiijjj $ is to be replaced by
\begin{eqnarray}
  \label{eq:externalpertinsertion}
h^\ext_{22'}\,  \langle 
\chbG{}_{i2'} \, \bmone \, \chbG{}_{2j} \rangle_{K, \nns} \; 
\quad \left[
\to {i {\bm{E}} (\omega_0)\cdot \bmj_{22'} \over \omega_0}
\langle  \chbG{}_{i2'} \, \bmone \,
\chbG{}_{2j} \rangle_{K, \nns} \right] \!\! , \phantom{.}
\end{eqnarray}
where the subscript ``ns'' denotes ``no (external) sources'', and
the term in brackets indicates the form which  $h^\ext_{22'}$
assumes under Fourier transformation, if we use
the gauge of \Eq{eq:representing-Eb}.

For any expectation value of the form $\langle \hat O (t) \rangle_{K}$,
the interaction term $\hat H_{iI}$ in $\hat S_{ci}$
can be decoupled using the Hubbard-Stratonovitch transformation of
Eqs.~(\ref{eq:Hubbard-Stratonovitch-all}), just
as in \Sec{sec:hs}, using the fields $V_F$ and $V_B$
for the forward and backward branches
of the Keldysh contour, respectively. One then readily finds
that $\langle \hat O (t) \rangle_{K}$
can be expressed as follows as a functional
average over all fields $V_{F/B}$:
\bsubequations
  \label{eq:ave-V-K}
  \begin{eqnarray}
    \label{eq:K-ave=V-ave}
    \langle \hat O_I (t) \rangle_K & = &
    \langle O^V (t,t_0) \rangle_V
\\
 O^V (t,t_0) &\equiv & { \langle {\cal T}_c \,
\hat {\cal S}_{cV} \, \hat {\cal S}_{c \tilde v} \,
\hat O_I (t)
    \rangle_0
\over  Z(t, t_0) }\; ,
\\ \label{eq?defineZ-K}
Z (t, t_0) & \equiv &
\langle {\cal T}_c \, \hat {\cal S}_{cV}  \hat {\cal S}_{c \tilde v}
    \,
\rangle_0 \; ,
\\
\hat {\cal S}_{cV}  & = &
{\cal T}_c \, e^{-{i\over \hbar}  \int_{t_0}^\infty dt_3 \int d x_3 \,
 \hat \unV_3  } \; , \rule[0.6cm]{0cm}{0cm}
 \\
 \hat \unV_3  & \equiv & e \bigl[ \hat n^F_{33'} V_{F} (\bmr_3)
 - \hat n^B_{33'} V_{B} (\bmr_3) \bigr] % \qquad \phantom{.}
 \rule[0.5cm]{0cm}{0cm}
% \\   & = & 
\; = \; 
e \bigl[  \unbmpsi^\dagger_{3'}
(\bmone V_{+3} + {\textstyle {1 \over 2}} \btau^1 V_{-3} )
\unbmpsi_3 \bigr] \; . \qquad \phantom{.} \qph 
  \end{eqnarray}
\esubequations
Here the functional average $\langle \; \rangle_V$
over all field configurations is defined, as before,
by \Eqs{eq:V-average}, where the functional
$Z$ occuring in \Eq{eq:S[V_a]} is now given by \Eq{eq?defineZ-K}.

\begin{figure}[t]
{\includegraphics[clip,width=0.98\linewidth]{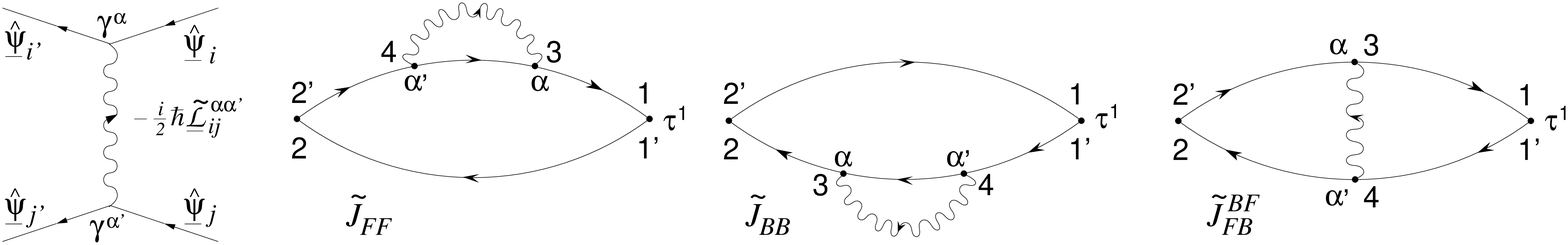}}%
\caption{
  Feynman diagrams for the interaction propagator
  [\Eq{eq:Feynman-vertex}] and the correlators $\tilde
  J_{FF}$, $\tilde J_{BB}$ and $\tilde J_{FB + BF}$
  [\Eqs{eq:Jpertubationtheory}] that give
  the leading correction to the conductivity due to electron-electron
  interaction. Solid lines denote matrix Green's functions $
  \cbG{}^0_{\iiijjj}$, wavy lines interaction propagators $
  \bcL_{\iiijjj}^{\alpha \alpha'} $, and the symbols $\alpha$ and
  $\alpha'$ the vertices $\bmgamma^\alpha$ and $\bmgamma^{\alpha'}$.
  Arrows point from the second to the first index of propagators. }
\label{fig:Feynman}
\end{figure}

\begin{figure}[htbp]%[h!!]
\center{\includegraphics[width=\linewidth]{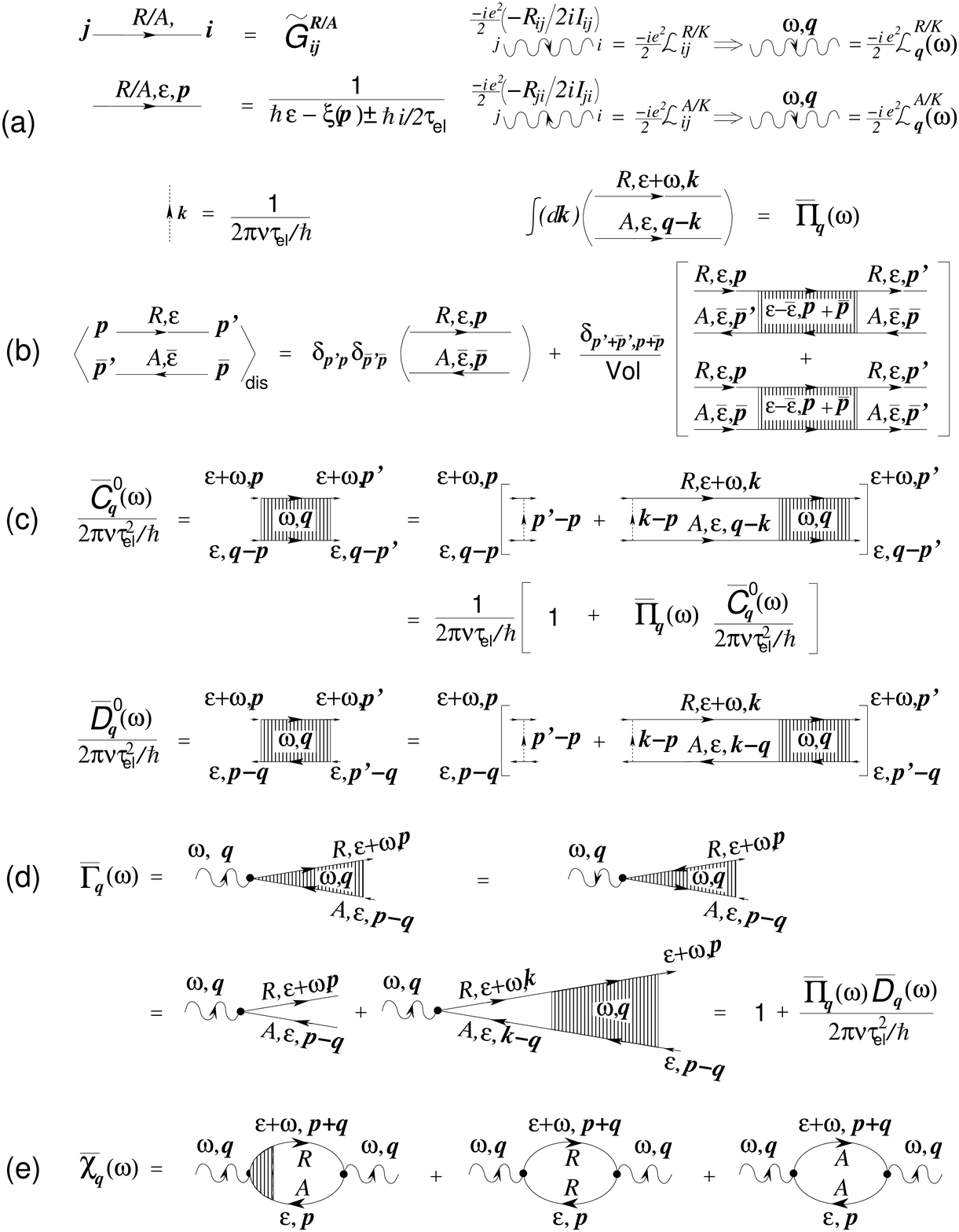}}
\caption{The building blocks of diagrammatic perturbation theory:
  (a) Basic definitions for the electron lines $\tilde G^{R/A}_\iiijjj$
  and 
  $\bcG^{R/A}_{\bmp} (\ve)$, impurity lines, the function
  $\bPiqw$ of \Eq{eq:Pidefinition}, and the interaction lines $\tilde
  {\cal L}_\iiijjj$ or $\overline {\cal L}_\bmq (\omega)$ of
  \Eq{eq:RIvsLRA}.  For all correlators, arrows point from 
  the second to the first indix.   creation to annihilation operators]
Internal impurity momenta are to be integrated over
  with $\int (d \bmk)$, as in $\bPiqw$.  (b)
  \Eq{eq:GGdisorderaverage}. (c) The bare Cooperon $\bcCqw$ 
[\Eq{eq:Cooperonexplicit}] and bare diffuson $ \bcDqw$ 
[\Eq{eq:diffusonexplicit}];    (d) the diffuson-dressed
  vertex $\bGqw$ [\Eq{eq:dressedvertex-result}] and  (e)
polarization bubble $\bar \chi_\bmq (\omega)$ [\Eq{eq:chi12omegaq-c}].
  For each of $\bPiqw$, $\bcCqw$
  and $\bGqw$, the frequency argument $\omega$ is defined 
as the frequency of the corresponding retarded
  Green's function minus that of the advanced one.  
}
  \label{fig:sigma-cooperon}
\end{figure}

To obtain an perturbation expansion
within the Keldysh approach, one expands $\hat {\cal S}_{cV}$ in
powers of $(-i/\hbar) \hat \unV_3$, which thus serves as basic interaction
vertex, and then applies Wick's theorem to the fermion fields.
In the $n$-th order term, there are $n!$ equivalent ways to connect
the $n$ vertices with $n$ fermion lines of the type
$\langle \unbmpsi_i^\pdag \unbmpsi^\dag_j \rangle_0
= i \hbar \cbG^0_\iiijjj$, yielding a combinatorical
factor of $(i \hbar )^n n!$
which cancels the $(-i / \hbar)^n /(n!)$ from the expansion of the
exponent of $\hat {\cal S}_{cV}$. Next,
the average $\langle \; \rangle_V$ over all field
configurations is to be performed, which yields
contractions between the interaction fields pairs
of vertices. These contractions have the form
\begin{eqnarray}  \label{eq:startvertex} \label{eq:contractVV}
\label{eq:Feynman-vertex}
%\langle \hat \unV_i \hat \unV_j \rangle_V
%& =&  - \toh i \hbar  \sum_{\alpha \alpha'}
%\unbmpsi^\dag_{3'} \bmgamma^\alpha \unbmpsi_3 \,
%\bcL_{34}^{\alpha \alpha'}
%\unbmpsi^\dag_{4'} \bmgamma^{\alpha'} \unbmpsi_4  \; ,
%\qquad \phantom{.}\\
\langle \hat \unV_i \hat \unV_j \rangle_V
& =&  - \toh i \hbar  \sum_{\alpha \alpha'}
\unbmpsi^\dag_{i'} \bmgamma^\alpha \unbmpsi_i \,
\bcL_{\iiijjj}^{\alpha \alpha'}
\unbmpsi^\dag_{j'} \bmgamma^{\alpha'} \unbmpsi_j  \; ,
\qquad \phantom{.}
\end{eqnarray}
where we introduced the ``vertex matrices''
 $\bmgamma^+ = \bmone$ and $\bmgamma^- = \btau^1$,
the field propagator in the Keldysh representation
$ \bcL_{\iiijjj}^{\alpha \alpha'}$ is given by
\Eqs{eq:tridiagonalKeldysh-LL},
and the Feynman diagram corresponding to Eq.~(\ref{eq:Feynman-vertex})
is the leftmost graph in Fig.~\ref{fig:Feynman}.
\Eq{eq:Feynman-vertex} implies the following Dyson equation (cf.\ 
Eq.~(5.6) of AAG\cite{AAG98}),
  \begin{eqnarray}
 \label{eq:KeldyshDyson}
    \cbG{}^\full_\iiijjj & = &
\cbG{}^0_\iiijjj + \int_{t_0}^\infty \! \! dt_3 \,
 dt_4 \int \! \! dx_3 \, dx_4  \, \cbG{}^0_{i3} \, \cbSigma_{34} \, 
\cbG{}^\full_{4j}  \; ,
\qquad \phantom{.}
\end{eqnarray}
where, to lowest order in the interaction, the self-energy is
given by
\begin{eqnarray}
  \label{eq:selfenergyKeldysh}
  \cbSigma_{34} & = & - \toh i \hbar
\bmgamma^\alpha \cbG{}^0_{34} \bmgamma^{\alpha'} \bcL_{34}^{\alpha
  \alpha'}
\; .
\end{eqnarray}

\subappendix{Conductivity}
\label{sec:conductivity-Keldysh}

In this section  we derive  a general expression for conductivity
$\sigma_\DC$ in the Keldysh approach and expand it to leading order
in the interaction propagator. This will allow us to check the
perturbative expansion (\ref{subeq:timesaa'}) 
of our influence functional $\tilde
J_{12',21'}$ of \Sec{app:1storderperturbation}.
%The goal is to reproduce the perturbative expression for the DC
%conductivity (Eq.~(5.12) of Ref.~\cite{AAG98}), and

We start by using  \Eq{eq:reddenmat-K}
to express the quantum-statistical
average of the current density operator $\hat \JJbm_{\!\! H} (t_1,
\bmr_1)$ of \Eq{eq:current-define}
as follows,
\begin{eqnarray}
%\nonumber
  \langle \hat \JJbm_{\!\! H} (t_1, \bmr_1) \rangle_K
& = &   \sum_{\sigma_1}
\left[  \bmj_{11'} - {e^2 \over m} \bmA(t_1,\bmr_1) \right]
  \label{eq:currentKeldysh}
( -i \hbar) \, \Tr_K \left[ \toh \btau^1  \cbG{}^\full_{11'} \right]
\; .
\end{eqnarray}
Next we expand \Eq{eq:currentKeldysh} to first order in $\hat h_\ext$
[using \Eq{eq:externalpertinsertion}], and then use
\Eq{eq:DCconductivity} to calculate $\sigma_\DC$; the result has the
form of \Eq{eq:sigmaAAG}, where $\tilde J_{12',21'} (\omega_0)$
therein is given by the
Fourier transform w.r.t.\ $t_{12}$ of the following expression:
  \begin{eqnarray}
    \label{eq:GRKeldyshVVgeneral}
    \tilde J^\Keldysh_{12',21'} & = & \hbar
\Tr_K \left[ \toh \btau^1
 \langle \chbG{}_{12'} \chbG{}_{21'}\rangle_{K,\nns} \right] .
\end{eqnarray}
In the absence of electron-electron interactions, this readily
reduces to 
 \begin{eqnarray}
    \label{eq:C1221t}
\tilde J^{(0), \Keldysh}_{12',21'} & = &
\toh \hbar  \left( \tilde G^R_{12'} \tilde G^K_{21'} +
\tilde G^K_{12'}  \tilde G^A_{21'} \right) \; = \; 
 \hbar  \left( \tilde G^R_{12'} \tilde G^<_{21'} +
\tilde G^<_{12'}  \tilde G^A_{21'} \right) \, . \qqph 
\end{eqnarray}
The second equality follows from \Eq{eq:GKeldyshdef} (with $\tilde
G^{R/A}_\iiijjj \tilde G^{A/R}_{ji} = 0$) and confirms
\Eq{eq:J01221freeF+B}.

Let us now obtain the leading correction to $\sigma_\DC$ due to the
electron-electron interaction. To this end, we have to expand
\Eq{eq:GRKeldyshVVgeneral} for $\tilde J^\Keldysh_{12',21'}$ to second
order in $ \hat \unV_3 $. One readily arrives at the following result
[which can also be obtained by starting directly from
\Eq{eq:currentKeldysh}, expanding $ \cbG{}^\full_{11'}$ therein to
first order in $ \cbSigma_{34}$ using \Eq{eq:KeldyshDyson}, and then
expanding each $\cbG{}^0_\iiijjj$ in the latter equation to first order in
$\hat h_\ext$ using \Eq{eq:externalpertinsertion}]:
\bsubequations
\label{eq:Jpertubationtheory}
\begin{eqnarray}
    \label{eq:C1122KeldyshJVV}
 \tilde J^{(2), \Keldysh}_{12',21'} &  = &
 { - \toh i \hbar^2}
\int_{t_0}^\infty \!\! dt_3 \, dt_4
\!\! \int \!\! dx_3 \, dx_4 \, 
%\sum_{\alpha \alpha'}
\Bigl( \tilde J_{FF} + \tilde J_{BB} + \tilde J^{BF}_{FB}  \Bigr) \; , 
\qquad \phantom{.}
\nonumber
\\
    \label{eq:C1122KeldyshVV-b}
\phantom{.} \hspace{-1cm}
\tilde J_{FF}
&  = &  \sum_{\alpha \alpha'}   \Tr_K \Bigl[ \toh
\btau^1 \cbG{}^0_{13}  \bmgamma^\alpha \cbG{}^0_{34}
\bmgamma^{\alpha'} \cbG{}^0_{42'} \cbG{}^0_{21'}
\,\bcL_{34}^{\alpha \alpha'}  \Bigr] ,
\qquad \phantom{.}
\\
    \label{eq:C1122KeldyshVV-c}
\tilde J_{BB}
 & = &   \sum_{\alpha \alpha'}  \Tr_K \Bigl[ \toh \btau^1
\cbG{}^0_{12'} \cbG{}^0_{23} \bmgamma^\alpha \cbG{}^0_{34}
\bmgamma^{\alpha'} \cbG{}^0_{41'} \,
\bcL_{34}^{\alpha \alpha'} \Bigr] ,
\qquad \phantom{.}
\\
    \label{eq:C1122KeldyshVV-d}
\tilde J_{FB}^{BF}
 & = & \sum_{\alpha \alpha'}  \Tr_K \Bigl[ \toh \btau^1
\cbG{}^0_{13}  \bmgamma^\alpha \cbG{}^0_{32'}  \cbG{}^0_{24}
\bmgamma^{\alpha'} \cbG{}^0_{41'} \, \bcL_{34}^{\alpha \alpha'}
\Bigr] \; .
\qquad \phantom{.}
  \end{eqnarray}
\esubequations
The correlators $\tilde J_{FF}$, $\tilde J_{BB}$ and $\tilde
J_{FB}^{BF}$ are illustrated in Fig.~\ref{fig:Feynman}, and correspond
to self-energy insertions in the upper and lower Keldysh contours, and
a vertex correction, respectively. 
  Multiplying out the Keldysh matrices explicitly, taking the trace
and omitting all terms involving the combinations
$\GR_{34} \tilde {\cal L}_{34}^A$ or $\GA_{34} \tilde {\cal
  L}_{34}^R$, which vanish (since $\theta_{34} \theta_{43} = 0$), 
we obtain:
\bsubequations
    \label{eq:sumJABC-full}
  \begin{eqnarray}
    %\sum_{\alpha \alpha'} 
\tilde J_{FF} & = &
\phantom{+} 
\toh \GR_{13} \bigl[
\GR_{34} \tilde {\cal L}_{34}^K +
\GK_{34} \tilde {\cal L}_{34}^R  \bigr]
\bigl( \GR_{42'} \GK_{21'} +  \GK_{42'} \GA_{21'} \bigr)
\\ \nonumber
 & & 
+ \toh \bigl[ \GR_{13} \bigl(
\GK_{34} \tilde {\cal L}_{34}^K +
\GR_{34} \tilde {\cal L}_{34}^R +
\GA_{34} \tilde {\cal L}_{34}^A \bigr)
+ \GK_{13} \bigl( 
\GA_{34} \tilde {\cal L}_{34}^K +
\GK_{34} \tilde {\cal L}_{34}^A
\bigr) \bigr]
\GA_{42'} \GA_{21'}  \; , 
\rule[-3mm]{0mm}{0mm}
\\
% \sum_{\alpha \alpha'} 
\tilde J_{BB} & = & \phantom{+}
\toh \bigl( \GR_{12'} \GK_{23} +  \GK_{12'} \GA_{23} \bigr)
\bigl[ \GA_{34} \tilde {\cal L}_{34}^K +
\GK_{34} \tilde {\cal L}_{34}^A  \bigr]
\GA_{41'}
\\ \nonumber
& & + 
\toh  \GR_{12'} \GR_{23} 
\bigl[ 
\bigl( \GR_{34} \tilde {\cal L}_{34}^K +
\GK_{34} \tilde {\cal L}_{34}^R \bigr) \GK_{41'}
+
\bigl( \GK_{34} \tilde {\cal L}_{34}^K +
\GR_{34} \tilde {\cal L}_{34}^R +
\GA_{34} \tilde {\cal L}_{34}^A 
\bigr) \GA_{41'} \bigr] \; ,
\rule[-3mm]{0mm}{0mm}
\\
%\sum_{\alpha \alpha'} 
\tilde J_{FB}^{BF} & = & \phantom{+}
\toh \bigl( \GR_{32'} \GK_{24} +  \GK_{32'} \GA_{24} \bigr)
\bigl[ \GA_{41'} \GR_{13} \tilde {\cal L}_{34}^K +
\GK_{41'} \GR_{13} \tilde {\cal L}_{34}^R  +
\GA_{41'} \GK_{13} \tilde {\cal L}_{34}^A  \bigr] 
\\ \nonumber
& & +
\toh  \GR_{32'} \GR_{24}
\bigl[ \GK_{41'} \GR_{13} \tilde {\cal L}_{34}^K +
\GA_{41'} \GR_{13} \tilde {\cal L}_{34}^R +
\bigl(\GK_{41'} \GK_{13} + \GR_{41'} \GA_{13} \bigr)
\tilde {\cal L}_{34}^A  \bigr]
%\rule[-3mm]{0mm}{0mm}
\\ \nonumber
& & +
\toh  \GA_{32'} \GA_{24}
\bigl[ \GA_{41'} \GK_{13} \tilde {\cal L}_{34}^K +
\GA_{41'} \GR_{13} \tilde {\cal L}_{34}^A +
\bigl(\GK_{41'} \GK_{13} + \GR_{41'} \GA_{13} \bigr)
\tilde {\cal L}_{34}^R  \bigr] \; . 
\end{eqnarray}
\esubequations
Now, terms that involve the combination $\GR_{i2'} \GR_{2j}$ or
$\GA_{i2'} \!  \GA_{2j}$ contribute to the so-called interaction
corrections, and do not contribute to ``decoherence''. Hence, we
retain only the first lines of \Eqs{eq:sumJABC-full} henceforth.  For
these, we use the identity [cf.\ (\ref{eq:GKeldyshdef})]
\begin{eqnarray}
  \label{eq:RK+KA}
  \toh (\GR \GK + \GK \GA ) = \GR \G^< + \G^< \GA + 
\toh (\GR \GR - \GA \GA) 
\end{eqnarray}
and drop the last term, for the same reason. The
remaining terms then take the following form:
\bsubequations
    \label{eq:sumJABC}
  \begin{eqnarray}
 \tilde J_{FF} & \!\! = \!\!  & 
\GR_{13} \bigl[
\GR_{34} \tilde {\cal L}_{34}^K +
\GK_{34} \tilde {\cal L}_{34}^R  \bigr]
\bigl( \GR_{42'} \G^<_{21'} +  \G^<_{42'} \GA_{21'} \bigr)
\\
\tilde J_{BB} & \!\! = \!\!  &  
\bigl( \GR_{12'} \G^<_{23} +  \G^<_{12'} \GA_{23} \bigr)
\bigl[ \GA_{34} \tilde {\cal L}_{34}^K +
\GK_{34} \tilde {\cal L}_{34}^A  \bigr]
\GA_{41'}
\\
\tilde J^{BF}_{FB} & \!\! = \!\!  & 
\bigl( \GR_{32'} \G^<_{24} +  \G^<_{32'} \GA_{24} \bigr)
\bigl[ \GA_{41'} \GR_{13} \tilde {\cal L}_{34}^K +
\GK_{41'} \GR_{13} \tilde {\cal L}_{34}^R  +
\GA_{41'} \GK_{13} \tilde {\cal L}_{34}^A  \bigr]  . \qqph \qph
\end{eqnarray}
\esubequations
These expressions agree with the expansion (\ref{subeq:timesaa'}) we
obtained from the influence functional approach, as can be
seen by relabelling $3 \leftrightarrow 4$ in some terms.
[$\tilde J_{FB}^{BF}$ here accounts for both  
$\tilde J_{BF}$ \emph{and} $\tilde J_{FB}$ there.]

%\appeqn
\appendix{Diagrammatic Disorder Averaging}
\label{sec:diagrammatics}

In this appendix we summarize, for 
reference purposes, some standard and well-known
conventions and results used for diagrammatically
performing disorder averages, using notations summarized
at the beginning of App.~\ref{sec:rederivation}.

\subappendix{Definitions, Standard Results and Useful Tricks}
\label{sec:buildingblocks}

To perform the disorder averages, we take the impurity potential to be
short-ranged, $ V_\imp (\bmr) = v_\imp \sum_i \delta (\bmr - \bmR_i)$,
%\begin{eqnarray}
%  \label{eq:Uimp}
%  V_\imp (\bmr) = v_\imp \sum_i \delta (\bmr - \bmR_i) \; ,
%\end{eqnarray}
($v_\imp$ has units of energy times volume),
 represent the fermion fields as
$\hat \psi_\sigma (t, \bmr) =  \Vol^{-1/2}
\sum_\bmp e^{i \bmp \cdot \bmr} c_\bmp (t)$, and Fourier transform
as follows:
\begin{eqnarray}
  \label{eq:GijFourier}
\tilde G^{R/A}_{\iiijjj}  =   \tilde G^{R/A} (t_\iiijjj, x_i, x_j)
 \equiv
\delta_{\sigma_i \sigma_j}
\int {d \ve \over 2 \pi} e^{-i \ve t_\iiijjj}
{1 \over \Vol } \sum_{\bmp_i \bmp_j}
\,
e^{i (\bmp_i \cdot \bmr_i - \bmp_j \cdot \bmr_j)}
\bar G^{R/A}_{\bmp_i \bmp_j} ( \ve) \; .  \qph
\end{eqnarray}

%\begin{eqnarray}
%  \label{eq:fermionfieldfourier}
%\hat \psi_\sigma (t, \bmr) = {1 \over \Vol^{1/2}}
%\sum_\bmp e^{i \bmp \cdot \bmr} c_\bmp (t) \; .
%\end{eqnarray}
%we evidently have
%\begin{eqnarr<ay}
%  \label{eq:Gp1p2=}
%  \bar G^{R/A}_{\bmp_i \bmp_j} (t_\iiijjj) & = &
%  \mp {i \over \hbar} \theta(t_\iiijjj) \bigl \langle \{ c_{\bmp_i} (t_i) ,
%  c^\dag_{\bmp_j} (t_j)\} \bigr \rangle \; .
%\end{eqnarray}
Using standard diagrammatic techniques,
the disorder-averaged single-particle propagator is found
to have the form [Fig.~\ref{fig:sigma-cooperon}(a)]:
% Book 6, pages 94 to 97
\bsubequations
  \label{eq:GRAdisavepe}
\begin{eqnarray}
  \label{eq:GRAdisavepe-a}
\bigl \langle \bar G^{R/A}_{\bmp' \bmp} (\ve) \bigr \rangle_\dis
& = &
\delta_{\bmp \bmp'} \,  \bcG^{R/A}_{\bmp} (\ve) \; ,
\\
  \label{eq:GRAdisavepe-b}
 \bcG^{R/A}_{\bmp} (\ve) & = & \int d t_{\iiijjj} \int
d \bmr_{\iiijjj} e^{-i \epsilon t_\iiijjj} e^{i \bmp \cdot \bmr_\iiijjj}
\bigl\langle  \tilde G^{R/A}_\iiijjj \bigr\rangle_\dis
={ 1 \over \hbar \ve - \xi_\bmp \pm  i \hbar / 2 \tauel} \; .
\qqph \qph
\end{eqnarray}
Here $\tauel = \hbar / ( 2 \pi \nud \, c_\imp  \, v_\imp^2) $ is the elastic
scattering time, $c_\imp$ the impurity concentration, 
$\xi_\bmp = \bmp^2 \hbar^2 /2m - \eF$, and calligraphic
symbols will be used throughout for disorder-averaged quantities.
The corresponding position-time expression, found
by inverse Fourier transforming, is:
\begin{eqnarray}
    \label{eq:shortrangeGtime}
        \tcG^{R/A}_\iiijjj (t) & = & 
%    \bigl\langle
%        \tilde G^{R/A}_\iiijjj (t)
%    \bigr \rangle_\dis     =  
\int (d \bmp) \, e^{i \bmp \cdot \bmr_\iiijjj} 
\int (d \ve) \, e^{-i \ve t } \,
\bcG^{R/A}_\bmp (\ve) \; , 
%\\
%    \label{eq:shortrangeGtimeresult-a}
%& = &
%\left[ \tilde G^{R/A}_\iiijjj (t)  \right]_{\rm   clean} e^{- |t|/2 \tauel} 
\\
    \label{eq:shortrangeGtimeresult-b}
& = & \mp {i\over \hbar} \theta(\pm t) \, 
\left({ m \over i 2\pi \hbar t } \right)^{d/2} \exp \left[ {i
   m \bmr_\iiijjj^2 \over 2 \hbar t}  \right] \, e^{i  \eF t / \hbar} 
\, e^{- |t|/2 \tauel} \; . 
\end{eqnarray}
\esubequations
The disorder-averaged products $ \langle \tilde G^R \tilde
G^A \rangle_\dis $ 
have the form [Fig.~\ref{fig:sigma-cooperon}(b)],
% Book 5, p. 65,
% or better: Book 6, page 97
\bsubequations
  \label{subeq:GGdisorderaverageRRorAA}
\begin{eqnarray}
  \label{eq:GGdisorderaverageRRorAA}
 \bigl\langle
\bar G^{R/A}_{\bmp' \bmp} (\ve ) \,
\bar G^{R/A}_{\bar \bmp' \bar \bmp} (\bar \ve) \bigr\rangle_\dis
& = &
\delta_{\bmp', \bmp} \, \delta_{\bar \bmp', \bar \bmp}  \;
\bcG^{R/A}_{\bmp} (\ve)\,
\bcG^{R/A}_{\bar \bmp} (\bar \ve) \; ,
\\
\nonumber
 \bigl\langle
\bar G^R_{\bmp' \bmp} (\ve ) \,
\bar G^A_{\bar \bmp' \bar \bmp} (\bar \ve) \bigr\rangle_\dis
& = &
\delta_{\bmp', \bmp} \, \delta_{\bar \bmp', \bar \bmp}  \;
\bcG^R_{\bmp} (\ve)\,
\bcG^A_{\bar \bmp} (\bar \ve)
\\
  \label{eq:GGdisorderaverage}
& & \phantom{.} \hspace{-4cm} \; + \;
 \delta_{\bmp' + \bar \bmp' , \bmp + \bar \bmp} \;
\bcG^R_{\bmp'} ( \ve) \,
\bcG^R_{\bmp} ( \ve ) \,
\bcG^A_{\bar \bmp'} (\bar \ve) \,
\bcG^A_{\bar \bmp} (\bar \ve) \,
{\bcD^0_{\bmp + \bar \bmp'} (\ve - \bar \ve )
+  \bcC^0_{\bmp + \bar \bmp} (\ve - \bar \ve )
\over  \Vol \, 2 \pi \nud \tauel^2 /
  \hbar} \; ,
\qquad \phantom{.}
\end{eqnarray}
\esubequations
$\bcCqw$ and $\bcDqw$ being the bare (\ie\ without
interactions) Cooperon and diffuson, respectively.
Fig.~\ref{fig:sigma-cooperon} summarizes the standard calculations 
of $\bcCqw$ and $\bcDqw$, and of the diffuson-dressed interaction 
vertex  $\bGqw$ and polarization bubble $\bar \chi_\bmq (\omega)$,
which is defined as the Fourier transform of \Eq{eq:chieta}:
\bsubequations
\label{subeq:chieexplicit}
\begin{eqnarray}
    \bigl \langle \bar \chi_\bmq (\omega) \bigl \rangle_\dis
 & = &
     - i \, 2 e^2  \hbar \int (d \varepsilon) (d \bmp)
\Bigl \langle 
\bcG^R_{\bmp + \bmq} (\varepsilon + \omega)
\bcG^<_{\bmp} (\varepsilon) + \bcG^<_{\bmp + \bmq} (\varepsilon + \omega)
\bcG^A_{\bmp} (\varepsilon) \Bigr \rangle_\dis \qqph 
\\ \nonumber 
 & \simeq &
     - i \, 2 e^2  \hbar \int (d \varepsilon) (d \bmp)
\Bigl \langle [- \omega \, n'_0 (\varepsilon)] 
\bcG^R_{\bmp + \bmq} (\varepsilon + \omega)
\bcG^A_{\bmp} (\varepsilon) \Bigr.
\\ \label{eq:chietaRRAA}
& & 
\qquad 
\Bigl.  - n_0 (\varepsilon) 
\Bigl[ \bcG^R_{\bmp + \bmq} (\varepsilon + \omega)
\bcG^R_{\bmp} (\varepsilon) 
- \bcG^A_{\bmp + \bmq} (\varepsilon + \omega)
\bcG^A_{\bmp} (\varepsilon) \Bigr]
\Bigr \rangle_\dis  \; . 
\end{eqnarray}
\esubequations
The results are:
\bsubequations
  \label{subeq:maindiagrammatics}
\begin{eqnarray}
  \label{eq:Pidefinition}
\bPiqw & = & \!\!\!  \int (d \bmk) \, \bcG^R_{\bmk} (\ve
+ \omega) \bcG^A_{\bmq - \bmk} (\ve) \; = 
{2 \pi \nud \tauel \over \hbar} \left[ 1 - \tauel ( \Dd \bmq^2 - i
 \omega) + \dots \right]  , \qqph 
\\
    \label{eq:Cooperonexplicit}
\bcCqw
 & = & {\tauel \over 1 - \bPiqw / ( 2 \pi \nud \tauel / \hbar)}
\; = \; {1 \over \Dd  \bmq^2 - i \omega + \gammaH} + \dots \; ,
\\
    \label{eq:diffusonexplicit}
\bcDqw
 & = & {\tauel \over 1 - \bPiqw / ( 2 \pi \nud \tauel / \hbar)}
\; = \; {1 \over \Dd  \bmq^2 - i \omega} + \dots \; ,
\\
\label{eq:dressedvertex-result}
  \bGqw & = & 1 + {\bPiqw \bcDqw \over 2 \pi \nud \tauel^2 / \hbar}
= {1 \over \tauel ( \Dd \bmq^2 - i \omega )} + \dots \; ,
\\
    \label{eq:chi12omegaq-c}
\bar \chi_\bmq (\omega) & =  &
 - i \, 2  e^2 \left[ {  \omega \, \nud \over \Dd \bmq^2 - i \omega}  -
     i\, \nud
     \right] \; = \;
- \, { \bmq^2 \, \sigma^\Drude_\DC \over \Dd \bmq^2 - i \omega} + \dots \;  .
\end{eqnarray}
\esubequations
Here $\Dd = v_F^2 \tauel/d$ is the diffusion constant in $d = 3$
or 2 dimensions, $\gammaH$ is a magnetic-field cutoff 
and the dots indicate subleading
terms that are small in $\omega \tauel \ll 1$ and
$q \lel \ll 1$.

\label{sec:diagramtricks}

For convenience, we also summarize here some results that are
useful for evaluating momentum integrals that arise in diagrammatic
perturbation theory.  Usually, the energy parameter $\hbar
\ve$ of the disorder-averaged Greens' functions
$\bcG^{R/A}_\bmp (\ve)$ is confined to the vicinity of $\eF$,
typically by the presence of a factor $- \partial_\ve n_0
(\hbar \ve)$ in an $\int d \ve$ integration, so that
terms of order $\hbar \ve / \eF$ can be neglected.  
[The second term of \Eq{eq:chietaRRAA} does not 
contain a factor  $- \partial_\ve n_0$, but one can
be generated by integrating by parts.]
The explicit form (\ref{eq:GRAdisavepe}) for $\bcG^{R/A}_{\bmp}
(\ve)$ then implies the following ``identities'':
\bsubequations
\label{eq:integrationidentities}
  \begin{eqnarray}
  \label{eq:integrationidentities2}
& &  \phantom{.} \hspace{-1.5cm}
\int \!\! {(d \bmp) \over 2 \pi \nud \tauel / \hbar} \;
\bcG^{R}_{\bmp} (\ve) \,
\bcG^{A}_{\bmp} (\ve)   =  1 \; , \qquad
 \int \!\! {(d \bmp) \over 2 \pi \nud \tauel / \hbar} \;
\bcG^{R/A}_{\bmp} (\ve) \,
\bcG^{R/A}_{\bmp} (\ve)
= 0 \; , \qqph 
\\
  \label{eq:integrationidentitiesgeneral}
& & \phantom{.} \hspace{-1.5cm}
\int \!\! {(d \bmp) \over 2 \pi \nud \tauel / \hbar} \;
\left[\bcG^{R/A}_{\bmp} (\ve) \right]^m \,
\left[\bcG^{A/R}_{\bmp} (\ve) \right]^n
 =  \Bigl( { - i \tau \over \hbar} \Bigr)^{m-1} 
    \Bigl({i \tau \over \hbar} \Bigr)^{n-1}
{ m+n-2 \choose m-1} . \qqph 
  \end{eqnarray}
\esubequations
Furthermore, in the limit of small frequencies ($\omega , \bomega \ll
1/ \tauel)$ and wavenumbers ($\bmq^2, \bbmq^2 \ll 1/ \Dd \tauel$),
integrals of the following kind can be evaluated by a systematic
expansion in the small paramters, combined with repeated use of
\Eqs{eq:integrationidentities}:
%{\bf (These integrals can of course be done exactly, see AAKL (3.2.11))}
\bsubequations
  \label{eq:IRARAARRAA}
\begin{eqnarray}
    \label{eq:IRA}
%I^{R/A} (\omega, \bmq) & \equiv &
& & \int  {(d\bmp)  \over 2 \pi \nud \tauel/ \hbar}
 \, \bcG^{R/A}_{\bmp} (\ve)
\, \bcG^{A/R}_{\bmp + \bmq} (\ve+ \omega)
\; =  \; 1 - \tauel \bigl[ \Dd \, \bmq^2 \pm i \omega \bigr] + \dots  \; ,
\rule[-5mm]{0mm}{0mm}
\\
    \label{eq:IRAA}
% I^{RAA/ARR} (\omega, \bmq;  \bar \omega, \bar \bmq) & \equiv & 
& & {\hbar \over \tauel}
\int  {(d\bmp)  \over 2 \pi \nud \tauel/ \hbar}
 \, \bcG^{R/A}_{\bmp} (\ve)
\, \bcG^{A/R}_{\bmp + \bmq} (\ve+ \omega)
\, \bcG^{A/R}_{\bmp + \bar \bmq} (\ve+ \bar \omega)
\\ \nonumber 
& & \qquad =  \pm i \Bigl\{ 1 - \tauel \bigl[ \Dd
(\bmq + \bar \bmq)^2 \pm i (\omega + \bar \omega) \bigr]
\Bigr\} + \dots \; ,
\rule[-5mm]{0mm}{0mm}
\\
    \label{eq:IR^2AA}
% I^{R^2AA/A^2RR} (\omega, \bmq;  \bar \omega, \bar \bmq) & \equiv & 
& & {\hbar^2 \over \tauel^2}
\int  {(d\bmp)  \over 2 \pi \nud \tauel/ \hbar}
 \, \left[ \bcG^{R/A}_{\bmp} (\ve) \right]^2
\, \bcG^{A/R}_{\bmp + \bmq} (\ve+ \omega)
\, \bcG^{A/R}_{\bmp + \bar \bmq} (\ve+ \bar \omega)
\\ \nonumber
& & \qquad =  2 - \tauel \bigl[4
\Dd (\bmq + \bar \bmq)^2
\pm 3 i  (\omega + \bar \omega) \bigr]
 + \dots \; ,
\\
    \label{eq:IRRAA}
% I^{RRAA/AARR} (\omega, \bmq;  \bar \omega, \bar \bmq; \omega', \bmq')
% & \equiv & 
& & {\hbar^2 \over \tauel^2}
\int  {(d\bmp)  \over 2 \pi \nud \tauel/ \hbar}
 \, \bcG^{R/A}_{\bmp} (\ve)
\, \bcG^{R/A}_{\bmp + \bmq'} (\ve + \omega')
\, \bcG^{A/R}_{\bmp + \bmq} (\ve+ \omega)
\, \bcG^{A/R}_{\bmp + \bar \bmq} (\ve+ \bar \omega)
\\ \nonumber 
& & \qquad =  2 - \tauel \bigl[
\Dd [4 (\bmq + \bar \bmq)^2 + 4 (\bmq')^2  - 6 \bmq' \cdot (\bmq + \bbmq)]
\pm 3 i  (\omega + \bar \omega - \omega') \bigr]
 + \dots \; .
\qqph
\end{eqnarray}
\esubequations

\subappendix{Cooperon Self Energy}
\label{sec:cooperonselfenergy-app}

%\begin{figure}[h!]
\begin{figure}[!ht]
  \centering
  \includegraphics[width=0.95\linewidth]{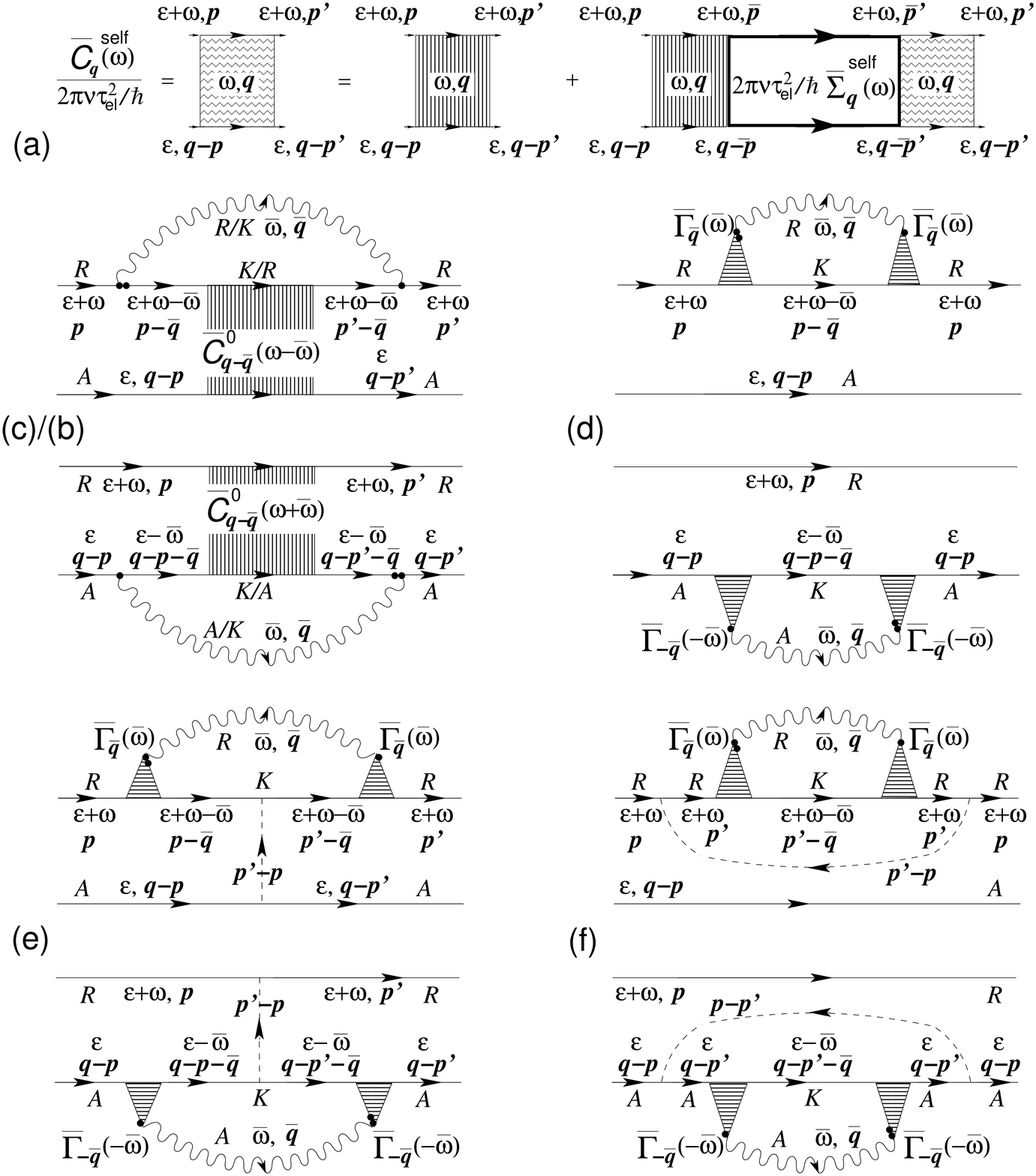}
  \caption{(a) Dyson equation [\Eq{CooperonDysona}] 
    for the Cooperon, for the case that the Cooperon self energy
    contains only self-energy contributions.  The latter are shown in
    diagrams (b) to (f), which depict how to perform the disorder
    average of the various contributions $\bSigma^{I/R}_{aa'} $ to the
    Cooperon self energy, leading to \Eqs{eq:selfenergyFFBBresults}
    and (\ref{eq:defineYgiveY}).  Diagrams (b) depict
    $\bSigma^I_{FF/BB}$; the diagrams (c) + (d) + (e) + (f) depict
    $\bSigma^R_{FF/BB}$, the four contributions corresponding to the
    terms in \Eqs{eq:selfenergySRFF}
    that contain $Y^{(1)}_{a}$, $Y^{(2)}_{a}$, $Y^{(3)}_{a}$ and
    $Y^{(4)}_{a}$, respectively.  [To avoid cluttering the figure with
    factors of $\toh$, the energy and momentum labels $\ve$, $\omega$
    and $\bmq$ used here were assigned in a less symmetrical way
    between upper and lower lines than in
    Fig.~\ref{fig:cooperonrealspace}(c); to transcribe the expressions
    used in this section into the notation used there, make the
    replacements $(\ve + \omega)_\here \to (\ve + \toh \omega)$,
    $\bmp'_\here \to (\bmp_1 + \toh \bmq)$, $\bmp_\here \to (\bmp_2 +
    \toh \bmq)$, and identify $({\cal E} + \toh \Omega_1)_\there =
    ({\cal E} + \toh \Omega_2)_\there = \varepsilon + \toh \omega$.] }
  \label{fig:Hikamiboxes}
\end{figure}

In this section, we provide some details for how the Cooperon self
energy can be calculated by performing the disorder average
diagrammatically.  As starting point we use
\Eqs{eq:selfenergies-explicit}, which we derived in
Sec.~\ref{sec:selfenergy} from the influence functional approach, but
which are equvalent to the standard Keldysh expressions following from
\Eq{eq:selfenergyKeldysh}.  According to
\Eqs{eq:selfenergies-explicit}, there are four self-energy
contributions to the Cooperon self energy, which we write as:
\begin{eqnarray}
  \label{eq:Cooperon-selfdefine}
 \bSigma_\bmq^\self (\omega) \equiv {1 \over \hbar}
\int {d \bomega   \over 2 \pi} \int (d \bbmq)
\Bigl[  \bSigma^I_{FF} + \bSigma^I_{BB} +  \bSigma^R_{FF} +
 \bSigma^R_{BB} \Bigr] 
\; .
\end{eqnarray}
The diagrams for $\bSigma^I_{aa}$ are depicted in
Fig.~\ref{fig:Hikamiboxes}(b), those for $\bSigma^R_{aa}$ in
Fig.~\ref{fig:Hikamiboxes}(c) to \ref{fig:Hikamiboxes}(f)
(which correspond one-to-one to Fig.~2(b) to 2(f) of
AAV\cite{AAV01}).   Starting
from \Eq{eq:selfenergies-explicit}, the corresponding algebraic
expressions can be written as:
  \begin{eqnarray}
 \label{eq:selfenergyFFBBresults}
%   \label{eq:selfenergySIFF}
 \bSigma^I_{FF} \!\! &  =  & \!\!
%\int {d \bomega  d (\bbmq) \over 2 \pi}
 - \toh i \bLKbbqw \, \bcC^0_{\bmq - \bbmq}
(\omega - \bomega) \, Y^{(1)}_{F}
%\; \stackrel{\omega = \bmq = 0}{\longrightarrow}
%\int {d \bomega  d (\bbmq) \over 2 \pi}
%\coth (\hbar \bomega / 2T) \, { {\rm Im}
%[\bLRbbqw]  \over \Dd \, \bq^2 + i \bomega} \; ,
\rule[-4.5mm]{0mm}{0mm}
\\
\nonumber    \label{eq:selfenergySIBB}
 \bSigma^I_{BB} 
\!\! &  =  & \!\!
% \int {d \bomega  d (\bbmq) \over 2 \pi}
- \toh i \bLKbbqw \, \bcC^0_{\bmq - \bbmq}
(\omega + \bomega) \, Y^{(1)}_{B} \; ,
\rule[-4.5mm]{0mm}{0mm}
\\
\nonumber    \label{eq:selfenergySRFF}
\bSigma^{R}_{FF}
\!\! &  =  & \!\!
- \toh i \bLRbbqw \, \tanh [\hbar (\ve + \omega - \bomega)/2T]
 \Bigl\{ \bcC^0_{\bmq - \bbmq} (\omega - \bomega) \, Y^{(1)}_{F}
-  \tauel \overline \Gamma^2_\bbmq (\bomega) 
%\bigl[ Y^2_{FF} + Y^3_{FF} + Y^4_{FF} \bigr]
\sum_{n=2}^4 Y^{(n)}_{F} 
 \Bigr\}  , 
\rule[-4.5mm]{0mm}{0mm}
\\
\nonumber    \label{eq:selfenergySRBB}
 \bSigma^{R}_{BB} 
\!\! &  =  & \!\!
-  \toh  i \bLAbbqw \, \tanh [\hbar (\ve  - \bomega)/2T]
 \Bigl\{- \bcC^0_{\bmq - \bbmq} (\omega + \bomega) \, Y^{(1)}_{B}
+  \tauel \overline \Gamma^2_{- \bbmq} (- \bomega) 
%\bigl[ Y^2_{BB} + Y^3_{BB} + Y^4_{BB} \bigr] 
\sum_{n=2}^4 Y^{(n)}_{B} 
\Bigr\} .
      \end{eqnarray}
      In the expressions for $ \bSigma^{R}_{aa} $, the minus signs
      before $Y^{(2)}_{F}$, $Y^{(3)}_{F}$, $Y^{(4)}_{F}$ and $Y^{(1)}_{B}$ arise
      from the minus sign in $\bcG^K = \tanh (\;)$ $ [\bcG^R -
      \bcG^A]$, and the $Y$'s represent integrals over internal
      momenta, that can be performed using variations of
      \Eqs{eq:IRARAARRAA}:
  \begin{eqnarray}
  \label{eq:defineYgiveY}
Y^{(1)}_{F}  & = &
{\hbar \over \tauel} \int
{ (d \bmp)  \over 2 \pi \nud \tauel/\hbar} \,
\bcG^A_{\bmq- \bmp} (\ve ) \,
\bcG^R_{\bmp} (\ve + \omega ) \,
\bcG^R_{\bmp - \bbmq} (\ve + \omega - \bomega) \,
\\
& & \times {\hbar \over \tauel} \int
{ (d \bmp')  \over 2 \pi \nud \tauel/\hbar} \,
\bcG^A_{\bmq- \bmp'} (\ve )
\bcG^R_{\bmp' - \bbmq} (\ve + \omega - \bomega) \,
\bcG^R_{\bmp'} (\ve + \omega )
\qquad \phantom{.} \nonumber
\\ \nonumber
& = & (-i)^2 \left\{ 1 - \tauel \Bigl[ \Dd (2 \bmq - \bbmq)^2 
- i (2 \omega - \bomega) \Bigr] + \dots \right\}^2 
\; ,
%= (-i)^2 \Bigl[ 1 - \tauel \bigl[
%\Dd (q^2 + \bar q^2 + \bmq \cdot \bar \bmq) +
% i (2 \omega - \bomega)\bigr]  + \dots \Bigr]^2
\rule[-4.5mm]{0mm}{0mm}
\\ 
Y^{(2)}_{F}  & = & {\hbar^2 \over \tauel^2} \int
{ (d \bmp)  \over 2 \pi \nud \tauel/\hbar} \,
\bigl[\bcG^R_{\bmp} (\ve + \omega ) \bigr]^2 \,
\bcG^A_{\bmp - \bbmq} (\ve + \omega - \bomega) \,
\bcG^A_{\bmq- \bmp} (\ve )
\qquad \phantom{.} \nonumber
\\ \nonumber
    \label{eq:Y2FF}
& = &  2 - \tauel \Bigl[ 4 \Dd (\bmq + \bbmq)^2 - 3 i  (\omega +
\bomega) \Bigr] \; , 
%= 2 - \tauel \bigl[
%4 \Dd (q^2 + \bar q^2 + \bmq \cdot \bar \bmq) -
% 3 i (\omega + \bomega)\bigr] + \dots
\rule[-4.5mm]{0mm}{0mm}
\\ \nonumber
Y^{(3)}_{F}  & = &
{\hbar \over \tauel} \int
{ (d \bmp)  \over 2 \pi \nud \tauel/\hbar} \,
\bcG^R_{\bmp} (\ve + \omega ) \,
\bcG^A_{\bmp - \bbmq} (\ve + \omega - \bomega) \,
\bcG^A_{\bmq- \bmp} (\ve ) \,
\\ &  & \nonumber
{\hbar \over \tauel} \int
{ (d \bmp')  \over 2 \pi \nud \tauel/\hbar} \,
\bcG^R_{\bmp'} (\ve + \omega ) \,
\bcG^A_{\bmp' - \bbmq} (\ve + \omega - \bomega) \,
\bcG^A_{\bmq- \bmp'} (\ve )
\nonumber \\
\label{eq:Y3FF}
& = &  \nonumber 
(i)^2 \left\{ 1 - \tauel \Bigl[ \Dd (\bmq + \bbmq)^2 
- i  (\omega + \bomega) \Bigr] + \dots \right\}^2 
%=  (i)^2 \Bigl[ 1 - \tauel \bigl[
%\Dd (q^2 + \bar q^2 + \bmq \cdot \bar \bmq) -
% i ( \omega + \bomega)\bigr]  + \dots \Bigr]^2
\rule[-4.5mm]{0mm}{0mm}
\\ \nonumber
Y^{(4)}_{F}  & = &
{\hbar^2 \over \tauel^2} \int
{ (d \bmp) (d \bmp') \over (2 \pi \nud \tauel/\hbar)^2} \,
\bcG^A_{\bmq- \bmp} (\ve ) \,
\bigl[\bcG^R_{\bmp} (\ve + \omega ) \bigr]^2\,
\bcG^A_{\bmp' - \bbmq} (\ve + \omega - \bomega) \,
\bigl[\bcG^R_{\bmp'} (\ve + \omega ) \bigr]^2\,
 \nonumber \\
\label{eq:Y4FF}
& = &   \nonumber
(-i)^2 
 \left\{ 1 - \tauel \Bigl[ \Dd (2 \bmq )^2 
- i 2 \omega \Bigr] + \dots \right\}^2 
 \left\{ 1 - \tauel \Bigl[ \Dd (2 \bbmq )^2 
- i 2 \bomega \Bigr] + \dots \right\}^2 
%\rule[-4mm]{0mm}{0mm}
%\\
\end{eqnarray}
Performing a similar set of integrals for the $Y_B^{(n)}$'s, we readily
find that $Y^{(n)}_B (\bomega) = Y^{(n)}_F (- \bomega) $. Note that
the sums $\sum_{n=2}^4 Y^{(n)}_{F/B}$, which are associated with the
so-called ``Hikami-box'' diagrams of Fig.~\ref{fig:Hikamiboxes}(d) to
\ref{fig:Hikamiboxes}(f), add up to zero in leading order, which is
why the next order had to be included.  Finally, the results for
$\overline \Sigma^{I,\self} $ and $\overline \Sigma^{R,\self }$,
given by \Eqs{subeq:selfenergygorydetail} in the main text, are
obtained by inserting \Eqs{eq:defineYgiveY} into
\Eqs{eq:selfenergyFFBBresults} and (\ref{eq:Cooperon-selfdefine}), and
making the replacement $\ve_\here \to \ve - \toh \omega$ (cf.\ caption
of Fig.\ref{fig:Hikamiboxes}).


\section*{References}

\begin{thebibliography}{1}

%\setcounter{enumiv}{-1}

\bibitem{Granada} Jan von Delft, "Influence functional for Decoherence
  of Interacting Electrons in Disordered Conductors", in
  \emph{Fundamental Problems of Mesoscopic Physics}, ed. I. V. Lerner,
  B.L. Altshuler, and Y. Gefen (eds.), Kluwer, London (2004), pp.\
  115-138.  This publication contains, modulo minor revisions, the
  main text and appendices A.1 to A.3 of the present review.



\bibitem{GZ1} D. S. Golubev and A. D. Zaikin, 
\emph{Phys.\ Rev.\ Lett.} 
{\bf 81}, 1074, (1998).

\bibitem{GZ2}
D. S. Golubev and A. D. Zaikin, 
\emph{Phys.\ Rev.} 
{\bf B59}, 9195, (1999).

\bibitem{GZ3}
D. S. Golubev and A. D. Zaikin,
\emph{Phys.\ Rev.}
{\bf B62}, 14061 (2000).

\bibitem{GZS}
D. S. Golubev, A. D. Zaikin, and G. Sch\"on,
\emph{J. Low Temp.\ Phys.} 
{\bf 126}, 1355 (2002) [cond-mat/0110495].

\bibitem{GolubevZaikin02}
D. S. Golubev and A. D. Zaikin, 
\emph{J. Low Temp.\ Phys.} {\bf 132}, 
11 (2003) [cond-mat/0208140].

\bibitem{GolubevHereroZaikina02}
D. S. Golubev, C. P. Herrero, Andrei D. Zaikin,
\emph{Europhys.\ Lett.} {\bf 63}, 426 (2003) [cond-mat/0205549].


\bibitem{MW}
P. Mohanty, E. M. Q. Jariwala, and R. A. Webb,
\emph{Phys.\ Rev.\ Lett.} {\bf 78}, 3366 (1997);
\emph{Fortschr.\ Phys.} {\bf 46}, 779 (1998).

\bibitem{controversy} Most relevant references can be found in
  the review\cite{GZS} by Golubev, Zaikin and Sch\"on, 
 which  gives a useful overview of the controversy
  from GZ's point of view.

\bibitem{AAK82}
B. L. Altshuler, A. G. Aronov, and D. E. Khmelnitskii,
\emph{J. Phys.} {\bf C15}, 7367 (1982).

\bibitem{EriksenHedegard}
K. A. Eriksen and P. Hedegard, cond-mat/9810297;
GZ replied in   cond-mat/9810368.

\bibitem{VavilovAmbegaokar98}
M. Vavilov and V. Ambegaokar, cond-mat/9902127.

\bibitem{KirkpatrickBelitz01}
T.R. Kirkpatrick, D. Belitz, cond-mat/0112063, cond-mat/0111398.

\bibitem{Imry02}
Y. Imry, cond-mat/0202044.

\bibitem{vonDelftJapan02}
 J. von Delft,
\emph{J. Phys. Soc. Jpn.}
{\bf 72} (2003), Suppl. A pp. 24-29 [cond-mat/0210644].

\bibitem{Marquardt02}
F. Marquardt, cond-mat/0207692.


\bibitem{AAG98}
I. Aleiner, B. L. Altshuler, and M. E. Gershenzon,
\emph{Waves and Random Media} {\bf 9}, 201 (1999) [cond-mat/9808053];
Phys.\ Rev.\ Lett.\ {\bf 82}, 3190 (1999).



\bibitem{AAV01}
 I. L. Aleiner, B. L. Altshuler, M. G. Vavilov,
\emph{J. Low Temp.\ Phys.} {\bf 126}, 1377 (2002) [cond-mat/0110545].





\bibitem{AAV02}
 I. L. Aleiner, B. L. Altshuler, M. G. Vavilov,
cond-mat/0208264.


\bibitem{MarquardtAmbegaokar04}
(MDSA-I) F. Marquardt, J. von  Delft, R. Smith and V. Ambegaokar,
\emph{Phys.\ Rev.} {\bf B76}, 195331 (2007) [cond-mat/0510556], and 
 (DMSA-II) J. von  Delft, F. Marquardt, R. Smith and V. Ambegaokar,
\emph{Phys.\ Rev.} {\bf B76}, 195332 (2007) [cond-mat/0510557].

\bibitem{ChakravartySchmid86}
S. Chakravarty and A. Schmid,
\emph{Phys.\ Rep.} {\bf 140}, 193 (1983).


\bibitem{FukuyamaAbrahams83}
H. Fukuyama and E. Abrahams,
\emph{Phys.\ Rev.}  {\bf B27}, 5976 (1983).


\bibitem{erratum} The expressions for $\tilde \Sigma$ that we
  published in \cite{vonDelftJapan02}, Eqs.~(A.16), contain incorrect
  signs and missing factors of $\toh$, and should be replaced by
  \Eqs{eq:selfenergies-explicit-main} of this review.



\bibitem{GZ05}
D. S. Golubev and A. D. Zaikin, cond-mat/0512411.


\end{thebibliography}
\end{document}